\documentclass [12pt]{article}

\usepackage{graphicx}
\usepackage{amsmath,amssymb}
\usepackage{amsthm,amscd,mathrsfs}
\usepackage{multirow}

\usepackage{color}
\input{colordvi.tex}

\usepackage{ulem}

\def\l{\left}
\def\r{\right}
\def\be{\begin{eqnarray}}
\def\ne{\nonumber\end{eqnarray}}
\def\nn{\nonumber}
\def\ee{\end{eqnarray}}
\def\c{\cdot}
\def\d{\partial}
\def\pd#1{{\partial~\over\partial{#1}}}
\def\hf{{1\over2}}
\def\qt{{1\over4}}
\def\Tr{{\rm tr}}
\def\vol{{\rm vol}}
\def\Det{{\cal D}{\rm et}}
\def\det{{\rm Det}}
\def\dets{{\rm det}}
\def\u{\underline}
\def\m{\mu}
\def\n{\nu}
\def\p{\rho}
\def\s{\sigma}
\def\t{\tau}
\def\om{\omega}
\def\Om{\Omega}
\def\e{\epsilon}
\def\ve{\varepsilon}
\def\Th{\Theta}
\def\th{\theta}
\def\vp{\varphi}
\def\a{\alpha}
\def\b{\beta}
\def\g{\gamma}
\def\dl{\delta}
\def\la{\lambda}
\def\La{\Lambda}
\def\da{\dot{\alpha}}
\def\db{\dot{\beta}}
\def\dg{\dot{\gamma}}
\def\ddl{\dot{\delta}}
\def\ta{\tilde{\alpha}}
\def\tb{\tilde{\beta}}
\def\tg{\tilde{\gamma}}
\def\tdl{\tilde{\delta}}
\def\ca{\check{\alpha}}
\def\cb{\check{\beta}}
\def\cg{\check{\gamma}}
\def\cdl{\check{\delta}}
\def\G{\Gamma}
\def\cD{{\cal D}}
\def\N{{\cal N}}
\def\cd{{\nabla}}
\def\RS{{\Sigma}}

\def\tr{\tilde{r}}
\def\done{\dot{1}}
\def\dtwo{\dot{2}}
\def\tchi{\tilde{\chi}}
\def\txi{\tilde{\xi}}
\def\teta{\tilde{\eta}}
\def\tkappa{\tilde{\kappa}}
\def\tvp{\tilde{\varphi}}
\def\ts{\tilde{\sigma}}
\def\tm{\tilde{m}}
\def\sa{\left(\sigma\cdot\alpha\right)}

\def\Vox#1{
\medskip
\begin{center}
\framebox(300,35){
\parbox{\textwidth}{#1}}
\end{center}
}
\def\BBox#1#2#3{
\begin{center}
\framebox({#1},{#2}){
\parbox{\textwidth}{#3}}
\end{center}
}

\begin{document}

\vspace*{-0.5cm}\hspace*{15.5cm}
\hfill\parbox{3cm}
{\normalsize 
{UT-15-16}\\
}\\

\vskip 0.5cm
\centerline{\Large\bf 
              5D SYM on 3D Deformed Spheres
}

\vskip .3in

\centerline{\sc Teruhiko Kawano}
\vskip 0in
\centerline{\it Department of Physics, University of Tokyo}
\centerline{\it Hongo, Tokyo 113-0033, Japan}
\vskip 0.5mm
\centerline{ and}
\vskip 0.5mm
\centerline{\sc Nariaki Matsumiya}
\vskip 0in
\centerline{\it Sumitomo Heavy Industries, Ltd. }
\centerline{\it 19 Natsushima-cho, Yokosuka-shi, Kanagawa 237-8555, Japan}

\vskip .5in

{
We reconsider the relation of superconformal indices of 
superconformal field theories of class ${\cal S}$ with 
five-dimensional $\N=2$ supersymmetric Yang-Mills theory 
compactified on the product space of a round three-sphere and 
a Riemann surface. We formulate the five-dimensional theory in 
supersymmetric backgrounds preseving $\N=2$ and $\N=1$ supersymmetries 
and discuss a subtle point in the previous paper 
concerned with the partial twisting on the Riemann surface.  
We further compute the partition function by localization 
of the five-dimensional theory on a squashed three-sphere in 
$\N=2$ and $\N=1$ supersymmetric backgrounds and on an ellipsoid 
three-sphere in an $\N=1$ supersymmetric background. 
}

\vskip .5in

\setcounter{footnote}{0}
\tableofcontents
\section{Introduction}

In the previous papers \cite{FKM,KM}, we have attempted to give 
a physical proof for the conjecture of \cite{Schur}. 
The conjecture states that the Schur limit of the superconformal index 
\cite{Index} of a four-dimensional $\N=2$ superconformal theory of class 
${\cal S}$ \cite{Gaiotto,WKB} can be computed by two-dimensional $q$-deformed 
Yang-Mills theory \cite{Aganagic}. 
The $\N=2$ superconformal theory of class ${\cal S}$ is defined 
in \cite{WKB,Gaiotto} as the infrared limit of M5-branes wrapped on 
a Riemann surface\footnote{The Riemann surface is commonly denoted by 
${\cal C}$ in the recent literature and is often refered to as a UV curve.} 
$\RS$, and according to the conjecture, one may compute the Schur index of 
the theory by making use of the $q$-deformed Yang-Mill theory on $\RS$ 
in the zero area limit.  

The superconformal index may be captured by the partition function of 
the four-dimensional theory compactified on $S^1\times{S}^3$. 
(See \cite{TachikawaNonzero} for the extension to $\RS$ with nonzero area.)  
Since the infrared limit of M5-branes gives rise to 
the putative six-dimensional ${\N=(2,0)}$ superconformal theory, 
it is conceivable to obtain the index by computing the parition 
function of the $\N=(2,0)$ theory compactified on $S^1\times{S}^3\times\RS$ 
with a partial twisiting on $\RS$. 
The idea\footnote{
Similar contructions have been used in \cite{WKB,finitearea} 
to explore four-dimensional $\N=2$ superconformal field theories of class 
${\cal S}$. See also \cite{Yonekura} for $\N=1$ supersymmetric theories. 
See \cite{3d-3d,Yamazaki} (also \cite{Yagi}) 
for related works on three-dimensional Chern-Simons theory from M5-branes.} 
has been argued in \cite{2dTQFT} as a ``top-down'' approach 
to uncover the relation of a generic superconformal index 
of the theories of class ${\cal S}$ with a topological field theory.
For a review of the superconformal indices of theories of class ${\cal S}$, 
see \cite{RastelliRazamat}.

In the previous papers \cite{FKM,KM}, we put the idea into pracitce by 
exchanging the order of the compactifications; regarding M5-branes 
wrapped on a circle as D4-branes, we compacitified 
five-dimensional $\N=2$ supersymmetric Yang-Mills theory on $\RS$ 
with a partial twisting and further on the round $S^3$, in a somewhat 
ad hoc way. Computing the partition fucntion of the 
compacitifed theory by localization, we have found that the fixed points 
give the fields of the $q$-deformed Yang-Mills theory, and that 
the one-loop contributions and the classical action at the fixed points 
yield the measure of the partition integral of it; namely, 
the partition function of the five-dimensional compactified theory 
is reduced to that of the two-dimensional $q$-deformed Yang Mill theory.

However, there was a confusion about the partial twisting in 
\cite{FKM}. The supersymmetric background used in \cite{FKM} 
- especially, the partial twisting - 
preserved only $\N=1$ supersymmetry in four dimensions. 
Since the conjecture in \cite{Schur} is concerned with the four-dimensional 
$\N=2$ superconformal theories, the results in \cite{FKM} seem to have 
nothing to do with the conjecture. 

The construction of the $\N=2$ superconformal theories by 
the twisted compactification of the $\N=(2,0)$ theory on $\RS$ 
has been generalized for $\N=1$ supersymmetic theories in four dimensions 
\cite{Sicilian,BahWecht,BahI,BahII}, which we will refer to as $\N=1$ 
theories of class ${\cal S}$. We can see that the twisting used in \cite{FKM} 
is identical to what is called\footnote{
See Appendix D of \cite{Sicilian} about the embedding of the spin connection 
on $\RS$ to the $R$-symmetry group. 
The $\N=1$ twist corresponds to the case of $l_1=l_2$ in their Calabi-Yau 
construction of \cite{BahII}. } the $\N=1$ twist in \cite{Sicilian}. 

Superconformal indices of the $\N=1$ superconformal theories of class 
${\cal S}$ have been calculated in four dimensions in 
\cite{GaddeBeem}. A simple comparision shows that 
the result in \cite{FKM} is in good agreement with the Schur limit of 
the mixed Schur index in \cite{GaddeBeem}, as we will see later. 

The questions we raise are two-fold; first, when the $\N=(2,0)$ theory 
compactified on $\RS$ with the $\N=2$ twisting so that $\N=2$ 
supersymmetry remains unbroken in four dimensions, 
whether will we obtain the $q$-deformed Yang-Mills theory on $\RS$ via 
localization? This was the original motivation 
in the previous paper \cite{FKM,KM}. 

Second, when replacing the round $S^3$ by a deformation of the $S^3$, 
such as a squashed $S^3$ and an ellipsoid $S^3$, as discussed in 
\cite{Hosomichi,Imamura}, whether will we obtain a deformation of 
the Schur index for the round $S^3$, like the mixed Schur index 
in \cite{GaddeBeem}? 

We will make an attempt to answer both of the questions in this paper, 
which is organized as follows: 
in sections \ref{5DSUGRA} and \ref{TensorSUGRA}, 
we will begin with the construction of the five-dimensional supersymmetric 
Yang-Mills theory on a curved space, based on the idea of \cite{FS} that 
the fields of an off-shell supergravity multiplet 
are utilized as background fields to preserve supersymmetries of 
the field theory on a curved space. 
In fact, through the dimensional reduction of 
the six-dimensional $\N=(2,0)$ conformal supergravity in \cite{BSvP2}, 
on-shell supersymmetry transformations and an on-shell action 
of the five-dimensional theory compactified on a curved space 
have been derived in \cite{CJ}, following the idea \cite{FS}. 

Therefore, sections \ref{5DSUGRA}, \ref{TensorSUGRA}, 
and \ref{NonAbelianTensor} are essentially devoted to a review of \cite{CJ}, 
up to a few points that we perform the dimensional reduction in the time 
direction of the six-dimensional theory, 
instead of the spatial direction as in \cite{CJ}. 
And we obtain off-shell supersymmetry transformations and an off-shell action 
of the five-dimensional theory on a curved space 
in section \ref{OffshellSUSY}, which are necessary to carry 
out localization. 

In section \ref{Backgrounds}, we will discuss the partial twistings mentioned 
above -the $\N=1$ twisting and the $\N=2$ twisting - in more details, 
in the language of the background gauge field of the $R$-symmetry group, 
and we will decribe the supersymmetric background 
on a round $S^3$ in \cite{FKM,KM} in terms of supergravity background fields 
for the $\N=1$ twisting in subsection \ref{oldR3Sbg}, 
and give a supersymmetric background on the round $S^3$ for the $\N=2$ 
twisting in subsection \ref{newR3Sbg}. 

We will proceed to consider 
two supersymmetric backgrounds on a squashed $S^3$ - the analog of 
the background in \cite{Hosomichi} and of the one in \cite{Imamura} -
in subsections \ref{HosomichiSquash} and \ref{ImamuraSquash}, respectively. 
Especially, for the former, we will give supersymmetry backgrounds 
for both of the twisitings. 

In subsection \ref{HosomichiEllipsoid}, we will discuss a supersymmetric 
background for the $\N=1$ twisting on an ellipsoid $S^3$, in an analogous way 
to \cite{Hosomichi}. 

After the discussions about the off-shell formulation of the five-dimensional 
theory in section \ref{OffshellSUSY}, as mentioned above, 
we will explain our localization method in depth in section \ref{localization}. 
We will compute the partition functions by localization 
on the round and squashed $S^3$'s in section \ref{SquashLocalization} 
for the background in section \ref{HosomichiSquash}
and that on the ellipsoid $S^3$ in section \ref{EllipsoidLocalization} 
for the background in section \ref{HosomichiEllipsoid}.

However, the computation of the partition function on the squashed $S^3$ 
for the background in section \ref{ImamuraSquash} somewhat doesn't seem 
straightforward to be done by localization, and we will leave it as an open 
question. Finally, section \ref{SummaryDiscussion} is devoted to the summary 
and discuusions of this paper. 

Appendix \ref{conventions} is a simple collection of our conventions 
about the (anti-)symmetrization of various indices and about differential 
forms, used in this paper, and the gamma matrices of the Lorentz groups 
in five and six dimensions are shown in our repsentation 
in appendix \ref{LorentzGamma}. 
The $R$-symmetry group of the six- and five-dimensional theories are 
commonly $Spin(5)_R\simeq{Sp(2)}_R$ and the associating gamma matrices 
in our represenation are given in appendix \ref{RGamma}. 
The spinors in the theories are symplectic Majorana-Weyl spinors and 
in appendix \ref{SMW}, our convetions about those spinors are explained. 

After the dimensional reduction of the conformal supergravity, 
supersymmetry transforms of the fermionic fields in the supergravity 
multiplet (the Weyl multiplet) yield supersymmetry conditions on the 
background fields to preserve supersymmetries on the curved background. 
Besides the supersymmetry condition derived from the gravitino field, 
there is another supersymmetry condition from the fermionic auxiliuary 
field in the Weyl multiplet and it is too long to write down 
explicitly in the text. Therefore, the explicit form of the supersymmetry 
condition is written in appendix \ref{SUSYchi}. 

In appendix \ref{SUSYMassPhi}, a few formulas which we think are useful 
to verify the invariance of the actions in sections \ref{TensorSUGRA} and 
\ref{NonAbelianTensor} under the supersymmetry transformations 
are given. 

In appendix \ref{3-sphere}, Killing spinors and metrics are discussed on 
the round, squashed, and ellipsoid $S^3$, following \cite{Hosomichi,Imamura}. 

Appendix \ref{Dictionary} explains the difference among the notations 
used in \cite{BSvP2}, in \cite{CJ}, and in this paper, and further 
the difference between the notations used here and in the previous paper 
\cite{FKM}.

\section{Euclidean 5D $\N=2$ SYM in SUGRA Backgrounds}
\label{5DSUGRA}

In this section, the dimensional reduction along the time direction 
will be performed for the six-dimensional $\N=(2,0)$ conformal supergravity 
derived in \cite{BSvP2}. 
This section, the sections \ref{TensorSUGRA} and \ref{NonAbelianTensor} 
are essentially a review of \cite{CJ}, but the spatial dimensional 
reduction was carried out there. 

In subsection \ref{ConformalTensorCalculus}, we will recapitulate the main 
results of \cite{BSvP2}, which we will need in this paper about 
the supergravity mulitplet called the Weyl multiplet in the conformal tensor 
calculus.

In subsection \ref{WeylConformalTensorCalculus}, 
we will discuss the dimensional reduction of 
the Weyl multiplet, which play roles of supersymmetric background fields to 
retain supersymmetreis of the five-dimensional Yang-Mills theory 
on a curved space. 
Subsection \ref{KillingVector&Spinor} is just a small digression about 
the relaton of Killing spinors with Killing vectors.


\subsection{Weyl Multiplet in 6D $\N=(2,0)$ Conformal Supergravity}
\label{ConformalTensorCalculus}

In this paper, following \cite{CJ}, we will carry out dimensional reduction 
of  the six-dimensional $\N=(2,0)$ supergravity in \cite{BSvP2} 
to obtain a five-dimensional Euclidean maximally supersymmetric Yang-Mills 
theory in supergravity backgrounds. It has been discussed in \cite{FS} that 
the supergravity backgrounds provide a systematic method for supersymmetric 
compactifications of supersymmetric field theories. 
The construction of the supergravity in \cite{BSvP2} is 
based on the conformal tensor calculus. (See the textbook \cite{Supergravity} 
for the conformal tensor calculus and references therein.)

In this approach, one starts with a gauge field theory by gauging 
the six-dimensional $\N=(2,0)$ superconformal symmetry group $OSp(2,6|4)$,
whose bosonic part consists of the conformal group $SO(2,6)$ and the $R$-symmetry 
group ${Spin(5)}$.
The symmetry group $OSp(2,6|4)$ is generated by 
\begin{eqnarray*}
P_{\u{a}}: {\rm translation}, 
\quad
D: {\rm translation}, 
\quad
M_{\u{ab}}: {\rm Lorentz}, 
\quad
K_{\u{a}}: {\rm special~conformal}, 
\nn\\
R_{IJ}: {\rm {\it R-}symmetry}, 
\quad
Q^{\a}: {\rm supersymmetry}, 
\quad
S^{\a}: {\rm conformal~supersymmetry}, 
\end{eqnarray*} 
whose corresponding gauge fields are shown in Table \ref{ConfGaugeField}. 

Let us list the notations of the various indices on the generators 
and the gauge fields;
\begin{itemize}
\item
$\u{a},\u{b}=0,1,\cdots,5$; the Lorentz indices,
\item
$\u{\m},\u{\n}=0,1,\cdots,5$; the coordinate frame indices,
\item
$I,J=1,\cdots,5$; the vector indices of the Spin(5)$_R$ symmetry,
\item
$\a,\b=1,\cdots,4$; the spinor indices of the Spin(5)$_R$ symmetry.
\end{itemize}

The fermionic fields $\u{\psi}{}_{\u{\m}}{}^\a$ and $\u{\phi}{}_{\u{\m}}{}^\a$ 
are the gauge fields of the supersymmetry and the conformal supersymmetry, 
respectively. 
They are symplectic Majorana-Weyl spinors of positive and negative 
chirality, respectively. 
See Appendix \ref{SMW} for our conventions about symplectic Majorana-Weyl 
spinors.

\begin{table}[h]
\begin{center}
\begin{tabular}{|c|c|c|c|c|}\hline
\makebox[25mm]{gauge fields}&\makebox[30mm]{transformations}
&\makebox[2cm]{restrictions}&\makebox[1.5cm]{Spin(5)${}_{R}$}
&\makebox[1cm]{weight}
\\ \hline \hline
{boson}&\multicolumn{4}{c|}{} 
\\ \hline
\multirow{2}{*}{$\u{E}^{\u{a}}{}_{\u{\m}}$}
&\multirow{2}{*}{$P_{\u{a}}$: translations}
&\multirow{2}{*}{sechsbein} 
&\multirow{2}{*}{{\bf 1}}
&\multirow{2}{*}{-1}
\\ 
& & & & \\ \hline
\multirow{2}{*}{$\u{b}_{\u{\m}}$}&\multirow{2}{*}{$D$: dilatation}
&\multirow{2}{*}{} 
&\multirow{2}{*}{{\bf 1}}
&\multirow{2}{*}{0}
\\
& & & & \\ \hline
\multirow{2}{*}{$\u{V}_{\u{\m}}{}^{IJ}$}
&\multirow{2}{*}{$R_{IJ}$: $R$-symmetry}
&\multirow{2}{*}{$\u{V}_{\u{\m}}{}^{IJ}=-\u{V}_{\u{\m}}{}^{JI}$} 
&\multirow{2}{*}{{\bf 10}}
&\multirow{2}{*}{0}
\\ 
& & & & \\ \hline 
{fermion}&\multicolumn{4}{c|}{} 
\\ \hline
\multirow{3}{*}{$\u{\psi}^{\a}{}_{\u{\m}}$}
&\multirow{3}{*}{$Q^{\a}$: supersymmetry}
&\multirow{1}{*}{gravitini} 
&\multirow{3}{*}{{\bf 4}}
&\multirow{3}{*}{-1/2}
\\ 
& &\multirow{2}{*}
{$\u{\G}{}^7\u{\psi}^{\a}{}_{\u{\m}}=\u{\psi}^{\a}{}_{\u{\m}}$}  
& & \\ 
& & & & \\ \hline\hline
\multicolumn{5}{|c|}
{dependent gauge fields} \\ \hline\hline
{boson}&\multicolumn{4}{c|}{} 
\\ \hline
\multirow{2}{*}{$\u{\Om}_{\u{\m}}{}^{\u{ab}}$}
&\multirow{2}{*}{$M_{\u{ab}}$: local Lorentz }
&\multirow{2}{*}{spin connection}  
&\multirow{2}{*}{{\bf 1}}
&\multirow{2}{*}{0}
\\
& & & & \\ \hline
\multirow{2}{*}{$\u{f}^{\u{a}}{}_{\u{\m}}$}
&\multirow{2}{*}{$K_{\u{a}}$: special conformal}
&\multirow{2}{*}{} 
&\multirow{2}{*}{{\bf 1}}
&\multirow{2}{*}{+1}
\\ 
& & & & \\ \hline
{fermion}&\multicolumn{4}{c|}{} 
\\ \hline
\multirow{2}{*}{$\u{\phi}^{\a}{}_{\u{\m}}$}
&\multirow{2}{*}{$S^{\a}$: conformal supersymmetry}
&\multirow{2}{*}
{$\u{\G}{}^7\u{\phi}^{\a}{}_{\u{\m}}=-\u{\phi}^{\a}{}_{\u{\m}}$} 
&\multirow{2}{*}{{\bf 4}}
&\multirow{2}{*}{+1/2}
\\ 
& & & & \\ \hline
\end{tabular}
\end{center}
\caption{The gauge fields of the 6D $\N=(2,0)$ superconformal symmetry.}
\label{ConfGaugeField}
\end{table}

A straightforward manner of gauging translations doesn't lead to 
general coordinate transformations which is indispensable to 
a theory of gravity. To gain general coordinate transformations from 
translations in the conformal tensor calculus approach, 
auxiliary fields\footnote{
They are referred to as `matter fields' in \cite{BSvP2,CJ}. 
It is not always necessary to introduce auxiliuary fields for the deformation, 
and it depends on the numbers of supersymmetries and the dimensions of spacetime, 
{\it i.e.}, superconformal algebras. 
See \cite{Supergravity} for more details. 
} 
in Table \ref{auxTensor} are introduced 
and the transformation laws of the gauge fields 
are deformed by imposing some constraints on 
the gauge field strengths and the auxiliary fields 
such that the resulting transformation laws give a closed algebra, 
as explained in \cite{Supergravity}. 

\begin{table}[h]
\begin{center}
\begin{tabular}{|c|c|c|c|}\hline
\makebox[25mm]{auxiliuary fields}&\makebox[30mm]{symmetries}
&\makebox[1.5cm]{Spin(5)${}_{R}$}&\makebox[1cm]{weight}
\\ \hline\hline
\multicolumn{4}{|c|}{bosonic fields} 
\\ \hline
\multirow{4}{*}{$\u{T}^{\a\b}{}_{\u{abc}}$}
&\multirow{2}{*}
{$\u{T}^{\a\b}{}_{\u{abc}}
=-{1\over3!}\ve_{\u{abc}}{}^{\u{def}}\u{T}^{\a\b}{}_{\u{def}}$,}
&\multirow{4}{*}{{\bf 5}}
&\multirow{4}{*}{1} \\ 
& & & \\ 
&\multirow{2}{*}{$\u{T}^{\a\b}{}_{\u{abc}}=-\u{T}^{\b\a}{}_{\u{abc}}$,\quad
$\Om_{\a\b}\u{T}^{\a\b}{}_{\u{abc}}=0$.}& & \\ 
& & & \\ \hline
\multirow{4}{*}{$\u{M}^{\a\b}{}_{\g\dl}$}
&\multirow{2}{*}{$\u{M}{}^{\a\b,\g\dl}=\u{M}{}^{\g\dl,\a\b}
=-\u{M}{}^{\b\a,\g\dl}=-\u{M}{}^{\a\b,\dl\g}$,}
&\multirow{4}{*}{{\bf 14}}
&\multirow{4}{*}{2} \\ 
& & & \\ 
&\multirow{2}{*}{$\Om_{\a\b}\u{M}{}^{\a\b,\g\dl}
=\Om_{\g\dl}\u{M}{}^{\a\b,\g\dl}=\Om_{\a\g}\Om_{\b\dl}\u{M}{}^{\a\b,\g\dl}=0$.}
 & & \\ 
& & & \\ \hline\hline
\multicolumn{4}{|c|}{fermionic field} 
\\ \hline
\multirow{4}{*}{$\u{\chi}^{\a\b}{}_{\g}$}
&\multirow{2}{*}{$\u{\G}^7\u{\chi}^{\a\b}{}_{\g}=\u{\chi}^{\a\b}{}_{\g}$,
\ 
$\u{\chi}{}^{\a\b}{}_{\g}=-\u{\chi}{}^{\b\a}{}_{\g}$,
\ 
$\Om_{\a\b}\u{\chi}{}^{\a\b}{}_{\g}=\u{\chi}{}^{\g\b}{}_{\g}=0$,}
&\multirow{4}{*}{{\bf 16}}
&\multirow{4}{*}{3/2} \\ 
& & & \\ 
&\multirow{2}{*}
{$({\u{\chi}}^{\a\b}{}_{\g}){}^\dag\u{\G}^0
=({\u{\chi}}^{\a'\b'}{}_{\g'}){}^T\u{C}\,\Om_{\a'\a}\Om_{\b'\b}
(\Om^{-1})^{\g'\g}$.}
 & & \\ 
& & & \\ \hline
\end{tabular}
\end{center}
\caption{The auxiliuary fields for the deformation of the superconformal 
symmtry.}
\label{auxTensor}
\end{table}

Furthermore, one requires the invertibility of the gauge field 
$\u{E}^{\u{a}}{}_{\u{\m}}$ of translations to solve the constraints, 
which allows us to regard it as the sechsbein. 
Solving the constraints makes the gauge fields $\u{\Om}_{\u{\m}}{}^{\u{ab}}$, 
$\u{f}^{\u{a}}{}_{\u{\m}}$, and $\u{\phi}^{\a}{}_{\u{\m}}$ dependent fields 
given in terms of the other gauge fields and the auxiliuary fields. 
In fact, they are given by  
\begin{eqnarray}
&&\u{\Om}_{\u{\m}}{}^{\u{ab}}=\u{\om}_{\u{\m}}{}^{\u{ab}}
+\u{E}_{\u{\m}}{}^{\u{a}}\,\u{b}^{\u{b}}
-\u{E}_{\u{\m}}{}^{\u{b}}\,\u{b}^{\u{a}}+\cdots,
\qquad
\u{\phi}^{\a}{}_{\u{\m}}=\cdots,
\label{defOm}\\
&&\u{f}^{\u{a}}{}_{\u{\m}}={1\over8}\u{R}_{\u{\m}}{}^{\u{a}}(\u{\Om})
-{1\over80}\u{E}_{\u{\m}}{}^{\u{a}}\u{R}_{\u{c}}{}^{\u{c}}(\u{\Om})
+{1\over32}\u{T}^{\u{\a\b}}{}_{\u{\m{cd}}}\,\u{T}_{\u{\a\b}}{}^{\u{{acd}}}
+\cdots,
\label{deff}
\end{eqnarray}
where the ellipses $\cdots$ denote the contributions from 
the fermionic fields. One can see that 
the spin connection $\u{\Om}_{\u{\m}}{}^{\u{ab}}$ is a generalization 
of the Levi-Civita spin connection $\u{\om}_{\u{\m}}{}^{\u{ab}}$ satisfying 
$$
d\u{E}{}^{\u{a}}+\u{\om}^{\u{a}}{}_{\u{b}}\wedge\u{E}{}^{\u{b}}=0,
\qquad
\u{E}{}^{\u{a}}=\u{E}_{\u{\m}}{}^{\u{a}}\,d\u{X}^{\u{\m}},
$$
and ${R}_{\u{\m}}{}^{\u{a}}(\u{\Om})$ 
is the Ricci tensor 
$$
\u{R}_{\u{\m}}{}^{\u{a}}(\u{\Om})=\u{\Th}{}^{\u{\n}}{}_{\u{b}}\,
\u{R}_{\u{\n\m}}{}^{\u{ba}}(\u{\Om}), 
$$
of the curvature tensor of the spin connection $\u{\Om}^{\u{a}}{}_{\u{b}}$, 
$$
\u{R}^{\u{a}}{}_{\u{b}}(\u{\Om})
={1\over2}\u{R}_{\u{cd}}{}^{\u{a}}{}_{\u{b}}(\u{\Om})\,\u{E}{}^{\u{c}}
\wedge\u{E}{}^{\u{d}}
=d\u{\Om}^{\u{a}}{}_{\u{b}}
+\u{\Om}^{\u{a}}{}_{\u{c}}\wedge\u{\Om}^{\u{c}}{}_{\u{b}},
$$
where $\u{\Th}{}^{\u{\n}}{}_{\u{b}}$ denotes the inverse of the sechsbein
$\u{E}^{\u{a}}{}_{\u{\m}}$, {\it i.e.}, coframe.

After the deformation, one finds a closed algebra with 
the (covariant) general coordinate transformations. The remaining 
independent gauge fields and auxiliary fields form a multiplet called 
the Weyl multiplet including the graviton, the gravitini and the others. 
We show the resulting bosonic transformations of the independent gauge fields, 
except for the (covariant) general coordinate transformations, 
\begin{eqnarray*}
&&\dl\u{E}_{\u{\m}}{}^{\u{a}}=-\La_D\,\u{E}_{\u{\m}}{}^{\u{a}}+
\La^{\u{a}}{}_{\u{b}}\,\u{E}_{\u{\m}}{}^{\u{b}},
\qquad
\dl\u{\psi}^{\a}{}_{\u{\m}}
=-\hf\La_D\,\u{\psi}^{\a}{}_{\u{\m}}
+\qt\La_R{}^{IJ}\l(\p_{IJ}\r)^{\a}{}_{\b}\,\u{\psi}^{\b}{}_{\u{\m}}
+\qt\La_{\u{ab}}\,\u{\G}^{\u{ab}}\,\u{\psi}^{\a}{}_{\u{\m}},
\nn\\
&&\dl\u{V}_{\u{\m}}{}^{IJ}=\d\La_R{}^{IJ}
+\La_R{}^{I}{}_{K}\,\u{V}_{\u{\m}}{}^{KJ}
+\La_R{}^{J}{}_{L}\,\u{V}_{\u{\m}}{}^{IL},
\qquad
\dl\u{b}_{\u{\m}}=\d_{\u{\m}}\La_D-2\La_{K}{}_{\u{a}}\,\u{E}_{\u{\m}}{}^{\u{a}},
\end{eqnarray*}
where $\La_D$, $\La_{ab}$, $\La_K{}_a$, and $\La_R{}^{IJ}$, are the 
parameters of dilatation, the Lorentz, special conformal, 
and $R$-symmetry transformations, respectively, 
and under the first four transformations, the auxiliuary fields transform as 
\begin{eqnarray*}
\dl\u{T}^{\a\b}{}_{\u{\m\n\p}}=\La_D\,\u{T}^{\a\b}{}_{\u{\m\n\p}},
\qquad
\dl\u{M}^{\a\b}{}_{\g\dl}=2\La_D\,\u{M}^{\a\b}{}_{\g\dl},
\qquad
\dl\u{\chi}^{\a\b}{}_{\g}={3\over2}\La_D\,\u{\chi}^{\a\b}{}_{\g}
+\qt\La_{\u{ab}}\,\u{\G}^{\u{ab}}\,\u{\chi}^{\a\b}{}_{\g}. 
\end{eqnarray*}
Under the $R$-symmetry transformations, they transform 
in the representations shown in the Table \ref{auxTensor}, 
respectively. 

The resulting supersymmetry ($Q$-) transformations and superconformal ($S$-)
transformations on the gauge fields and the auxiliuary fields are given by
\begin{eqnarray}
&&\dl\u{E}_{\u{\m}}{}^{\u{a}}
={i\over2}\Om_{\a\b}\l(\u{\e}^\a\r)^T
\u{C\G^{\u{a}}}\,\u{\psi}^{\b}{}_{\u{\m}},
\qquad
\dl\u{\psi}^{\a}{}_{\u{\m}}
=\u{\cD}_{\u{\m}}\u{\e}^{\a}
+{1\over4!}\,T^{\a\b}{}_{\u{abc}}\,\u{\G}^{\u{abc}}\u{\G}_{\u{\m}}\,\u{\e}_{\b}
+\u{\G}_{\u{\m}}\,\u{\eta}^{\a},
\nn\\
&&\dl\u{b}_{\u{\m}}=\hf\Om_{\a\b}\l(\u{\e}^\a\r)^T\u{C}\u{\phi}^{\b}{}_{\u{\m}}
-\hf\Om_{\a\b}\l(\u{\eta}^\a\r)^T\u{C}\u{\psi}^{\b}{}_{\u{\m}},
\nn\\
&&\dl\u{V}_{\u{\m}}{}^{IJ}=\big(\Om\p^{IJ}\big)_{\a\b}
\l[\l(\u{\e}^{\a}\r)^T\u{C}\,\u{\phi}^{\b}{}_{\u{\m}}
+\l(\u{\eta}^{\a}\r)^T\u{C}\,\u{\psi}^{\b}{}_{\u{\m}}\r]
-{1\over15}\l(\p^{IJ}\r)^{\a}{}_{\b}\Om_{\g\dl}
\l(\u{e}^{\g}\r)^T\u{\G}_{\u{\m}}\,\u{\chi}^{\dl\b}{}_{\a},
\label{QStransf}\\
&&\dl\u{T}^{\a\b}{}_{\u{abc}}={1\over16}\l(\u{\e}^{[\a}\r)^T\u{C}\u{\G}^{\u{de}}
\u{\G}_{\u{abc}}\,\u{R}^{\b]}{}_{\u{de}}(Q)
-{1\over15}\l(\u{\e}^\g\r)^T\u{C}\u{\G}_{\u{abc}}\chi^{\a\b}{}_{\g}
-({\rm trace}),
\nn\\
&&\dl\u{M}^{\a\b}{}_{\g\dl}=
-\l(\u{\e}^{[\a}\r)^T\u{C}\u{\G}^{\u{\m}}\u{\cD}_{\u{\m}}
\u{\chi}_{\g\dl}{}^{\b]}
+2\l(\u{\eta}^{[\a}\r)^T\u{C}\u{\chi}_{\g\dl}{}^{\b]}
-({\rm trace}),
\nn\\
&&\dl\u{\chi}^{\a\b}{}_{\g}=
{5\over32}\,\u{\G}^{\u{abc}}\u{\G}^{\u{\m}}\u{\e}_{\g}\,
\u{\cD}_{\u{\m}}\u{T}^{\a\b}{}_{\u{abc}}
+{15\over32}\,\u{\G}^{\u{\m\n}}\u{\e}^{[\a}\,\u{R}_{\u{\m\n}}{}_{\g}{}^{\b]}
-\qt{}D^{\a\b}{}_{\g\dl}\,\u{\e}^{\dl}
+{5\over8}\,\u{\G}^{\u{abc}}\u{\eta}_{\g}\,\u{T}^{\a\b}{}_{\u{abc}}
-({\rm trace}),
\nn
\end{eqnarray}
with (trace) denoting necessary terms to give the same 
irreducible representations of the $R$-symmetry group 
as the fields on the left hand sides. 
The parameter $\u{\e}^{\a}$ of a supersymmetry transformation and 
$\u{\eta}^{\a}$ of a superconformal transformation 
are symplectic Majorana-Weyl spinors of positive and 
negative chirality, respectively; 
\begin{eqnarray*}
\u{\G}{}^7\u{\e}^{\a}=\u{\e}^{\a},
\qquad
\u{\G}{}^7\u{\eta}^{\a}=-\u{\eta}^{\a}. 
\end{eqnarray*}
The operation $T$ denotes transpose, and so 
$\l(\u{\e}^{\a}\r)^T$ and $\l(\u{\eta}^{\a}\r)^T$ are 
the tranposes of $\u{\e}^{\a}$ and $\u{\eta}^{\a}$, respectively. 
The curvature $\u{R}^{\a}{}_{\u{ab}}(Q)$ is the field strength of 
the supersymmetry gauge field (graitini) $\u{\psi}^{\a}{}_{\u{\m}}$, 
whose exact form can be seen in \cite{BSvP2}, but it will not be necessary 
in this paper. 

Here the covariant derivatives of $\u{\e}^{\a}$ and $\u{T}^{\a\b}{}_{\u{abc}}$ 
are given by
\begin{eqnarray*}
&&\u{\cD}_{\u{\m}}\u{\e}^{\a}=\d_{\u{\m}}\u{\e}^{\a}
+\hf\u{b}_{\u{\m}}\u{\e}^{\a}
+\qt\big(\u{\Om}_{\u{\m}}\big)^{\u{ab}}\u{\G}_{\u{ab}}\u{\e}^{\a}
-\qt\big(\u{V}_{\u{\m}}\big)^{IJ}\big(\u{\p}_{IJ}\big)^{\a}{}_{\b}\u{\e}^{\b},
\nn\\
&&\u{\cD}_{\u{\m}}\u{T}^{\a\b}{}_{\u{abc}}
={\d}_{\u{\m}}\u{T}^{\a\b}{}_{\u{abc}}
+\big(\u{\Om}_{\u{\m}}\big){}_{[\u{a}}{}^{\u{d}}\u{T}^{\a\b}{}_{\u{bc]d}}
-\u{b}_{\u{\m}}\u{T}^{\a\b}{}_{\u{abc}}
+\qt\u{V}_{\u{\m}}{}^{IJ}\big(\p_{IJ}\big){}^{[\a}{}_{\g}
\u{T}^{\b]\g}{}_{\u{abc}}.
\end{eqnarray*}
Here, the field strength of the $R$-symmetry gauge field $\u{V}_{\u{\m}}{}^{IJ}$ 
is given by
\begin{eqnarray*}
\u{R}_{\u{\m\n}}{}^{\a}{}_{\b}
=\hf\u{R}_{\u{\m\n}}{}^{IJ}\big(\p_{IJ}\big)^{\a}{}_{\b}
=\hf\l[\d_{\u{\m}}\u{V}_{\u{\n}}{}^{IJ}-\d_{\u{\n}}\u{V}_{\u{\m}}{}^{IJ}
-\u{V}_{\u{\m}}{}^{I}{}_{K}\u{V}_{\u{\n}}{}^{KJ}
+\u{V}_{\u{\n}}{}^{I}{}_{K}\u{V}_{\u{\m}}{}^{KJ}
\r]\big(\p_{IJ}\big)^{\a}{}_{\b}.
\end{eqnarray*}

\subsection{Temporally Dimensional Reduction of the Weyl Multiplet}
\label{WeylConformalTensorCalculus}

In this subsection, the dimensional reduction of the Weyl multiplet 
along the time direction will be considered in the same way 
as the dimensional reduction along one spatial direction was performed in 
\cite{CJ}, where the strategy in \cite{KO} was followed.

For the usual ansatz for the metric 
$$
ds^2_6=-{1\over\a^2}\l(dt+C\r)^2+ds^2_5
=-\u{E}^0\u{E}^0+ds_5^2=\sum_{\u{a}=0}^5\u{E}^{\u{a}}\u{E}^{\u{a}},
$$
where
$$
\u{E}^0={1\over\a}\l(dt+C\r),
\qquad
ds_5^2=\sum_{{a}=1}^5\u{E}^{{a}}\u{E}^{{a}},
$$
as a gauge-fixing condition, 
the six-dimensional coframe $\u{E}^{\u{a}}{}_{\u{\m}}$ ($\u{\m}=t,1,\cdots,5$; 
$\u{a}=0,1,2,\cdots,5$.) can be taken by a local Lorentz transformation to be 
$$
\l(\u{E}^{\u{a}}{}_{\u{\m}}\r)=
\left(\begin{array}{cc}
e^0{}_{t}&e^0{}_\m\\ e^a{}_t&e^a{}_\m
\end{array}\right)
=\left(\begin{array}{cc}
{\a}^{-1}&{\a}^{-1}C_\m\\0&e^a{}_\m
\end{array}\right), 
$$
where $\m=1,2,\cdots,5$; $a=1,2,\cdots,5$. 
Here, $\a$ is a scalar field (a.k.a. dilaton), which is sometimes denoted by 
$\sim\exp(-\varphi)$, but we will follow \cite{KO,CJ} to denote it 
by $\a$. 
Therefore, one can see that
$$
\u{E}^{0}={1\over\a}\l(dt+C\r), 
\qquad
\u{E}^{a}=e^a=e^a{}_{\m}dx^{\m}.
$$
For the gauge field $C=C_{\m}dx^\m$, we define the field strength 
$$
G=dC=\hf{G}_{\m\n}dx^\m\wedge{dx}^\n. 
$$

Since the six-dimensional coframe $\u{\Th}^{\u{\m}}{}_{\u{a}}$ 
is inverse to the sechsbein $\u{E}^{\u{a}}{}_{\u{\m}}$, 
it takes the form
$$
\l(\u{\Th}^{\u{\m}}{}_{\u{a}}\r)=
\left(\begin{array}{cc}
\th^t{}_{0}&\th^t{}_a\\ \th^\m{}_0&\th^\m{}_a
\end{array}\right)
=\left(\begin{array}{cc}
{\a}&-C_a\\0&\th^\m{}_a
\end{array}\right),
$$
under the gauge-fixing condition, 
with $C_a=\th^\n{}_aC_\n$, 
where the funfbein $e^a{}_\m$ and the five-dimensional 
coframe $\th^\m{}_b$ satisfy 
$$
e^a{}_\m\,\th^\m{}_b=\dl^a{}_b,
\qquad
\th^\m{}_a{e}^a{}_\n=\dl^\m{}_\n.
$$

One then finds the Levi-Civita spin connection 
$\u{\om}^{\u{ab}}=\u{\om}_{\u{c}}{}^{\u{ab}}e^{\u{c}}$ satisfying 
$de^{\u{a}}+\u{\om}^{\u{a}}{}_{\u{b}}\wedge{e}^{\u{b}}=0$,  
$$
\l(\u{\om}_0\r)^{0}{}_a=-{1\over\a}\th^\m{}_a\,\d_\m\a,
\qquad
\l(\u{\om}_0\r){}_{ab}=\l(\u{\om}_a\r){}_{0b}={1\over2\a}G_{ab},
\qquad
\l(\u{\om}_a\r)^{bc}=\l({\om}_a\r)^{bc},
$$
with the five-dimensional Levi-Civita spin connection 
$\om^{ab}=\l(\om_c\r)^{ab}\,e^c$ 
satisfying $de^{a}+\om^{a}{}_{b}\wedge{e}^b=0$, 
and $G_{ab}=\th^\m{}_{a}\th^{\n}{}_{b}G_{\m\n}$.

As in \cite{KO,CJ}, we will continue the partial gauge-fixing by using 
the conformal supersymmetry transformation $S^\a$ 
to set $\u{\psi}^{\a}{}_{\u{0}}=0$; 
the special conformal tansformations $K_0$ to set $\u{b}_{\u{0}}=0$, and 
$K_a$ to 
$$
\u{b}_{{\m}}={1\over\a}\d_\m\a, 
\qquad
(\m=1,2,\cdots,5).
$$
The latter condition makes the dilaton field $\a$ covariant constant \cite{KO};
$$
\cD_\m\a=\d_\mu\a-b_\m\a=0,
$$
which will be convenient for the calculations below. 

The partial gauge fixing conditions are summarized as 
\begin{center}
\framebox(420,40){\parbox{\textwidth}{
\begin{eqnarray}
~~~~~
\u{E}^{\u{a}}{}_{{t}}=0, 
\qquad
\u{\psi}^{\a}{}_{{0}}=0, 
\qquad
\u{b}_{{0}}=0,
\qquad
\u{b}_{{\m}}={\a}^{-1}\d_\m\a, \quad(\m=1,2,\cdots,5).
\label{gaugefixing}
\end{eqnarray}
}}
\end{center}
We will use $b_\m$ as shorthand for ${\a}^{-1}\d_\m\a$ and 
$b_a=\th_a{}^\m{b}_\m$.

Therefore, under the gauge fixing condition, one has 
the dependent gauge field $\u{\Om}_{\u{\m}}{}^{\u{ab}}$ in (\ref{defOm}) 
\begin{eqnarray*}
\l(\u{\Om}_{t}\r){}^{0}{}_{a}=0,
\qquad
\l(\u{\Om}_{t}\r){}_{ab}={1\over\a^2}G_{ab},
\qquad
\l(\u{\Om}_{\m}\r){}^{0}{}_{a}=-{1\over2\a}G_{\m{a}},
\nn\\
\l(\u{\Om}_{\m}\r){}_{ab}=\l(\om_\m\r){}_{ab}+{1\over2\a^2}C_\m{G}_{ab}
+\l(e_{a\m}\th^{\n}{}_{b}-e_{b\m}\th^{\n}{}_{a}\r){1\over\a}\d_\n\a.
\end{eqnarray*}
Among them, after the dimensional reduction, the component
$$
\l(\u{\Om}_c\r)_{ab}=\u{\Th}^{\u{\m}}{}_{c}\big(\u{\Om}{}_{\u{\m}}\big){}_{ab}
=\l(\om_c\r)_{ab}+\dl_{ac}\,b_b-\dl_{bc}\,b_a
\equiv\th^\m{}_c\l(\Om_\m\r){}_{ab}
$$
often appears in the covariant derivatives, and we refer to it as 
$\l(\Om_c\r){}_{ab}$. 

The auxiliuary fields $\u{V}_{\u{a}}{}^{IJ}$, $\u{T}{}^{\a\b}{}_{\u{abc}}$ 
are decomposed into five-dimensional fields $S^{IJ}$, $V_a{}^{IJ}$, 
$t^I{}_{ab}$ by
\begin{eqnarray*}
\u{V}_{\u{a}}{}^{IJ} 
=\begin{cases}
\displaystyle
\u{V}_0{}^{IJ}\equiv{S}^{IJ},
\\
\displaystyle
\u{V}_{a}{}^{IJ}\equiv{}A_{a}{}^{IJ},
\end{cases}
\qquad
\u{T}{}^{\a\b}{}_{\u{abc}}
=\begin{cases}
\displaystyle
\u{T}{}^{\a\b}{}_{{0ab}}\equiv{}t^{I}{}_{ab}\l(\p_I\Om^{-1}\r)^{\a\b},
\\
\displaystyle
\u{T}{}^{\a\b}{}_{{abc}}=-(1/2)\ve_{abc}{}^{de}\u{T}{}^{\a\b}{}_{{0de}},
\end{cases}
\end{eqnarray*}
with $\ve_{12345}=\ve^{12345}=1$. 
Note that the gauge field $A_\m{}^{IJ}$ is given by
$$
A_\m{}^{IJ}=e^a{}_{\m}A_a{}^{IJ}=e^a{}_{\m}\u{\Th}^{\u{\n}}{}_a
\u{V}{}_{\u{\n}}{}^{IJ}=\u{V}{}_\m{}^{IJ}-C_\m\,\u{V}{}_t{}^{IJ}
=\u{V}{}_\m{}^{IJ}-{1\over\a}C_\m\,S{}^{IJ}. 
$$
Let us remove the underline $\u{\quad}$ from $\u{M}^{\a\b}{}_{\g\dl}$ 
to denote its reduced one as ${M}^{\a\b}{}_{\g\dl}$. 
It is sometimes convenient to replace the spinor indices $\a,\b$ of 
${M}^{\a\b}{}_{\g\dl}$ by the vector indices $I,J$ as
$$
{M}^{\a\b}{}_{\g\dl}
=-{M}^{IJ}\l(\p_I\Om^{-1}\r){}^{\a\b}\l(\Om\p_J\r){}_{\g\dl}.
$$
The field $M^{IJ}$ is in the reprensentation {\bf 14} of the Spin(5)$_R$ 
group and enjoys the symmetry properties
$$
M^{IJ}=M^{JI}, 
\qquad
\dl_{IJ}\,M^{IJ}=0.
$$

The time component of the gravitini is set to zero by the gauge fixing 
condition (\ref{gaugefixing}); $\u{\psi}^{\a}{}_{t}=0$, and 
we will denote 
the remaining components $\u{\psi}^{\a}{}_{\m}$ ($\m=1,\cdots,5$)
simply as 
\begin{eqnarray*}
\u{\psi}^{\a}{}_{\m}
=\begin{pmatrix}
{\psi}^{\a}{}_{\m}\\0
\end{pmatrix},
\end{eqnarray*}
since it is of positive chirality, and our convetion of the chirality is 
found in Appendix \ref{SMW}. 

Since the auxiliuary spinor $\u{\chi}^{\a\b}{}_{\g}$ is also of positive 
chirality, we will take 
\begin{eqnarray*}
\u{\chi}^{\a\b}{}_{\g}
={15\over16}\begin{pmatrix}
~{\chi}^{\a\b}{}_{\g}\\0
\end{pmatrix},
\end{eqnarray*}
with the convenient coefficient $15/16$ in \cite{CJ}.

The parameters $\u{\e}^\a$ and $\u{\eta}^\a$ 
of supersymmetry and conformal supersymmetry transformations 
are of positive and negative chirality, respectively, and 
we will take 
\begin{eqnarray*}
\u{\e}^{\a}
=\begin{pmatrix}
~{\e}^{\a}\\0
\end{pmatrix},
\qquad
\u{\eta}^{\a}
=\begin{pmatrix}
0\\{\eta}^{\a}
\end{pmatrix}. 
\end{eqnarray*}

The gauge fixing condition (\ref{gaugefixing}) 
is changed under the supersymmetry ($Q$-) 
tranformation (\ref{QStransf}). In particular, the zeroth component 
of the gravitino transforms under the supersymmetry ($Q$) and the conformal 
supersymmetry ($S$) as 
\begin{eqnarray}
\dl\u{\psi}^{\a}{}_{0}
={1\over8\a}G_{ab}\u{\G}^{ab}\u{\e}^{\a}
-{1\over4}{S}_{IJ}\l(\p^{IJ}\r){}^{\a}{}_{\b}\u{\e}^\b
+{1\over4}\,t^{\a\b}{}_{{ab}}\,\u{\G}^{{ab}}\,\u{\e}_{\b}
+\u{\G}_{0}\,\u{\eta}^{\a},
\label{delpsi}
\end{eqnarray}
under the gauge fixing condition (\ref{gaugefixing}). 
However, combining the supersymmetry ($Q$-) 
and the conformal supersymmetry ($S$-) transformations, 
one can find that one linear combination of them leaves 
the condition $\u{\psi}^{\a}{}_{0}=0$ unchanged. 
For any $\u{\e}^\a$, one can see that 
the conformal supersymmetry transformation with the parameter 
\begin{equation}
{\eta}^{\a}
={1\over8\a}G_{ab}{\g}^{ab}{\e}^{\a}
-{1\over4}{S}_{IJ}\l(\p^{IJ}\r){}^{\a}{}_{\b}{\e}^\b
+{1\over4}\,t^{\a\b}{}_{{ab}}\,\g^{{ab}}\,{\e}_{\b},
\label{superconfeta}
\end{equation}
compensates for the deviation (\ref{delpsi}) from the gauge fixing 
condition on the gravitini. 

Among the other gauge fixing conditions in (\ref{gaugefixing}), 
the condition $\u{E}^{\u{a}}{}_{t}=0$ remains unchanged under the 
supersymmetry ($Q$-) and the conformal supersymmetry ($S$-) transformaions. 
But, the remaining gauge fixing conditions $\u{b}_0=0$ and 
$\u{b}_{\m}=\a^{-1}\d_\m\a$ are changed under those transformations. 
However, the deviations can be canceled by the 
special conformal ($K$-) transformations with appropriate paramemters 
$\La_K{}_{\u{a}}$. Note here that $\u{E}^{\u{a}}{}_{t}$ and 
$\u{\psi}^{\a}{}_{0}$ are left invariant under the special conformal ($K$-) 
transformations. Thus, one may define a supersymmetry transformation 
in the reduced five-dimensional theory as the linear combination of 
supersymmetry ($Q$-), conformal supersymmetry ($S$-), and  
special conformal ($K$-) transformations.

Following the ideas in \cite{FS}, we are seeking for supersymmetric 
backgrounds of the reduced theory to obtain supersymmetric compactifications 
of the $\N=2$ supersymmetric Yang-Mills theory in five dimensions. 
Since we would like to consider bosonic backgrounds, we will turn off 
background spinor fields, and we will find the supersymmetric bosonic 
backgrounds leaving the spinor fields $\psi^\a{}_\m$, $\chi^{\a\b}{}_{\g}$ 
unchanged under some of supersymmetry transformations in the reduced theory.

From the supersymmetry transformation of the gravitini 
\begin{eqnarray*}
\dl_\e\psi^\a{}_{a}=\th^\m{}_a\,\dl\psi^\a{}_\m
=\cD_a\e^\a-\qt{S^{IJ}}(\p_{IJ}){}^\a{}_\b\g_a\e^\b+{1\over2\a}G_{ab}\g^b\e^\a
+{1\over8\a}G_{bc}\g_a{}^{bc}\e^\a
-\hf{t^I}{}_{bc}(\p_{I}){}^{\a}{}_{\b}\g_a{}^{bc}\e^\b,
\end{eqnarray*}
with the covariant derivative of the supersymmetry parameter
$$
\cD_a\e^\a=\th^\m{}_a\cD_\m\e^\a
=\th^\m{}_a\d_\m\e^\a+\hf{b}_a\e^\a+\qt(\Om_a){}^{bc}\g_{bc}\e^\a
-\qt{A}_a{}^{IJ}\l(\p_{IJ}\r){}^\a{}_\b\e^\b,
$$
one can see that the supersymmetric bosonic backgrounds should obey 
\begin{center}
\framebox(420,40){\parbox{\textwidth}{
\begin{eqnarray}
~~~~~\cD_a\e^\a=\qt{S^{IJ}}(\p_{IJ}){}^\a{}_\b\g_a\e^\b
-{1\over2\a}G_{ab}\g^b\e^\a
-{1\over8\a}G_{bc}\g_a{}^{bc}\e^\a
+\hf{t^I}{}_{bc}(\p_{I}){}^{\a}{}_{\b}\g_a{}^{bc}\e^\b.
\label{KSE}
\end{eqnarray}
}}
\end{center}

Under a supersymmetry transformation, the auxiliuary spinor $\chi^{\a\b}{}_{\g}$ 
transforms as 
\begin{eqnarray}
\dl_\e\chi^{\a\b}{}_{\g}&=
&-2\c{1\over4!}\ve^{abcde}\cD^f{t}^{\a\b}{}_{fa}\g_{bcde}\e_\g
+{1\over4!}\ve^{abcde}\cD_a{S}_\g{}^{[\a}\g_{bcde}\e^{\b]}
-t^{\a\b}{}_{ab}t{}_{\g\dl}{}_{cd}\g^{abcd}\e^{\dl}
\nn\\
&&-\hf\ve^{abcde}\cD_at^{\a\b}{}_{bc}\g_{de}\e_\g
+2{1\over\a}G_a{}^c{t}^{\a\b}{}_{bc}\g^{ab}\e_\g
-{1\over2\a}G^{ab}S_\g{}^{[\a}\g_{ab}\e^{\b]}
\nn\\
&&-\hf{S}^{[\a}{}_{\dl}t^{\b]\dl}{}_{ab}\g^{ab}\e_\g
-2\,t^{\a\b}{}_{ab}S_{\g\dl}\g^{ab}\e^{\dl}
+4t^{\a\b}{}_{ac}t_{\g\dl}{}_b{}^c\,\g^{ab}\e^{\dl}
-\hf{F}^{ab}{}_{\g}{}^{[\a}\g_{ab}\e^{\b]}
\nn\\
&&-{4\over15}M^{\a\b}{}_{\g\dl}\e^{\dl}
-{1\over\a}G^{ab}t^{\a\b}{}_{ab}\e_\g
+2t^{\a\b}{}_{ab}t_{\g\dl}{}^{ab}\e^{\dl}
+\cdots,
\label{chiKSE}
\end{eqnarray}
with $t^{\a\b}{}_{ab}=t^I{}_{ab}(\p_I\Om^{-1})^{\a\b}$, and 
$S^{\a}{}_{\b}=(1/2)S^{IJ}(\p_{IJ})^\a{}_\b$, 
where the ellipse $\cdots$ denotes the necessary terms\footnote{
In \cite{CJ,BSvP2}, they are denoted as (trace).
} to leave the right hand side in the representation {\bf 16} of the Spin(5)$_R$ 
symmetry, since $\chi^{\a\b}{}_{\g}$ is in the reprensentation {\bf 16}. 
Here, the two covariant derivatives are given by
\begin{eqnarray*}
&&\cD_\m{t}^{\a\b}{}_{ab}=\d_\m{t}^{\a\b}{}_{ab}
+\l(\Om_\m\r){}_a{}^c{t}^{\a\b}{}_{cb}+\l(\Om_\m\r){}_b{}^c{t}^{\a\b}{}_{ac}
-b_\m{t}^{\a\b}{}_{ab}-\hf{A}_\m{}^{\a}{}_{\g}{t}^{\g\b}{}_{ab}
-\hf{A}_\m{}^{\b}{}_{\g}{t}^{\a\g}{}_{ab},
\nn\\
&&\cD_\m{S}^{\a\b}=\d_\m{S}^{\a\b}-b_\m{S}^{\a\b}
-\hf{A}_\m{}^\a{}_\g{S}^{\g\b}-\hf{A}_\m{}^\b{}_\g{S}^{\a\g},
\end{eqnarray*}
with $A_\m{}^\a{}_\b=(1/2)A_\m{}^{IJ}(\p_{IJ}){}^\a{}_\b$,
whose curvature tensor 
$F_{\m\n}{}^{\a}{}_{\b}=(1/2)F_{\m\n}{}^{IJ}\l(\p_{IJ}\r)^\a{}_\b$ 
is defined by
$$
F_{\m\n}{}^I{}_J=\d_\m{A}_{\n}{}^I{}_J-\d_\n{A}_{\m}{}^I{}_J
-{A}_{\m}{}^{I}{}_{K}{A}_{\n}{}^{K}{}_{J}
+{A}_{\n}{}^{I}{}_{K}{A}_{\m}{}^{K}{}_{J}.
$$
Therefore, the other condition for the supersymmetric backgrounds 
is that the right hand side of (\ref{chiKSE}) should vanish. 
The explicit form (\ref{SUSYchieq}) of the supersymmetry condition is given 
in Appendix \ref{SUSYchi}, because the equation is very lengthy to write it 
here. 

Thus, (\ref{KSE}) gives the Killing spinor equation, and supersymmetric 
backgrounds have to allow the existence of 
the solutions (the Killing spinors) to the equation. 
One may interpret that (\ref{SUSYchieq}) determines the background field 
$M^{\a\b}{}_{\g\dl}$, which will appear in the mass term of the scalar 
fields in the five-dimensional $\N=2$ supersymmetric theroy, as 
will be seen below.

\subsection{The Killing Vectors and the Killing Spinors}
\label{KillingVector&Spinor}

The Killing spinors $\e^\a$, $\eta^\a$ obeying the equation (\ref{KSE}) 
form the bilinear 
$$
\xi^a
=\l(\eta^\a\r)^TC\g^a\e^\b\Om_{\a\b}
\equiv\bar\eta\c\g^a\e,
$$
and its covariant derivative 
\begin{eqnarray*}
\cD_\m\xi^a&\equiv&\d_\m\xi^a+b_\m\xi^a+\Om_\m{}^{a}{}_{b}\xi^b
=\bigg[\l(\cD_\m\eta^\a\r)^T{\c}C\g^a\e^\b
+\l(\eta^\a\r)^TC\g^a\cD_\m\e^\b\bigg]\Om_{\a\b}
\nn\\
&=&-\hf{S}_{IJ}\l(\bar\eta\c\p^{IJ}\g_\m{}^a\e\r)
-{1\over\a}G_\m{}^a\l(\bar\eta\c\e\r)
+\ve_\m{}^{abcd}\bigg[
{1\over4\a}G_{bc}\l(\bar\eta\c\g_d\e\r)
-t^I{}_{bc}\l(\bar\eta\c\p_I\g_d\,\e\r)\bigg],
\end{eqnarray*}
satisfies $\cD_a\xi_b+\cD_b\xi_a=0$. 
See Appendix \ref{SMW} for the notations for the bilinears 
$\l(\bar\eta\c\p_{I_1\cdots{I}_n}\g^{a_1\cdots{a}_m}\e\r)$.

The vector field $\xi^a$ obeys the conformal Killing vector equation 
\begin{eqnarray}
\cd_a\,\xi_b+\cd_b\,\xi_a={2\over5}\eta_{ab}\l(\cd_c\,\xi^c\r),
\label{confKVE}
\end{eqnarray}
with the covariant derivative 
$
\cd_\m\xi^a\equiv\d_\m\xi^a+\om_\m{}^a{}_b\xi^b,
$
which is related to the previous covariant derivative as
$$
\cD_a\xi_b=\cd_a\xi_b+\eta_{ab}\l(b_c\xi^c\r)+b_{[a}\xi_{b]}. 
$$
In fact, the equation $\cD_a\xi_b+\cD_b\xi_a=0$ leads to
$$
\cd_a\xi_b+\cd_b\xi_a=-{2}\eta_{ab}\l(b_c\xi^c\r),
$$
which gives the conformal Killing vector equation (\ref{confKVE}).

\section{Tensor Multiplet in the Supergravity Theory}
\label{TensorSUGRA}

To the conformal supergravity, tensor multiplets can be added as matters, 
and after the dimensional reduction, they give rise to 
$\N=2$ gauge multiplets in five dimensions. 
It therefore yields a five-dimensional $\N=2$ supersymmetric Abelian theory 
in the supergravity background. It is the topic of this section.

A tensor multiplet 
($\u{B}{}_{\u{\m\n}}$, $\u{\phi}{}^{\a\b}$, $\u{\chi}{}^{\a}$) 
of the $\N=(2,0)$ supergravity is listed in Table \ref{mattertensor},
and the field strength of the two-form $\u{B}$ is given by 
$$
\u{H}={1\over3!}\,\u{H}{}_{\u{abc}}\,\u{E}^a\wedge\u{E}^b\wedge\u{E}^c
=d\u{B}.
$$
The transformation rules and the equations of motion of the tensor multiplet 
were derived in \cite{BSvP2}.

\begin{table}[h]
\begin{center}
\begin{tabular}{|c|c|c|c|}\hline
\makebox[30mm]{tensor multiplet}&\makebox[30mm]{symmetries}
&\makebox[1.5cm]{Spin(5)${}_{R}$}&\makebox[1cm]{weight}
\\ \hline\hline
\multicolumn{4}{|c|}{bosonic fields} 
\\ \hline
\multirow{2}{*}{$\u{B}{}_{\u{\m\n}}$}
&\multirow{2}{*}
{$\u{B}{}_{\u{\m\n}}=-\u{B}{}_{\u{\n\m}}$,}
&\multirow{2}{*}{{\bf 1}}
&\multirow{2}{*}{0} \\ 
& & & \\ \hline
\multirow{2}{*}{$\u{\phi}^{\a\b}$}
&\multirow{2}{*}{$\u{\phi}^{\a\b}=-\u{\phi}^{\b\a}$, 
~$\Om_{\a\b}\u{\phi}^{\a\b}=0$,}
&\multirow{2}{*}{{\bf 5}}
&\multirow{2}{*}{2} \\ 
& & & \\ \hline\hline
\multicolumn{4}{|c|}{fermionic field} 
\\ \hline
\multirow{2}{*}{$\u{\chi}^{\a}$}
&\multirow{2}{*}{$\u{\G}^7\u{\chi}^{\a}=-\u{\chi}^{\a}$,}
&\multirow{2}{*}{{\bf 4}}
&\multirow{2}{*}{5/2} \\ 
& & & \\ \hline\end{tabular}
\end{center}
\caption{The tensor multiplet in the six-dimensional supergravity.}
\label{mattertensor}
\end{table}

Under a fermionic transformation (supersymmetry$+$ conformal supersymmetry), 
the tensor multiplet transforms as  
\begin{equation}
\begin{split}
&\dl\u{B}_{\u{\m\n}}=i\l(\u{\e}^\a\r)^\dag\u{\G}^0\u{\G}_{\u{\m\n}}\u{\chi}^\a,
\qquad
\dl\u{\phi}^{\a\b}=-2i\l(\u{\e}_\a\r)^\dag\u{\G}^0\u{\chi}^\b
+2i\l(\u{\e}_\b\r)^\dag\u{\G}^0\u{\chi}^\a
-i\Om^{\a\b}\l(\u{\e}^\g\r)^\dag\u{\G}^0\u{\chi}^\g,
\\
&\dl\u{\chi}^\a
={1\over8}\c{1\over3!}\,{\u{H}^+}{}_{\u{\m\n\p}}\,
\u{\G}{}^{\u{\m\n\p}}\u{\e}{}^\a
+{1\over4}\,\u{\cD}{}_{\u{\m}}\u{\phi}{}^{\a\b}\,\u{\G}{}^{\u{\m}}\u{\e}{}_\b
-\u{\phi}{}^{\a\b}\u{\eta}{}_\b,
\end{split}
\label{6DtensorSUSY}
\end{equation}
where $\u{H}{}^{\pm}=(1/2)\l(\u{H}\pm\u{*}\u{H}\r)$. 
(See the definition of the Hodge dual $\u{*}$ in Appendix \ref{conventions}.)
The covariant derivative of the scalar field $\u{\phi}^{\a\b}$ is 
$$
\u{\cD}{}_{\u{\m}}\,\u{\phi}{}^{\a\b}=
\d{}_{\u{\m}}\,\u{\phi}^{\a\b}-2\u{b}{}_{\u{\m}}\u{\phi}^{\a\b}
-\qt{}\u{V}{}_{\u{\m}}{}^{IJ}\l(\p_{IJ}\r){}^\a{}_\g\,\u{\phi}^{\g\b}
-\qt{}\u{V}{}_{\u{\m}}{}^{IJ}\l(\p_{IJ}\r){}^\b{}_\g\,\u{\phi}^{\a\g}. 
$$

The equations of motion of the tensor multiplet are given by
\begin{eqnarray}
&&\u{H}^{-}-\hf\,\u{\phi}{}_{\a\b}\,\u{T}^{\a\b}=0,
\label{6DHeom}\\
&&\u{\cD}{}^{\u{a}}\u{\cD}{}_{\u{a}}\u{\phi}{}_{\a\b}
-{1\over15}\,\u{M}{}_{\a\b}{}^{\g\dl}\u{\phi}{}_{\g\dl}
+{1\over3}\,{\u{H}^{+}}{}_{\u{abc}}\u{T}{}_{\a\b}{}^{\u{abc}}=0,
\label{6Dphieom}\\
&&\u{\G}{}^{\u{a}}\u{\cD}{}_{\u{a}}\,\u{\chi}{}^\a
-{1\over12}\,\u{T}{}^{\a\b}{}_{\u{abc}}\,\u{\G}{}^{\u{abc}}\u{\chi}{}_{\b}=0,
\label{6Dchieom}
\end{eqnarray}
with the covariant derivatives 
\begin{eqnarray*}
&&\u{\cD}{}^{\u{a}}\u{\cD}{}_{\u{a}}\u{\phi}{}^{\a\b}
=\u{\Th}{}^{\u{\m}}{}_{\u{a}}\l(\d_{\u{\m}}-3\u{b}{}_{\u{\m}}\r)
\u{\cD}{}^{\u{a}}\u{\phi}{}^{\a\b}
+\l(\u{\Om}{}_{\u{a}}\r){}^{\u{ab}}\u{\cD}{}_{\u{b}}\,\u{\phi}{}^{\a\b}
\nn\\
&&\qquad\qquad\qquad
-\qt\,\u{V}{}_{\u{a}}{}^{IJ}\l(\p_{IJ}\r){}^{\a}{}_{\g}
\u{\cD}{}^{\u{a}}\,\u{\phi}{}^{\g\b}
-\qt\,\u{V}{}_{\u{a}}{}^{IJ}\l(\p_{IJ}\r){}^{\b}{}_{\g}
\u{\cD}{}^{\u{a}}\,\u{\phi}{}^{\a\g}
-{1\over5}\u{R}\l(\u{\Om}\r)\u{\phi}{}^{\a\b},
\nn\\
&&\u{\cD}{}_{\u{\m}}\,\u{\chi}{}^\a
=\l(\d_{\u{\m}}-{5\over2}\,\u{b}{}_{\u{\m}}
+\qt\,\big(\u{\Om}{}_{\u{\m}}\big){}^{\u{ab}}\u{\G}{}_{\u{ab}}\r)\u{\chi}{}^\a
-\qt\,\u{V}{}_{\u{\m}}{}^{IJ}\l(\p_{IJ}\r){}^\a{}_\b\u{\chi}{}^\b.
\end{eqnarray*}

\subsection{Dimensional Reduction of the Tensor Multiplet}

From the six-dimensional Minkowski space to the five-dimensional 
Euclidean space, 
the dimensional reduction of the tensor multiplet gives rise to 
the five-dimensional abelian gauge multiplet ($A_\m$, $\phi^I$, $\chi^\a$), 
\begin{eqnarray*}
&&\u{B}{}_{\u{ab}} \quad\longrightarrow\quad 
\u{B}{}_{{a0}}\equiv\a\,{A}_a=\a\,\th^\m{}_a{A}_\m, 
\qquad (a=1,2,\cdots,5)
\nn\\
&&\u{\phi}{}^{\a\b} \quad\longrightarrow\quad \a\,{\phi}{}^{\a\b}
=\a\,{\phi}{}^I\l(\p_I\Om^{-1}\r){}^{\a\b},
\nn\\
&&\u{\chi}{}^\a \quad\longrightarrow\quad 
{\a\over4}\begin{pmatrix}0\\\chi^\a\end{pmatrix}.
\end{eqnarray*}
The remaining components $\u{B}_{ab}$ are described by $A_\m$ and $\phi^I$ 
through the equation (\ref{6DHeom}) of motion of $\u{H}^{-}$, which is 
reduced to 
$$
\u{H}{}^{-}{}_{ab0}=\hf\,\u{\phi}{}_{\a\b}\,\u{T}{}^{\a\b}{}_{ab0}
\quad\longrightarrow\quad
2\a\,\phi_I\,{t}^I{}_{ab}.
$$
Since the components $\u{H}{}_{ab0}$ reduce to the field strength $F_{\m\n}$ 
of $A_\m$, 
$$
\u{H}{}_{\m\n{t}}=\d_\m B_{\n{t}}+\d_\n B_{{t}\m}+\d_t B_{\m\n}
\longrightarrow\d_\m A_\n-\d_\n A_\m=F_{\m\n},
\quad
\u{H}{}_{ab0}\rightarrow\a\th^\m{}_a\th^\n{}_bF_{\m\n}=\a{F}_{ab},
$$
one can see that the components $\u{H}{}_{abc}$ are reduced as 
\begin{eqnarray*}
&&\u{H}{}{}_{abc}=\u{H}{}^+{}_{abc}+\u{H}{}^-{}_{abc}
=\hf\ve_{abc}{}^{de}\l(\u{H}{}^+{}_{de0}-\u{H}{}^-{}_{de0}\r)
=\hf\ve_{abc}{}^{de}\l(\u{H}{}{}_{de0}-2\u{H}{}^-{}_{de0}\r)
\nn\\
&&\qquad\longrightarrow\quad
\a\,\hf\,\ve_{abc}{}^{de}\l(F_{de}-4\,\phi_i\,{t}^i{}_{de}\r). 
\end{eqnarray*}

We have previously seen that a six-dimensional supersymmetry transformation 
with a transformation parameter $\u{\e}{}^\a$ combined with 
the superconformal transformation with $\u{\eta}{}^\a$ in (\ref{superconfeta}) 
is reduced to a five-dimensional supersymmetry transformation. 
Substituting the parameter $\u{\eta}{}^\a$ in (\ref{superconfeta}) into 
the fermionic transformation rules in (\ref{6DtensorSUSY}) 
of the tensor multiplet, one can see that their reduction gives 
the supersymmetry transformation of the abelian gauge multiplet, 
\begin{eqnarray}
&&\dl_\e{A}_\m=-{i\over4}\Om_{\a\b}\l(\e^\a\r)^TC\g_\m\chi^\a,
\qquad
\dl_\e\phi^I={i\over4}\l(\Om\p^I\r){}_{\a\b}\l(\e^\a\r)^TC\chi^\b,
\nn\\
&&\dl_\e\chi^\a=-{1\over2}F_{ab}\g^{ab}\e^\a
-\g^\m\cD_\m\phi^I\l(\p_I\r){}^\a{}_\b\e^\b
+{1\over2\a}G_{ab}\phi^I\l(\p_I\r){}^\a{}_\b\g^{ab}\e^\b
+S^I{}_J\phi^J\l(\p_I\r)^\a{}_\b\e^\b
\label{5DabelianSUSY}\\
&&\qquad\qquad
+{1\over2}\ve_{IJKLM}S^{IJ}\phi^M\l(\p^{KL}\r){}^\a{}_\b\e^\b
+t^I{}_{ab}\phi^J\l(\p_{IJ}\r){}^\a{}_\b\g^{ab}\e^\b,
\nn\end{eqnarray}
with the covariant derivative of $\phi^I$,
$$
\cD_\m\phi^I=\d_\m\phi^I-b_\m\phi^I-A_\m{}^I{}_J\phi^J.
$$

The reduction of the external derivative of the equation (\ref{6DHeom}) 
$$
d\l(\u{*}\,\u{H}+\u{\phi}{}_{\a\b}\,\u{T}{}^{\a\b}
\r)=0,
$$
yields the equation of motion of the gauge field $A_\m$, 
\Vox{
\begin{equation}
d\l[
\a{*}\l(F-4\,\phi_I\,t^I\r)
\r]+F\wedge{G}=0,
\label{5Dgaugeeom}
\end{equation}
}
where 
$$
F=\hf\,F_{\m\n}dx^\m\wedge{dx}^\n, 
\qquad
t^I=\hf\,t^I{}_{\m\n}dx^\m\wedge{dx}^\n, 
\qquad
G=\hf\,G_{\m\n}dx^\m\wedge{dx}^\n. 
$$

The equations (\ref{6Dphieom},\ref{6Dchieom}) of motion are reduced into
\BBox{400}{95}{
\begin{eqnarray}
&&\g^\m\cD_\m\chi^\a
-{1\over8\a}G_{ab}\g^{ab}\chi^\a-{1\over4}S_{IJ}(\p^{IJ}){}^\a{}_\b\chi^\b
+{1\over2}t^I{}_{ab}(\p_I){}^\a{}_\b\g^{ab}\chi^\b=0,
\nn\\
&&\cD^a\cD_a\phi^{I}
-S^I{}_J\,S^J{}_K\,\phi^{K}
-{1\over5}{R}\l({\Om}\r)\phi^{I}
-{4\over15}M^{I}{}_{J}\phi^{J}
\label{5Dchiphieom}\\
&&\hskip3.5cm
-{1\over20\a^2}G_{ab}G^{ab}\phi^{I}
+4\,t^{I}{}_{ab}\,t_{J}{}^{ab}\,\phi^{J}
-2\,t^{I}{}_{ab}\,F^{ab}
=0,
\nn
\end{eqnarray}
}
with 
$$
\cD^a\cD_a\phi^{I}
=\th^{\m{a}}\l[
\l(\d_\m-2b_\m\r)\cD_a\phi^{I}+\Om_\m{}_a{}^b\cD_b\phi^{I}
-\qt{A}_\m{}^{I}{}_{J}\cD_a\phi^{J}\r],
$$
where the covariant derivative of $\chi^\a$ 
$$
\cD_\m\chi^\a=\d_\m\chi^\a-{3\over2}b_\m\chi^\a
+{1\over4}\Om_{\m}{}^{bc}\g_{bc}\chi^\a
-{1\over4}A_\m{}^{IJ}\l(\p_{IJ}\r){}^\a{}_\b\chi^\b,
$$
with the spin connection 
$\Om_\m{}^{ab}=\om_\m{}^{ab}+(e^a{}_\m\th^{\n{b}}-e^b{}_\m\th^{\n{a}})b_\n$,
and the scalar curvature $R(\Om)$ of $\Om_\m{}^{ab}$ is defined by 
$R(\Om)=\th^\m{}_a\th^\n{}_b(\d_{[\m}\Om_{\n]}{}^{ab}
+\Om_{[\m}{}^{ae}\Om_{\n]}{}_{e}{}^{b})$, which comes from 
$$
\u{R}\l(\u{\Om}\r)=R\l(\Om\r)+{1\over4\a^2}G_{ab}G^{ab}.
$$

From the equations of motion (\ref{5Dgaugeeom},\ref{5Dchiphieom}), 
one obtains the bosonic part of 
the action of the abelian gauge multiplet
\begin{eqnarray}
L_B&=&-\hf\int \l[\a\l(F-4\phi^I\,t_I\r)\wedge*\l(F-4\phi^J\,t_J\r)
+C\wedge{F}\wedge{F}\r]
\nn\\
&&+\hf\int dx^5\sqrt{g}\,\a\l[
\cD_a\phi^I\cD^a\phi_I
+{\cal M}_{B}{}_{IJ}\phi^I\phi^J
\r],
\label{5DabelianLB}
\end{eqnarray}
with 
$$
{\cal M}_{B}{}_{IJ}
={1\over5}\,\dl_{IJ}\l(R\l(\Om\r)+{1\over4\a^2}G_{ab}G^{ab}\r)
+{4\over15}M_{IJ}+4t_{I}{}^{ab}t_J{}_{ab}-S_I{}^KS_{JK},
$$
and the fermionic part
\begin{eqnarray}
&&L_F={i\over8}\int dx^5\sqrt{g}\,\a\l(\chi^\a\r)^TC\bigg[
\g^a\cD_a\chi^\b\Om_{\a\b}
-{1\over8\a}G_{ab}\g^{ab}\chi^\b\Om_{\a\b}
\nn\\
&&\hskip6cm
-{1\over4}S_{IJ}\chi^\b\l(\Om\p^{IJ}\r)_{\a\b}
+{1\over2}t^I{}_{ab}\g^{ab}\chi^\b\l(\Om\p_I\r)_{\a\b}
\bigg].
\label{5DabelianLF}
\end{eqnarray}

One can verify that 
the total action $L=L_F+L_B$ is left invariant under the supersymmetry 
transformation (\ref{5DabelianSUSY}). However, it is a lengthy calculation 
to verify the supersymmetry invariance of the action $L$. 
Although we do not intend to pause for a detailed demonstration of it, 
we will discuss a supersymmetry transformaion of the mass term of 
the scalar fields $\phi^I$ in the action in Appendix \ref{SUSYMassPhi},  
which we think is one of the keys to verify the supersymmetry invariance 
of the action.

\section{The Generalization for a Non-Abelian Gauge Group}
\label{NonAbelianTensor}

The reduced theory of the six-dimensional tensor multiplet gives rise to 
the abelian gauge theory in five dimensions. 
We will extend the abelian gauge multiplet 
($A_\m$, $\phi^I$, $\chi^\a$) to the adjoint representation 
of a non-abelian gauge group $G$ and replace the partial derivatives
by covariant ones;
$$
\d_\m \phi^I  \quad\longrightarrow\quad 
\d_\m \phi^I +ig\l[A_\m,\,\phi^I\r],
\qquad
\d_\m \chi^\a  \quad\longrightarrow\quad 
\d_\m \chi^\a +ig\l[A_\m,\,\chi^\a\r]. 
$$
We will henceforth denote the covariant derivatives as 
\begin{eqnarray*}
&&\cD_\m\phi^I=\d_\m\phi^I-b_\m\phi^I-A_\m{}^I{}_J\phi^J+ig\l[A_\m,\,\phi^I\r],
\nn\\
&&
\cD_\m\chi^\a=\d_\m\chi^\a-{3\over2}b_\m\chi^\a
+{1\over4}\Om_{\m}{}^{bc}\g_{bc}\chi^\a
-{1\over4}A_\m{}^{IJ}\l(\p_{IJ}\r){}^\a{}_\b\chi^\b
+ig\l[A_\m,\,\chi^\a\r]. 
\end{eqnarray*}

For the non-abelian extension of 
the supersymmetry transformations (\ref{5DabelianSUSY}) 
and the equations of motion (\ref{5Dgaugeeom},\ref{5Dchiphieom}), 
there are two conditions to be satisfied. 
In the flat limit where all the backgrounds go to zero, 
they should be reduced to the ones
in the $\N=2$ supersymmetric Yang-Mills theory on a flat space, 
and in the abelian limit $g\to0$,  the extension has to go back to 
(\ref{5DabelianSUSY},\ref{5Dgaugeeom},\ref{5Dchiphieom}). 
Our ansatz for the non-abelian extension of the supersymmetry 
tranformations is 
\BBox{410}{115}{
\begin{eqnarray}
&&\dl_\e{A}_\m=-{i\over4}\Om_{\a\b}\l(\e^\a\r)^TC\g_\m\chi^\a,
\qquad
\dl_\e\phi^I={i\over4}\l(\Om\p^I\r){}_{\a\b}\l(\e^\a\r)^TC\chi^\b,
\nn\\
&&\dl_\e\chi^\a=-{1\over2}F_{ab}\g^{ab}\e^\a
-\g^\m\cD_\m\phi^I\l(\p_I\r){}^\a{}_\b\e^\b
+{1\over2\a}G_{ab}\phi^I\l(\p_I\r){}^\a{}_\b\g^{ab}\e^\b
\label{SUSY}\\
&&\qquad\qquad
+S^I{}_J\phi^J\l(\p_I\r)^\a{}_\b\e^\b
+{1\over2}\ve_{IJKLM}S^{IJ}\phi^M\l(\p^{KL}\r){}^\a{}_\b\e^\b
\nn\\
&&\qquad\qquad
+t^I{}_{ab}\phi^J\l(\p_{IJ}\r){}^\a{}_\b\g^{ab}\e^\b
+{i\over2}g\l[\phi^I,\,\phi^J\r]\l(\p_{IJ}\r){}^\a{}_\b\e^\b,
\nn\end{eqnarray}
}
with the field strength of the non-abelian gauge field $A_\m$ 
\begin{eqnarray*}
F_{ab}=\th^\m{}_a\th^\n{}_bF_{\m\n}
=\th^\m{}_a\th^\n{}_b\Big(\d_{\m}A_{\n}-\d_{\n}A_{\m}+ig\l[A_\m,\,A_\n\r]\Big). 
\end{eqnarray*}

In the abelian gauge theory, the algebra of the supersymmetry 
transformaions (\ref{5DabelianSUSY}) is closed on-shell, and 
in the flat limit of the non-abelian gauge theory, it is also closed 
on-shell. Therefore, in order to see the closure of the algebra of 
the supersymmetry transformations (\ref{SUSY}), we make 
an ansatz for the equation of motion of the spinor $\chi^\a$, 
\BBox{400}{55}{
\begin{eqnarray}
&&\g^\m\cD_\m\chi^\a+ig\l(\p_I\r){}^\a{}_\b\l[\phi^I,\,\chi^\b\r]
\nn\\
&&\qquad
-{1\over8\a}G_{ab}\g^{ab}\chi^\a-{1\over4}S_{IJ}(\p^{IJ}){}^\a{}_\b\chi^\b
+{1\over2}t^I{}_{ab}(\p_I){}^\a{}_\b\g^{ab}\chi^\b=0. 
\label{chieom}
\end{eqnarray}
}

The supersymmetry transforms of $\cD_\m\phi^I$ and 
($F_{ab}-2\phi_I\,{t}^I{}_{ab}$) may be useful to see that 
the algebra of the supersymmetry transformations is closed on-shell; 
\begin{eqnarray*}
&&\dl_\e\cD_\m\phi^I
={i\over4}\bigg[
\bar\e\c\p^I\cD_\m\chi
+{1\over8\a}G_{bc}(\bar\e\c\p^I\g_\m{}^{bc}\chi)
-{1\over2\a}G_{\m{b}}(\bar\e\c\p^I\g^{b}\chi)
\nn\\
&&\hskip6cm
-{1\over4}S_{KL}(\bar\e\c\p^{KL}\p^I\g_\m\chi)
-{1\over2}t^J{}_{bc}(\bar\e\c\p_J\p^I\g_\m{}^{bc}\chi)
\bigg],
\nn\\
&&\dl_\e\l(F_{ab}-2\phi_I\,{t}^I{}_{ab}\r)
={i\over4}\bigg[
\bar\e\c\g_{[a}\cD_{b]}\chi
-{1\over4\a}G_{cd}
\bar\e\c\l(\g^{cd}{}_{ab}-3\dl^{c}{}_{[a}\g^{d}{}_{b]}
-4\dl^{c}{}_{a}\dl^{d}{}_{b}\r)\chi 
\nn\\
&&\hskip4.5cm
+{1\over2}S_{IJ}(\bar\e\c\p^{IJ}\g_{ab}\chi)
+t^I{}_{cd}
\bar\e\c\p_I\l(\g^{cd}{}_{ab}-\dl^{c}{}_{[a}\g^{d}{}_{b]}
-2\dl^{c}{}_{a}\dl^{d}{}_{b}\r)\chi 
\bigg].
\end{eqnarray*}

Using the equation of motion (\ref{chieom}) and 
the Killing spinor equation (\ref{KSE}), 
one can verify that the algebra of the supersymmetry transformations (\ref{SUSY}) 
is closed on-shell. 
\begin{eqnarray}
&&\l[\dl_\e,\,\dl_\eta\r]A_\m
={i\over2}\l[
F_{\m\n}\xi^\n
+\cD_\m\l(\phi^I\l(\bar\eta\c\p_I\e\r)\r)
\r]
=-{i\over2}\bigg[
\xi^\n\d_\n{A}_\m+\d_\m\xi^\n\c{A}_\n\bigg]
-\cD_\m\La_G,
\nn\\
&&\l[\dl_\e,\,\dl_\eta\r]\phi^I
=-{i\over2}\bigg(
\xi^\m\d_\m\phi^I-\xi^a{b}_a\phi^I\bigg)
+ig\l[\La_G,\,\phi^I\r]
-\La^I{}_J\phi^J,
\label{onshellSUSYalg}\\
&&\l[\dl_\e,\,\dl_\eta\r]\chi^\a
=-{i\over2}\bigg[
\xi^\m\d_\m\chi^\a-{3\over2}\xi^a{b}_a\chi^\a\bigg]
+ig\l[\La_G,\,\chi^\a\r]
+{1\over4}\La^{ab}\g_{ab}\chi^\a
-{1\over4}\La_{IJ}\l(\p^{IJ}\r){}^\a{}_\b\chi^\b,
\nn
\end{eqnarray}
with the Killing vector $\xi^a=(\bar\eta\c\g^a\e)$, 
where the parameters are given by 
\begin{eqnarray*}
&&\La_{IJ}
=-{i\over2}\bigg[
A_a{}_{IJ}\xi^a
+S_{IJ}(\bar\eta\c\e)
-\ve_{IJKLM}S^{KL}(\bar\eta\c\p^M\e)
\nn\\
&&\hskip6cm
+{1\over2\a}G_{ab}(\bar\eta\c\p_{IJ}\g^{ab}\e)
-t^K{}_{ab}(\bar\eta\c\p_{IJK}\g^{ab}\e)
\bigg],
\nn\\
&&\La^{ab}=-{i\over2}\l[\cD^a\xi^b+\xi^c\Om_c{}^{ab}\r],
\qquad
\La_G
=-{i\over2}\l[\xi^a{A}_a+\phi^I(\bar\eta\c\p_I\e)\r],
\end{eqnarray*}
(See Appendix \ref{SMW} for the abbreviation 
$\l(\bar\eta\c\p_{I_1\cdots{I}_n}\g^{a_1\cdots{a}_m}\e\r)$ ), 
and the covariant derivative of $\phi^I\l(\bar\eta\c\p_I\e\r)$ is
$$
\cD_\m\l(\phi^I\l(\bar\eta\c\p_I\e\r)\r)
=\d_\m\l(\phi^I\l(\bar\eta\c\p_I\e\r)\r)
+ig\l[A_\m,\,\phi^I\l(\bar\eta\c\p_I\e\r)\r]. 
$$

Since we have seen that the supersymmetry transformation (\ref{SUSY}) gives 
an on-shell closed algebra with the equation of motion (\ref{chieom}), 
we will proceed with (\ref{SUSY}) and (\ref{chieom}) to obtain the non-abelian 
extension of the action (\ref{5DabelianLB},\ref{5DabelianLF}).

A simple calculation shows that the equation of motion (\ref{chieom}) 
may be derived from the fermionic part of the non-abelian action
\begin{eqnarray}
&&S_F={i\over8}\int dx^5\sqrt{g}\,\a\,\Tr\bigg[
\bar\chi\c\g^a\cD_a\chi
-{1\over8\a}G_{ab}\bar\chi\c\g^{ab}\chi
-{1\over4}S_{IJ}\bar\chi\c\p^{IJ}\chi
\nn\\
&&\hskip6cm
+{1\over2}t^I{}_{ab}\bar\chi\c\p_I\g^{ab}\chi
+(ig)\bar\chi\c\p_I\l[\phi^I,\,\chi\r]
\bigg],
\label{5DLF}
\end{eqnarray}
where the symbol $\Tr$ denotes a trace in the adjoint representation 
of the gauge group $G$.

In the abelian limit $g\to0$, 
the non-abelian action should go to (\ref{5DabelianLB}), 
-- more precisely, the abelian action of the $|G|$ abelian gauge multiplets 
with $|G|$ denoting the dimension of the adjoint representation of $G$ --, 
and in the flat limit, we must regain the familiar non-abelian action in 
the $\N=2$ supersymmetric Yang-Mills theory. It therefore seems natural 
to take the ansatz 
\begin{eqnarray}
S^{(0)}_B&=&-\hf\int \Tr\Big[\a\l(F-4\phi^I\,t_I\r)\wedge*\l(F-4\phi^J\,t_J\r)
+C\wedge{F}\wedge{F}\Big]
\nn\\
&&
+\hf\int dx^5\sqrt{g}\,\a\, \Tr\Big[
\cD_a\phi^I\cD^a\phi_I
+{\cal M}_{B}{}_{IJ}\phi^I\phi^J
-{1\over2}(ig)^2[\phi^I,\,\phi^J][\phi_I,\,\phi_J]\Big],
\label{5DLBtree}
\end{eqnarray}
where 
\begin{equation}
{\cal M}_{B}{}_{IJ}
={1\over5}\,\dl_{IJ}\l(R\l(\Om\r)+{1\over4\a^2}G_{ab}G^{ab}\r)
+{4\over15}M_{IJ}+4t_{I}{}^{ab}t_J{}_{ab}-S_I{}^KS_{JK}.
\label{onshellscalarmass}
\end{equation}

In order to examine the supersymmetry invariance of the sum $S_F+S^{(0)}_B$, 
one needs to perform a similar calculation to what is done 
for the abelian action $L$. 
The calculation may be painful, especially in the mass term of the scalar 
fields $\phi^i$, of which the details is shown in Appendix \ref{SUSYMassPhi}. 

However, it turns out that the variation of the sum $S_F+S^{(0)}_B$ 
under the supersymmetry transformation (\ref{SUSY}) doesn't vanish 
at the order ${\cal O}(g)$. 
Therefore, in order to obtain a supersymmetric action, 
as discussed in \cite{CJ}, one needs the additional term
\begin{eqnarray}
S^{(1)}_B&=&
-{1\over6}\int dx^5\sqrt{g}\,\a\, \Tr\Big[
(ig)\,\ve^{IJKLM}S_{IJ}\phi_K\l[\phi_L,\,\phi_M\r]
\Big],
\label{5DLB}
\end{eqnarray}
to cancel the supersymmetry variation of $S^{(0)}_B+S_F$. 
Thus, one may see that $S=S_B+S_F=S^{(0)}_B+S^{(1)}_B+S_F$ 
yields a supersymmetric non-abelian action. 


\section{Supersymmetric Backgrounds}
\label{Backgrounds}

In this section, we will discuss the supersymmetric solutions to 
the Killing spinor equation (\ref{KSE}) and the condition (\ref{SUSYchieq}) 
from the spinor variation $\dl\chi^{\a\b}{}_{\g}$, 
which gives rise to supersymmetric backgrounds for the five-dimensional 
supersymmetric Yang-Mills theory. 

In this paper, we will make an assumption 
\BBox{400}{30}{
\begin{equation}
b_\m=0,
\quad
t^{i}{}_{ab}=0, 
\quad
S_{i5}=-S_{5i}=0, 
\quad
A_\m{}^{i5}=-A_\m{}^{5i}=0, 
\quad
(i=1,\cdots,4)
\label{assumption}
\end{equation}}
which is satisfied by the background in the previous papers \cite{FKM,KM}, 
as will be seen below. 
In \cite{FKM,KM}, we have considered the product space 
of a round $S^3$ and a Riemann surface $\RS$. 
In this paper, we are especially interested in supersymmetric backgrounds 
for deformed 3-spheres - a squashed and an ellipsoid $S^3$. 
We will find supersymmetric backgrounds on the product spaces of 
those 3-spheres and $\RS$, which turn out to satisfy 
the assumption (\ref{assumption}). 

It is convenient under the assumption (\ref{assumption}) 
to decompose the supersymmetry parameter $\e^\a$ as
$$
\p^5\e^\a=\e^\a \quad\longrightarrow\quad 
\e^\a=\begin{pmatrix}\e^{\ta}\\ 0 \end{pmatrix}, 
\qquad
\p^5\e^\a=-\e^\a \quad\longrightarrow\quad 
\e^\a=\begin{pmatrix} 0 \\ \ve^{\da}\end{pmatrix},
$$
in the representation with $\p^5={\rm diag.}(+{\bf 1}_2,\,-{\bf 1}_2)$.
While the Killing spinor equation (\ref{KSE}) in a generic background 
gives a differential equation of $\e^{\ta}$ and $\ve^{\da}$ coupled to 
each other, the assumption (\ref{assumption}) splits them into 
\begin{eqnarray}
\cD_\m\e^{\ta}=
\qt{S}_{ij}\l(\s^{ij}\r){}^{\ta}{}_{\tb}\,\g_\m\e^{\tb}
-{1\over2\a}G_{\m\n}\g^\n\e^{\ta}
-{1\over8\a}G_{bc}\g_\m{}^{bc}\e^{\ta}
+\hf\,{t}_{bc}\,\g_\m{}^{bc}\e^{\ta},
\label{tildeN=1KSE}
\\
\cD_\m\ve^{\da}=
\qt{S}_{ij}\l(\bar{\s}^{ij}\r){}^{\da}{}_{\db}\,\g_\m\ve^{\db}
-{1\over2\a}G_{\m\n}\g^\n\ve^{\da}
-{1\over8\a}G_{bc}\g_\m{}^{bc}\ve^{\da}
-\hf\,{t}_{bc}\,\g_\m{}^{bc}\ve^{\da},
\label{N=1KSE}
\end{eqnarray}
with 
$t_{ab}\,{\equiv}\,t^5{}_{ab}$,
where the covariant derivatives are defined by 
\begin{eqnarray*}
\cD_\m\e^{\ta}\equiv\d_\m\e^{\ta}+\hf{b}_\m\e^{\ta}
+\qt\Om_\m{}^{bc}\g_{bc}\e^{\ta}
-\qt{A}_\m{}^{ij}\l(\s_{ij}\r){}^{\ta}{}_{\tb}\e^{\tb},
\nn\\
\cD_\m\ve^{\da}\equiv\d_\m\ve^{\da}+\hf{b}_\m\ve^{\da}
+\qt\Om_\m{}^{bc}\g_{bc}\ve^{\da}
-\qt{A}_\m{}^{ij}\l(\bar{\s}_{ij}\r){}^{\da}{}_{\db}\ve^{\db}.
\end{eqnarray*}

We will further make an ansatz for the Killing spinors, 
\BBox{450}{40}{
\begin{eqnarray}
\ve^{\da=1}=\e\otimes\zeta_{+},
\quad
\ve^{\da=2}=C_3\e^*\otimes\zeta_{-};
\qquad
\e^{\ta=1}=\tilde{\e}\otimes\zeta_{\pm},
\quad
\e^{\ta=2}=C_3{\tilde{\e}}^*\otimes\zeta_{\mp},
\label{KSansatz}
\end{eqnarray}
}
with two-dimensional spinors $\e$, $\tilde{\e}$ on the $S^3$ 
and constant two-dimensional spinors 
$$
\zeta_{\pm}={1\over\sqrt{2}}\begin{pmatrix}1\\ \pm{i}\end{pmatrix}, 
$$
on $\RS$, obeying that $\t_2\zeta_{\pm}=\pm\zeta_{\pm}$, 
with the Pauli matrix $\t_2$.
Note that they satisfy 
\begin{equation}
\g_{45}\,\ve^{\da}=-i(\t_3)^{\da}{}_{\db}\,\ve^{\db}, 
\qquad
\g_{45}\,\e^{\ta}=\mp{i}(\t_3)^{\ta}{}_{\tb}\,\e^{\tb}.
\label{RSchirality}
\end{equation}

For later convenience, let us consider the commutation relation of the 
covariant derivatives acting on $\ve^{\da}$, which by definition gives
\begin{eqnarray*}
\l[\cD_a,\,\cD_b\r]\ve^{\da}
=\qt\,R_{ab}{}^{cd}(\Om)\g_{cd}\ve^{\da}-\qt\,F_{ab}{}^{ij}
\l(\bar{\s}_{ij}\r){}^{\da}{}_{\db}\ve^{\db},
\end{eqnarray*}
and acting $\g^{ab}$ on this, one obtains 
\begin{equation}
\g^{ab}\cD_a\cD_b\,\ve^{\da}
=\hf\,\g^{ab}\l[\cD_a,\cD_b\r]\ve^{\da}
=-\qt\,R\l(\Om\r)\ve^{\da}
-{1\over8}{F}_{ab}{}^{ij}\l(\bar{\s}_{ij}\r)^{\da}{}_{\db}\g^{ab}\ve^{\db}.
\label{CommRelN=1KSE}
\end{equation}
On the other hand, using the Killing spinor equation (\ref{N=1KSE}) twice 
for $\g^{ab}\cD_a\cD_b\,\ve^{\da}$ and equating it and the right hand side 
of (\ref{CommRelN=1KSE}), one finds that 
\begin{eqnarray}
&&
-\qt\,R\l(\Om\r)\ve^{\da}
-{1\over8}{F}_{ab}{}^{ij}\l(\bar{\s}_{ij}\r)^{\da}{}_{\db}\g^{ab}\ve^{\db}
\nn\\
&&=
\l(\bar{\s}^{ij}\r)^{\da}{}_{\db}\,\cD_{a}S_{ij}\g^{a}\ve^{\db}
-\cD_{a}t_{bc}\g^{abc}\ve^{\da}
-3\c\cD^{b}t_{ba}\g^{a}\ve^{\da}
-5\cdot{1\over4\a}\cD^{b}G_{ba}\g^a\ve^{\da}
\nn\\
&&\quad
+{5\over4}\c{}S_{ij}S_{kl}\l(\bar{\s}^{ij}\bar{\s}^{kl}\r)^{\da}{}_{\db}
\ve^{\db}
+\l[4\c\l({1\over4\a}\r)^2G_{ab}G^{ab}-3\c\l(t_{ab}+{1\over4\a}G_{ab}\r)
\l(t^{ab}+{1\over4\a}G^{ab}\r)\r]\ve^{\da}
\nn\\
&&\quad
-{9\over2}\c\l(\bar{\s}^{ij}\r)^{\da}{}_{\db}{1\over4\a}G_{ab}S_{ij}\g^{ab}
\ve^{\db}
-{3\over2}\c\l(\bar{\s}^{ij}\r)^{\da}{}_{\db}t_{ab}S_{ij}\g^{ab}
\ve^{\db}
-8\c{1\over4\a}G_a{}^{c}t_{bc}\g^{ab}\ve^{\da}
\nn\\
&&\quad
-\l[t_{ab}t_{cd}-5\c\l({1\over4\a}\r)^2G_{ab}G_{cd}\r]\g^{abcd}\ve^{\da}.
\label{RFtN=1KSE}
\end{eqnarray}

Decompsing the fields $\chi^\a$, $\phi^I$ 
in the representation with $\p^5={\rm diag.}(+{\bf 1}_2,\,-{\bf 1}_2)$ 
as 
$$
\chi^{\a}\to\begin{pmatrix}\psi^{\ta}\\\la^{\da}\end{pmatrix}, 
\qquad
\phi^I \to (\phi^{{i=1,\cdots,4}},\,\phi^5=\s), 
$$
one can see that the supersymmetry transformation under 
the assumption (\ref{assumption}) becomes 
\begin{eqnarray}
\dl_\e{A}_\m
&=&-{i\over4}\ve_{\ta\tb}\l(\e^{\ta}\r)^TC\g_\m\psi^{\tb}
-{i\over4}\ve_{\da\db}\l(\ve^{\da}\r)^TC\g_\m\la^{\db},
\nn\\
\dl_\e{\s}
&=&{i\over4}\ve_{\ta\tb}\l(\e^{\ta}\r)^TC\psi^{\tb}
-{i\over4}\ve_{\da\db}\l(\ve^{\da}\r)^TC\la^{\db},
\nn\\
\dl_\e{\phi}^i
&=&{i\over4}\l(\ve\s^i\r)_{\ta\db}\l(\e^{\ta}\r)^TC\la^{\db}
+{i\over4}\l(\ve\bar{\s}^i\r)_{\da\tb}\l(\ve^{\da}\r)^TC\psi^{\tb},
\label{N=1SUSY}\\
\dl_\e\psi^{\ta}
&=&-\l(\hf{F}_{ab}\g^{ab}+\g^a\cD_a\s-{1\over2\a}G_{ab}\s\g^{ab}\r)\e^{\ta}
-\bigg(S^{ij}\s
-{i\over2}g\l[\phi^i,\,\phi^j\r]\bigg)\l(\s_{ij}\r){}^{\ta}{}_{\tb}\e^{\tb}
\nn\\
&&-\bigg(\g^a\cD_a\phi^i
-{1\over2\a}G_{ab}\phi^i\g^{ab}
-\l(S^{ij}+\ve^{ijkl}S_{kl}\r)\phi_j
-t_{ab}\phi^i\g^{ab}
-ig\l[\s,\,\phi^i\r]\bigg)
\l(\s_i\r)^{\ta}{}_{\db}\ve^{\db},
\nn\\
\dl_\e\la^{\da}
&=&-\l(\hf{F}_{ab}\g^{ab}-\g^a\cD_a\s+{1\over2\a}G_{ab}\s\g^{ab}\r)\ve^{\da}
+\bigg(S^{ij}\s
+{i\over2}g\l[\phi^i,\,\phi^j\r]\bigg)\l(\bar{\s}_{ij}\r){}^{\da}{}_{\db}\ve^{\db}
\nn\\
&&-\bigg(\g^a\cD_a\phi^i
-{1\over2\a}G_{ab}\phi^i\g^{ab}
-\l(S^{ij}-\ve^{ijkl}S_{kl}\r)\phi_j
+t_{ab}\phi^i\g^{ab}
+ig\l[\s,\,\phi^i\r]\bigg)
\l(\bar{\s}_i\r)^{\da}{}_{\tb}\e^{\tb}. 
\nn
\end{eqnarray}

The equations of motion of the spinors $\psi^{\ta}$, $\la^{\da}$ under the 
assumption (\ref{assumption}) give 
\begin{eqnarray*}
\g^\m\cD_\m\psi^{\ta}+ig\l[\s,\,\psi^{\ta}\r]
+ig\l(\s_i\r){}^{\ta}{}_{\db}\l[\phi^i,\,\la^{\db}\r]
={1\over8\a}G_{ab}\g^{ab}\psi^{\ta}
+\qt{S}_{ij}\l(\s^{ij}\r){}^{\ta}{}_{\tb}\psi^{\tb}
-\hf\,t_{ab}\g^{ab}\psi^{\ta},
\nn\\
\g^\m\cD_\m\la^{\da}-ig\l[\s,\,\la^{\da}\r]
+ig\l(\bar{\s}_i\r){}^{\da}{}_{\tb}\l[\phi^i,\,\psi^{\tb}\r]
={1\over8\a}G_{ab}\g^{ab}\la^{\da}
+\qt{S}_{ij}\l(\bar{\s}^{ij}\r){}^{\da}{}_{\db}\la^{\db}
+\hf\,t_{ab}\g^{ab}\la^{\da},
\end{eqnarray*}
with the covariant derivatives 
\begin{eqnarray*}
\cD_\m\psi^{\ta}&=&\d_\m\psi^{\ta}-{3\over2}b_\m\psi^{\ta}
+\qt\Om_\m{}^{ab}\g_{ab}\psi^{\ta}
-\qt{A}_\m{}^{ij}\l(\s_{ij}\r)^{\ta}{}_{\tb}\psi^{\tb}
+ig\l[A_\m,\,\psi^{\ta}\r],
\nn\\
\cD_\m\la^{\da}&=&\d_\m\la^{\da}-{3\over2}b_\m\la^{\da}
+\qt\Om_\m{}^{ab}\g_{ab}\la^{\da}
-\qt{A}_\m{}^{ij}\l(\bar{\s}_{ij}\r)^{\da}{}_{\db}\la^{\db}
+ig\l[A_\m,\,\la^{\da}\r].
\end{eqnarray*}

\subsection{The $\N=1$ SUSY Background in the Previous Paper}
\label{oldR3Sbg}

We start with the background in the previous paper \cite{FKM,KM}, 
where the compactification on the product space of a unit round $S^3$ 
and a Riemann surface, $S^3\times\Sigma$ was considered, 
and we will reinterpret it as a supersymmetric background 
in terms of $A_\m{}^i{}_j$, $S_{ij}$, $G_{ab}$, $t{}_{ab}\equiv{t}^5{}_{ab}$.  
See Appendix \ref{oldbackground} for the differences of the old notations used 
in \cite{FKM} from the ones in this paper. 

The background in \cite{FKM} can be read in the notations of this paper as 
\begin{equation}
t_{45}={1\over4r}, 
\qquad
S_{12}=S_{34}={1\over2r},
\qquad
{1\over4\a}G_{45}=-{1\over{4r}},
\label{oldbg}
\end{equation} 
in the Lorentz frame $t_{ab}$, $G_{ab}$, 
where we have replaced the unit radius of the $S^3$ by $r$.

On the Riemann surface $\Sigma$ with local coordinates ($x^4$, $x^5$), 
the twisting is required to preserve supersymmetries by turning on 
the background gauge field $A^i{}_j$ as 
\begin{equation}
A^{12}=A^{34}=-\hf\om^{45}, 
\label{oldRgauge}
\end{equation}
with the spin connection $\om^{45}$ on the surface $\Sigma$. 
This together with $S_{12}=S_{34}$ break the $Spin(5)_R$ $R$-symmetry 
group to $SU(2)_l\times{U(1)}_r\subset{SU(2)_l}\times{SU(2)}_r$, 
when regarding the subgroup $Spin(4)$ of the $Spin(5)_R$ as 
$SU(2)_l\times{SU(2)}_r$. We refer to it as the $\N=1$ twisting, 
following \cite{Sicilian}. 

The supersymmetry condition (\ref{SUSYchieq}) determines the background 
$M_{IJ}$
$$
{4\over15}M_{55}={1\over5}\l[{1\over{r}^2}-R(\Sigma)\r],
\qquad
{4\over15}M_{ij}=-{1\over20}\l[{1\over{r}^2}-R(\Sigma)\r]\dl_{ij},
\quad
(i,j=1,\cdots,4)
$$
where the scalar curvature $R(\RS)$ is derived 
from the spin connection $\om^{45}$,
$$
\hf\,R(\RS)\,e^4\wedge{e}^5=d\om^{45},
$$
and substituting these into (\ref{onshellscalarmass}) gives\footnote{
In the previous paper \cite{FKM} (v3 on the arXiv), 
the scalar curvature $R(\RS)$ was dropped from 
the mass terms ${\cal M}_{B\,ij}$ of the $\N=1$ hypermultiplet scalars.
} 
$$
{\cal M}_{B\,55}={2\over{r}^2}, 
\qquad
{\cal M}_{B\,ij}=\qt\l[R(\RS)+{4\over{r}^2}\r]\dl_{ij}, 
\quad
(i,j=1,\cdots 4).
$$

The Killing spinor equation (\ref{N=1KSE}) in the background (\ref{oldbg}) 
is identical to the one in \cite{FKM},  
$$
\cD_\m\ve^{\da}=-{1\over2r}\g_\m{}^{45}\ve^{\da}, 
$$
with the ansatz (\ref{KSansatz}).

The scalar curvature $R(\Om)$ on the $S^3\times\RS$ is given by 
$$
R(\Om)=R(S^3)+R(\RS)={6\over{r}^2}+R(\RS), 
$$
for the round $S^3$ of radius $r$. 
Since the gauge field $A^i{}_j$ is minus the half of 
the spin connection $\om^{45}$ on the surface $\RS$, the field strength of 
$A^{ij}$ results in 
$$
F_{45}{}^{12}=F_{45}{}^{34}
=-\qt{R}(\RS).
$$

The equation (\ref{RFtN=1KSE}) identically holds for 
the curvatures and the background fields, and  
it is consistent with the existence of the Killing spinor $\ve^{\da}$. 
In fact, as explained in \cite{FKM,KM} and in Appendix \ref{3-sphere}, 
the Killing spinor is given by
$$
\ve^{\da=1}=\e_0\otimes\zeta_{+}, 
\qquad
\ve^{\da=2}=C_{3}^{-1}{\e_0}^*\otimes\zeta_{-}, 
$$
with $\e_0$ a constant spinor on the $S^3$, which is consistent with our ansatz 
(\ref{KSansatz}).

For the other supersymmetry paramter $\e^{\ta}$, the Killing spinor 
equation (\ref{tildeN=1KSE}) in the same background gives 
$$
\cD_a\e^{\ta}=-{1\over2\a}G_{ab}\g^b\e^{\ta}+{1\over2r}\g_a{}^{45}\e^{\ta}.
$$
Note that $S^{12}=S^{34}$ obeys $S^{ij}\s_{ij}=0$. 
With $A^{12}=A^{34}$, we have $A^{ij}\s_{ij}=0$, and the twisting 
of the background $A^{ij}$ have no effects inside the covariant derivative 
$\cD_a\e^{\ta}$. 
In a generic Riemann surface $\RS$, we don't have a solution to the 
above Killing spinor equation. 
In fact, the calculation of $\g^{ab}\cD_a\cD_b\e^{\ta}$ shows that 
the scalar curvature $R(\RS)$ is an obstacle to the existence of 
a Killing spinor for $\e^{\ta}$. 

We can see from (\ref{oldbg}), (\ref{oldRgauge}) that the background 
breaks the $Spin(5)_R$ group of the $R$-symmetry into 
$SU(2)_l\times{U(1)}_r$, which is a subgroup of 
$SU(2)_l\times{SU(2)}_r\simeq{Spin(4)}_R\subset{Spin(5)}_R$. 
The symmetry breaking is caused by the twisting $A^{12}=A^{34}$ 
(and also $S^{12}=S^{34}$). As we have seen just above, 
the twisting only retains the half of the supersymmetries. 
Therefore, it is consistent with the fact that the $SU(2)_l$ symmetry 
doesn't give rise to the $SU(2)_R$ $R$-symmetry in four-dimensional 
$\N=2$ supersymmetric theories \cite{WKB,Sicilian}.

The background (\ref{oldbg}) is not a unique solution\footnote{It has been pointed out in \cite{ImamuraMatsuno} in the context of five-dimensional 
$\N=1$ supersymmetric theories.} 
to yield an $\N=1$ supersymmetric background on the round $S^3$. 
Even under the ansatz
\begin{eqnarray*}
S_{12}=S_{34}=\hf{S}, 
\end{eqnarray*}
with only non-zero components $G_{45}$ and $t_{45}$, 
there exists a Killing spinor for $\ve^{\da}$, if 
\begin{eqnarray*}
\hf{S}+2\c{1\over4\a}G_{45}=0,
\qquad
\hf{S}+{1\over4\a}G_{45}+t_{45}={1\over2r},
\end{eqnarray*}
which can be read from the Killing spinor equation (\ref{N=1KSE}). 
They may therefore be parametrized by $S$;
\begin{eqnarray*}
{1\over4\a}G_{45}=-{1\over4}S, 
\qquad
t_{45}=-{1\over4}S+{1\over2r}. 
\end{eqnarray*}
The other supersymmetric condition (\ref{SUSYchieq}) gives one more 
constraint - the backgrounds are constant on $\RS$, 
$$
\cD_4S=\cD_5S=0,
$$
and determines the remaining background $M_{IJ}$, 
\begin{eqnarray*}
{4\over15}\,M_{55}=-{1\over5}R(\RS)+{4\over5r^2}-{3\over5}S^2,
\qquad
{4\over15}\,M_{ij}=\l[{1\over20}R(\RS)-{1\over5r^2}+{3\over20}S^2\r]\dl_{ij},
\end{eqnarray*}
for $i,j=1,\cdots,4$. The scalar mass parameters ${\cal M}_{B\,IJ}$ are given 
by 
\begin{eqnarray*}
{\cal M}_{B\,55}={2\over{r}}\l({2\over{r}}-S\r),
\qquad
{\cal M}_{B\,11}=\cdots={\cal M}_{B\,44}=\qt{R}(\RS)+{1\over{r}^2}.
\end{eqnarray*}
When $S=1/r$, it certainly retains the mass term of the scalar $\s$ 
in the previous papers \cite{FKM,KM}.

\subsection{${\cal N}=2$ SUSY Backgrounds on the Round $S^3\times\RS$}
\label{newR3Sbg}

While the background in \cite{FKM,KM} preserves half of the supersymmetries, 
we will find a new supersymmetric background preserving both of 
$\ve^{\da}$ and $\e^{\ta}$ on the $S^3\times\RS$. 


Taking the breaking of the $R$-symmetry group $Spin(5)_R$ into account, 
we will turn on $A^{12}$ and $S^{12}=S$ only, 
and it would break the $Spin(5)_R$ group down to $U(1)_R\times{SU(2)}_R$. 
We could instead turn on $A^{34}$ or $S^{34}$ only, but it is just 
a matter of convention. We refer to this partial twisting as the $\N=2$ 
twisting. 

Since we have the covariant derivatives with the ansatz (\ref{KSansatz}),
\begin{eqnarray*}
\cD_\m\e^{\ta}
&=&\d_\m\e^{\ta}-{i\over2}\l(A_\m{}^{12}\pm\om_\m{}^{45}\r)
\l(\t_3\r)^{\ta}{}_{\tb}\e^{\tb},
\quad
\cD_\m\ve^{\da}
=\d_\m\ve^{\da}-{i\over2}\l(A_\m{}^{12}+\om_\m{}^{45}\r)
\l(\t_3\r)^{\da}{}_{\db}\ve^{\db},
\end{eqnarray*}
in order to cancel the spin connection $\om^{45}$ by $A_{12}$ in both of the 
covariant derivatives, 
the chirality of $\e^{\ta}$ on the surface $\RS$ should be the same as 
the one of $\ve^{\da}$; $i\g_{45}\ve^{\ta}=\l(\t_3\r)^{\ta}{}_{\tb}\e^{\tb}$. 
Therefore, the twisting 
$$
A^{12}=-\om^{45}
$$
works for both of $\ve^{\da}$ and $\e^{\ta}$. 
When we turn on the components $G_{45}$ and $t_{45}$ only, 
the Killing spinor equations (\ref{tildeN=1KSE}), (\ref{N=1KSE}) become 
\begin{eqnarray*}
\cD_a\e^{\ta}&=&{i\over2}S\l(\t_3\r)^{\ta}{}_{\tb}\g_a\e^{\tb}
-{1\over2\a}G_{ab}\g^b\e^{\ta}
-\l({1\over4\a}G_{45}-t_{45}\r)\g_a{}^{45}\e^{\ta},
\nn\\
\cD_a\ve^{\da}&=&{i\over2}S\l(\t_3\r)^{\da}{}_{\db}\g_a\ve^{\db}
-{1\over2\a}G_{ab}\g^b\ve^{\da}
-\l({1\over4\a}G_{45}+t_{45}\r)\g_a{}^{45}\ve^{\da}.
\end{eqnarray*}
For $a=4,5$, the Killing spinor equation is satisfied with $\e^{\ta}$ 
and $\ve^{\da}$ constant on $\RS$, if 
$$
S+4\c{1\over4\a}G_{45}=0. 
$$

With the ansatz (\ref{KSansatz}), the Killing spinors on the round $S^3$ 
(See Appendix \ref{KSroundS}) are lifted to 
\begin{eqnarray*}
\cD_a\ve^{\da}=-{1\over2r}\g_a{}^{45}\ve^{\da},
\qquad
\cD_a\e^{\ta}=\mp{1\over2r}\g_a{}^{45}\e^{\ta},
\end{eqnarray*}
and the comparison of this with the above Killing spinor equations for 
$a=1,2,3$ leads to 
\begin{eqnarray*}
\hf{S}+{1\over4\a}G_{45}+t_{45}={1\over2r},
\qquad
\hf{S}+{1\over4\a}G_{45}-t_{45}=\pm{1\over2r},
\end{eqnarray*}
Depending upon the sign, there are two solutions;
$$
S=-4\c{1\over4\a}G_{45}={2\over{r}}, \quad {1\over4\a}G_{45}=-{1\over{2r}}, 
\qquad t_{45}=0. 
$$
and
$$
S={1\over4\a}G_{45}=0, \qquad t_{45}={1\over2r}. 
$$
We will call the former background type B and the latter type A, 
respectively. 

Let us begin with the type A background;
$$
S_{ij}=0,
\qquad
{1\over4\a}G_{ab}=0,
\quad
t_{45}={1\over2r}.
$$

In the background, since 
the Killing spinor equation (\ref{KSE}) is reduced into 
\begin{eqnarray*}
\cD_\m\e^{\ta}={1\over2r}\g_\m{}^{45}\e^{\ta},
\qquad 
\cD_\m\ve^{\da}=-{1\over2r}\g_\m{}^{45}\ve^{\da},
\end{eqnarray*}
one obtains the solution to them,
\begin{eqnarray*}
\l(\e^{\ta=1},\e^{\ta=2}\r)
=\l(U^{-1}\tilde{\e}_0\otimes\zeta_{+},\,
C_3^{-1}U^{T}\tilde{\e}^*_0\otimes\zeta_{-}\r),
\qquad
\l(\ve^{\da=1},\ve^{\da=2}\r)
=\l(\e_0\otimes\zeta_{+},\, C_3^{-1}\e^*_0\otimes\zeta_{-}\r), 
\end{eqnarray*}
with $\e_0$ and $\tilde{\e}_0$ constant spinors 
and $U$ the mapping of the 3-sphere to 
the $SU(2)$ group given in Appendix \ref{3-sphere} and with $C_3$ 
the three-dimensional charge conjugation matrix explained 
in Appendix \ref{LorentzGamma}.

The supersymmetry condition (\ref{SUSYchieq}) determines the background 
$M_{IJ}$;
\begin{eqnarray*}
&&
{4\over15}M_{55}={4\over5}{1\over{r}^2}-{1\over5}R(\RS),
\nn\\
&&{4\over15}M_{11}={4\over15}M_{22}
=-{1\over5}{1\over{r}^2}+{3\over10}R(\RS),
\quad
{4\over15}M_{33}={4\over15}M_{44}
=-{1\over5}{1\over{r}^2}-{1\over5}R(\RS),
\end{eqnarray*}
which gives rise to the masses ${\cal M}_{B\,IJ}$ of the scalar fields $\phi^I$,
$$
{\cal M}_{B\,55}={4\over{r}^2},
\quad
{\cal M}_{B\,11}={\cal M}_{B\,22}
={1\over{r}^2}+{1\over2}R(\RS),
\quad
{\cal M}_{B\,33}={\cal M}_{B\,44}
={1\over{r}^2}. 
$$

Turning on the field $t_{45}=t^{I=5}_{45}$ breaks the 
$Spin(5)_R$ symmetry group into $Spin(4)_R$ and with the twisting 
by $A^{12}=-\om^{45}$ into $U(1)\times{U(1)}$. Thus, the background 
doesn't repsect the $R$-symmetry of the four-dimensional $\N=2$ 
conformal algebra, but it retains the $\N=2$ supersymmetry. 

Let us move on to the type B background;
$$
S={2\over{r}}, \qquad {1\over4\a}G_{45}=-{1\over{2r}}, 
\qquad t_{45}=0.
$$
It gives rise to 
the Killing spinor equation
\begin{eqnarray*}
\cD_\m\e^{\ta}=-{1\over2r}\g_\m{}^{45}\e^{\ta},
\qquad 
\cD_\m\ve^{\da}=-{1\over2r}\g_\m{}^{45}\ve^{\da},
\end{eqnarray*}
and one gives the same constant solution 
for the both $\e^{\ta}$ and $\ve^{\da}$;
\begin{eqnarray*}
\l(\e^{\ta=1},\e^{\ta=2}\r)
=\l(\tilde{\e}_0\otimes\zeta_{+},\,\tilde{\e}^*_0\otimes\zeta_{-}\r),
\qquad
\l(\ve^{\da=1},\ve^{\da=2}\r)
=\l(\e_0\otimes\zeta_{+},\, C_3^{-1}\e^*_0\otimes\zeta_{-}\r), 
\end{eqnarray*}
with $\e_0$ and $\tilde{\e}_0$ constant spinors as above. 

The supersymmetric condition (\ref{SUSYchieq}) is obeyed by the background, 
if the background fields $M_{IJ}$ satisfy  
\begin{eqnarray*}
{4\over15}M_{55}={4\over15}M_{33}={4\over15}M_{44}
=-{8\over5}{1\over{r}^2}-{1\over5}R(\RS),
\qquad
{4\over15}M_{11}={4\over15}M_{22}
={12\over5}{1\over{r}^2}+{3\over10}R(\RS),
\end{eqnarray*}
which surely respects the $R$-symmetry group $U(1)_R\times{SU(2)}_R$. 
The scalars $\s$, $\phi^3$, $\phi^4$ remain massless, while 
the remaining $\phi^1$, $\phi^2$, are lifted by a half of the scalar curvature 
$R(\RS)$;
$$
{\cal M}_{B\,55}={\cal M}_{B\,33}={\cal M}_{B\,44}=0,
\qquad
{\cal M}_{B\,11}={\cal M}_{B\,22}=\hf\,{R}(\RS).
$$
Thus, they respect the remaining $R$-symmetry group $U(1)_R\times{SU(2)}_R$. 

Turning on either of $A^{12}$ or $A^{34}$ without $t_{45}\not=0$ 
breaks the $R$-symmetry group 
$SO(5)_R$ into $SO(2)\times{SO(3)}$$\simeq{U(1)_R}\times{SU(2)}_R$, 
which can be identified with the $R$-symmetry group of the ${\cal N}=2$ 
superconformal group, if the theory flows into an infrared fixed point. 
On the other hand, as in the previous papers 
\cite{FKM,KM}, turning on both $A^{12}$ and $A^{34}$ such that 
$A^{12}=A^{34}$, the $SO(5)_R$ group is broken to 
${SU(2)}_l\times{U(1)_r}$, which is the subgroup of 
${SU(2)_l}\times{SU(2)}_{r}\simeq{SO(4)}\subset{SO(5)}_R$. 
The subgroup ${SU(2)}_l$ cannot be identified to the $R$-symmetry group 
$SU(2)_R$, because the above results shows that such a background preserves 
only half of the supersymmetries\footnote{We thank Yuji Tachikawa for 
clarification on this point.}. This is consistent with the result 
in \cite{WKB,Sicilian}.

\subsection{A Squashed 3-Sphere with constant Killing spinors}
\label{HosomichiSquash}

A squashed 3-sphere is a deformation of a round $S^3$, 
and regarding it as a circle fibration over a round 2-sphere, 
{\it i.e.}, the Hopf fibration, the radius of the fiber differs 
from the radius of the base. See Appendix \ref{3-sphere} for more details. 
In \cite{Hosomichi}, three-dimensional supersymmetric field theories on 
the squashed 3-sphere has beed discussed, and we will make use of 
their construction for the five-dimensional theory.

The constant solution $\ve^{\da}$ on the round $S^3$; 
$
\l(\ve^1,\ve^2\r)=\l(\e_0\otimes\zeta_{+},\, C_3^{-1}\e^*_0\otimes\zeta_{-}\r)
$
with 
\begin{eqnarray}
\e_0=\begin{pmatrix}1\\0\end{pmatrix}, 
\qquad
C_3^{-1}\e^*_0=\begin{pmatrix}0\\1\end{pmatrix}, 
\qquad
\g^3\ve^{\da}=\ve^{\da}.
\label{constsquashKS}
\end{eqnarray}
solves the differential equation
\begin{eqnarray}
\l(d+\qt\om^{ab}\g_{ab}\r)\ve^{\da}
-{i\over\tr}\l(1-{\tr^2\over{r}^2}\r)(\t_3)^{\da}{}_{\db}e_3\ve^{\db}
=-\hf{\tr\over{r}^2}e^a\g_a{}^{45}\ve^{\da},
\label{HosomichiSqKSE}
\end{eqnarray}
where $\om^{ab}$ is the spin connection of the squashed $S^3$ with the 
fiber radius $\tr$ and the base radius $r$. 
See Appendix \ref{3-sphere} for the squashed $S^3$. 

We will begin with the $\N=2$ twisting by turning on $A^{12}$ only. 
A comparison of (\ref{HosomichiSqKSE}) with the Killing spinor equation 
(\ref{N=1KSE}) suggests that 
\begin{eqnarray}
A^{12}={2\over\tr}\l(1-{\tr^2\over{r}^2}\r)e^3-\om^{45}, 
\qquad
{1\over4\a}G_{45}=-\hf{\tr\over{r}^2},
\qquad
S_{12}={2\tr\over{r}^2}.
\label{N=2Sqbg}
\end{eqnarray}

For the other supersymmetry parameter $\e^{\ta}$, it is easy to 
find a Killing spinor on the squashed $S^3$, if we make the same ansatz 
as for $\ve^{\da}$; it is a constant spinor 
$
\l(\e^1,\e^2\r)
=\l(\tilde{\e}_0\otimes\zeta_{+},\, C_3^{-1}\tilde{\e}^*_0\otimes\zeta_{-}\r) 
$
obeying $\tilde{\e}_0=(1,0)^T$. One then see that it obeys the same 
differential equation (\ref{HosomichiSqKSE}), and thus the background 
(\ref{N=2Sqbg}) preserves the both Killing spinors $\ve^{\da}$, $\e^{\ta}$. 

The other supersymmetry condition (\ref{SUSYchieq}) determines the background 
fields $M_{IJ}$,
$$
{4\over15}M_{11}={4\over15}M_{22}={3\over10}R(\RS)+{12\over5}{1\over{r}^2},
\qquad
{4\over15}M_{33}={4\over15}M_{44}={4\over15}M_{55}
=-{1\over5}R(\RS)-{8\over5}{1\over{r}^2},
$$
and plugging them into (\ref{onshellscalarmass}), one obtains the scalar masses 
${\cal M}_{B\,IJ}$,
$$
{\cal M}_{B\,11}={\cal M}_{B\,22}=\hf{R}(\RS)
+{4\over{r}^2}\l(1-{\tr^2\over{r}^2}\r),
\qquad
{\cal M}_{B\,33}={\cal M}_{B\,44}={\cal M}_{B\,55}=0.
$$

Let us proceed to the $\N=1$ twisting so that 
we will turn on the gauge field $A^{ij}$ of only one $SU(2)$ 
subgroup of the $Spin(5)_R$ group by requiring that $A^{12}=A^{34}$, 
and then a comparison with the Killing spinor equation (\ref{KSE}) 
identifies the background $R$-symmetry gauge field
$$
A^{12}=A^{34}={1\over\tr}\l(1-{\tr^2\over{r}^2}\r)e^3-\hf\om^{45}, 
$$
and for the other background fields, taking account of (\ref{SUSYchieq}), 
one finds that 
\begin{eqnarray}
&&
S_{12}=S_{34}=\hf{S},
\qquad
{1\over4\a}G_{45}=-\qt{S},
\qquad
t_{45}=-\qt{S}+\hf{\tr\over{r}^2},
\label{constsquashbg}\\
&&{4\over15}M_{55}={4\over5}{\tr^2\over{r}^4}-{1\over5}R(\RS)
-{3\over5}S^2
-{8\over5}{1\over{r}^2}\l(1-{\tr^2\over{r}^2}\r),
\nn\\
&&{4\over15}M_{ij}=\l[-{1\over5}{\tr^2\over{r}^4}+{1\over20}R(\RS)
+{3\over20}S^2
+{2\over5}{1\over{r}^2}\l(1-{\tr^2\over{r}^2}\r)\r]\dl_{ij},
\nn
\end{eqnarray}
for $i,j=1,\cdots,4$. In the limit $\tr\to{r}$, one regains the $\N=1$ 
supersymmetric background on the round $S^3$ in the previous subsection. 
It follows from (\ref{onshellscalarmass}) that 
$$
{\cal M}_{B\,55}={2\tr\over{r}^2}\l({2\tr\over{r}^2}-S\r), 
\qquad
{\cal M}_{B\,11}={\cal M}_{B\,22}={\cal M}_{B\,33}={\cal M}_{B\,44}=
\qt{R}(\RS)+{\tr^2\over{r}^4}+{2\over{r}^2}\l(1-{\tr^2\over{r}^2}\r). 
$$

\subsection{A Squashed 3-Sphere with non-constant Killing Spinors}
\label{ImamuraSquash}

Upon the Kaluza-Klein compactification on the time circle 
to the round $S^3\times\RS$, the periodic boundary condition $t\to{t+2\pi}$ 
was assumed in the previous papers \cite{FKM,KM}. The partition function 
is supposed to give the index of the six-dimensional theory. 
Let us generalize this by considering a slant boundary condition
$$
t \quad\sim\quad t+2\pi,
\qquad
\psi \quad\sim\quad \psi+{4u\pi\over{r\a}},
$$
where $\psi$ is the fiber coodinate in the Hopf fibration of the 
3-sphere. See Appendix \ref{3-sphere} for more details. 

It has been explained in \cite{Imamura} that 
the Kaluza-Klein reduction along this slant circle gives rise 
to a squashed $S^3$. Changing the local coodinates ($t,\,\psi$) into 
($\tilde{t},\,\tilde{\psi}$) by
\begin{eqnarray*}
\tilde{t}=t \quad\sim\quad \tilde{t}+2\pi,
\qquad
\tilde{\psi}=\psi-{2u\over{r\a}}t \quad\sim\quad \tilde{\psi},
\end{eqnarray*}
the ordinary reduction in the $\tilde{t}$ direction will be carried out. 
Then, the mapping $U(\psi,\th,\phi)$ in (\ref{mappingU})
from the 3-sphere to the $SU(2)$ group is given 
in terms of the new coordinates by 
$$
U(\psi,\th,\phi)=e^{{i\over2}\phi\t_3}e^{{i\over2}\th\t_3}e^{{i\over2}\psi\t_3}
=U(\tilde{\psi},\th,\phi)e^{i{u\tilde{t}\over{r\a}}\t_3},
$$
and the vielbein $\tilde{\m}^{(0)}$ in the new coordinates of the 3-sphere, 
$$
\tilde{\m}^{(0)}
=\l({1\over{i}}\r)U^{-1}(\tilde{\psi},\th,\phi)dU(\tilde{\psi},\th,\phi)
$$
is related to ${\m}^{(0)}$ in the original coordinates by
\begin{eqnarray*}
\tilde{\m}^{(0)}_{1}=\cos\l({2u\over{r}\a}t\r){\m}^{(0)}_{1}
+\sin\l({2u\over{r}\a}t\r){\m}^{(0)}_{2},
~~
\tilde{\m}^{(0)}_{2}=\cos\l({2u\over{r}\a}t\r){\m}^{(0)}_{2}
-\sin\l({2u\over{r}\a}t\r){\m}^{(0)}_{1},
~~
\tilde{\m}^{(0)}_{3}={\m}^{(0)}_{3}
-{u\over{r}\a}dt.
\qquad
\end{eqnarray*}
Under the change of coordinates, in the six-dimensional metric 
\begin{eqnarray*}
ds_6^2=ds_{\RS}^2
+r^2\l[\l(\m^{(0)}_1\r)^2+\l(\m^{(0)}_2\r)^2+\l(\m^{(0)}_3\r)^2\r]
-{1\over\a^2}dt^2,
\end{eqnarray*}
in addition to the trivial change of the base part in the Hopf fibration
$$
r^2\l[\l(\m^{(0)}_1\r)^2+\l(\m^{(0)}_2\r)^2\r]
=r^2\l[\l(\tilde{\m}^{(0)}_1\r)^2+\l(\tilde{\m}^{(0)}_2\r)^2\r],
$$
the last two terms are changed into
\begin{eqnarray*}
&&-{1\over\a^2}dt^2+r^2\l(\m_3^{(0)}\r)^2
~\rightarrow~
-{1-u^2\over\a^2}\l(d\tilde{t}-\a{ru\over1-u^2}\tilde{\m}^{(0)}_3\r)^2
+{r^2\over1-u^2}\l(\tilde{\m}_3^{(0)}\r)^2
=-{1\over\tilde{\a}^2}\l(d\tilde{t}+{C}\r)^2+e_3^2,
\end{eqnarray*}
where 
\begin{eqnarray*}
&&\tilde{\a}={1\over\sqrt{1-u^2}}\a,
\qquad
{C}=-\a{ru\over1-u^2}\tilde{\m}^{(0)}_3=-\tilde{\a}ue_3,
\qquad
e_3={r\over\sqrt{1-u^2}}\tilde{\m}^{(0)}_3
=\tr\l({d\tilde{\psi}+\cos\th{d\phi}\over2}\r),
\end{eqnarray*}
with 
$
\tr={r/\sqrt{1-u^2}}.
$
Therefore, the slant boundary condition turns on the graviphoton field $C$ 
and deforms the radius of the circle fiber, which results in a squashed 
3-sphere.

Upon the reduction to five dimensions, one has the metric 
\begin{eqnarray*}
ds_5^2=ds^2_{\RS}+\l(e_1^2+e_2^2+e_3^2\r), 
\qquad e_1=r\tilde{\m}^{(0)}_1,\quad e_2=r\tilde{\m}^{(0)}_2,
\end{eqnarray*}
and the field strength of the graviphoton,  
$$
{1\over\tilde{\a}}G={1\over\tilde{\a}}dC=-{2\over{r}}{u\over\sqrt{1-u^2}}
\,{e}_1\wedge{e}_2.
$$
Note that $\tilde{\a}$ is the radius of the circle in the $\tilde{t}$ direction, 
while $\tilde{r}$ is the radius of the fiber in the Hopf fibration of the 
squashed $S^3$.

If we started with a non-vanishing graviphoton field $C$ in 
the round $S^3\times\RS$, we wouldn't gain a simple squashed 3-sphere. 
Therefore, let us consider the supersymmetric backgrounds 
in subsection \ref{newR3Sbg},
where we have $C=0$ on a round $S^3\times\RS$. 
One can now see that the above change of coordinates leads the backgrounds 
in subsection \ref{newR3Sbg} to supersymmetric backgrounds on 
the squashed $S^3\times\RS$. We will begin with the $\N=1$ supersymmetric 
background in \ref{newR3Sbg} by turning on 
$$
A^{12}=A^{34}=-\hf\om^{45},
$$
breaking the $Spin(5)_R$ symmetry down to $SU(2)_l\times{U}(1)_r$.
Besides the $R$-symmetry gauge field $A^i{}_j$, the only auxiliuary field 
${t}_{45}$ is turned on in subsection \ref{newR3Sbg}. 

Returning to six dimensions, the sechsbein
$$
\l(e^1, e^2, e^3, e^0={1\over\tilde{\a}}\l(d\tilde{t}+C\r)\r)
$$
is related to the sechsbein ($\m^1$, $\m^2$, $\m^3$, $\m^0=(1/\a)dt$), 
where $\m_a=r\m^{(0)}_a$ ($a=1,2,3$), by a local Lorentz transformation, 
\begin{eqnarray}
e^a=\sum_{b=0}^{3}\La^{a}{}_{b}\,\m^{b}, \quad (a=0,\cdots,3), 
\label{LorentzSquashed}
\end{eqnarray}
where 
$$
\l(\La^{a}{}_{b}\r)
=\left(\begin{array}{cccc}
\cos(2ut/r\a)&\sin(2ut/r\a)& 0 & 0
\\
-\sin(2ut/r\a)&\cos(2ut/r\a)& 0 & 0
\\
0 & 0 & \cosh\xi & -\sinh\xi 
\\
0 & 0 & -\sinh\xi & \cosh\xi
\end{array}\right),
$$
with $\cosh\xi=1/\sqrt{1-u^2}$ and $\sinh\xi=u/\sqrt{1-u^2}$. 

From the six-dimensional view point, the background $t_{45}$ may be regarded as 
$$
\u{T}^{\a\b}{}_{045}=-\u{T}^{\a\b}{}_{123}=t_{45}\l(\p^5\Om^{-1}\r)^{\a\b}.
$$
Recall that $\u{T}^{\a\b}{}_{\u{abc}}$ is anti-self under the Hodge duality. 
The field $\u{T}^{\a\b}{}_{\u{abc}}$ is transformed under the Lorentz 
transformation (\ref{LorentzSquashed}), and one obtains 
$$
\u{\tilde{T}}^{\a\b}{}_{045}=t_{45}\cosh\xi, 
\qquad
\u{\tilde{T}}^{\a\b}{}_{012}=-t_{45}\sinh\xi.
$$
Therefore, on the squashed $S^3\times\RS$, besides the $R$-symmetry gauge field 
$A^{12}=A^{34}$ and the graviphoton field $G$, the auxiliuary fields 
$$
\tilde{t}_{45}=t_{45}\cosh\xi, 
\qquad
\tilde{t}_{12}=-t_{45}\sinh\xi.
$$
are turned on.

Under the Lorentz transformation (\ref{LorentzSquashed}), 
a six-dimensional supersymmetry parameter $\u{\e}^{\a}$ transforms as
\begin{eqnarray}
\u{\e}^{\a}\quad\to\quad
\u{\tilde{\e}}^{\a}=
\exp\l({ut\over{r\a}}\,\u{\G}^{12}\r)\exp\l(\hf\xi\,\u{\G}^{03}\r)\u{\e}^{\a},
\label{LorentzSquashedSpinor}
\end{eqnarray}
and recalling that it is of positive chirality,
$$
\u{\e}^{\a}=\l(\begin{array}{c}\e^{\a}\\{0}\end{array}\r),
$$
one can see that the five-dimensional spinor $\e^\a$ transforms as
$$
\e^{\a}\quad\to\quad
\tilde{\e}^{\a}=
\exp\l({ut\over{r\a}}\,{\g}^{12}\r)\exp\l(\hf\xi\,{\g}^{3}\r){\e}^{\a}.
$$

The Lorentz transform (\ref{LorentzSquashedSpinor}) of the Killing spinors 
in the subsection \ref{newR3Sbg} also gives the Killing spinors 
in the background on the squashed $S^3\times\RS$. 
In fact in the subsection \ref{newR3Sbg}, 
depending on the sign of the background $t_{45}=\pm1/2r$, 
one has the Killing spinors
\begin{eqnarray*}
&&\l(\ve^{\da=1},\ve^{\da=2}\r)
=\l(\e_0\otimes\zeta_{+},\, C_3^{-1}\e^*_0\otimes\zeta_{-}\r), 
\quad {\rm for~} t_{45}=1/2r, 
\nn\\
&&\l(\ve^{\da=1},\ve^{\da=2}\r)
=\l(U^{-1}\e_0\otimes\zeta_{+},\,C_3^{-1}U^{T}\e^*_0\otimes\zeta_{-}\r), 
\quad {\rm for~} t_{45}=-1/2r, 
\end{eqnarray*}
and they are transformed under the Lorentz transformation into
\begin{eqnarray*}
\ve^{1}&&\quad\rightarrow\quad
\exp\l(i{ut\over{r\a}}\,\t_3\r)\exp\l(\hf\xi\,\t_3\r)\e_0\otimes\zeta_{+},
\hskip 4cm {\rm for~} t_{45}=1/2r, 
\nn\\
\ve^{1}&&\quad\rightarrow\quad
\exp\l(i{ut\over{r\a}}\,\t_3\r)\exp\l(\hf\xi\,\t_3\r)\c
\exp\l(-i{ut\over{r\a}}\,\t_3\r)
U^{-1}(\tilde{\psi},\th,\phi)\e_0\otimes\zeta_{+}
\nn\\
&&\quad=\quad
\exp\l(\hf\xi\,\t_3\r)U^{-1}(\tilde{\psi},\th,\phi)\e_0\otimes\zeta_{+},
\hskip 4.5cm {\rm for~} t_{45}=-1/2r. 
\end{eqnarray*}
One can thus see that the former never survive the Kaluza-Klein reduction, 
and the latter yields the Killing spinor on the squashed $S^3\times\RS$,
$$
\tilde{\ve}^{\da=1}=
\exp\l(\hf\xi\,\t_3\r)U^{-1}(\tilde{\psi},\th,\phi)\e_0\otimes\zeta_{+},
\qquad
\tilde{\ve}^{\da=2}=
\exp\l(-\hf\xi\,\t_3\r)C^{-1}_3U^{T}(\tilde{\psi},\th,\phi)
\e^{*}_0\otimes\zeta_{-},
$$
which is the solution to 
\begin{eqnarray*}
\cD_a\ve^{\da}
&=&-{1\over2{\ta}}G_{12}\l(\dl_{a}{}^1\dl_{b}{}^2-\dl_{a}{}^2\dl_{b}{}^1\r)
\g^b\ve^{\da}
-\l[{1\over4\ta}G_{12}+\tilde{t}_{12}\r]\g_a{}^{12}\ve^{\da}
-\tilde{t}_{45}\,\g_a{}^{45}\ve^{\da}
\nn\\
&=&-{1\over{r}}\sinh\xi\l(\dl_{a}{}^1\dl_{b}{}^2-\dl_{a}{}^2\dl_{b}{}^1\r)
\g^b\ve^{\da}
+{1\over2r}\cosh\xi\,\g_a{}^{45}\ve^{\da}. 
\end{eqnarray*}
which agrees with the Killing spinor equation (\ref{N=1KSE}) with 
the background obtained in this subsection. 

Let us turn to the remaining supersymmetry condition (\ref{SUSYchieq}) 
from $\dl_\e{\chi}^{\a\b}{}_{\g}=0$, which determines the auxiliuary field 
$M_{IJ}$. Substituting the background fields into (\ref{SUSYchieq}) and 
noticing that\footnote{
The latter formula is written for later use. 
} 
$$
\cD_2t_{23}=\cD_1t_{13}=2t_{12}\c{t}_{45},
\qquad
{1\over\a}\cD_2G_{23}={1\over\a}\cD_1G_{13}=4{\tr\over{r}^2}t_{12},
$$
one obtains 
\begin{eqnarray*}
&&{4\over15}M_{55}
={4\over5}{1\over{r}^2}-{1\over5}R\l(\RS\r),
\nn\\
&&{4\over15}M_{11}={4\over15}M_{22}
=-{1\over5}{1\over{r}^2}-{1\over5}R\l(\RS\r)-F_{45}{}^{12}
=-{1\over5}{1\over{r}^2}+{1\over20}R\l(\RS\r),
\nn\\
&&{4\over15}M_{33}={4\over15}M_{44}
=-{1\over5}{1\over{r}^2}-{1\over5}R\l(\RS\r)-F_{45}{}^{34}
=-{1\over5}{1\over{r}^2}+{1\over20}R\l(\RS\r),
\end{eqnarray*}
and the scalar masses ${\cal M}_{B\,IJ}$
$$
{\cal M}_{B\,55}
=4\,{\tr^2\over{r}^4}, 
\qquad
{\cal M}_{B\,11}={\cal M}_{B\,22}={\cal M}_{B\,33}={\cal M}_{B\,44}
={1\over{r}^2}+{1\over4}R(\RS).
$$

In summary, we have found the supersymmetric background on 
the squashed $S^3\times\RS$,
\begin{eqnarray*}
\tilde{t}_{12}=-{1\over4\ta}G_{12}={1\over{2r}}\sinh\xi,
\qquad
\tilde{t}_{45}=-{1\over2{r}}\cosh\xi,
\qquad
A^{12}=A^{34}=-\hf\om^{12}, 
\end{eqnarray*}
with the above scalar masses ${\cal M}_{B\,IJ}$. 




\subsection{An Ellipsoid 3-Sphere}
\label{HosomichiEllipsoid}

As explained in \cite{Hosomichi}, an ellipsoid 3-sphere is 
defined by the set of solutions $(x_1,x_2,x_3,x_4)\in{\bf R}^4$ to 
$$
{x_1^2+x_2^2\over{r}^2}+{x_3^2+x_4^2\over{\tr}^2}=1,
$$
which is solved by polar coordinates ($\phi$, $\chi$, $\th$)
$$
x_1=r\cos\th\cos\vp, 
\quad
x_2=r\cos\th\sin\vp, 
\qquad
x_3=\tr\sin\th\cos\chi, 
\quad
x_4=\tr\sin\th\sin\chi, 
$$
with $0\le\phi\le2\pi$, $0\le\chi\le2\pi$, $0\le\th\le\pi/2$. 
The metric is induced by embedding it into a flat ${\bf R}^4$,
$$
ds^2=dx_1^2+dx_2^2+dx_3^2+dx_4^2.
$$
For more details, see \cite{Hosomichi} or Appendix \ref{EllipsoidS^3}. 

The Killing spinors $\e$ and $\e^c=C^{-1}_3\e^*$ on the ellipsoid $S^3$ 
are given\footnote{The Killing spinors are the same as in \cite{Hosomichi}.} 
in Appendix \ref{EllipsoidS^3}. Using them, we form five-dimensional 
Killing spinors, $\ve^1=\e\otimes\zeta_+$, $\ve^2=\e^c\otimes\zeta_-$, 
with the same $\zeta_{\pm}$ as before, and see that they obey 
\begin{equation}
\cD_\m\ve^{\da}=-{1\over2f}\g_\m{}^{45}\ve^{\da},
\label{5DKSEellipsoid}
\end{equation}
where the $R$-symmetry gauge field $A^{i}{}_{j}$ in  
the covariant derivative $\cD\ve^{\da}$ is given by 
$$
\hf\l(A^{12}+A^{34}\r)=-V-\hf\om^{45}, 
$$
with $V$ the background $U(1)$ gauge field 
on the ellisoid $S^3$ 
given in (\ref{bgV}) of Appendix \ref{EllipsoidS^3}, 
and with $\om^{45}$ the spin connection on the Riemann surface $\RS$. 

We will consider an $\N=1$ supersymmetric background by taking 
$A^{12}=A^{34}$, and break the $Spin(5)_R$ symmetry group to 
$SU(2)_l\times{U}(1)_r$.

For the other background fields $S_{ij}$, $G_{ab}$, and $t_{ab}$, 
the Killing spinor equation (\ref{N=1KSE}) in the background satisfying
\begin{eqnarray}
S+2\cdot{1\over4\a}G_{45}=0,
\qquad
S+{1\over4\a}G_{45}+t_{45}={1\over{2f}},
\label{onshellSUSYellisoid}
\end{eqnarray}
where we have assumed that $S=S_{12}=S_{34}$, 
is reduced into (\ref{5DKSEellipsoid}). 
However, substituting them into the other supersymmetry condition 
(\ref{SUSYchieq}), we find that the background 
\begin{equation}
S=0, \qquad {1\over4\a}G_{45}=0, \qquad t_{45}={1\over{2f}},
\label{ellipsoidbg}
\end{equation}
is the only solution to (\ref{SUSYchieq}), and that the background fields 
$M_{IJ}$ are given by
$$
{4\over15}M_{55}={2\over{f}^2}-{1\over5}{R(\Om)}
={2\over5}\l[{4f^2-r^2-\tr^2\over{f}^4}-{\hf}R(\RS)\r],
\qquad
{4\over15}M_{ij}=-\dl_{ij}\,{1\over15}M_{55}, 
\quad
(i,j=1,\cdots,4). 
$$
It means that the scalar masses ${\cal M}_{B\,IJ}$ are
$$
{\cal M}_{B\,55}={4\over{f}^2},
\qquad
{\cal M}_{B\,ij}=\hf\l[{r^2+\tr^2\over{f}^2}+\hf{R}(\RS)\r]\dl_{ij},
\quad
(i,j=1,\cdots,4).
$$


\section{The Off-Shell Formulation of the Reduced Theory}
\label{OffshellSUSY}

In order to implement the localization technique in calculating 
the partition function of the five-dimensional theory in a 
supergravity background, a supersymmetry transformation for 
the localization must be defined off-shell. 
To this end, the half of the supersymmetry transformations (\ref{SUSY}) will 
be extended off-shell by introducing auxiliuary fields. 
The supersymmetry parameters $\e^{\ta}$, $\ve^{\da}$ are decoupled 
in the supersymmetry transformations, if the background fields obey 
the conditions
$$
t^i{}_{ab}=0, 
\qquad
S_{i5}=0,
\qquad
A_\m{}^{i5}=0, 
\qquad
(i=1,\cdots,4). 
$$
The backgrounds we would like to consider in this paper, obey 
these conditions. Therefore, we will content ourselves with the 
construction of an off-shell formulation of the theory in 
this restricted type of backgrounds. 
In addition, the backgrounds in this paper also satisfy 
the condition $b_\m=0$, and we will add it to the above conditions.

We would like to use one of the supersymmetry transformations for 
the localization. Since the supersymmetry parameters $\e^{\ta}$, $\ve^{\da}$ 
are decoupled in the supersymmetry transformations in the restricted type of 
background, we will turn off the parameter $\e^{\ta}$ and focus only on 
$\ve^{\da}$. It is then convenient to regard the $\N=2$ gauge multiplet 
as the sum of an $\N=1$ gauge multiplet ($\s$, $A_\m$, $\la^{\da}$) and 
$\N=1$ hypermultiplets ($\phi^i$, $\psi^{\ta}$). 

We will introduce an auxiliuary field $D^{\da}{}_{\db}$ in the adjoint 
representation of the $SU(2)_r$ subgroup of the $Spin(5)_R$ $R$-symmetry group;
$$
D^{\da\db}=\ve^{\db\dg}D^{\da}{}_{\dg}=D^{\db\da},
$$
and replace 
\begin{eqnarray}
\l(\bar{\s}^{ij}\r)^{\da}{}_{\db}\,S_{ij}\s
+{i\over2}g\l[\phi^i,\,\phi^j\r]\l(\bar{\s}_{ij}\r)^{\da}{}_{\db}
\quad\rightarrow\quad
D^{\da}{}_{\db},
\label{replacement}
\end{eqnarray}
in (\ref{N=1SUSY}), to which the supersymmetry tranformation (\ref{SUSY}) 
is reduced in the restricted type of backgrounds.  

In a consistent way to the above replacement, the supersymmetry transformation 
of $D^{\da}{}_{\db}$ is determined by using the equation of motion of 
$\la^{\da}$, and one obtains 
\begin{eqnarray}
&&\dl_{\e}A_{\m}=-{i\over4}(\bar{\ve}\c\g_\m\la),
\qquad
\dl_{\e}\s=-{i\over4}(\bar{\ve}\c\la),
\nn\\
&&\dl_{\e}\la^{\da}
=-\hf{F}_{ab}\g^{ab}\ve^{\da}+\g^a\cD_a\s\c\ve^{\da}
+D^{\da}{}_{\db}\ve^{\db}-{1\over2\a}G_{ab}\s\g^{ab}\ve^{\da},
\nn\\
&&\dl_{\e}D^{\da\db}
=-{i\over4}\bar{\ve}^{(\da}
\l[\g^a\cD_a\la^{\db)}-ig\l[\s,\,\la^{\db)}\r]
-{1\over8\a}G_{ab}\g^{ab}\la^{\db)}
-\qt{S}_{ij}\l(\bar{\s}^{ij}\r){}^{\db)}{}_{\dg}\la^{\dg}
-\hf{t}_{ab}\g^{ab}\la^{\db)}\r]
\nn\\
&&\hskip1.6cm
+{i\over4}S_{ij}\l(\bar{\s}^{ij}\r){}^{\da}{}_{\dg}\ve^{\dg\db}
\l(\bar{\eta}\c\la\r).
\label{VectorSUSY}
\end{eqnarray}


The off-shell supersymmetry transformations (\ref{VectorSUSY}) are  
closed into the other bosonic transformations. 
Using the Killing spinor equation (\ref{N=1KSE}) in this type of backgrounds, 
one obtains 
\begin{eqnarray}
&&\l[\dl_{\e},\,\dl_{\eta}\r]A_\m
={i\over2}\xi^\n{F}_{\m\n}-{i\over2}\l(\bar{\eta}\c\ve\r)\cD_\m\s
+{i\over2\a}\xi^\n{G}_{\m\n}\s
=-{i\over2}\l(\xi^\n\d_\n{A}_\m+\d_\m\xi^\n\c{A}_\n\r)
-\d_\m\La_{G}-ig\l[A_\m,\,\La_{G}\r],
\nn\\
&&\l[\dl_{\e},\,\dl_{\eta}\r]\s
=-{i\over2}\xi^a\cD_a\s
=-{i\over2}\xi^a\d_a\s+\La_D\,\s+ig\l[\La_G,\,\s\r],
\nn\\
&&\l[\dl_{\e},\,\dl_{\eta}\r]\la^{\da}
=-{i\over2}\xi^a\cD_a\la^{\da}
+{i\over2}\c(ig)\l[\l(\bar{\eta}\c\ve\r)\s,\,\la^{\da}\r]
-{i\over8}\l(\cD_a\xi_b\r)\g^{ab}\la^{\da}
-{1\over4}\La^{(0)}_{ij}\l(\bar{\s}^{ij}\r){}^{\da}{}_{\db}\la^{\db}
\nn\\
&&\hskip1.8cm
=-{i\over2}\xi^\m\d_\m\la^{\da}+{3\over2}\La_D\,\la^{\da}
+ig\l[\La_G,\,\la^{\da}\r]
+{1\over4}\La_{ab}\g^{ab}\la^{\da}
-{1\over4}\La^{ij}\l(\bar{\s}_{ij}\r){}^{\da}{}_{\db}\la^{\db},
\label{OffShellAlgGauge}\\
&&\l[\dl_{\e},\,\dl_{\eta}\r]D^{\da\db}
=-{i\over2}\xi^a\cD_aD^{\da\db}
+{i\over2}\c(ig)\l[\l(\bar\eta\c\e\r)\s,\,D^{\da\db}\r]
-\qt\La^{(0)}_{ij}\l[\l(\bar{\s}^{ij}\r){}^{\da}{}_{\dg}D^{\dg\db}
+\l(\bar{\s}^{ij}\r){}^{\db}{}_{\dg}D^{\da\dg}\r]
\nn\\
&&\hskip2.1cm
=-{i\over2}\xi^\m\d_\m{D}^{\da\db}
+2\La_D\,{D}^{\da\db}
+ig\l[\La_G,\,D^{\da\db}\r]
-\qt\La_{ij}\l[\l(\bar{\s}^{ij}\r){}^{\da}{}_{\dg}D^{\dg\db}
+\l(\bar{\s}^{ij}\r){}^{\db}{}_{\dg}D^{\da\dg}\r],
\nn
\end{eqnarray}
with the transformation parameters
\begin{eqnarray}
&&\xi^a=\bar{\eta}\c\g^a\ve, 
\qquad
\La_D={i\over2}\xi^a{b}_a, 
\qquad
\La_{ab}=-{i\over2}\l[\cD_a\xi_b+\xi^\m\l(\Om_\m\r)_{ab}\r],
\qquad
\La_{G}=-{i\over2}\l[\xi^a{A}_a-\l(\bar{\eta}\c\ve\r)\s\r],
\nn\\
&&\La^{(0)}_{ij}=-{i\over2}\l[3S_{ij}\l(\bar{\eta}\c\e\r)
+\l(t_{ab}+{1\over2\a}G_{ab}\r)\l(\bar{\eta}\c\bar{\s}_{ij}\g^{ab}\e\r)\r],
\qquad
\La_{ij}=\La^{(0)}_{ij}-{i\over2}\l(\xi^a{A}_a\r)_{ij}.
\label{bosonicparamter}
\end{eqnarray}

Let us proceed to the hypermultiplets ($\phi^i,\,\psi^{\ta}$). 
In writing the off-shell transformation for them, it will turn out 
that the spinor index notation of the scalars $\phi^{\ta}{}_{\db}$, 
$$
\phi^{\ta}{}_{\db}=\phi^i\l(\s_i\r)^{\ta}{}_{\db},
\qquad
\bar{\phi}^{\da}{}_{\tb}=\phi^i\l(\bar{\s}_i\r)^{\da}{}_{\tb},
$$
will be convenient. 
In order to formulate an off-shell supersymmetry transformation\footnote{
We will follow the prescription in \cite{S^5,FKM,KM}, where we won't 
expect to have a full-fledged off-shell formulation. However, 
it will be sufficient to carry out the localization.}
of the hypermultiplets, 
we will introduce auxiliuary fields $F^{\ta}{}_{\cb}$ with the index $\cb$ 
labeling a doublet of a new SU(2) flavor group, which is not a subgroup of 
the $Spin(5)_R$ $R$-symmetry group, following \cite{S^5}. 

Further, we will also introduce different supersymmetry parameters $\ve^{\ca}$ 
from $\ve^{\da}$ and $\e^{\ta}$, the former of which span 
the whole four-dimensional spinor space with $\ve^{\da}$. 

The auxiliuary fields $F^{\ta}{}_{\cb}$ and the new parameters $\ve^{\ca}$ 
are expected to play the role to impose the equation of motion of the spinors 
in the off-shell supersymmetry formulation. To this end, requiring that 
$\dl_{\e}F^{\ta}{}_{\cb}$ be proportional to the equation of motion of 
the spinor $\psi^{\ta}$, one obtains an off-shell supersymmetry transformation 
\begin{eqnarray}
&&\dl_{\e}\phi^{\ta}{}_{\db}={i\over2}(\bar{\ve}_{\db}\c\psi^{\ta}),
\nn\\
&&\dl_{\e}\psi^{\ta}
=-\g^a\cD_a\phi^{\ta}{}_{\db}\c\ve^{\db}+ig\l[\s,\,\phi^{\ta}{}_{\db}\r]\ve^{\db}
+\l(t_{ab}+{1\over2\a}G_{ab}\r)\phi^{\ta}{}_{\db}\g^{ab}\ve^{\db}
\nn\\
&&\hskip1.5cm
-\qt{S}_{ij}\l[3\phi^{\ta}{}_{\dg}\l(\bar{\s}^{ij}\r)^{\dg}{}_{\db}
+\l(\s^{ij}\r){}^{\ta}{}_{\tg}\phi^{\tg}{}_{\db}
\r]\ve^{\db}
+F^{\ta}{}_{\cb}\ve^{\cb},
\label{HyperSUSY}\\
&&\dl_{\e}F^{\ta}{}_{\cb}
={i\over2}\bar{\ve}_{\cb}\l[\g^a\cD_a\psi^{\ta}+ig\l[\s,\,\psi^{\ta}\r]
+ig\l[\phi^{\ta}{}_{\dg},\,\la^{\dg}\r]
+\hf\l(t_{ab}-{1\over4\a}G_{ab}\r)\g^{ab}\psi^{\ta}
-{1\over4}S_{ij}\l(\s^{ij}\r){}^{\ta}{}_{\tg}\psi^{\tg}\r].
\nn
\end{eqnarray}
In terms of the vector notation of the scalars $\phi^i$, it gives 
\begin{eqnarray*}
&&\dl_{\e}\phi^i
={i\over4}(\bar{\s}^i)^{\da}{}_{\tb}(\bar{\ve}_{\da}\c\psi^{\tb}),
\nn\\
&&\dl_{\e}\psi^{\ta}
=(\s_i)^{\ta}{}_{\db}\bigg[
-\g^a\cD_a\phi^i\c\ve^{\db}+ig\l[\s,\,\phi^i\r]\ve^{\db}
+\l(t_{ab}+{1\over2\a}G_{ab}\r)\phi^i\g^{ab}\ve^{\db}
+\l({S}^{ij}+\ve^{ijkl}S_{kl}\r)\phi_j\ve^{\db}\bigg]
\nn\\
&&\hskip1.5cm
+F^{\ta}{}_{\cb}\ve^{\cb},
\nn\\
&&\dl_{\e}F^{\ta}{}_{\cb}
={i\over2}\bar{\ve}_{\cb}\l[\g^a\cD_a\psi^{\ta}+ig\l[\s,\,\psi^{\ta}\r]
+ig(\s_i)^{\ta}{}_{\dg}\l[\phi^i,\,\la^{\dg}\r]
+\hf\l(t_{ab}-{1\over4\a}G_{ab}\r)\g^{ab}\psi^{\ta}
-{1\over4}S_{ij}\l(\s^{ij}\r){}^{\ta}{}_{\tg}\psi^{\tg}\r],
\end{eqnarray*}
where the conditions are assumed;
\begin{eqnarray}
\l(\ve^{\ca}\r)^T{}C\eta^{\db}-\l(\eta^{\ca}\r)^T{}C\ve^{\db}=0,
\qquad
(\bar{\eta}_{\db}\,\ve^{\db})=(\bar{\eta}_{\cb}\,\ve^{\cb}),
\qquad
(\bar{\eta}_{\db}\g^a\ve^{\db})=-(\bar{\eta}_{\cb}\g^a\ve^{\cb}),
\label{orthogonaldotcheck}
\end{eqnarray}
which will be necessary for the off-shell closure of the supersymmetry 
transformations on $\phi^{\ta}{}_{\db}$ and $\psi^{\ta}$. 
Note that in terms of the spinor index notation, the covariant derivative 
$\cD_\m\phi^{\ta}{}_{\db}$ can be read as
$$
\cD_\m\phi^{\ta}{}_{\db}=\d_\m\phi^{\ta}{}_{\db}-b_\m\phi^{\ta}{}_{\db}
+ig\l[A_\m,\,\phi^{\ta}{}_{\db}\r]
-\qt{A}_\m{}^{ij}\l[\l(\s_{ij}\r){}^{\ta}{}_{\tg}\phi^{\tg}{}_{\db}
-\phi^{\ta}{}_{\dg}\l(\bar{\s}_{ij}\r){}^{\dg}{}_{\db}\r].
$$

Making use of (\ref{orthogonaldotcheck}), one can verify that 
\begin{eqnarray}
\l[\dl_{\e},\,\dl_{\eta}\r]\phi^{\ta}{}_{\db}
&=&-{i\over2}\xi^a\cD_a\phi^{\ta}{}_{\db}
+{i\over2}\c(ig)\l[(\bar\eta\c\ve)\s,\,\phi^{\ta}{}_{\db}\r]
+\qt\La^{(0)}_{ij}\phi^{\ta}{}_{\dg}\l({\bar\s}^{ij}\r){}^{\dg}{}_{\db}
-{1\over4}\tilde{\La}^{(0)}_{ij}\l(\s^{ij}\r){}^{\ta}{}_{\tg}\phi^{\tg}{}_{\db},
\nn\\
&=&-{i\over2}\xi^\m\d_\m\phi^{\ta}{}_{\db}+\La_D\,\phi^{\ta}{}_{\db}
+ig\l[\La_G,\,\phi^{\ta}{}_{\db}\r]
+\qt\La_{ij}\phi^{\ta}{}_{\dg}\l({\bar\s}^{ij}\r){}^{\dg}{}_{\db}
-{1\over4}\tilde{\La}_{ij}\l(\s^{ij}\r){}^{\ta}{}_{\tg}\phi^{\tg}{}_{\db},
\nn\\
\l[\dl_{\e},\,\dl_{\eta}\r]\psi^{\ta}
&=&-{i\over2}\xi^a\cD_a\psi^{\ta}
-{i\over8}\l(\cD_a\xi_b\r)\g^{ab}\psi^{\ta}
+{i\over2}\c(ig)\l[(\bar\eta\c\ve)\s,\,\phi^{\ta}{}_{\db}\r]
-{1\over4}\tilde{\La}^{(0)}_{ij}\l(\s^{ij}\r){}^{\ta}{}_{\tg}\psi^{\tg},
\label{OffShellAlgHyper}\\
&=&-{i\over2}\xi^\m\d_\m\psi^{\ta}+{3\over2}\La_D\,\psi^{\ta}
+{1\over4}\La_{ab}\g^{ab}\psi^{\ta}
+ig\l[\La_G,\,\phi^{\ta}{}_{\db}\r]
-{1\over4}\tilde{\La}_{ij}\l(\s^{ij}\r){}^{\ta}{}_{\tg}\psi^{\tg},
\nn
\end{eqnarray}
with the parameters in (\ref{bosonicparamter}), 
where the transformation parameters of the other SU(2) subgroup of 
the Spin(5)$_R$ are given by 
$$
\tilde{\La}^{(0)}_{ij}={i\over2}(\bar\eta\c\ve)S_{ij},
\qquad
\tilde{\La}_{ij}=\tilde{\La}^{(0)}_{ij}-{i\over2}\xi^\m{A}_\m{}_{ij}.
$$

So far, we have seen that the supersymmetry transformations (\ref{HyperSUSY}) 
on $\phi^{\ta}{}_{\db}$, $\psi^{\ta}$ are closed off-shell, if we require 
the condition (\ref{orthogonaldotcheck}) on the supersymmetry parameters. 
However, we will see that the supersymmetry transformations (\ref{HyperSUSY}) 
on the auxiliuary fields $F^{\ta}{}_{\cb}$ are not automatically closed. 
In order to achieve an off-shell supersymmetry transformation, 
it seems that one has to require the supergravity backgrounds to obey 
additional conditions. 

Let us look at the supersymmetry transformations on $F^{\ta}{}_{\cb}$. 
Using the condition (\ref{orthogonaldotcheck}), one obtains 
\begin{eqnarray}
&&\l[\dl_{\e},\,\dl_{\eta}\r]F^{\ta}{}_{\cb}
=-{i\over2}\xi^\m\d_\m{F}^{\ta}{}_{\cg}
+\La_D{F}^{\ta}{}_{\cg}
+ig\l[\La_G,\,F^{\ta}{}_{\cg}\r]
+{1\over4}\tilde{\La}_{ij}\l(\s_{ij}\r){}^{\ta}{}_{\tg}F^{\tg}{}_{\cb}
+\check{\La}{}_{\cb}{}^{\cg}F^{\ta}{}_{\cg}
\nn\\
&&\hskip2.6cm
+\qt\l(\Th^{ij}\r){}_{\cb}{}^{\dg}\l(\s_{ij}\r)^{\ta}{}_{\tdl}
\,\phi^{\tdl}{}_{\dg}
+\Delta_{\cb}{}^{\dg}\phi^{\ta}{}_{\dg},
\label{OffShellAlgF}
\end{eqnarray}
where the parameter $\check{\La}{}_{\ca}{}^{\cb}$ of the new $SU(2)$ 
transformation is given by 
\begin{eqnarray*}
&&\check{\La}{}_{\ca}{}^{\cb}
={i\over2}\bigg[
\l({\bar\eta}_{\cb}\g^a\cD_a\ve^{\cg}-{\bar\ve}_{\cb}\g^a\cD_a\eta^{\cg}\r)
+\hf\l(t_{ab}-{1\over4\a}G_{ab}\r)
\l({\bar\eta}_{\cb}\g^{ab}\ve^{\cg}-{\bar\ve}_{\cb}\g^{ab}\eta^{\cg}\r)
\bigg],
\end{eqnarray*}
and the last two terms on the right hand side suggest that the supersymmetry 
transformations fail to be closed off-shell, where the parameters
$\l(\Th^{ij}\r){}_{\cb}{}^{\dg}$, and $\Delta_{\cb}{}^{\dg}$ are given by 
\begin{eqnarray*}
&&\l(\Th^{ij}\r){}_{\cb}{}^{\dg}
={i}
\l(\bar{\eta}_{\cb}\g^{ab}\ve^{\dg}-\bar{\ve}_{\cb}\g^{ab}\eta^{\dg}\r)
\l(\qt\,F_{ab}{}^{ij}+{1\over4\a}G_{ab}S^{ij}\r)
-{i\over2}
\l(\bar{\eta}_{\cb}\g^{a}\ve^{\dg}
-\bar{\ve}_{\cb}\g^{a}\eta^{\dg}\r)\cD_a{S}^{ij},
\nn\\
&&\Delta_{\cb}{}^{\dg}
={i\over2}\bigg[
\l({1\over4\a}\r)^2G_{ab}G_{cd}
\l(\bar{\eta}_{\cb}\g^{abcd}\ve^{\dg}-\bar{\ve}_{\b}\g^{abcd}\eta^{\dg}\r)
-\l(\cD^at_{ab}+{1\over4\a}\cD^aG_{ab}\r)
\l(\bar{\eta}_{\cb}\g^{b}\ve^{\dg}-\bar{\ve}_{\b}\g^{b}\eta^{\dg}\r)
\bigg]
\nn\\
&&\hskip1.5cm
+{i\over2}\l(\bar{\s}^{ij}\r){}^{\dg}{}_{\ddl}\bigg[
\qt\cD_aS_{ij}
\l(\bar{\eta}_{\cb}\g^{a}\ve^{\ddl}-\bar{\ve}_{\b}\g^{a}\eta^{\ddl}\r)
-\hf\l(t_{ab}+2{1\over4\a}G_{ab}\r)S_{ij}
\l(\bar{\eta}_{\cb}\g^{ab}\ve^{\ddl}-\bar{\ve}_{\b}\g^{ab}\eta^{\ddl}\r)
\bigg]. 
\end{eqnarray*}

Therefore, in order to gain an off-shell closed algebra of the supersymmetry 
transformations, the conditions 
\begin{eqnarray}
\l(\Th^{ij}\r){}_{\cb}{}^{\dg}\,\l(\s_{ij}\r)^{\ta}{}_{\tdl}=0,
\qquad
\Delta_{\cb}{}^{\dg}=0,
\label{Offshellcondition}
\end{eqnarray}
are required. 
Although the implications of the condition (\ref{Offshellcondition}) 
have been unclear for the authors, the backgrounds in the section 
\ref{Backgrounds} satisfy the condition (\ref{Offshellcondition}). 
So henceforth, we will assume that the backgrounds satisfy 
the condition (\ref{Offshellcondition}).

Let us make sure that the supersymmetric backgrounds 
in section \ref{Backgrounds} obey the conditions (\ref{orthogonaldotcheck}) 
and (\ref{Offshellcondition}). 
With the ansatz (\ref{KSansatz}) and as explained in Appendix \ref{3-sphere}, 
the supersymmetry parameters 
$\ve^{\da}$, $\ve^{\ca}$ are given by 
\begin{eqnarray*}
\ve^{\da=1}=\e\otimes\zeta_{+}, 
\quad
\ve^{\da=2}=C_{3}^{-1}\e^*\otimes\zeta_{-}; 
\qquad\quad
\ve^{\ca=1}=\e\otimes\zeta_{-}, 
\quad
\ve^{\ca=2}=C_{3}^{-1}\e^*\otimes\zeta_{+}, 
\end{eqnarray*}
obeying that $\ve^{\ca}=\g_5\ve^{\ca}$. It follows from this that 
\begin{eqnarray*}
\l(\ve^{\ca}\r)^TC\eta^{\db}-\l(\eta^{\ca}\r)^TC\ve^{\db}=0,
\qquad
\bar\eta_{\ca}\ve^{\ca}=\bar\eta_{\da}\ve^{\da},
\qquad
\bar\eta_{\ca}\g^a\ve^{\ca}=-\bar\eta_{\da}\g^a\ve^{\da},
\end{eqnarray*}
and the condition (\ref{orthogonaldotcheck}) is satisfied by 
the supersymmetry parameters $\ve^{\da}$, $\ve^{\ca}$. 

Since the field $S_{ij}$ in all the $\N=1$ supersymmetric backgrounds 
in section \ref{Backgrounds} satisfies $S^{ij}\,\s_{ij}=0$, 
and the field strength $F_{ab}{}^{ij}$ satisfies 
$F_{ab}{}^{ij}\,\s_{ij}=0$, it is easy to see that they satisfy 
the former condition in (\ref{Offshellcondition}). 
In the $\N=2$ background in subsection \ref{HosomichiSquash}, 
the non-zero components of $F_{ab}{}^{ij}$ are $F_{45}{}^{ij}$ and 
$F_{12}{}^{ij}$, and $G_{ab}$ has only nonzero component $G_{45}$. 
At the first sight, the nonzero $F_{45}{}^{ij}$ and $G_{45}$ 
seems to give the contributions to the former condition of 
(\ref{Offshellcondition}), but, since we have the formula
\begin{eqnarray}
\bar\eta_{\ca}\g^{45}\ve^{\db}-\bar\ve_{\ca}\g^{45}\eta^{\db}=0, 
\label{etacag45vedb}
\end{eqnarray}
they yield no contributions to the condition. 
Furthermore, the nonzero $F_{12}{}^{ij}$ appears on the left hand side 
of the condition with the term 
\begin{eqnarray}
\bar\eta_{\ca}\g^{12}\ve^{\db}-\bar\ve_{\ca}\g^{12}\eta^{\db}
\quad\propto\quad
\bar\eta_{\ca}\g^{3}\ve^{\db}-\bar\ve_{\ca}\g^{3}\eta^{\db}, 
\end{eqnarray}
but, the conditions $\g^3\ve^{\da}=\ve^{\da}$ and $\g^3\eta^{\da}=\eta^{\da}$ 
reduce $\bar\eta_{\ca}\g^{3}\ve^{\db}-\bar\ve_{\ca}\g^{3}\eta^{\db}$ to 
the left hand side of the first condition in (\ref{orthogonaldotcheck}). 
Therefore, the former condition in (\ref{Offshellcondition}) is obeyed 
also for the $\N=2$ supersymmetric background. 

The covariant derivatives $\cD_a{t}_{bc}$, $\cD_a{G}_{bc}$, and $\cD_a{S}_{ij}$ 
are vanishing except for the background in subsection \ref{ImamuraSquash}.
However, ever for the background, $\cD^a{t}_{ab}+\cD^a{G}_{ab}/(4\a)=0$. 
Further, it is obvious that $G_{ab}G_{cd}\ve^{abcde}=0$. Therefore, 
taking (\ref{etacag45vedb}) into account, one can see that all the 
backgrounds in section \ref{Backgrounds} also satisfy the latter condition 
in (\ref{Offshellcondition}). Now we can see that all the supersymmetry 
backgrounds in (\ref{Offshellcondition}) allow the off-shell supersymmetry.

Let us proceed to the construction of an off-shell supersymmetric action. 
In order to perform the replacement (\ref{replacement}) with $D^{\da}{}_{\db}$ 
within the on-shell invariant action $S=S_F+S^{(0)}_B+S^{(1)}_B$ in 
(\ref{5DLF}-\ref{5DLB}), we will add the term 
$$
\hf\int dx^5\sqrt{g}\,\a\, 
\Tr\Big[-\hf\hat{D}^{\da}{}_{\db}\hat{D}^{\db}{}_{\da}\Big]
\quad
{\rm with}
\quad
\hat{D}^{\da}{}_{\db}\equiv D^{\da}{}_{\db}-\l(\bar{\s}^{ij}\r)^{\da}{}_{\db}
\l(S_{ij}\s+\hf(ig)\l[\phi_i,\,\phi_j\r]\r)
$$
to the on-shell action $S$. 

For the hypermultiplets, the off-shell supersymmetry transformation of 
$\psi^{\ta}$ has an additional term $F^{ta}{}_{\cg}\ve^{\cg}$, compared 
to the on-shell supersymmetry transformations. 
Therefore, under the off-shell supersymmetry transformation, 
the on-shell fermionic action $S_F$ in (\ref{5DLF}) gains 
an additional term 
$$
-\hf\ve_{\ta\tb}F^{\ta}{}_{\cg}\l(\ve^{\cg}\r)^TC
\Big(\g^a\cD_a\psi^{\tb}+\cdots\Big)
=-\hf\ve_{\ta\tb}F^{\ta}{}_{\cg}\c\ve^{\tg\tdl}\dl_\e{F}^{\tb}{}_{\cdl}
=-\qt\dl_\e\l(\ve_{\ta\tb}\ve^{\cg\cdl}F^{\ta}{}_{\cg}{F}^{\tb}{}_{\cdl}\r)
$$
with the ellipsis denoting the other terms of the equation of motion for 
$\psi^{\ta}$, from the term 
$$
-{i\over4}\ve_{\ta\tb}\l(\dl_{\e}\psi^{\ta}\r)^TC
\Big(\g^a\cD_a\psi^{\tb}+\cdots\Big)+\cdots
$$
in $\dl_\e{S}_F$. Thus, it is necessary to add the term
$$
-\int\,\sqrt{g}\,d^5x\,{\a\over4}\Tr
\l[\ve_{\ta\tb}\ve^{\cg\cdl}F^{\ta}{}_{\cg}{F}^{\tb}{}_{\cdl}\r]
$$
to cancel the additional term from $\dl_\e{S}_F$. 

Finally, the construction of an off-shell action is achived as 
\begin{eqnarray}
{\cal S}
=
-\int \Tr\l[\hf\,C\wedge{F}\wedge{F}
+{\a\over2}\l(F-4t\s\r)\wedge*\l(F-4t\s\r)\r]
-\int \sqrt{g}\,d^5x\,\a\Tr\l[\,{\cal L}\,\r],
\label{OffShellAction}
\end{eqnarray}
where the `matter' Lagrangian ${\cal L}$ is given by
\begin{eqnarray}
&&{\cal L}=-\hf\cD_a\s\cD^a\s-\hf\cD_a\phi^i\cD^a\phi^i
-\hf{\cal M}_{\s}\s^2-\hf{\cal M}_{ij}\phi^i\phi^j
+\qt{}D^{\da}{}_{\db}D^{\db}{}_{\da}
+\qt\ve_{\ta\tb}\ve^{\cg\cdl}F^{\ta}{}_{\cg}F^{\tb}{}_{\cdl}
\nn\\
&&\qquad
-\hf{}D^{\da}{}_{\db}(\bar{\s}^{ij})^{\db}{}_{\da}
\l(S_{ij}\s+\hf(ig)\l[\phi_i,\,\phi_j\r]\r)
+\hf(ig)^2\l[\s,\,\phi^i\r]\l[\s,\,\phi^i\r]
-igS_{ij}\s\l[\phi^i,\,\phi^j\r]
\nn\\
&&\qquad
-{i\over8}\bar{\la}\c\g^a\cD_a\la
-{i\over8}\bar{\psi}\c\g^a\cD_a\psi
-{i\over16}\l(t_{ab}-{1\over4\a}G_{ab}\r)\bar{\psi}\c\g^{ab}\psi
+{i\over16}\l(t_{ab}+{1\over4\a}G_{ab}\r)\bar{\la}\c\g^{ab}\la
\nn\\
&&\qquad
+{i\over32}S_{ij}\bar{\psi}\c\s^{ij}\psi
+{i\over32}S_{ij}\bar{\la}\c\bar{\s}^{ij}\la
-{i\over8}(ig)\bar{\psi}_{\ta}\c\l[\s,\,\psi^{\ta}\r]
+{i\over8}(ig)\bar{\la}_{\da}\c\l[\s,\,\la^{\da}\r]
\nn\\
&&\qquad
-{i\over8}(ig)\bar{\psi}_{\ta}\c(\s_i)^{\ta}{}_{\db}\l[\phi^i,\,\la^{\db}\r]
-{i\over8}(ig)\bar{\la}_{\da}\c(\bar{\s}_i)^{\da}{}_{\tb}
\l[\phi^i,\,\psi^{\tb}\r],
\label{OffShellMatterLagrangian}
\end{eqnarray}
with the `mass' parameters
\begin{eqnarray*}
{\cal M}_{\s}&=&
\l({4\over15}M_{55}+{1\over5}R(\Om)+{1\over20\a^2}G_{ab}G^{ab}
+4t_{ab}t^{ab}-\hf\Tr(\bar{\s}^{ij}\bar{\s}^{kl})S_{ij}S_{kl}\r),
\nn\\
{\cal M}_{ij}&=&
\l[{4\over15}M_{ij}+{1\over5}\l(R(\Om)
+{1\over4\a^2}G_{ab}G^{ab}\r)\dl_{ij}-S_i{}^kS_{jk}\r]. 
\end{eqnarray*}

\section{Localization and Twistings}
\label{localization}

In this section, let us proceed to compute the partition function of 
the theory by using the localization. 
Before going on, there is a subtle point that 
the kinectic terms of the fields $\s$, $D^{\da}{}_{\db}$, $\phi^i$, and 
$F^{\ta}{}_{\cb}$ have the negative sign in the Lagrangian 
(\ref{OffShellMatterLagrangian}). In order to circumvent it, 
we would like to follow the same strategy for $\s$, $D^{\da}{}_{\db}$, 
and $F^{\ta}{}_{\cb}$ as in \cite{FKM,KM}. Recall that the scalars 
$\phi^i$ had the positive kinetic terms in \cite{FKM,KM}, where 
the five-dimensional theory was obtained by the dimensional reduction 
from the six-dimensional maximally supersymmetric Yang-Mills theory. 

To this end, we will perform the `analytic continuation' for the scalars, 
\begin{eqnarray*}
\s\quad\to\quad{i\s}, 
\qquad
\phi^i\quad\to\quad-{i\phi^i}. 
\end{eqnarray*}

For the auxiliuary fields $D^{\da}{}_{\db}$ and $F^{\ta}{}_{\cb}$, 
let us carefully recall what we have done in the previous papers \cite{FKM,KM}. 
First, we have shifted $D^{\done}{}_{\done}$ as\footnote{
In the previous papers \cite{FKM,KM}, although we haven't considered the 
supergravity backgrounds such as $G_{ab}$, we can interpret what was done 
as the shift (\ref{Dshift}) in terms of supergravity backgrounds.
}
\begin{equation}
D^{\done}{}_{\done} \quad\to\quad 
D^{\done}{}_{\done}-iF_{45}-{i\over\a}G_{45}\s,
\label{Dshift}
\end{equation}
and we then impose the reality condition
$$
\l(D^{\da}{}_{\db}\r)^*=D^{\db}{}_{\da}.
$$

In previous papers, we implicitly left the sign of the kinetic term 
of $F^{\ta}{}_{\cb}$ negative. Since the integration over the auxiliuary 
fields $F^{\ta}{}_{\cb}$ is trivial; there is no dependence on the 
vacuum expectation value of $\s$, we have just ignored the divergence from it. 
Therefore, we assumed that
$$
\qquad
\l(F^{\ta}{}_{\cb}\r)^*=-F^{\tg}{}_{\cdl}\,\ve^{\cdl\cb}\,\ve_{\tg\ta}.
$$
Although we don't have any rationale for the prescriptions, 
it seems to work well, and we will also follow the same prescriptions 
in this paper.

In order to carry out the localization, we will define a BRST transformation 
out of the supersymmetry transformation by setting both of $\ve^{\dot{2}}$ and 
$\ve^{\check{2}}$ to be zero, following the strategy in \cite{FKM,KM}. 
Note that this is possible, because $\ve^{\dot{2}}$ decouples from 
$\ve^{\dot{1}}$ in the Killing spinor equation. 
This is also the case for $\ve^{\ca}$. 
Furthermore, introducing bosonic Killing spinors 
$\ve$ and $\check{\ve}$, we take 
$$
\ve^{\dot{1}}=\Upsilon\,\ve, 
\qquad
\ve^{\check{1}}=\Upsilon\,\check{\ve},
$$
where $\Upsilon$ is a Grassmann odd number. 
For a generic field $\Phi$, then we define the BRST transformation of $\Phi$ by 
$$
\dl_\e\Phi=\Upsilon\,\dl_Q\Phi.
$$

Before the shift (\ref{Dshift}), it follows from (\ref{VectorSUSY}) that 
the BRST transformation on the gauge multiplet is given by
\begin{eqnarray}
&&\dl_QA_\m={i\over4}\bar{\la}^{\dtwo}\g_\m\ve,
\qquad
\dl_Q\s=\qt\bar{\la}^{\dtwo}\ve
\nn\\
&&\dl_Q\la^{\done}=-\hf{F}_{ab}\g^{ab}\ve+i\cD_a\s\,\g^a\ve
+D^{\done}{}_{\done}\ve-{i\over2\a}G_{ab}\,\s\g^{ab}\ve,
\qquad
\dl_Q\la^{\dtwo}=D^{\dtwo}{}_{\done}\,\ve,
\nn\\
&&\dl_QD^{\done}{}_{\dtwo}=-{i\over2}\l[
\bar\ve\g^a\cD_a\la^{\dot{1}}+g\l[\s,\,\bar\ve\la^{\dot{1}}\r]
-\l({1\over8\a}G_{ab}+\hf{t}_{ab}\r)\bar\ve\g^{ab}\la^{\dot{1}}
-\qt{S}_{ij}\l(\bar{\s}^{ij}\r){}^1{}_1\bar\ve\la^{\dot{1}}
\r],
\label{VectorBRST}\\
&&\dl_QD^{\done}{}_{\done}=-{i\over4}\l[
\cD_a\bar\la^{\dtwo}\g^a\ve+g\l[\s,\,\bar\la^{\dtwo}\ve\r]
+{1\over8\a}G_{ab}\bar\la^{\dtwo}\g^{ab}\ve+\hf{t}_{ab}\bar\la^{\dtwo}\g^{ab}\ve
-{3\over4}S_{ij}\l(\bar{\s}^{ij}\r)^1{}_1\bar\la^{\dtwo}\ve\r],
\nn\end{eqnarray}
and from (\ref{HyperSUSY}) that for the hypermultiplets, 
\begin{eqnarray}
&&\dl_Q\phi^i=\qt\l(\bar\s^i\r)^2{}_{\ta}\,\bar{\psi}^{\ta}\ve,
\qquad
\dl_QF^{\ta}{}_{1}=0,
\nn\\
&&\dl_QF^{\ta}{}_{2}={i\over2}\check\ve^TC\l[
\g^a\cD_a\psi^{\ta}-g\l[\s,\,\psi^{\ta}\r]
+g\l(\s_i\r)^{\ta}{}_{\dg}\l[\phi^i,\,\la^{\dg}\r]
+\hf\l(t_{ab}-{1\over4\a}G_{ab}\r)\g^{ab}\psi^{\ta}
\r],
\label{HyperBRST}\\
&&\dl_Q\psi^{\ta}=i\l(\s_i\r)^{\ta}{}_{1}
\l[\cD_a\phi^i\g^a+g\l[\s,\,\phi^i\r]-{1\over2\a}G_{ab}\phi^i\g^{ab}
-t_{ab}\phi^i\g^{ab}-\l(S_{ij}+\ve_{ijkl}S^{kl}\r)\phi^j\r]\ve
+F^{\ta}{}_{1}\check\ve,
\nn
\end{eqnarray}
where $\bar{\la}^{\da}$ is an abbreviation for $(\la^{\da})^TC$, 
$\bar\ve$ is for $\ve^TC$.

The algebra of the supersymmetry transformations in (\ref{OffShellAlgGauge}), 
(\ref{OffShellAlgHyper}) and (\ref{OffShellAlgF}) may be used to check 
the nilpotency of the BRST transformation, 
assuming that (\ref{orthogonaldotcheck}) and (\ref{Offshellcondition}) are 
satisfied. 
Substituting 
$$
\dl_\e\Phi=\Upsilon\,\dl_Q\Phi, 
\qquad
\dl_\eta\Phi=\Upsilon'\,\dl_Q\Phi, 
$$
into (\ref{OffShellAlgGauge}-\ref{OffShellAlgF}), 
through the relation 
$$
\l[\dl_\e,\,\dl_\eta\r]\Phi=2\Upsilon\Upsilon'\,(\dl_Q)^2\Phi,
$$
one can compute $(\dl_Q)^2\Phi$.
Since $\ve^{\dot{2}}=\eta^{\dot{2}}=0$, 
$$
\bar\eta\c\ve
=(\eta^{\dot{1}})^TC\ve^{\dot{2}}-(\eta^{\dot{2}})^TC\ve^{\dot{1}}
=0,
\qquad
\xi^a=\bar\eta\c\g^a\ve
=(\eta^{\dot{1}})^TC\g^a\ve^{\dot{2}}-(\eta^{\dot{2}})^TC\g^a\ve^{\dot{1}}
=0,
$$
so that 
$\bar\eta\c\ve$ and $\xi^a$ on the right hand sides 
of (\ref{OffShellAlgGauge}-\ref{OffShellAlgF}) are zero. 
Recalling that $\ve$ is chiral - $i\g^{45}\ve=\ve$ - on $\RS$, one can see that 
$$
\bar\eta\c\bar{\s}_{ij}\g^{ab}\e
=-\Upsilon\Upsilon'\,\l(\bar{\s}_{ij}\r){}^2{}_1\,\ve^TC\g^{ab}\ve
\quad
(a,b=1,2,3),
\qquad
\bar\eta\c\bar{\s}_{ij}\g^{45}\e
=-\Upsilon\Upsilon'\,\l(\bar{\s}_{ij}\r){}^2{}_1\,\ve^TC\g^{45}\ve=0.
$$
However, since $\bar\eta\c\bar{\s}_{ij}\g^{a4}\e$ and 
$\bar\eta\c\bar{\s}_{ij}\g^{a5}\e$ are not necessarily zero 
for a generic background, the BRST transformation isn't always nilpotent.  
But, for the backgrounds of our interest in this paper, 
since there are no fields carring the mixed components 
tangent to the 3-sphere and to the Riemann surface 
at the same time, one can find that it is nilpontent. 

Let us now take the shift (\ref{Dshift}) into account. Although it never 
affects the nilpotency of the BRST transformation $\dl_Q$, it does affect 
$\dl_Q\la^{\done}$ and $\dl_Q{D}^{\done}{}_{\done}$, 
\begin{eqnarray*}
&&\dl_Q\la^{\done}
=-\hf\sum_{(a,b)\not=(4,5),(5,4)}
\l({F}_{ab}+{i\over\a}{G}_{ab}\s\r)\g^{ab}\ve
+i\cD_a\s\,\g^a\ve+D^{\done}{}_{\done}\ve,
\nn\\
&&\dl_QD^{\done}{}_{\done}=-{i\over4}\l[
\sum_{m=1}^{3}\cD_m\bar\la^{\dtwo}\g^m\ve+g\l[\s,\,\bar\la^{\dtwo}\ve\r]
-{i\over\a}G_{45}\bar\la^{\dtwo}\ve
+\hf\l({t}_{ab}+{1\over4\a}G_{ab}\r)\bar\la^{\dtwo}\g^{ab}\ve
-{3\over4}S_{ij}\l(\bar{\s}^{ij}\r)^1{}_1\bar\la^{\dtwo}\ve\r],
\end{eqnarray*}
where we have assumed that the Killing spinor $\ve$ obeys  
$i\g_{45}\ve=\ve$ and $\cD_4\ve=\cD_5\ve=0$. These conditions are 
satisfied by the Killing spinors in section \ref{Backgrounds}. 

The partition function of the theory with the action ${\cal S}$ 
in (\ref{OffShellAction}) 
$$
Z=\int\l[d\Phi\r]\, \exp\l({\cal S}\r)
$$
is invariant under a deformation of the action 
$$
Z=\int\l[d\Phi\r]\, \exp\l({\cal S}+{t}{\cal S}_Q\r)
$$
by the `regulator' action 
$$
{\cal S}_Q=\dl_Q\Psi,
$$
which is the BRST transform of a functional $\Psi$ of the fields, 
with a parameter $t$. 
More explicitly, we will choose the regulator action to be
$$
{\cal S}_Q=-\int{d^5x\sqrt{g}}\,\dl_Q\Tr\l[
\l(\dl_Q\la^{\da}\r)^\dag\c\la^{\da}+\l(\dl_Q\psi^{\ta}\r)^\dag\c\psi^{\ta}
\r].
$$

Since the partition function $Z$ never depends on the parameter $t$, 
one can take a large $t$ limit, while leaving the value of $Z$ intact.  
In the large $t$ limit, the main contributions to $Z$ comes from 
the fixed points of the fields given by $\dl_Q\la^{\da}=0$ and 
$\dl_Q\psi^{\ta}=0$. Then, writing the fields $\Phi$ in terms of 
quantum fluctuations $\tilde{\Phi}$ about one of the fixed points $\Phi_0$ 
as 
$$
\Phi=\Phi_0+{1\over\sqrt{t}}\tilde{\Phi},
$$
and interating over the fluctuations to carry out the one-loop computation, 
one may compute the partition function $Z$ exactly.

In order to carry out the localization, it is convenient to 
convert spinor and vector fields to scalar fields\footnote{
They are scalar fields on the 3-spheres, but not necessarily scalar 
fields on $\RS$. We will see the spin content of them on $\RS$, below.
} on the 3-spheres, and then 
there is no need to introduce spinor or vector spherical harmonics 
on the three-spheres. 

For the $\N=1$ hypermultiplet, 
\begin{eqnarray*}
&&\psi^{\tilde{1}}
=\l(\chi\otimes\e+\xi\otimes\e^c\r)\otimes\zeta_+
+\l(\eta\otimes\e+\kappa\otimes\e^c\r)\otimes\zeta_-,
\nn\\
&&\psi^{\tilde{2}}
=\l(\tchi\otimes\e+\txi\otimes\e^c\r)\otimes\zeta_+
+\l(\teta\otimes\e+\tkappa\otimes\e^c\r)\otimes\zeta_-,
\nn\\
&&\tilde{H}={1\over\sqrt{2}}\l(\s_i\r)^1{}_{1}\phi^i
=-{i\over\sqrt{2}}\l(\phi^3+i\phi^4\r),
\qquad
{H}={1\over\sqrt{2}}\l(\bar{\s}_i\r)^2{}_{1}\phi^i
={i\over\sqrt{2}}\l(\phi^1+i\phi^2\r).
\nn\\
&&\tilde{H}^*={1\over\sqrt{2}}\l(\bar{\s}_i\r)^1{}_{1}\phi^i
={i\over\sqrt{2}}\l(\phi^3-i\phi^4\r),
\qquad
{H}^*={1\over\sqrt{2}}\l({\s}_i\r)^1{}_{2}\phi^i
=-{i\over\sqrt{2}}\l(\phi^1-i\phi^2\r).
\end{eqnarray*}
Note that $\e$ and $\e^c$ are the Killing spinors on each of the $S^3$s  
and that they are linearly independent as two-component vectors, 
as discussed in Appendix \ref{3-sphere}. 
The fields ($\chi$, $\xi$, $\eta$, $\kappa$) and 
($\tchi$, $\txi$, $\teta$, $\tkappa$) are scalar fields on the three-spheres.

For the $\N=1$ vector multiplet, 
\begin{eqnarray*}
&&\la^1
=\l(\xi\,\e+\eta\,\e^c\r)\otimes\zeta_+
+\l(\vp\,\e+\chi\,\e^c\r)\otimes\zeta_-,
\nn\\
&&\l(\la^2\r)^T=-\l[
\l(\tvp\,\e^\dag+\tchi\,{\e^c}^\dag\r)\otimes\zeta_+
+\l(\txi\,\e^\dag+\teta\,{\e^c}^\dag\r)\otimes\zeta_-
\r]C^{-1},
\nn\\
&&A_m=\l(\e^\dag\t_m\e\r)V_0+\l({\e^c}^\dag\t_m\e\r)V_-
+\l(\e^\dag\t_m\e^c\r)V_+,
\end{eqnarray*}
for $m=1,2,3$. The three-component vectors $\l(\e^\dag\t_m\e\r)$, 
$\l({\e^c}^\dag\t_m\e\r)$, and $\l(\e^\dag\t_m\e^c\r)$ are orthogonal 
among them, and are normalized by $\e^\dag\e={\e^c}^\dag\e^c=1$. 
See Appendix \ref{3-sphere} in more detail. 
Let us recall that the background on the squashed $S^3$ in subsection 
\ref{ImamuraSquash} is not up to our mind here, because we will leave 
the calculation of the partition function for the background undone 
in this paper, as explained in Introduction. 

In order to denote the scalar fields in the gauge multiplet, 
we use the same Greek letters $\chi$, $\xi$, $\eta$, as for the ones 
in the hypermultiplet. 
But, we never mean that they are the same fields. What it really means is 
the shortage of the Greek letters we can assign to each of the fields. 
We will compute the one-loop contributions from the gauge multiplet and 
the hypermultiplet, separately. Therefore, we believe and hope that it 
wouldn't cause any confusion. 

\subsection{The $\N=2$ Twisting and the $\N=1$ Twisting}

As we have seen in section \ref{Backgrounds}, 
the $\N=2$ twisting by turing on $A^{12}$ only gives rise to 
the $\N=2$ supersymmetric backgrounds on a round and a squashed $S^3$, 
and on the other hand, the $\N=1$ twisting by turning on $A^{12}$ and 
$A^{34}$ with $A^{12}=A^{34}$ gives rise to the $\N=1$ supersymmetric 
backgrounds on a round, a squashed and an ellipsoid $S^3$. 
The difference between the $\N=2$ twisting and the $\N=1$ twisting 
has no effects on the BRST transformation of the $\N=1$ gauge multiplet, 
but affects the transformation of the $\N=1$ hypermultiplet. 
Therefore, the one-loop 
contributions from the $\N=1$ gauge multiplet don't depend on which 
twisting is done and yield the same results on an identical sphere. 

Therefore, before proceeding to the one-loop calculations, 
let us see how the spin content of the two-dimensional fields 
in the hypermultiplet is changed upon each of the twistings. 
Then, we will see the spin content of the $\N=1$ gauge multiplet 
after the twistings, too. 

The spin content of the two-dimensional fields on $\RS$ from the hypermultiplet 
can be read from the covariant derivatives of the component fields of 
the hypermultiplet 
along the surface $\RS$ with the local coordinates $(x^4,x^5)$. 

For the $\N=2$ twisting,
\begin{eqnarray*}
\cD_z\psi^{\ta}&=&\hf\l(\cD_4\psi^{\ta}-i\cD_5\psi^{\ta}\r)
=\d_z\psi^{\ta}+\hf\om_z{}^{45}\g_{45}\psi^{\ta}
+\hf\om_z{}^{45}\l(\s_{12}\r)^{\ta}{}_{\tb}\psi^{\tb}
+ig\l[A_z,\,\psi^{\ta}\r],
\nn\\
\cD_z\phi^1&=&\d_z\phi^1+\om_z{}^{45}\phi^2+ig\l[A_z,\,\phi^1\r],
\qquad
\cD_z\phi^2=\d_z\phi^2-\om_z{}^{45}\phi^1+ig\l[A_z,\,\phi^2\r],
\nn\\
\cD_z\phi^3&=&\d_z\phi^3+ig\l[A_z,\,\phi^3\r],
\qquad
\cD_z\phi^4=\d_z\phi^4+ig\l[A_z,\,\phi^4\r],
\end{eqnarray*}
where the complex coordinate $z$ is defined by $z=x^4+ix^5$ and 
$\d_z=(\d_4-i\d_5)/2$.

Therefore, we can see that
the bosonic field $\tilde{H}\sim(\phi^3+i\phi^4)$ 
remains a scalar under the $\N=2$ twisting, 
but $H\sim(\phi^1+i\phi^2)$ becomes a $(0,1)$-form; $H\to{H}_{\bar{z}}$. 
The fermionic fields $(\chi,\xi)$ are changed to be a scalar, 
and $(\eta, \kappa)$ to be $(1,0)$-forms; 
$(\eta, \kappa)\to(\eta_z, \kappa_z)$. 
On the other hand, the fermionic fields $(\tchi,\txi)$ give $(0,1)$-forms; 
$(\tchi,\txi)\to(\tchi_{\bar{z}},\txi_{\bar{z}})$, 
and $(\eta, \kappa)$ become scalars.

For the $\N=1$ twisting,
\begin{eqnarray*}
\cD_z\psi^{\ta}&=&\hf\l(\cD_4\psi^{\ta}-i\cD_5\psi^{\ta}\r)
=\d_z\psi^{\ta}+\hf\om_z{}^{45}\g_{45}\psi^{\ta}+ig\l[A_z,\,\psi^{\ta}\r],
\nn\\
\cD_z\phi^i&=&\d_z\phi^i+\hf\om_z{}^{45}\e^{ij}\phi^i+ig\l[A_z,\,\phi^i\r],
\end{eqnarray*}
for $i,j=1,\cdots,4$, 
where $\e^{ij}$ is an antisymmetric tensor with non-zero components 
$\e^{12}=-\e^{21}=\e^{34}=-\e^{43}=1$ only.

The bosonic fields ($\tilde{H}$, $H$) and the fermionic fields 
($\chi$, $\xi$), ($\tchi$, $\txi$)
become two-dimensional Weyl spinors of positive chirality, and 
($\eta$, $\kappa$) and ($\teta$, $\tkappa$) are Weyl spinors 
of negative chirality\footnote{In a two-dimensional Euclidean space, 
the complex conjugate of a Weyl spinor of negative chirality is 
of positive chirality. Therefore, essentially, after the $\N=1$ twisting, 
all the fields in the hypermultiplet carry the same spin on $\RS$. }. 

The results are summarized in Table \ref{twistingtable}.
The notation $(k,l)$ in the table denotes a $(k,l)$-form for 
integers $k$, $l$. For half integers $k$, $l$, $(\hf,0)$ denotes 
a Weyl spinor of positive chirality, and $(0,\hf)$ of negative chirality. 
Whichever $k$ and $l$ are integer or half-integer, 
the covariant derivative of a field $\Phi$ of $(k,l)$ carries 
the spin connection 
$\om^{45}$ of $\RS$ as
$$
\cD_z\Phi=\d_z\Phi+i(k-l)\om_z{}^{45}\Phi+ig\l[A_z,\,\Phi\r].
$$

\begin{table}[h]
\begin{center}
\begin{tabular}{|c|c|c|c|c|c|}\hline
 & &\multicolumn{2}{|c|}{the $\N=2$ twisting}
&\multicolumn{2}{|c|}{the $\N=1$ twisting}
\\ \hline \hline
{5D fields}&{scalars}&{spin} $(k,l)$&{charge} $q$&{spin} $(k,l)$&{charge} $q$ 
\\ \hline \hline
\multirow{4}{*}{$\phi^i$}
&$\tilde{H}$
& $(0,0)$ & 0 & $(\hf,0)$ & -1~
\\ \cline{2-6}
& $H$ & $(0,1)$ & -2~ & $(\hf,0)$ & -1~
\\ \cline{2-6}
& ${\tilde{H}}^*$ &  $(0,0)$ & 0 & $(0,\hf)$ & 1
\\ \cline{2-6}
& $H^*$ & $(1,0)$ & 2 & $(0,\hf)$ & 1
\\ \hline
\multirow{4}{*}{$\psi^{\tilde{1}}$}
&$\chi$
& $(0,0)$ & 0 & $(\hf,0)$ & -1~
\\ \cline{2-6}
& $\xi$ & $(0,0)$ & 2 & $(\hf,0)$ & 1
\\ \cline{2-6}
& $\eta$ & $(1,0)$  & 0 & $(0,\hf)$ & -1~
\\ \cline{2-6}
& $\kappa$ & $(1,0)$ & 2 & $(0,\hf)$ & 1
\\ \hline 
\multirow{4}{*}{$\psi^{\tilde{2}}$}
&$\tchi$ & $(0,1)$ & -2~ & $(\hf,0)$ & -1~
\\ \cline{2-6}
& $\txi$ & $(0,1)$ & 0 & $(\hf,0)$ & 1
\\ \cline{2-6}
& $\teta$ & $(0,0)$ & -2~ & $(0,\hf)$ & -1~
\\ \cline{2-6}
& $\tkappa$ & $(0,0)$ & 0 & $(0,\hf)$ & 1
\\ \hline
\end{tabular}
\end{center}
\caption{The scalar fields on the $S^3$'s of the hypermultiplet 
upon the twistings.}
\label{twistingtable}
\end{table}

On a squashed and an ellipsoid $S^3$, as we have seen in subsection 
\ref{HosomichiSquash} and \ref{HosomichiEllipsoid}, we have also turned on 
the background field $A^i{}_j$ along the $S^3$. 
When the Killing spinors are reduced to the three-dimensional ones 
$\e$ and $\e^c$ on the spheres, we refer to the background $R$-symmetry field 
as $V$ so that the covariant derivatives of $\e$ and $\e^c$ are given by 
$$
\cD\e=\l(d+\qt\om_{mn}\t^{mn}+iV\r)\e,
\qquad
\cD\e^c=\l(d+\qt\om_{mn}\t^{mn}-iV\r)\e^c. 
$$

Then, on the squashed $S^3$ in subsection \ref{HosomichiSquash}, we have 
for the $\N=2$ twisting,
$$
A^{12}|_{S^3}={2\over\tr}\l(1-{\tr^2\over{r}^2}\r)e^3=-2V,
$$
and for the $\N-1$ twisting, 
$$
A^{12}|_{S^3}=A^{34}|_{S^3}={1\over\tr}\l(1-{\tr^2\over{r}^2}\r)e^3=-V,
$$
where $|_{S^3}$ denotes the components along the $S^3$. 

On an ellipsoid $S^3$ in subsection \ref{HosomichiEllipsoid}, 
the $\N=1$ twisting causes along the $S^3$ the background field 
$$
A^{12}|_{S^3}=A^{34}|_{S^3}=-V,
$$
which is identical to $V$ given in (\ref{bgV}).

When a two-dimensional field $\Phi$ has the covariant derivative
$$
\cD\Phi|_{S^3}=d\Phi|_{S^3}+iq{V}\Phi, 
$$
we will say that the field $\Phi$ carries charge $q$ 
under the background field $V$. 
The charges of the two-dimensional fields from the hypermultiplet are 
listed in Table \ref{twistingtable}. 

\begin{table}[h]
\begin{center}
\begin{tabular}{|c||c|c||c|c||c|c||c|c||c|c|c|c|c|c|}\hline
{2D fields}& $\vp$ & $\chi$ & $\tvp$ & $\tchi$ & $\xi$ & $\eta$ & 
$\txi$ & $\teta$ & $V_0$ & $V_+$ & $V_-$ 
\\ \hline
{spin} $(k,l)$ & \multicolumn{2}{|c||}{$(1,0)$} 
& \multicolumn{2}{|c||}{$(0,1)$}
& \multicolumn{2}{|c||}{$(0,0)$}
& \multicolumn{2}{|c||}{$(0,0)$}
& \multicolumn{3}{|c|}{$(0,0)$}
\\ \hline 
{charge} $q$ & 0 & 2 & 0 & -2 & 0 & 2 & 0 & -2 & 0 & 2 & -2
\\ \hline 
\end{tabular}
\end{center}
\caption{The charges $q$ of the two-dimensional fields 
from the gauge multiplet under the background $V$.}
\label{ChargeGaugeMultiplet}
\end{table}

Let us turn to the two-dimensional fields on the three-spheres 
in the $\N=1$ gauge multiplet and 
see how the spin content of them is changed under the twisting. 
As mentioned before, both of the $\N=2$ and $\N=1$ twistings affect 
the spin content of them in the same way.

The two-dimensional fields 
from the $\N=1$ gauge multiplet are also charged 
under the gauge field $V$. But, the charges of them don't depend upon 
which twisting we perform. The charges under $V$ which two-dimensional fields 
from the $\N=1$ gauge multiplet carry are also listed 
in Table \ref{ChargeGaugeMultiplet}.

\section{Localization on the Round and Squashed $S^3$} 
\label{SquashLocalization}

In this section, we will compute the partition function by localization for 
the backgrounds on the squashed $S^3$ discussed in subsection 
\ref{HosomichiSquash}. In the round limit $\tr\to{r}$, 
we will see that the previous results in \cite{FKM,KM} are regained for 
the $\N=1$ twisting, and we will obtain new results for the $\N=2$ twisting 
on the round $S^3$ in subsection \ref{newR3Sbg} and on the squashed $S^3$ in 
subsection \ref{HosomichiSquash}. 

As mentioned before, in order to carry out localization, 
we need to find fixed points of the regulator action 
${\cal S}_Q$, which are given by $\dl_Q\la^{\da}=0$ and $\dl_Q\psi^{\ta}=0$. 
In the squashed $S^3$ background in subsection \ref{HosomichiSquash}, 
the former conditions gives 
\begin{eqnarray*}
&&\dl_Q\la^{\done}
=-\hf\sum_{(a,b)\not=(4,5),(5,4)}{F}_{ab}\g^{ab}\ve
+i\cD_a\s\,\g^a\ve+D^{\done}{}_{\done}\ve=0,
\qquad
\dl_Q\la^{\dtwo}=D^{\dtwo}{}_{\done}\,\ve=0,
\end{eqnarray*}
which are reduced to
\begin{eqnarray*}
\hf\e_{mkl}F_{kl}=\cD_m\s, 
\qquad
\l[F_{mz}\t^m-i\cD_z\s\r]\e=0, 
\qquad
D^{\da}{}_{\db}=0,
\end{eqnarray*}
with the complex coordinate $z=x^4+ix^5$ combining the local coordinates 
$(x^4,x^5)$ of $\RS$. 

The first equation means that 
$$
A_m=0, \qquad e^m\d_m\s=0 \quad\to\quad \s=\s(z,\bar{z}),
$$
and the second equation in turn implys that 
$$
\cD_z\s=\d_z\s+ig\l[A_z,\,\s\r]=0,
\qquad
e^m\d_mA_z=0 \quad\to\quad A_z=A_z(z,\bar{z}).
$$
We will `diagonalize' the scalar field $\s$ at the fixed point 
by partial gauge fixing,
$$
\s(z,\bar{z})=\sum_{i=1}^{r}\s^i(z,\bar{z})\,H_i,
$$
where $H_i$ ($i=1,\cdots,r$) are the generators of the Cartan subalgebra 
of the gauge group $G$ with $r$ the rank of $G$. It then follows from 
$\cD_z\s=0$ that the gauge field $A_z$ takes values in the Cartan subalgebra, 
too,
$$
A_z(z,\bar{z})=\sum_{i=1}^{r}A_z^i(z,\bar{z})\,H_i,
$$
and that $\d_z\s^i=0$, {\it i.e.}, the solution $\s^i$ is a constant.

As for the latter conditions $\dl_Q\psi^{\ta}=0$, a simple examination 
shows that the solution to $\dl_Q\psi^{\ta}=0$ is given by 
$$
\tilde{H}=H=0, 
\qquad 
F^1{}_1=2\sqrt{2}\cD_z\tilde{H}=0,
\qquad
F^2{}_1=-2\sqrt{2}\cD_z{H}=0,
$$
for both of the $\N=2$ and $\N=1$ twistings.

We will proceed to calculate the one-loop contributions about the fixed points 
of the regulator action ${\cal S}_Q$ in the next two subsections.

\subsection{One-Loop Contributions from the $\N=1$ Gauge Multiplet}
\label{SquashOne-LoopGaugeMultiplet}

The BRST transformations of the $\N=1$ gauge multiplet are the same 
for both of the $\N=2$ and $\N=1$ twistings. 
As discussed in section \ref{localization}, we would like to 
reduce all the component fields in the gauge multiplet into scalar fields 
on the $S^3$. 

In particular, when we will convert the gauge field $A_m$ to 
$V_0$ and $V_\pm$, the field strength 
$F_{mn}=\cD_mA_n-\cD_nA_m+ig\l[A_m,\,A_n\r]$ may be rewritten in terms of them 
as
\begin{eqnarray*}
\hf\e_{mkl}\l({\e}^\dag\t_m\e\r)F_{kl}=\e_{mkl}\l({\e}^\dag\t_m\e\r)\cD_{k}A_{l}
+\cdots
={2\tr\over{r}^2}V_0+i\l({\e^c}^\dag\t_m\e\r)\cD^{(-2)}_mV_{-}
-i\l({\e}^\dag\t_m\e^c\r)\cD^{(2)}_mV_+
+\cdots,
\nn\\
\hf\e_{mkl}\l({\e^c}^\dag\t_m\e\r)F_{kl}
=\e_{mkl}\l({\e^c}^\dag\t_m\e\r)\cD_{k}A_{l}+\cdots
={4\tr\over{r}^2}V_++2i\l({\e}^\dag\t_m\e\r)\cD^{(2)}_mV_{+}
-i\l({\e^c}^\dag\t_m\e\r)\cD^{(0)}_mV_0
+\cdots,
\end{eqnarray*}
where the ellipsis stands for the gauge interaction terms, which 
gives no contributions to the partition function in the large $t$ limit, 
and we will omit them. Note that the formulas (\ref{squashcross}) were used 
to derive these. Also omitting the gauge interaction terms, the field strength 
$F_{mz}$ is given in terms of this language by 
\begin{eqnarray*}
F_{mz}
&=&
\l({\e^c}^\dag\t_m\e\r)\l[\hf\l({\e}^\dag\t^n\e^c\r)\cD^{(0)}_nA_z-\cD_zV_-\r]
+\l({\e}^\dag\t_m\e^c\r)\l[\hf\l({\e^c}^\dag\t^n\e\r)\cD^{(0)}_nA_z-\cD_zV_+\r]
\nn\\
&&+\l({\e}^\dag\t_m\e\r)\Big[\l({\e}^\dag\t^n\e\r)\cD^{(0)}_nA_z-\cD_zV_0\Big].
\end{eqnarray*}

After the conversion, we can see that the BRST transformation of 
the bosonic fields is given by 
\begin{eqnarray*}
&&\dl_Q\ts=-\qt\txi,
\qquad
\dl_QV_0=-{i\over4}\txi,
\quad
\dl_QV_-=-{i\over4}\teta,
\quad
\dl_QV_+=0,
\qquad
\dl_QA_{\bar{z}}=\qt\tvp, 
\quad
\dl_QA_{z}=0,
\nn\\
&&\dl_Q{D}^1{}_1={i\over4}\l[
\l({\e}^\dag\t^m\e\r)\cD^{(0)}_m\txi-2i{\tr\over{r}^2}\txi+g\l[\s,\,\txi\r]
+\l({\e^c}^\dag\t^m\e\r)\cD^{(-2)}_m\teta\r],
\nn\\
&&\dl_Q{D}^1{}_2={i\over4}\l[
-\l({\e}^\dag\t^m\e\r)\cD^{(2)}_m\chi+2i{\tr\over{r}^2}\chi-g\l[\s,\,\chi\r]
+\l({\e^c}^\dag\t^m\e\r)\cD^{(0)}_m\vp-2i\cD_z\eta
\r],
\end{eqnarray*}
where we denote a fixed point of the scalar field $\s$ as the same letter $\s$, 
and the fluctuation about this fixed point $\s$ as $\ts$. 
Henceforth, we will keep this notation until the end of this section. 

The BRST tranformation of the fermionic fields is given by
\begin{eqnarray*}
&&\dl_Q\txi=0, \qquad \dl_Q\teta=0, 
\qquad 
\dl_Q\tvp=0,\qquad \dl_Q\tchi=-D^2{}_1,
\nn\\
&&\dl_Q\xi
=-2i{\tr\over{r}^2}V_0+g\l[\s,\,V_0\r]+i\l({\e}^\dag\t^m\e\r)\d_m\ts
+\l({\e^c}^\dag\t^m\e\r)\cD^{(-2)}_mV_-
-\l({\e}^\dag\t^m\e^c\r)\cD^{(2)}_mV_++D^1{}_1,
\nn\\
&&\dl_Q\eta
=-4i{\tr\over{r}^2}V_++2g\l[\s,\,V_+\r]
+2\l({\e}^\dag\t^m\e\r)\cD^{(2)}_mV_+
+i\l({\e^c}^\dag\t^m\e\r)\d_m\ts
-\l({\e^c}^\dag\t^m\e\r)\d_mV_0,
\nn\\
&&\dl_Q\vp=2i\l[\l({\e^\dag}\t^m\e\r)\d_mA_z+g\l[\s,\,A_z\r]
-\cD_zV_0+i\cD_z\ts\r],
\qquad
\dl_Q\chi=2i\l[\l({\e^c}^\dag\t^m\e\r)\d_mA_z-2\cD_zV_+\r]. 
\end{eqnarray*}

Taking account of $\l(V_\pm\r)^\dag=V_\mp$, we deduce that 
\begin{eqnarray}
&&
\dl_Q\l(\dl_Q\xi\r)^\dag
=\hf\l[
2{\tr\over{r}^2}\txi+ig\l[\s,\,\txi\r]+i\l({\e}^\dag\t^m\e\r)\d_m\txi
+i\l({\e^c}^\dag\t^m\e\r)\cD^{(-2)}_m\teta\r],
\nn\\
&&\dl_Q\l(\dl_Q\eta\r)^\dag
=-{i\over2}\l[
2i{\tr\over{r}^2}\teta-g\l[\s,\,\teta\r]
+\l({\e}^\dag\t^m\e\r)\cD^{(-2)}_m\teta
-\l({\e}^\dag\t^m\e^c\r)\cD^{}_m\txi
\r],
\label{dlQdagdlQfermigauge}\\
&&\dl_Q\l(\dl_Q\vp\r)^\dag
=-{i\over2}\l[\l({\e^\dag}\t^m\e\r)\d_m\tvp-g\l[\s,\,\tvp\r]
+2i\cD_{\bar{z}}\txi\r],
\quad
\dl_Q\l(\dl_Q\chi\r)^\dag=-{i\over2}
\l[\l({\e}^\dag\t^m\e^c\r)\d_m\tvp+2i\cD_{\bar{z}}\teta\r], 
\nn\\
&&\dl_Q\l(\dl_Q\tchi\r)^\dag=
-{i\over2}\l[
-\l({\e}^\dag\t^m\e\r)\cD_m\chi+2i{\tr\over{r}^2}\chi-g\l[\s,\,\chi\r]
+\l({\e^c}^\dag\t^m\e\r)\cD_m\vp-2i\cD_z\eta
\r].
\nn\end{eqnarray}

Each of these fluctuations is in the adjoint representation of the gauge 
group $G$, whose Cartan generators we denote as $H_i$ ($i=1,\cdots,r$) 
with r the rank of $G$, and the remaining generators as $E_\a$ with 
$\a$ a root of $G$. We assume that they obey 
$$
\l[H_i,\,E_\a\r]=\a_i E_\a,
\qquad
\l[E_\a,\,E_{-\a}\r]=\sum_{i=1}^{r}\a_iH_i\equiv\a\c{H},
$$
and are normalized as 
$$
\Tr\l[H_iH_j\r]=\dl_{i,j}, 
\qquad
\Tr\l[E_{-\a}E_\a\r]=1. 
$$

Since the fluctuations have no interactions in the large $t$ limit, 
the fluctuations taking values in the Cartan subalgebra are decoupled from 
the remaining sector, and they yield an overall constant to the partition 
function. We are interested in the dependence of the partition function 
on the value $\s$ at one of the fixed points, and therefore we will 
focus on the remaining sector, where the fluctuations are expanded in terms of 
the basis $\{E_{\a}\}_{\a\in\La}$ with $\La$ the set of all the roots of $G$. 

We then assume that $\sa=\sum_{i=1}^{r}\s_i\a^i$ is non-zero for a generic 
$(\s^1,\cdots,\s^r)$. It implys that the operator $\l[\s,\,\c\r]$ acting 
on the sector we are interested in is invertible, and the following 
shifts are allowed to be done:
\begin{eqnarray*}
V_0 \quad\to\quad V_0-i{1\over{g}\l[\s,\,\c\r]}\l({\e}^\dag\t^m\e\r)\d_m\ts,
\qquad
V_+ \quad\to\quad 
V_+-{i\over2}{1\over{g}\l[\s,\,\c\r]}\l({\e^c}^\dag\t^m\e\r)\d_m\ts,
\nn\\
V_- \quad\to\quad 
V_--{i\over2}{1\over{g}\l[\s,\,\c\r]}\l({\e}^\dag\t^m\e^c\r)\d_m\ts,
\qquad
A_z \quad\to\quad 
A_z-{i}{1\over{g}\l[\s,\,\c\r]}\d_z\ts,
\end{eqnarray*}
which enables us to `gauge away' the fluctuation $\ts$ in the above BRST 
transformation. In order to ensure this, we need to use 
(\ref{CommRelDifOpSquashS}) in Appendix \ref{constKSSquashS}.

We would now like to contemplate the relation of the scalar $\s$ with 
a parameter $\th$ of the gauge transformation. In order to elucidate 
the discussion, we will refer to the value $\s$ at one of the fixed points 
as $\s_0$. Before `diagonalizing' $\s_0$, the scalar field $\s$ is given 
by the sum 
$$
\s=\s_0(z,\bar{z})+\ts,
$$
where the fluctuaion $\ts$ is defined as the non-zero modes on the $S^3$ 
so that 
$$
\ts=\sum_{l=1/2}^{\infty}\sum_{m,\tm=-l}^{l}
\ts_{l,m,\tm}(z,\bar{z})\vp_{l,m,\tm},
$$
with the scalar spherical harmonics $\vp_{l,m,\tm}$ 
($l=0,1/2,1,3/2,\cdots; -l\le{m},\tm\le{l}$) on the $S^3$. 
The fixed point $\s_0(z,\bar{z})$ therefore corresponds to 
the zero mode $\vp_{0,0,0}$ on the $S^3$. 
With the parameter $\th$ of the gauge transformation, the scalar field $\s$ 
is transformed infinitesimally as 
$$
\s \quad\to\quad \s+ig\l[\th,\s\r].
$$
The parameter $\th$ may be expanded in terms of the harmonics $\vp_{l,m,\tm}$, 
$$
\th
=\sum_{l=0}^{\infty}\sum_{m,\tm=-l}^{l}
\th_{l,m,\tm}(z,\bar{z})\vp_{l,m,\tm}
=\th_0(z,\bar{z})+\sum_{l=1/2}^{\infty}\sum_{m,\tm=-l}^{l}
\th_{l,m,\tm}(z,\bar{z})\vp_{l,m,\tm}.
$$

Since there is the correspondence between $\ts_{l,m,\tm}$ and 
$\th_{l,m,\tm}$ for $l\not=0$, the measure of the non-zero modes 
$\prod_{l=1/2}\prod_{-l\le{m},\tm\le{l}}[d\ts_{l,m,\tm}]$ in the path integral 
can be cancelled by the gauge degrees of freedom, 
$\prod_{l=1/2}\prod_{-l\le{m},\tm\le{l}}[d\th_{l,m,\tm}]$, 
if the fluctuation $\ts$ never appear in the integrand of the path integral. 
This is indeed the case, as we have seen above in the large $t$ limit. 

The remaining gauge degrees of freedom $\th_0(z,\bar{z})$ are used to 
`diagonalize' $\s_0(z,\bar{z})$, as explained before. As we have elucidated 
in the previous paper \cite{FKM}, 
the ratio of the measures $[d\s_0]/[d\th_0]$ gives rise to the Fadeev-Popov 
determinant 
\begin{eqnarray}
Z_{{\rm FP}}=\prod_{\a\in\La}\det_{(0,0)}\l[ig\sa\r]
=\int \l[d\bar{c}(z,\bar{z})dc(z,\bar{z})\r] 
\exp\l[-ig\sum_{\a\in\La}\int_{\RS} d^2z \sqrt{g_{\RS}}
\l(\s_0\c\a\r)\bar{c}_{-\a}c_{\a}\r],
\label{FPdetSquash}
\end{eqnarray}
with the Fadeev-Popov ghost $c_{\a}(z,\bar{z})$, $\bar{c}_{\a}(z,\bar{z})$ 
$(\a\in\La)$, which scalar fields on $\RS$. 

Thus, we will set $\ts$ to zero in the BRST transformations, and 
let us proceed to the evaluation of the one-loop determinants in 
the partition function. 

From the bosonic part of the gauge multiplet in the regulator action 
${\cal S}_Q$, 
\begin{eqnarray}
-\int d^5x\sqrt{g}\,\Tr\l[
\l(\dl_Q\xi\r)^\dag\c\dl_Q\xi+\l(\dl_Q\eta\r)^\dag\c\dl_Q\eta
+\l(\dl_Q\vp\r)^\dag\c\dl_Q\vp+\l(\dl_Q\chi\r)^\dag\c\dl_Q\chi
+\l(\dl_Q\tchi\r)^\dag\c\dl_Q\tchi
\r],
\label{bosongaugeSQ}
\end{eqnarray}
we can see that the auxiliuary fields $D^1{}_2$ and $D^2{}_1$ show up in 
the last term $\l(\dl_Q\tchi\r)^\dag\c\dl_Q\tchi\sim\l|D^2{}_1\r|^2$, 
and we will immediately integrated them out in the path integral. 
Furthermore,  we will also integrate out the auxiliuary field $D^1{}_1$, 
since it appears only in the first term 
$\l(\dl_Q\xi\r)^\dag\c\dl_Q\xi\sim\l|D^1{}_1\r|^2+\cdots$, with 
no $D^1{}_1$ in the ellipsis.  Then, the sum of the first term and the second 
term is reduced to
\begin{eqnarray}
&&-\int d^5x\sqrt{g}\,\Tr\bigg[
\Big|
-2i{\tr\over{r}^2}V_0+g\l[\s,\,V_0\r]
+\l({\e^c}^\dag\t^m\e\r)\cD^{(-2)}_mV_-
-\l({\e}^\dag\t^m\e^c\r)\cD^{(2)}_mV_+\Big|^2
\nn\\
&&\hskip3cm
+\Big|
-4i{\tr\over{r}^2}V_++2g\l[\s,\,V_+\r]
+2\l({\e}^\dag\t^m\e\r)\cD^{(2)}_mV_+
-\l({\e^c}^\dag\t^m\e\r)\d_mV_0\Big|^2
\bigg],
\label{V0pmSQ}
\end{eqnarray}
and the third and fourth terms are summed to yield 
\begin{eqnarray*}
-4\int d^5x\sqrt{g}\,\Tr\l[
\l|\l({\e}^\dag\t^m\e\r)\d_mA_z+g\l[\s,\,A_z\r]-\cD_zV_0\r|^2
+\l|\l({\e^c}^\dag\t^m\e\r)\d_mA_z-2\cD_zV_+\r|^2
\r],
\end{eqnarray*}
which we will integrate by parts to obtain
\begin{eqnarray*}
&&-4\int d^5x\sqrt{g}\,\Tr\Big[
A_z\Delta_0A_{\bar{z}}
+A_z\cD_{\bar{z}}\l(\l({\e}^\dag\t^m\e\r)\d_mV_0+g\l[\s,\,V_0\r]
+2\l({\e^c}^\dag\t^m\e\r)\cD^{(-2)}_mV_-\r)
\nn\\
&&\hskip 1cm
+\cD_{{z}}\l(\l({\e}^\dag\t^m\e\r)\d_mV_0-g\l[\s,\,V_0\r]
+2\l({\e}^\dag\t^m\e^c\r)\cD^{(2)}_mV_+\r){\c}A_{\bar{z}}
+\cD_zV_0\cD_{\bar{z}}V_0+4\cD_zV_+\cD_{\bar{z}}V_-
\Big],
\end{eqnarray*}
where $\Delta_0$ denotes the differential operator
$$
-\l[\l({\e}^\dag\t^m\e\r)\d_m+g\l[\s,\,\c\r]\r]
\l[\l({\e}^\dag\t^n\e\r)\d_n-g\l[\s,\,\c\r]\r]
-\l({\e^c}^\dag\t^m\e\r)\cD^{(-2)}_m\l({\e}^\dag\t^n\e^c\r)\d_n.
$$

Since the three differential operators
\begin{eqnarray}
\l({\e}^\dag\t^m\e\r)\d_m={2i\over\tr}L_3,
\quad
\l({\e^c}^\dag\t^m\e\r)\d_m={2i\over{r}}L_+,
\quad
\l({\e}^\dag\t^m\e^c\r)\d_m={2i\over{r}}L_-,
\label{DiffOp0L}
\end{eqnarray}
with $d=e^m\d_m$ obey the Lie algebra of $SU(2)$, 
$$
\l[L_3,\,L_{\pm}\r]={\pm}L_{\pm}, 
\qquad
\l[L_+,\,L_{-}\r]=2L_{3}, 
$$ 
we can regard $\Delta_0$ as
$$
{4\over\tr^2}L_3^2+{4\over{r}^2}L_+L_-+g^2\l[\s,\,\l[\s,\,\c\r]\r],
$$
which is potitive in the root sector expanded in the basis $\{E_{\a}\}$, 
and the operator $\Delta_0$ is invertible in the sector. 
Using the inverse of it, we will shift $A_z$ and $A_{\bar{z}}$ in the above 
integrand to give
\begin{eqnarray}
&&-4\int d^5x\sqrt{g}\,\Tr\Big[
A_z\Delta_0A_{\bar{z}}
+\cD_{{z}}J_+\c{1\over\Delta_{-2}}\l(\cD_{{z}}J_+\r)^\dag
\Big],
\label{AzSQ}
\end{eqnarray}
after integrations by parts, 
with 
$$
J_+=2\l(\l({\e}^\dag\t^m\e\r)\d_mV_++g\l[\s,\,V_+\r]-{2i}{\tr\over{r}^2}V_+\r)
-\l({\e^c}^\dag\t^m\e\r)\d_mV_0, 
$$
where we have defined the operator $\Delta_{-2}$ by 
\begin{eqnarray*}
-\l[\l({\e}^\dag\t^m\e\r)\cD^{(-2)}_m+2i{\tr\over{r}^2}+g\l[\s,\,\c\r]\r]
\l[\l({\e}^\dag\t^n\e\r)\cD^{(-2)}_n+2i{\tr\over{r}^2}-g\l[\s,\,\c\r]\r]
-\l({\e}^\dag\t^n\e^c\r)\cD^{(0)}_n\l({\e^c}^\dag\t^m\e\r)\cD^{(-2)}_m.
\end{eqnarray*}
From the definition, it is obvious that 
\begin{eqnarray}
&&\l({\e^c}^\dag\t^m\e\r)\cD^{(-2)}_m=\l({\e^c}^\dag\t^m\e\r)\d_m
={2i\over{r}}L_+,
\nn\\
&&\l({\e}^\dag\t^m\e\r)\cD^{(-2)}_m+2i{\tr\over{r}^2}
=\l({\e}^\dag\t^m\e\r)\l(\d_m+2i{1\over\tr}\l(1-{\tr^2\over{r}^2}\r)e^3_m\r)
+2i{\tr\over{r}^2}={2i\over\tr}\l(L_3+1\r).
\label{DiffOp-2L}
\end{eqnarray}
and so the operator $\Delta_{-2}$ may be rewritten as 
$$
{4\over\tr^2}\l(L_3+1\r)^2+g^2\l[\s,\,\l[\s,\,\c\r]\r]+{4\over{r}^2}L_-L_+.
$$
Therefore, with the same reason as for $\Delta_0$, we can see that 
the operator $\Delta_{-2}$ is invertible. 

To achieve the above expression (\ref{AzSQ}) of the integrand, we have used 
the formulas (\ref{CommRelDifOpSquashS}), repeatly.
In particular,  from (\ref{CommRelDifOpSquashS}), we can deduce more customized
formulas for this purpose, 
\begin{eqnarray}
D^{(q+2)}_0D^{(q)}_+=D^{(q)}_+\l[D^{(q)}_0+2i{\tr\over{r}^2}\r],
\qquad
D^{(q)}_-D^{(q)}_0=\l[D^{(q-2)}_0+2i{\tr\over{r}^2}\r]D^{(q)}_-,
\nn\\
D^{(0)}_0\Delta_0=\Delta_0D^{(0)}_0,
\qquad
D^{(0)}_-\Delta_0=\Delta_{-2}D^{(0)}_-,
\qquad
\Delta_0D^{(-2)}_+=D^{(-2)}_+\Delta_{-2},
\label{operatorsalgSU(2)}
\end{eqnarray}
with the abbreviations,
$$
D^{(q)}_0=\l({\e}^\dag\t^m\e\r)\cD^{(q)}_m,
\qquad
D^{(q)}_+=\l({\e^c}^\dag\t^m\e\r)\cD^{(q)}_m, 
\qquad
D^{(q)}_-=\l({\e}^\dag\t^m\e^c\r)\cD^{(q)}_m.
$$

In the sum of (\ref{V0pmSQ}) and (\ref{AzSQ}), we will shift $V_\pm$ as 
$$
V_\pm \quad\to\quad V_\pm+\hf{1\over{D}^{(\pm2)}_0\mp{2i}{\tr\over{r}^2}
\pm{g}\l[\s,\,\c\r]}D^{(0)}_{\pm}V_0.
$$
This shift is possible, because the operators acting on the root $E_\a$ 
$$
{D}^{(\pm2)}_0\mp{2i}{\tr\over{r}^2}\pm{g}\l[\s,\,\c\r]
={2i\over\tr}\l(L_3\mp1\r)\pm{g}\sa
$$
have no zero-modes for a generic $\sa$. Since the term in $\dl_Q\xi$, 
$$
{D}^{(2)}_-V_+-{D}^{(-2)}_+V_-+\l(2i{\tr\over{r}^2}-g\l[\s,\,\c\r]\r)V_0
$$
is shifted to become
${D}^{(2)}_-V_+-{D}^{(-2)}_+V_--{\cal K}_0V_0$, with 
${\cal K}_0V_0$ denoting 
\begin{eqnarray*}
{g\l[\s,\,\c\r]\over
{\l(D^{(0)}_0+g\l[\s,\,\c\r]\r)\l(D^{(0)}_0-g\l[\s,\,\c\r]\r)}}
\l[{\l(D^{(0)}_0-2i{\tr\over{r}^2}+g\l[\s,\,\c\r]\r)
\l(D^{(0)}_0-g\l[\s,\,\c\r]\r)}+D^{(-2)}_+D^{(0)}_-\r]V_0,~~~
\end{eqnarray*}

We obtain the integrand of the resulting sum of (\ref{V0pmSQ}) and 
(\ref{AzSQ}), after integrations by parts, 
\begin{eqnarray*}
&&A_z\Delta_0A_{\bar{z}}
-V_+{1\over\Delta_{-2}}\l[\Delta_{-2}-4\cD_{{z}}\cD_{\bar{z}}\r]
\l(D^{(-2)}_0+2i{\tr\over{r}^2}+g\l[\s,\,\c\r]\r)
\l(D^{(-2)}_0+2i{\tr\over{r}^2}-g\l[\s,\,\c\r]\r)V_-
\nn\\
&&+\qt\l|{D}^{(2)}_-V_+-{D}^{(-2)}_+V_--{\cal K}_0V_0\r|^2,
\end{eqnarray*}
where we will shift $V_0$ appropriately to eliminate the term 
${D}^{(2)}_-V_+-{D}^{(-2)}_+V_-$. This is possible, since 
the operator ${\cal K}_0$ is invertible in the same sense as explained above.
The last term in the above integrand then gives 
$$
-\qt\l({\cal K}_0V_0\r)^2. 
$$
Note that ${\cal K}_0V_0$ is pure imaginary, which may be ensured by using 
\begin{eqnarray}
&&
\l(D^{(0)}_0-2i{\tr\over{r}^2}\mp{g}\l[\s,\,\c\r]\r)
\l(D^{(0)}_0\pm{g}\l[\s,\,\c\r]\r)+D^{(-2)}_+D^{(0)}_-
\nn\\
&&\qquad=
\l(D^{(0)}_0+2i{\tr\over{r}^2}\pm{g}\l[\s,\,\c\r]\r)
\l(D^{(0)}_0\mp{g}\l[\s,\,\c\r]\r)+D^{(2)}_-D^{(0)}_+.
\label{hermitianD0}
\end{eqnarray}

We now see that the resulting integrand is `diagonalized', and 
it is a simple of matter to compute the one-loop determinants from 
the bosonic fields of the gauge multiplet,
\begin{eqnarray*}
Z_{{\rm V,B}}^{{\rm 1-loop}}=Z^{{\rm 1-loop}}_{{\rm V},0}
{\Det_{(0,0)}\l[\Delta_{-2}\r]\over\Det_{(0,1)}\l[\Delta_0\r]
\Det_{(0,0)}\l[\l[\Delta_{-2}-4\cD_{{z}}\cD_{\bar{z}}\r]
\l(D^{(-2)}_0+2i{\tr\over{r}^2}+g\l[\s,\,\c\r]\r)
\l(D^{(-2)}_0+2i{\tr\over{r}^2}-g\l[\s,\,\c\r]\r)
\r]},
\end{eqnarray*}
where the determinant $\Det_{(k,l)}[D]$ for an operator $D$ is defined by 
$$
{1\over\Det_{(k,l)}\l[D\r]}
=\int [d\vp^\dag][d\vp]\exp\l[-\int d^5x \sqrt{g}\, \Tr[\vp^\dag{D}\vp]\r],
$$
for a bosonic $(k,l)$-form field $\vp$ on $\RS$ and its partner $\vp^\dag$, 
both of which are also scalar fields on the $S^3$. 
We denote the one-loop contribution from $V_0$ as $Z_{{\rm V},0}$. 
Since $V_0$ is a real field; $V_0^\dag=V_0$, 
some care is required to integrate over it. Upon expanding it 
in terms of the basis $\{E_{\a}\}_{\a\in\La}$, we have 
$$
V_0=\sum_{i=1}^{r}V_0^i\,H_i+\sum_{\a\in\La}V_0^{\a}\,E_{\a}, 
$$
and the reality condition implys that $\l(V_0^{\a}\r)^\dag=V_0^{-\a}$. 
Therefore, $Z_{V,0}$ is given by the path integral 
\begin{eqnarray*}
\int \prod_{\a\in\La}\l[dV_0^{\a}\r]\exp\l[\qt\int d^5x \sqrt{g} 
\Tr\l[\l({\cal K}_0V_0\r)^2\r]
\r]
\end{eqnarray*}
with the exponent $\Tr\l[\l({\cal K}_0V_0\r)^2\r]$ expanded as 
\begin{eqnarray*}
&&-2\sum_{\a\in\La_+}
\bigg|{g\sa\over\l(D^{(0)}_0+{g}\sa\r)
\l(D^{(0)}_0-{g}\sa\r)}
\nn\\
&&\hskip4cm\times
\l[\l(D^{(0)}_0-2i{\tr\over{r}^2}+g\sa\r)
\Big(D^{(0)}_0-g\sa\Big)+D^{(-2)}_+D^{(0)}_-\r]V_0^\a
\bigg|^2,
\end{eqnarray*}
up to the Cartan part, 
with $\La_+$ the set of all the positive roots of $\La$. 
Taking account of (\ref{hermitianD0}), we will integrate it to obtain 
\begin{eqnarray*}
Z^{{\rm 1-loop}}_{V,0}&=&\prod_{\a\in\La_+}\det_{(0,0)}\l[
{g\sa\over\l(D^{(0)}_0+{g}\sa\r)
\l(D^{(0)}_0-{g}\sa\r)}\r]^{-2}
\nn\\
&&\qquad\times
\det_{(0,0)}\l[\l(D^{(0)}_0-2i{\tr\over{r}^2}+g\sa\r)
\Big(D^{(0)}_0-g\sa\Big)+D^{(-2)}_+D^{(0)}_-\r]^{-2}, 
\end{eqnarray*}
where we defined the determinant $\det_{(k,l)}[D]$ for an operator $D$ as
$$
{1\over\det_{(k,l)}\l[D\r]}
=\int [d\l(\vp_\a\r)^\dag]
[d\vp_\a]\exp\l[-\int d^5x \sqrt{g}\,[\l(\vp_\a\r)^\dag{D}\vp_\a]\r],
$$
for a bosonic $(k,l)$-form field $\vp_{\a}$ on $\RS$ and its partner 
$\l(\vp_\a\r)^\dag$, 
both of which are also scalar fields on the $S^3$. 
They are just one components of an adjoint $\vp$ in the expansion 
$$
\vp=\sum_{i=1}^{r}\vp_i{H}_i+\sum_{\a\in\La}\vp_\a{E}_\a.
$$
Therefore, up to an overall constant including the Cartan part, 
$$
Z^{{\rm 1-loop}}_{{\rm V},0}={1\over{Z}_{{\rm FP}}}\cdot
{\l(\Det_{(0,0)}\l[D_0^{(0)}-g\l[\s,\,\c\r]\r]\r)^2
\over
\Det_{(0,0)}\l[\l(D^{(0)}_0-2i{\tr\over{r}^2}+g\l[\s,\,\c\r]\r)
\Big(D^{(0)}_0-g\l[\s,\,\c\r]\Big)+D^{(-2)}_+D^{(0)}_-\r]}.
$$

Let us proceed to the one-loop contributions from the fermionic part 
of the gauge multiplet, of which the part in the regulator action ${\cal S}_Q$ 
is 
\begin{eqnarray*}
-\int d^5x\sqrt{g}\l(
\dl_Q\l(\dl_Q\xi\r)^\dag\c\xi+\dl_Q\l(\dl_Q\eta\r)^\dag\c\eta
+\dl_Q\l(\dl_Q\vp\r)^\dag\c\vp+\dl_Q\l(\dl_Q\chi\r)^\dag\c\chi
+\dl_Q\l(\dl_Q\tchi\r)^\dag\c\tchi\r).
\end{eqnarray*}

Substituting (\ref{dlQdagdlQfermigauge}) into this and integrating by parts, 
the first two terms become 
\begin{eqnarray*}
-\l({i\over2}\r)\l[
\raise1.5ex\hbox{$\begin{pmatrix}\txi, \teta\end{pmatrix}$}
\begin{pmatrix}
D^{(0)}_0+2i{\tr\over{r}^2}+g\l[\s,\c\r]
&
D^{(2)}_-
\cr
D^{(0)}_+
&
-D^{(2)}_0+2i{\tr\over{r}^2}+g\l[\s,\c\r]
\end{pmatrix}
\begin{pmatrix}\xi\cr\eta\end{pmatrix}\r],
\end{eqnarray*}
and the remaining terms yield 
\begin{eqnarray*}
\l({i\over2}\r)\bigg[
\raise1.5ex\hbox{$\begin{pmatrix}\tvp, \tchi\end{pmatrix}$}
\begin{pmatrix}
D^{(0)}_0-g\l[\s,\c\r]
&
D^{(2)}_-
\cr
D^{(0)}_+
&
-D^{(2)}_0+2i{\tr\over{r}^2}-g\l[\s,\c\r]
\end{pmatrix}
\begin{pmatrix}\vp\cr\chi\end{pmatrix}
\bigg]
+
\raise1.5ex\hbox{$\begin{pmatrix}\cD_{\bar{z}}\txi, 
\cD_{\bar{z}}\teta\end{pmatrix}$}
\begin{pmatrix}\vp\cr\chi\end{pmatrix}
+
\raise1.5ex\hbox{$\begin{pmatrix}\tvp,&\tchi\end{pmatrix}$}
\begin{pmatrix}0\cr\cD_z\chi\end{pmatrix}.
\end{eqnarray*}

Integrating over $\vp,\chi,\tvp$, and $\tchi$ to give the one-loop determinant 
\begin{eqnarray*}
\Det_{(1,0)}\l[
\begin{pmatrix}
D^{(0)}_0-g\l[\s,\c\r]
&
D^{(2)}_-
\cr
D^{(0)}_+
&
-D^{(2)}_0+2i{\tr\over{r}^2}-g\l[\s,\c\r]
\end{pmatrix}
\r],
\end{eqnarray*}
the latter terms in the integrand are reduced to
\begin{eqnarray*}
-{2i}\bigg[
\raise1.5ex\hbox{$\begin{pmatrix}\txi, \teta\end{pmatrix}$}
\begin{pmatrix}
D^{(0)}_0-g\l[\s,\c\r]
&
D^{(2)}_-
\cr
D^{(0)}_+
&
-D^{(2)}_0+2i{\tr\over{r}^2}-g\l[\s,\c\r]
\end{pmatrix}^{-1}
\begin{pmatrix}0\cr\cD_{\bar{z}}\cD_{z}\eta\end{pmatrix}
\bigg],
\end{eqnarray*}
after integration by parts. Summing this and the first two terms in the 
integrand results in 
\begin{eqnarray*}
-{i\over2}\bigg[
\raise1.5ex\hbox{$\begin{pmatrix}\txi, \teta\end{pmatrix}$}
\begin{pmatrix}
D^{(0)}_0-g\l[\s,\c\r]
&
D^{(2)}_-
\cr
D^{(0)}_+
&
-D^{(2)}_0+2i{\tr\over{r}^2}-g\l[\s,\c\r]
\end{pmatrix}^{-1}
\begin{pmatrix}
D_1
&
0
\cr
D_3
&
D_4
\end{pmatrix}
\begin{pmatrix}\xi\cr\eta\end{pmatrix}
\bigg],
\end{eqnarray*}
where the operators $D_1$, $D_3$, and $D_4$ denote 
\begin{eqnarray*}
D_1&=&
\Big(D^{(0)}_0-g\l[\s,\,\c\r]\Big)\l(D^{(0)}_0+2i{\tr\over{r}^2}
+g\l[\s,\,\c\r]\r)+D^{(2)}_-D^{(0)}_+, 
\nn\\
D_3&=&D^{(0)}_+\l(D^{(0)}_0+2i{\tr\over{r}^2}+g\l[\s,\,\c\r]\r)
-\l(D^{(2)}_0-2i{\tr\over{r}^2}+g\l[\s,\c\r]\r)D^{(0)}_+,
\nn\\
D_4&=&
4\cD_{\bar{z}}\cD_z
+\l(D^{(2)}_0-2i{\tr\over{r}^2}+g\l[\s,\,\c\r]\r)
\l(D^{(2)}_0-2i{\tr\over{r}^2}-g\l[\s,\,\c\r]\r)
+D^{(0)}_+D^{(2)}_-,
\end{eqnarray*}
and the zero in the top right component of the matrix is seen from the 
calculation
\begin{eqnarray*}
\Big(D^{(0)}_0-g\l[\s,\,\c\r]\Big)D^{(2)}_--D^{(2)}_-
\l(D^{(2)}_0-2i{\tr\over{r}^2}-g\l[\s,\,\c\r]\r)=0,
\end{eqnarray*}
with help of (\ref{operatorsalgSU(2)}). 
Integrating over $\xi$, $\eta$, $\txi$, and $\teta$, we obtain the 
one-loop determinants
\begin{eqnarray*}
{\Det_{(0,0)}\l[
\begin{pmatrix}
D_1
&
0
\cr
D_2
&
D_4
\end{pmatrix}
\r]\over
\Det_{(0,0)}\l[
\begin{pmatrix}
D^{(0)}_0-g\l[\s,\c\r]
&
D^{(2)}_-
\cr
D^{(0)}_+
&
-D^{(2)}_0+2i{\tr\over{r}^2}-g\l[\s,\c\r]
\end{pmatrix}
\r]}.
\end{eqnarray*}

Thus, we compute the one-loop contributions from the fermionic fields of 
the gauge multiplet, 
\begin{eqnarray*}
Z_{V,F}^{{\rm 1-loop}}=
{\Det_{(1,0)}\l[
\begin{pmatrix}
D^{(0)}_0-g\l[\s,\c\r]
&
D^{(2)}_-
\cr
D^{(0)}_+
&
-D^{(2)}_0+2i{\tr\over{r}^2}-g\l[\s,\c\r]
\end{pmatrix}
\r]
\over
\Det_{(0,0)}\l[
\begin{pmatrix}
D^{(0)}_0-g\l[\s,\c\r]
&
D^{(2)}_-
\cr
D^{(0)}_+
&
-D^{(2)}_0+2i{\tr\over{r}^2}-g\l[\s,\c\r]
\end{pmatrix}
\r]}
\Det_{(0,0)}\l[
\begin{pmatrix}
D_1
&
0
\cr
D_2
&
D_4
\end{pmatrix}
\r],
\end{eqnarray*}
where the last factor is easily evaluated as 
\begin{eqnarray*}
&&\Det_{(0,0)}\l[
\begin{pmatrix}
D_1&0
\cr
D_2&D_4
\end{pmatrix}
\r]
=
\Det_{(0,0)}\l[D_1\r]\Det_{(0,0)}\l[D_4\r]
\nn\\
&&=
\Det_{(0,0)}\l[\Big(D^{(0)}_0-g\l[\s,\,\c\r]\Big)\l(D^{(0)}_0+2i{\tr\over{r}^2}
+g\l[\s,\,\c\r]\r)+D^{(2)}_-D^{(0)}_+\r]
\nn\\
&&\quad\times
\Det_{(0,0)}\l[
4\cD_{\bar{z}}\cD_z
+\l(D^{(2)}_0-2i{\tr\over{r}^2}+g\l[\s,\,\c\r]\r)
\l(D^{(2)}_0-2i{\tr\over{r}^2}-g\l[\s,\,\c\r]\r)
+D^{(0)}_+D^{(2)}_-\r].
\end{eqnarray*}

Let us evaluate the determinant 
$$
\Det_{(k,l)}\l[
\begin{pmatrix}
D^{(0)}_0-g\l[\s,\c\r]
&
D^{(2)}_-
\cr
D^{(0)}_+
&
-D^{(2)}_0+2i{\tr\over{r}^2}-g\l[\s,\c\r]
\end{pmatrix}
\r]. 
$$
For four operators $A$, $B$, $C$, and $D$, we have the formula 
(see for example, \cite{KugoText})
\begin{eqnarray*}
{\Det}_{(k,l)}\l[\begin{pmatrix}
A&B\cr{C}&D
\end{pmatrix}\r]={\Det}_{(k,l)}\l[A-B{1\over{D}}C\r]{\Det}_{(k,l)}\l[D\r],
\end{eqnarray*}
for an invertible $D$. If there is another differential operator ${D'}$ 
satisfying the relation 
$$
B{1\over{D}}={1\over{D'}}B, 
$$
we then obtain the formula 
\begin{eqnarray*}
{\Det}_{(k,l)}\l[\begin{pmatrix}
A&B\cr{C}&D
\end{pmatrix}\r]={{\Det}_{(k,l)}\l[D\r]\over{\Det}_{(k,l)}\l[D'\r]}
{\Det}_{(k,l)}\l[{D'}A-BC\r].
\end{eqnarray*}
When we regard 
$$
B=D^{(2)}_-,
\qquad
D=-D^{(2)}_0+2i{\tr\over{r}^2}-g\l[\s,\c\r], 
$$
using (\ref{operatorsalgSU(2)}), we find the operator 
$$
D'=-D^{(0)}_0-g\l[\s,\c\r],
$$
and the determinant gives 
\begin{eqnarray*}
{\Det_{(k,l)}\l[D^{(2)}_0-2i{\tr\over{r}^2}+g\l[\s,\c\r]\r]
\over
\Det_{(k,l)}\l[D^{(0)}_0+g\l[\s,\c\r]\r]}
\Det_{(k,l)}\l[-\l(D^{(0)}_0+g\l[\s,\c\r]\r)\l(D^{(0)}_0-g\l[\s,\c\r]\r)
-D^{(2)}_-D^{(0)}_+\r].
\end{eqnarray*}

On the other hand, we also have the formula 
(see for example, \cite{KugoText})
\begin{eqnarray*}
{\Det}_{(k,l)}\l[\begin{pmatrix}
A&B\cr{C}&D
\end{pmatrix}\r]={\Det}_{(k,l)}\l[A\r]
{\Det}_{(k,l)}\l[D-C{1\over{A}}B\r],
\end{eqnarray*}
for an invertible $A$. If there is another differential operator ${A'}$ 
satisfying the relation 
$$
C{1\over{A}}={1\over{A'}}C, 
$$
we then obtain the formula 
\begin{eqnarray*}
{\Det}_{(k,l)}\l[\begin{pmatrix}
A&B\cr{C}&D
\end{pmatrix}\r]={{\Det}_{(k,l)}\l[A\r]\over{\Det}_{(k,l)}\l[A'\r]}
{\Det}_{(k,l)}\l[{A'}D-CB\r].
\end{eqnarray*}
If we then identify 
$$
A=D^{(0)}_0-g\l[\s,\c\r], 
\qquad
C=D^{(0)}_+,
$$
using (\ref{operatorsalgSU(2)}), we can find the operator 
$$
A'=D^{(2)}_0-2i{\tr\over{r}^2}-g\l[\s,\c\r], 
$$
and therefore, the same determinant have another expression
\begin{eqnarray*}
{\Det_{(k,l)}\l[D^{(0)}_0-g\l[\s,\c\r]\r]
\over
\Det_{(k,l)}\l[D^{(2)}_0-2i{\tr\over{r}^2}-g\l[\s,\c\r]\r]}
\Det_{(k,l)}\l[
-\l(D^{(2)}_0-2i{\tr\over{r}^2}-g\l[\s,\c\r]\r)\l(D^{(2)}_0-2i{\tr\over{r}^2}+g\l[\s,\c\r]\r)-D^{(0)}_+D^{(2)}_-\r].
\end{eqnarray*}

For a $(k,l)$-form fermionic field $v$ on $\RS$ of charge $2$, 
which is also a scalar on the $S^3$, and its hermitian conjugate $v^*$, 
integration by part is used to deduce 
\begin{eqnarray*}
&&\int d^5x \sqrt{g} \Tr\l[v^*\l(
\l(D^{(2)}_0-2i{\tr\over{r}^2}-g\l[\s,\c\r]\r)
\l(D^{(2)}_0-2i{\tr\over{r}^2}+g\l[\s,\c\r]\r)+D^{(0)}_+D^{(2)}_-\r)v\r]
\nn\\
&&=
\int d^5x \sqrt{g} \Tr\l[v\l(\l(D^{(-2)}_0+2i{\tr\over{r}^2}-g\l[\s,\c\r]\r)
\l(D^{(-2)}_0+2i{\tr\over{r}^2}+g\l[\s,\c\r]\r)+D^{(0)}_-D^{(-2)}_+\r)v^*\r],
\end{eqnarray*}
which implys that 
\begin{eqnarray*}
&&\Det_{(k,l)}\l[
-\l(D^{(2)}_0-2i{\tr\over{r}^2}-g\l[\s,\c\r]\r)
\l(D^{(2)}_0-2i{\tr\over{r}^2}+g\l[\s,\c\r]\r)-D^{(0)}_+D^{(2)}_-\r]
\nn\\
&&=\Det_{(l,k)}\l[
-\l(D^{(-2)}_0+2i{\tr\over{r}^2}+g\l[\s,\c\r]\r)
\l(D^{(-2)}_0+2i{\tr\over{r}^2}-g\l[\s,\c\r]\r)-D^{(0)}_-D^{(-2)}_+\r]
=\Det_{(l,k)}\l[\Delta_{-2}\r].
\end{eqnarray*}

Similarly, we can see that 
\begin{eqnarray*}
&&\Det_{(k,l)}\l[-\l(D^{(0)}_0+g\l[\s,\c\r]\r)\l(D^{(0)}_0-g\l[\s,\c\r]\r)
-D^{(2)}_-D^{(0)}_+\r]
\nn\\
&&=
\Det_{(l,k)}\l[-\l(D^{(0)}_0+g\l[\s,\c\r]\r)\l(D^{(0)}_0-g\l[\s,\c\r]\r)
-D^{(-2)}_+D^{(0)}_-\r]
=\Det_{(l,k)}\l[\Delta_{0}\r],
\end{eqnarray*}
and that
\begin{eqnarray*}
\Det_{(k,l)}\l[D^{(2)}_0-2i{\tr\over{r}^2}\pm{g}\l[\s,\c\r]\r]
=
\Det_{(l,k)}\l[D^{(-2)}_0+2i{\tr\over{r}^2}\pm{g}\l[\s,\c\r]\r].
\end{eqnarray*}

Using these, we may rewrite the determinant 
\begin{eqnarray}
&&\Det_{(k,l)}\l[
\begin{pmatrix}
D^{(0)}_0-g\l[\s,\c\r] & D^{(2)}_-
\cr
D^{(0)}_+ & -D^{(2)}_0+2i{\tr\over{r}^2}-g\l[\s,\c\r]
\end{pmatrix}
\r]
\label{determinantfermi}\\
&&=
{\Det_{(l,k)}\l[D^{(-2)}_0+2i{\tr\over{r}^2}+g\l[\s,\c\r]\r]
\over
\Det_{(l,k)}\l[D^{(0)}_0+g\l[\s,\c\r]\r]}
\Det_{(l,k)}\l[\Delta_{0}\r]
=
{\Det_{(l,k)}\l[D^{(0)}_0-g\l[\s,\c\r]\r]
\over
\Det_{(l,k)}\l[D^{(-2)}_0+2i{\tr\over{r}^2}-g\l[\s,\c\r]\r]}
\Det_{(l,k)}\l[\Delta_{-2}\r],
\nn\end{eqnarray}
which also implys the formula
$$
{\Det_{(l,k)}\l[\Delta_{0}\r]\over\Det_{(l,k)}\l[\Delta_{-2}\r]}
=
{\Det_{(l,k)}\l[D^{(0)}_0-g\l[\s,\c\r]\r]
\Det_{(l,k)}\l[D^{(0)}_0+g\l[\s,\c\r]\r]
\over
\Det_{(l,k)}\l[D^{(-2)}_0+2i{\tr\over{r}^2}-g\l[\s,\c\r]\r]
\Det_{(l,k)}\l[D^{(-2)}_0+2i{\tr\over{r}^2}+g\l[\s,\c\r]\r]}.
$$

Using (\ref{determinantfermi}) twice in a bit tricky way, 
we obtain $Z_{{\rm V, F}}^{{\rm 1-loop}}$ 
\begin{eqnarray*}
{\Det_{(0,1)}\l[D^{(-2)}_0+2i{\tr\over{r}^2}+g\l[\s,\c\r]\r]
\Det_{(0,0)}\l[D^{(-2)}_0+2i{\tr\over{r}^2}-g\l[\s,\c\r]\r]
\over
\Det_{(0,1)}\l[D^{(0)}_0+g\l[\s,\c\r]\r]
\Det_{(0,0)}\l[D^{(0)}_0-g\l[\s,\c\r]\r]}
{\Det_{(0,1)}\l[\Delta_{0}\r]
\over
\Det_{(0,0)}\l[\Delta_{-2}\r]}
\Det_{(0,0)}\l[D_1\r]\Det_{(0,0)}\l[D_4\r].
\end{eqnarray*}

With the same reasoning as the argument about integration by parts in 
the integrand, it is easy to see that 
\begin{eqnarray*}
&&\Det_{(0,0)}\l[D_1\r]
=\Det_{(0,0)}\l[\l(D^{(0)}_0-2i{\tr\over{r}^2}+g\l[\s,\,\c\r]\r)
\Big(D^{(0)}_0-g\l[\s,\,\c\r]\Big)+D^{(-2)}_+D^{(0)}_-\r],
\nn\\
&&\Det_{(0,0)}\l[D_4\r]
=\Det_{(0,0)}\l[4\cD_z\cD_{\bar{z}}-\Delta_{-2}\r].
\end{eqnarray*}
Using this, we will combine the one-loop contributions 
$Z^{{\rm 1-loop}}_{{\rm V,B}}$, $Z^{{\rm 1-loop}}_{{\rm V,F}}$, 
and $Z_{{\rm FP}}$ from the gauge multiplet to yield 
\BBox{525}{60}{
\begin{eqnarray}
Z^{{\rm 1-loop}}_{{\rm V}}
=Z_{{\rm FP}}Z^{{\rm 1-loop}}_{{\rm V,B}}Z^{{\rm 1-loop}}_{{\rm V,F}}
={\Det_{(0,1)}\l[D^{(-2)}_0+2i{\tr\over{r}^2}+g\l[\s,\c\r]\r]
\over
\Det_{(0,0)}\l[D^{(-2)}_0+2i{\tr\over{r}^2}+g\l[\s,\c\r]\r]}
{\Det_{(0,0)}\l[D^{(0)}_0-g\l[\s,\c\r]\r]
\over
\Det_{(0,1)}\l[D^{(0)}_0-g\l[\s,\c\r]\r]},
\label{Z_V}
\end{eqnarray}
}
where we have made use of the invariance 
$$
\Det_{(0,0)}\l[D^{(0)}_0-g\l[\s,\c\r]\r]
=\Det_{(0,0)}\l[D^{(0)}_0+g\l[\s,\c\r]\r],
$$
under $\a\to-\a$ for all the roots $\a\in\La$.

The determinant $\Det_{(k,l)}$ can be evaluated by using the basis 
$\{\vp_{l,m,\tm}\otimes{v}\otimes{E}_{\a}, 
\vp_{l,m,\tm}\otimes{v}\otimes{H}_{i}\}$, for $v$ running over all the 
basis vectors of $\Om^{(k,l)}(\RS)$, the set of all $(k,l)$-forms on $\RS$, 
upon regarding $\Om^{(k,l)}(\RS)$ as a linear space. 
Here, $\vp_{l,m,\tm}$ ($l=0,1/2,1,3/2,\cdots; -l\le{m},\tm\le{l}$) denote 
the scalar spherical harmonics on the $S^3$, and through the relations 
(\ref{DiffOp0L}) of the differential operators with the generators of 
the Lie algebra of $SU(2)$, they provide the representations of the $SU(2)$ 
algebra;
$$
L_3\vp_{l,m,\tm}=m\vp_{l,m,\tm}, 
\qquad
L_{\pm}\vp_{l,m,\tm}=\sqrt{(l\mp{m})(l\pm{m}+1)}\vp_{l,m\pm{1},\tm}. 
$$

On the basis\footnote{Since the contributions from the basis vectors 
$\vp_{l,m,\tm}\otimes{v}\otimes{H}_{i}$ to the determinants are constant 
factors to the partition function, we will omit them.} 
$\{\vp_{l,m,\tm}\otimes{v}\otimes{E}_{\a}\}$, using (\ref{DiffOp-2L}), 
we deduce 
\begin{eqnarray*}
&&\Det_{(k,l)}\l[D^{(0)}_0-g\l[\s,\c\r]\r]
=\prod_{\a\in\La}\prod_{l\in\hf{\bf Z}_{\ge0}}\prod_{m,\tm=-l}^{l}
\dets_{(k,l)}\l[{2i\over\tr}m-g\sa\r],
\nn\\
&&\Det_{(k,l)}\l[D^{(-2)}_0+2i{\tr\over{r}^2}+g\l[\s,\c\r]\r]
=\prod_{\a\in\La}\prod_{l\in\hf{\bf Z}_{\ge0}}\prod_{m,\tm=-l}^{l}
\dets_{(k,l)}\l[{2i\over\tr}(m+1)+g\sa\r],
\end{eqnarray*}
with $\hf{\bf Z}_{\ge0}$ the set of non-negative half integers, 
where the determinant $\dets_{(k,l)}$ is defined over the space 
$\Om^{(k,l)}(\RS)$.

As explained in \cite{BlauThompson}, 
the Hodge decomposition implys that for the space $\Om^{k,l}(\RS)$ 
of all the ($k,l$)-forms on the Riemann surface $\RS$, 
$$
\Om^{1,0}(\RS)\oplus\Om^{0,1}(\RS)
=\l(\Om^{0,0}(\RS)\ominus{H}^{0}(\RS)\r)\oplus
\l(\Om^{0,0}(\RS)\ominus{H}^{0}(\RS)\r)\oplus{H}^1(\RS),
$$
where $H^p(\RS)$ is the space of all the harmonic $p$-forms on $\RS$. 
It follows from this that for a constant $D$, 
\begin{eqnarray}
{\dets_{(0,0)}\l[D\r]\over\dets_{(0,1)}\l[D\r]}=D^{b_0(\RS)-\hf{b}_1(\RS)}
=D^{\hf\chi(\RS)}
\label{Hodge}
\end{eqnarray}
with $b_i(\RS)=\dim{H}_i(\RS)$ the $i$-th Betti number, and 
with the Euler number $\chi(\RS)$ of the surface $\RS$; 
$$
\chi(\RS)=b_0(\RS)-b_1(\RS)+b_2(\RS)=2b_0(\RS)-b_1(\RS),
$$
where we have used the Hodge duality; $b_0(\RS)=b_2(\RS)$. 

Taking account of this, we can reduce $Z^{{\rm 1-loop}}_{{\rm V}}$ to
\begin{eqnarray*}
Z^{{\rm 1-loop}}_{{\rm V}}
=\prod_{\a\in\La}\prod_{l\in\hf{\bf Z}_{\ge0}}\prod_{m,\tm=-l}^{l}
\l({{2i\over\tr}m-g\sa
\over
{2i\over\tr}(m+1)+g\sa}\r)^{\hf\chi(\RS)}
=\prod_{\a\in\La_+}\l[{\tr{g}\sa}
\prod_{n=1}\l(n^2+\tr^2g^2\sa^2\r)\r]^{\chi(\RS)},
\end{eqnarray*}
where we have replaced $l$ by $n=2l$ ($n=0,1,2,\cdots$). 

From the formula 
$$
{1\over\pi}\sinh{\pi{x}}=x\prod_{m=1}^{\infty}\l(1+{x^2\over{m^2}}\r),
$$
together with the zeta regularization, 
$$
\prod_{m=1}^{\infty}m=e^{\sum_{m=1}^{\infty}\log{m}}
\quad\to\quad e^{-\zeta'(0)}=\sqrt{2\pi},
$$
it follows that 
\BBox{400}{45}{
\begin{eqnarray}
Z^{{\rm 1-loop}}_{{\rm V}}
=\prod_{\a\in\La_+}\l[2\sinh\big(\pi\tr{g}\sa\big)\r]^{\chi(\RS)}. 
\label{squashZ_V}
\end{eqnarray}
}

In the round limit $\tr\to{r}$, $Z_{{\rm V}}$ is in agreement with the 
previous result in \cite{FKM}.

\subsection{One-Loop Contributions from the $\N=1$ Hypermultiplet}

Let us proceed to compute the one-loop contributions from the hypermultiplet 
by localization. 
Since the BRST transfromation in the $\N=1$ twisting differs from the one 
in the $\N=2$ twisting, we will discuss them separately in the next 
two subsubsections. 

However, the BRST transformations of the scalar fields $H$, $H^\dag$ 
of the hypermultiplet are common in both the $\N=1$ and the $\N=2$ twistings.
From (\ref{HyperBRST}), the BRST transformation of the scalar fields 
is reduced to  
\begin{eqnarray*}
\dl_Q\tilde{H}=0, 
\quad
\dl_Q{H}=0, 
\qquad
\dl_Q\tilde{H}^\dag=-{1\over2\sqrt{2}}\tkappa, 
\quad
\dl_Q{H}^\dag=-{1\over2\sqrt{2}}\kappa. 
\end{eqnarray*}

\subsubsection{The $\N=1$ Twisting}
\label{LocalizationN=1Squash}

Let us begin with the $\N=1$ twisting to 
calculate the one-loop contributions from the hypermultiplet 
to the partition function by localization. 

The BRST transformation of the fermions in the hypermultiplet is 
given by
\begin{eqnarray*}
&&\dl_Q\chi=\sqrt{2}i\l[
\l({\e}^\dag\t^m\e\r)\cD^{(-1)}{}_m\tilde{H}+g\l[\s,\,\tilde{H}\r]
+i{\tr\over{r}^2}\tilde{H}\r],
\quad
\dl_Q\xi=\sqrt{2}i\l({\e^c}^\dag\t^m\e\r)\cD^{(-1)}{}_m\tilde{H},
\nn\\
&&\dl_Q\eta=F^1{}_{1}-2\sqrt{2}\cD_z\tilde{H},
\qquad
\dl_Q\kappa=0, 
\nn\\
&&\dl_Q\tchi=-\sqrt{2}i\l[
\l({\e}^\dag\t^m\e\r)\cD^{(-1)}{}_m{H}+g\l[\s,\,{H}\r]
+i{\tr\over{r}^2}{H}\r],
\quad
\dl_Q\txi=-\sqrt{2}i\l({\e^c}^\dag\t^m\e\r)\cD^{(-1)}{}_m{H},
\nn\\
&&\dl_Q\teta=F^2{}_{1}+2\sqrt{2}\cD_z{H},
\qquad
\dl_Q\tkappa=0, 
\end{eqnarray*}
and its hermitian conjugate by 
\begin{eqnarray*}
&&\l(\dl_Q\chi\r)^\dag=-\sqrt{2}i\l[
\l({\e}^\dag\t^m\e\r)\cD^{(1)}{}_m\tilde{H}^\dag-g\l[\s,\,\tilde{H}^\dag\r]
-i{\tr\over{r}^2}\tilde{H}^\dag\r],
\quad
\l(\dl_Q\xi\r)^\dag
=-\sqrt{2}i\l({\e}^\dag\t^m\e^c\r)\cD^{(1)}{}_m\tilde{H}^\dag,
\nn\\
&&\l(\dl_Q\eta\r)^\dag=-F^2{}_{2}-2\sqrt{2}\cD_{\bar{z}}\tilde{H}^\dag,
\qquad
\l(\dl_Q\kappa\r)^\dag=0, 
\nn\\
&&\l(\dl_Q\tchi\r)^\dag=\sqrt{2}i\l[
\l({\e}^\dag\t^m\e\r)\cD^{(1)}{}_m{H}^\dag-g\l[\s,\,{H}^\dag\r]
-i{\tr\over{r}^2}{H}^\dag\r],
\quad
\l(\dl_Q\txi\r)^\dag=\sqrt{2}i\l({\e}^\dag\t^m\e^c\r)\cD^{(1)}{}_m{H}^\dag,
\nn\\
&&\l(\dl_Q\teta\r)^\dag=F^1{}_{2}+2\sqrt{2}\cD_{\bar{z}}{H}^\dag,
\qquad
\l(\dl_Q\tkappa\r)^\dag=0.
\end{eqnarray*}

Using the BRST transformation of the auxiliuary fields $F^1{}_2$, $F^2{}_2$,
\begin{eqnarray*}
\dl_QF^1{}_2={i\over2}\l[
\l({\e^c}^\dag\t^m\e\r)\cD^{(-1)}{}_m{\chi}
-\l({\e}^\dag\t^m\e\r)\cD^{(1)}{}_m\eta-g\l[\s,\,\eta\r]
+i{\tr\over{r}^2}\eta-2i\cD_{\bar{z}}\kappa
\r],
\nn\\
\dl_QF^2{}_2={i\over2}\l[
\l({\e^c}^\dag\t^m\e\r)\cD^{(-1)}{}_m{\tchi}
-\l({\e}^\dag\t^m\e\r)\cD^{(1)}{}_m\teta-g\l[\s,\,\teta\r]
+i{\tr\over{r}^2}\teta-2i\cD_{\bar{z}}\tkappa
\r],
\end{eqnarray*}
where we have omitted the terms 
$g\l(\s^i\r)^{\ta}{}_{\dg}\l[\phi^i,\,\la^{\dg}\r]$ on the right hand sides 
of both the equations, because they vanish in the large $t$ limit, 
we find that 
\begin{eqnarray*}
&&\dl_Q\l(\dl_Q\chi\r)^\dag={i\over2}\l[
\l({\e}^\dag\t^m\e\r)\cD^{(1)}{}_m\tkappa-g\l[\s,\,\tkappa\r]
-i{\tr\over{r}^2}\tkappa\r],
\qquad
\dl_Q\l(\dl_Q\xi\r)^\dag={i\over2}\l({\e}^\dag\t^m\e^c\r)\cD^{(1)}{}_m\tkappa,
\nn\\
&&\dl_Q\l(\dl_Q\eta\r)^\dag
=-{i\over2}\l\{\l({\e^c}^\dag\t^m\e\r)\cD^{(-1)}{}_m\tchi
-\l[\l({\e}^\dag\t^m\e\r)\cD^{(1)}{}_m\txi+g\l[\s,\,\txi\r]
-i{\tr\over{r}^2}\txi\r]\r\},
\qquad
\dl_Q\l(\dl_Q\kappa\r)^\dag=0, 
\nn\\
&&\dl_Q\l(\dl_Q\tchi\r)^\dag=-{i\over2}\l[
\l({\e}^\dag\t^m\e\r)\cD^{(1)}{}_m\kappa-g\l[\s,\,\kappa\r]
-i{\tr\over{r}^2}\kappa\r],
\qquad
\dl_Q\l(\dl_Q\txi\r)^\dag=-{i\over2}\l({\e}^\dag\t^m\e^c\r)\cD^{(1)}{}_m\kappa,
\nn\\
&&\dl_Q\l(\dl_Q\teta\r)^\dag
={i\over2}\l\{\l({\e^c}^\dag\t^m\e\r)\cD^{(-1)}{}_m\chi
-\l[\l({\e}^\dag\t^m\e\r)\cD^{(1)}{}_m\xi+g\l[\s,\,\xi\r]
-i{\tr\over{r}^2}\xi\r]\r\},
\qquad
\dl_Q\l(\dl_Q\tkappa\r)^\dag=0.
\end{eqnarray*}

The system of $(\tilde{H},\tilde{H}^\dag,\chi,\xi,\teta,\tkappa,F^1{}_2)$ 
is identical to 
the one of $(H,H^\dag,\tchi,\txi,\eta,\kappa,F^2{}_2)$. If the former 
contributes the one-loop determinant $Z^{1-{\rm loop}}_{H}$ to the partition 
function, both of the systems contribute $(Z^{1-{\rm loop}}_{H})^2$. 
Therefore, we will focus on the former system only. 

From the fermionic part of the systerm 
$(\tilde{H},\tilde{H}^\dag,\chi,\xi,\teta,\tkappa,F^1{}_2)$ 
of the regulator action ${\cal S}_Q$,
\begin{eqnarray*}
&&-\int \sqrt{g} d^5x \l[
\dl_Q\l(\dl_Q\chi\r)^\dag\c\chi+\dl_Q\l(\dl_Q\xi\r)^\dag\c\xi
+\dl_Q\l(\dl_Q\teta\r)^\dag\c\teta+\dl_Q\l(\dl_Q\tkappa\r)^\dag\c\tkappa \r]
\nn\\
&&={i\over2}\int \sqrt{g} d^5x \l[
\raise1.5ex\hbox{$\begin{pmatrix}\chi,&\xi\end{pmatrix}$}
\begin{pmatrix}
\l({\e}^\dag\t^m\e\r)\cD^{(1)}{}_m-g\l[\s,\c\r]-i{\tr\over{r}^2}
&
\l({\e^c}^\dag\t^m\e\r)\cD^{(-1)}{}_m
\cr
\l({\e}^\dag\t^m\e^c\r)\cD^{(1)}{}_m
&
-\l({\e}^\dag\t^m\e\r)\cD^{(-1)}{}_m-g\l[\s,\c\r]-i{\tr\over{r}^2}
\end{pmatrix}
\begin{pmatrix}\tkappa\cr\teta\end{pmatrix}
\r],
\end{eqnarray*}
where we have performed integration by parts, 
the one-loop determinant from the fermions 
of the system $(\tilde{H},\tilde{H}^\dag,\chi,\xi,\teta,\tkappa,F^1{}_2)$ 
can be read as 
\begin{eqnarray*}
Z^{{\rm 1-loop}}_{{\rm H,F}}=\Det_{(0,\hf)}\l[
\begin{pmatrix}
\l({\e}^\dag\t^m\e\r)\cD^{(1)}{}_m-g\l[\s,\c\r]-i{\tr\over{r}^2}
&
\l({\e^c}^\dag\t^m\e\r)\cD^{(-1)}{}_m
\cr
\l({\e}^\dag\t^m\e^c\r)\cD^{(1)}{}_m
&
-\l({\e}^\dag\t^m\e\r)\cD^{(-1)}{}_m-g\l[\s,\c\r]-i{\tr\over{r}^2}
\end{pmatrix}\r], 
\end{eqnarray*}
where the determinant $\Det_{(0,\hf)}$ is defined such that 
the path integral over a fermionic $(k,l)$-form $\psi$ on $\RS$ 
and its partner $\la$ with a differential operator $D$ yields 
$$
\int\l[d\la\r]\l[{d\psi}\r]\exp\l[\int\sqrt{g}d^5x\,\la{D}\psi\r]
=\Det_{(k,l)}\l[D\r].
$$

For four differential operators $D_1,\cdots,D_4$, we have the formula
\begin{eqnarray*}
{\Det}_{(k,l)}\l[\begin{pmatrix}
D_1&D_2\cr{D}_3&D_4
\end{pmatrix}\r]={\Det}_{(k,l)}\l[D_1\r]
{\Det}_{(k,l)}\l[D_4-D_3{1\over{D}_1}D_2\r],
\end{eqnarray*}
for an invertible $D_1$. If there is another differential operator ${D'}_1$ 
satisfying the relation 
$$
D_3{1\over{D}_1}={1\over{D'}_1}D_3, 
$$
we obtain the formula 
\begin{eqnarray*}
{\Det}_{(k,l)}\l[\begin{pmatrix}
D_1&D_2\cr{D}_3&D_4
\end{pmatrix}\r]={{\Det}_{(k,l)}\l[D_1\r]\over{\Det}_{(k,l)}\l[D'_1\r]}
{\Det}_{(k,l)}\l[{D'}_1D_4-D_3D_2\r].
\end{eqnarray*}

In our case, we have
$$
D_1=\l({\e}^\dag\t^m\e\r)\cD^{(1)}{}_m-g\l[\s,\c\r]-i{\tr\over{r}^2},
\qquad
D_3=\l({\e}^\dag\t^m\e^c\r)\cD^{(1)}{}_m,
$$
which both act on the spinor $\tkappa$ of negative chirality on $\RS$ and 
of charge $q=1$. Using (\ref{CommRelDifOpRoundS}) in Appendix 
\ref{constKSSquashS}, we can find the operator ${D'}_1$,
\begin{eqnarray*}
D_3D_1
&=&\l({\e}^\dag\t^n\e^c\r)\cD^{(1)}{}_n\l[
\l({\e}^\dag\t^m\e\r)\cD^{(1)}{}_m-g\l[\s,\c\r]-i{\tr\over{r}^2}
\r]
\nn\\
&=&
\l[\l({\e}^\dag\t^m\e\r)\cD^{(-1)}{}_m
-g\l[\s,\c\r]+i{\tr\over{r}^2}\r]
\l({\e}^\dag\t^n\e^c\r)\cD^{(1)}{}_n
={D'}_1D_3.
\end{eqnarray*}
Therefore, we find that 
\begin{eqnarray*}
Z^{{\rm 1-loop}}_{{\rm H,F}}=
\Det_{(0,\hf)}\l[\Delta^{\N=1}_{{\rm H,B}}\r]
{\Det_{(0,\hf)}\l[
\l({\e}^\dag\t^m\e\r)\cD^{(1)}{}_m-g\l[\s,\c\r]-i{\tr\over{r}^2}\r]
\over
\Det_{(0,\hf)}\l[
\l({\e}^\dag\t^m\e\r)\cD^{(-1)}{}_m-g\l[\s,\c\r]+i{\tr\over{r}^2}\r]},
\end{eqnarray*}
where the differential operator $\Delta^{\N=1}_{{\rm H,B}}$ denotes 
\begin{eqnarray*}
\Delta^{\N=1}_{{\rm H,B}}
&=&
-\l(\l({\e}^\dag\t^m\e\r)\cD^{(-1)}{}_m
-g\l[\s,\c\r]+i{\tr\over{r}^2}\r)
\l(\l({\e}^\dag\t^m\e\r)\cD^{(-1)}{}_m
+g\l[\s,\c\r]+i{\tr\over{r}^2}\r)
\nn\\
&&
-\l({\e}^\dag\t^m\e^c\r)\cD^{(1)}{}_m\l({\e^c}^\dag\t^m\e\r)\cD^{(-1)}{}_m.
\end{eqnarray*}

Note that 
\begin{eqnarray*}
&&\Det_{(0,\hf)}\l[
\l({\e}^\dag\t^m\e\r)\cD^{(1)}{}_m-g\l[\s,\c\r]-i{\tr\over{r}^2}\r]
=\int\l[d\chi\r]\l[{d\tkappa}\r]
\exp\l[{i\over2}\int\sqrt{g}d^5x\,\chi{D_1}\tkappa\r]
\nn\\
&&=\int\l[d\chi\r]\l[{d\tkappa}\r]
\exp\l[{i\over2}\int\sqrt{g}d^5x\,\tkappa{D'}_1\chi\r]
=\Det_{(\hf,0)}\l[
\l({\e}^\dag\t^m\e\r)\cD^{(-1)}{}_m-g\l[\s,\c\r]+i{\tr\over{r}^2}\r].
\end{eqnarray*}
The differential operator ${D'}_1$ in the determinant 
$\Det_{(\hf,0)}\l[{D'}_1\r]$ on the most right hand side 
doesn't depend on the chirality of $\chi$, and therefore, 
$\Det_{(\hf,0)}\l[{D'}_1\r]=\Det_{(0,\hf)}\l[{D'}_1\r]$. 
It means that the ratio of the determinants is unity;
$$
{\Det_{(0,\hf)}\l[
\l({\e}^\dag\t^m\e\r)\cD^{(1)}{}_m-g\l[\s,\c\r]-i{\tr\over{r}^2}\r]
\over
\Det_{(0,\hf)}\l[
\l({\e}^\dag\t^m\e\r)\cD^{(-1)}{}_m-g\l[\s,\c\r]+i{\tr\over{r}^2}\r]}
=1,
$$
and we obtain 
$$
Z^{{\rm 1-loop}}_{{\rm H,F}}=
\Det_{(0,\hf)}\l[\Delta^{\N=1}_{{\rm H,B}}\r].
$$

In the bosonic part of the system 
$(\tilde{H},\tilde{H}^\dag,\chi,\xi,\teta,\tkappa,F^1{}_2)$ 
of the regulator action ${\cal S}_Q$,
\begin{eqnarray*}
-\int d^5x \sqrt{g} \l[
\l(\dl_Q\chi\r)^\dag\c\dl_Q\chi+\l(\dl_Q\xi\r)^\dag\c\dl_Q\xi
+\l(\dl_Q\teta\r)^\dag\c\dl_Q\teta+\l(\dl_Q\tkappa\r)^\dag\c\dl_Q\tkappa 
\r],
\end{eqnarray*}
we will shift the auxiliuary fields $F^{\ta}{}_{\cb}$ so that 
we can trivially integrate them out. 
We will integrate the remaining part of the action by parts to obtain 
\begin{eqnarray*}
-2\int d^5x \sqrt{g}\tilde{H}^\dag\Delta^{\N=1}_{{\rm H,B}}\tilde{H},
\end{eqnarray*}
and see that the one-loop determinant from the bosonic fields of 
the system $(\tilde{H},\tilde{H}^\dag,\chi,\xi,\teta,\tkappa)$ is given by
$$
Z^{1-{\rm loop}}_{{\rm H,B}}={1\over\Det_{(0,\hf)}
\l[\Delta^{\N=1}_{{\rm H,B}}\r]}.
$$

Therefore, the contributions from the hypermultiplet to the partition 
function are trivial; 
$$
\l(Z^{1-{\rm loop}}_{{\rm H}}\r)^2
=\l(Z^{1-{\rm loop}}_{{\rm H,F}}Z^{1-{\rm loop}}_{{\rm H,B}}\r)^2=1.
$$
In the round limit $\tr\to{r}$, the contributions from the hypermultiplet 
reproduce the previous results about the hypermultiplet in \cite{FKM}.

\subsubsection{The $\N=2$ Twisting}

Let us proceed to the $\N=2$ twising. 
In contrast to the $\N=1$ twisting, the system of the fluctuations 
$(\tilde{H},\tilde{H}^\dag,\chi,\xi,\teta,\tkappa,F^1{}_2)$ 
yields the different contribution to the partition function 
from the  one from 
$({H},{H}^\dag,\tchi,\txi,\eta,\kappa,F^2{}_2)$,  
and we will treat them separately below. 

The BRST transformation of the fermions in the hypermultiplet is 
given by
\begin{eqnarray*}
&&\dl_Q\chi=\sqrt{2}i\l[
\l({\e}^\dag\t^m\e\r)\cD^{(0)}{}_m\tilde{H}+g\l[\s,\,\tilde{H}\r]
+2i{\tr\over{r}^2}\tilde{H}\r],
\quad
\dl_Q\xi=\sqrt{2}i\l({\e^c}^\dag\t^m\e\r)\cD^{(0)}{}_m\tilde{H},
\nn\\
&&\dl_Q\eta=F^1{}_{1}-2\sqrt{2}\cD_z\tilde{H},
\qquad
\dl_Q\kappa=0, 
\nn\\
&&\dl_Q\tchi=-\sqrt{2}i\Big[
\l({\e}^\dag\t^m\e\r)\cD^{(-2)}{}_m{H}+g\l[\s,\,{H}\r]\Big],
\quad
\dl_Q\txi=-\sqrt{2}i\l({\e^c}^\dag\t^m\e\r)\cD^{(-2)}{}_m{H},
\nn\\
&&\dl_Q\teta=F^2{}_{1}+2\sqrt{2}\cD_z{H},
\qquad
\dl_Q\tkappa=0, 
\end{eqnarray*}
and its hermitian conjugate by 
\begin{eqnarray*}
&&\l(\dl_Q\chi\r)^\dag=-\sqrt{2}i\l[
\l({\e}^\dag\t^m\e\r)\cD^{(0)}{}_m\tilde{H}^\dag-g\l[\s,\,\tilde{H}^\dag\r]
-2i{\tr\over{r}^2}\tilde{H}^\dag\r],
\quad
\l(\dl_Q\xi\r)^\dag
=-\sqrt{2}i\l({\e}^\dag\t^m\e^c\r)\cD^{(0)}{}_m\tilde{H}^\dag,
\nn\\
&&\l(\dl_Q\eta\r)^\dag=-F^2{}_{2}-2\sqrt{2}\cD_{\bar{z}}\tilde{H}^\dag,
\qquad
\l(\dl_Q\kappa\r)^\dag=0, 
\nn\\
&&\l(\dl_Q\tchi\r)^\dag=\sqrt{2}i\Big[
\l({\e}^\dag\t^m\e\r)\cD^{(2)}{}_m{H}^\dag-g\l[\s,\,{H}^\dag\r]\Big],
\quad
\l(\dl_Q\txi\r)^\dag=\sqrt{2}i\l({\e}^\dag\t^m\e^c\r)\cD^{(2)}{}_m{H}^\dag,
\nn\\
&&\l(\dl_Q\teta\r)^\dag=F^1{}_{2}+2\sqrt{2}\cD_{\bar{z}}{H}^\dag,
\qquad
\l(\dl_Q\tkappa\r)^\dag=0.
\end{eqnarray*}

Using the BRST transformation of the auxiliuary fields $F^1{}_2$, $F^2{}_2$,
\begin{eqnarray*}
\dl_QF^1{}_2&=&{i\over2}\l[
\l({\e^c}^\dag\t^m\e\r)\cD^{(-1)}{}_m{\chi}
-\l({\e}^\dag\t^m\e\r)\cD^{(1)}{}_m\eta-g\l[\s,\,\eta\r]
-2i\cD_{\bar{z}}\kappa
\r],
\nn\\
\dl_QF^2{}_2&=&{i\over2}\l[
\l({\e^c}^\dag\t^m\e\r)\cD^{(-1)}{}_m{\tchi}
-\l({\e}^\dag\t^m\e\r)\cD^{(1)}{}_m\teta-g\l[\s,\,\teta\r]
+2i{\tr\over{r}^2}\teta-2i\cD_{\bar{z}}\tkappa
\r],
\end{eqnarray*}
where we have omitted the terms 
$g\l(\s^i\r)^{\ta}{}_{\dg}\l[\phi^i,\,\la^{\dg}\r]$ on the right hand sides 
of both the equations, because they vanish in the large $t$ limit, 
we find that 
\begin{eqnarray*}
&&\dl_Q\l(\dl_Q\chi\r)^\dag={i\over2}\l[
\l({\e}^\dag\t^m\e\r)\cD^{(0)}{}_m\tkappa-g\l[\s,\,\tkappa\r]
-2i{\tr\over{r}^2}\tkappa\r],
\qquad
\dl_Q\l(\dl_Q\xi\r)^\dag={i\over2}\l({\e}^\dag\t^m\e^c\r)\cD^{(0)}{}_m\tkappa,
\nn\\
&&\dl_Q\l(\dl_Q\eta\r)^\dag
=-{i\over2}\l\{\l({\e^c}^\dag\t^m\e\r)\cD^{(-2)}{}_m\tchi
-\l[\l({\e}^\dag\t^m\e\r)\cD^{(0)}{}_m\txi+g\l[\s,\,\txi\r]
-2i{\tr\over{r}^2}\txi\r]\r\},
\qquad
\dl_Q\l(\dl_Q\kappa\r)^\dag=0, 
\nn\\
&&\dl_Q\l(\dl_Q\tchi\r)^\dag=-{i\over2}\Big[
\l({\e}^\dag\t^m\e\r)\cD^{(2)}{}_m\kappa-g\l[\s,\,\kappa\r]\Big],
\qquad
\dl_Q\l(\dl_Q\txi\r)^\dag=-{i\over2}\l({\e}^\dag\t^m\e^c\r)\cD^{(2)}{}_m\kappa,
\nn\\
&&\dl_Q\l(\dl_Q\teta\r)^\dag
={i\over2}\l\{\l({\e^c}^\dag\t^m\e\r)\cD^{(0)}{}_m\chi
-\Big[\l({\e}^\dag\t^m\e\r)\cD^{(2)}{}_m\xi+g\l[\s,\,\xi\r]\Big]\r\},
\qquad
\dl_Q\l(\dl_Q\tkappa\r)^\dag=0.
\end{eqnarray*}

From the fermionic part of the systerm 
$(\tilde{H},\tilde{H}^\dag,\chi,\xi,\teta,\tkappa,F^1{}_2)$ 
of the regulator action ${\cal S}_Q$,
\begin{eqnarray*}
&&-\int \sqrt{g} d^5x \l[
\dl_Q\l(\dl_Q\chi\r)^\dag\c\chi+\dl_Q\l(\dl_Q\xi\r)^\dag\c\xi
+\dl_Q\l(\dl_Q\teta\r)^\dag\c\teta+\dl_Q\l(\dl_Q\tkappa\r)^\dag\c\tkappa \r]
\nn\\
&&={i\over2}\int \sqrt{g} d^5x \l[
\raise1.5ex\hbox{$\begin{pmatrix}\chi,&\xi\end{pmatrix}$}
\begin{pmatrix}
\l({\e}^\dag\t^m\e\r)\cD^{(0)}{}_m-g\l[\s,\c\r]-2i{\tr\over{r}^2}
&
\l({\e^c}^\dag\t^m\e\r)\cD^{(-2)}{}_m
\cr
\l({\e}^\dag\t^m\e^c\r)\cD^{(0)}{}_m
&
-\l({\e}^\dag\t^m\e\r)\cD^{(-2)}{}_m-g\l[\s,\c\r]
\end{pmatrix}
\begin{pmatrix}\tkappa\cr\teta\end{pmatrix}
\r],
\end{eqnarray*}
where we have performed integration by parts, 
the one-loop determinant from the fermions 
of the system $(\tilde{H},\tilde{H}^\dag,\chi,\xi,\teta,\tkappa,F^1{}_2)$ 
can be read as 
\begin{eqnarray}
Z^{{\rm 1-loop}}_{{\rm H,F}}=\Det_{(0,0)}\l[
\begin{pmatrix}
\l({\e}^\dag\t^m\e\r)\cD^{(0)}{}_m-g\l[\s,\c\r]-2i{\tr\over{r}^2}
&
\l({\e^c}^\dag\t^m\e\r)\cD^{(-2)}{}_m
\cr
\l({\e}^\dag\t^m\e^c\r)\cD^{(0)}{}_m
&
-\l({\e}^\dag\t^m\e\r)\cD^{(-2)}{}_m-g\l[\s,\c\r]
\end{pmatrix}\r], 
\label{N=2ZH}
\end{eqnarray}
with the determinant $\Det_{(0,0)}$ defined in the previous subsection 
\ref{LocalizationN=1Squash}. 

As in the $\N=1$ twisting in subsection \ref{LocalizationN=1Squash}, 
upon computing the one-loop determinant (\ref{N=2ZH}), 
we may identify the differential operators $D_1$ and $D_3$ with 
$$
D_1=\l({\e}^\dag\t^m\e\r)\cD^{(0)}{}_m-g\l[\s,\c\r]-2i{\tr\over{r}^2},
\qquad
D_3=\l({\e}^\dag\t^m\e^c\r)\cD^{(0)}{}_m,
$$
respectively, and using (\ref{CommRelDifOpRoundS}) in Appendix 
\ref{constKSSquashS}, we obtain the operator ${D'}_1$,
\begin{eqnarray*}
D_3D_1
&=&\l({\e}^\dag\t^n\e^c\r)\cD^{(0)}{}_n\l[
\l({\e}^\dag\t^m\e\r)\cD^{(0)}{}_m-g\l[\s,\c\r]-2i{\tr\over{r}^2}
\r]
\nn\\
&=&
\l[\l({\e}^\dag\t^m\e\r)\cD^{(-2)}{}_m
-g\l[\s,\c\r]\r]
\l({\e}^\dag\t^n\e^c\r)\cD^{(0)}{}_n
={D'}_1D_3.
\end{eqnarray*}
Using this relation, we can compute the one-loop determinant 
\begin{eqnarray*}
Z^{{\rm 1-loop}}_{{\rm H,F}}=
\Det_{(0,0)}\l[\Delta^{\N=2}_{{\rm H,B}}\r]
{\Det_{(0,0)}\Big[
\l({\e}^\dag\t^m\e\r)\cD^{(0)}{}_m-g\l[\s,\c\r]-2i{\tr\over{r}^2}\Big]
\over
\Det_{(0,0)}\Big[
\l({\e}^\dag\t^m\e\r)\cD^{(-2)}{}_m-g\l[\s,\c\r]\Big]},
\end{eqnarray*}
where the differential operator $\Delta^{\N=2}_{{\rm H,B}}$ denotes 
\begin{eqnarray*}
\Delta^{\N=2}_{{\rm H,B}}
&=&
-\Big(\l({\e}^\dag\t^m\e\r)\cD^{(-2)}{}_m-g\l[\s,\c\r]\Big)
\Big(\l({\e}^\dag\t^m\e\r)\cD^{(-2)}{}_m+g\l[\s,\c\r]\Big)
-\l({\e}^\dag\t^m\e^c\r)\cD^{(0)}{}_m\l({\e^c}^\dag\t^m\e\r)\cD^{(-2)}{}_m.
\end{eqnarray*}

In the bosonic part of the system 
$(\tilde{H},\tilde{H}^\dag,\chi,\xi,\teta,\tkappa)$ 
of the regulator action ${\cal S}_Q$,
\begin{eqnarray*}
-\int d^5x \sqrt{g} \l[
\l(\dl_Q\chi\r)^\dag\c\dl_Q\chi+\l(\dl_Q\xi\r)^\dag\c\dl_Q\xi
+\l(\dl_Q\teta\r)^\dag\c\dl_Q\teta+\l(\dl_Q\tkappa\r)^\dag\c\dl_Q\tkappa 
\r], 
\end{eqnarray*}
we will shift the auxiliuary fields $F^{\ta}{}_{\cb}$ so that 
we can trivially integrate them out. 
We will integrate the remaining part of the action by parts to obtain 
\begin{eqnarray*}
-2\int d^5x \sqrt{g}\tilde{H}^\dag\tilde{\Delta}^{\N=2}_{{\rm H,B}}\tilde{H},
\end{eqnarray*}
with the differential operator $\tilde{\Delta}^{\N=2}_{{\rm H,B}}$ given by 
\begin{eqnarray*}
-\Big(\l({\e}^\dag\t^m\e\r)\cD^{(0)}{}_m-g\l[\s,\c\r]+2i{\tr\over{r}^2}\Big)
\Big(\l({\e}^\dag\t^m\e\r)\cD^{(0)}{}_m+g\l[\s,\c\r]+2i{\tr\over{r}^2}\Big)
-\l({\e}^\dag\t^m\e^c\r)\cD^{(2)}{}_m\l({\e^c}^\dag\t^m\e\r)\cD^{(0)}{}_m.
\end{eqnarray*}
Therefore, we can read 
the one-loop determinant from the bosonic fields of 
the system $(\tilde{H},\tilde{H}^\dag,\chi,\xi,\teta,\tkappa,F^1{}_2)$ as 
$$
Z^{1-{\rm loop}}_{{\rm H,B}}
={1\over\Det_{(0,0)}\l[\tilde{\Delta}^{\N=2}_{{\rm H,B}}\r]}.
$$

Let us move onto the fermionic part of the systerm 
$({H},{H}^\dag,\tchi,\txi,\eta,\kappa,F^2{}_2)$ 
of the regulator action ${\cal S}_Q$,
\begin{eqnarray*}
&&-\int \sqrt{g} d^5x \l[
\dl_Q\l(\dl_Q\tchi\r)^\dag\c\tchi+\dl_Q\l(\dl_Q\txi\r)^\dag\c\txi
+\dl_Q\l(\dl_Q\eta\r)^\dag\c\eta+\dl_Q\l(\dl_Q\kappa\r)^\dag\c\kappa \r]
\nn\\
&&={i\over2}\int \sqrt{g} d^5x \l[
\raise1.5ex\hbox{$\begin{pmatrix}\tchi,&\txi\end{pmatrix}$}
\begin{pmatrix}
\l({\e}^\dag\t^m\e\r)\cD^{(2)}{}_m-g\l[\s,\c\r]
&
\l({\e^c}^\dag\t^m\e\r)\cD^{(0)}{}_m
\cr
\l({\e}^\dag\t^m\e^c\r)\cD^{(2)}{}_m
&
-\l({\e}^\dag\t^m\e\r)\cD^{(0)}{}_m-g\l[\s,\c\r]-2i{\tr\over{r}^2}
\end{pmatrix}
\begin{pmatrix}\kappa\cr\eta\end{pmatrix}
\r],
\end{eqnarray*}
upto an integration by parts. It gives rise to 
the one-loop determinant 
\begin{eqnarray}
\tilde{Z}^{{\rm 1-loop}}_{{\rm H,F}}=\Det_{(1,0)}\l[
\begin{pmatrix}
\l({\e}^\dag\t^m\e\r)\cD^{(2)}{}_m-g\l[\s,\c\r]
&
\l({\e^c}^\dag\t^m\e\r)\cD^{(0)}{}_m
\cr
\l({\e}^\dag\t^m\e^c\r)\cD^{(2)}{}_m
&
-\l({\e}^\dag\t^m\e\r)\cD^{(0)}{}_m-g\l[\s,\c\r]-2i{\tr\over{r}^2}
\end{pmatrix}\r], 
\label{N=2ZH}
\end{eqnarray}
with the determinant $\Det_{(1,0)}$ defined in the previous subsection 
\ref{LocalizationN=1Squash}. 

As we have done just above, 
identifying the differential operators $D_1$ and $D_3$ with 
$$
D_1=\l({\e}^\dag\t^m\e\r)\cD^{(2)}{}_m-g\l[\s,\c\r],
\qquad
D_3=\l({\e}^\dag\t^m\e^c\r)\cD^{(2)}{}_m,
$$
and using (\ref{CommRelDifOpRoundS}) in Appendix 
\ref{constKSSquashS}, the operator ${D'}_1$ is found to be
\begin{eqnarray*}
D_3D_1
&=&\l({\e}^\dag\t^n\e^c\r)\cD^{(2)}{}_n\Big[
\l({\e}^\dag\t^m\e\r)\cD^{(2)}{}_m-g\l[\s,\c\r]
\Big]
\nn\\
&=&
\l[\l({\e}^\dag\t^m\e\r)\cD^{(0)}{}_m
-g\l[\s,\c\r]+2i{\tr\over{r}^2}\r]
\l({\e}^\dag\t^n\e^c\r)\cD^{(2)}{}_n
={D'}_1D_3.
\end{eqnarray*}
It follows from this relation that the one-loop determinant is computed to give 
\begin{eqnarray*}
\tilde{Z}^{{\rm 1-loop}}_{{\rm H,F}}=
\Det_{(1,0)}\l[\tilde{\Delta}^{\N=2}_{{\rm H,B}}\r]
{\Det_{(1,0)}\Big[
\l({\e}^\dag\t^m\e\r)\cD^{(2)}{}_m-g\l[\s,\c\r]
\Big]
\over
\Det_{(1,0)}\Big[
\l({\e}^\dag\t^m\e\r)\cD^{(0)}{}_m-g\l[\s,\c\r]+2i{\tr\over{r}^2}\Big]},
\end{eqnarray*}
with $\tilde{\Delta}^{\N=2}_{{\rm H,B}}$ given just above.

In the bosonic part of the system 
$({H},{H}^\dag,\tchi,\txi,\eta,\kappa,F^2{}_2)$ 
of the regulator action ${\cal S}_Q$,
\begin{eqnarray*}
-\int d^5x \sqrt{g} \l[
\l(\dl_Q\tchi\r)^\dag\c\dl_Q\tchi+\l(\dl_Q\txi\r)^\dag\c\dl_Q\txi
+\l(\dl_Q\eta\r)^\dag\c\dl_Q\eta+\l(\dl_Q\kappa\r)^\dag\c\dl_Q\kappa 
\r], 
\end{eqnarray*}
we can trivially integrate $F^{\ta}{}_{\cb}$ out in the same way as above. 
We will integrate the remaining part of the action by parts to obtain 
\begin{eqnarray*}
-2\int d^5x \sqrt{g}{H}^\dag{\Delta}^{\N=2}_{{\rm H,B}}{H},
\end{eqnarray*}
with the same $\tilde{\Delta}^{\N=2}_{{\rm H,B}}$ as above.  
It immediately gives the one-loop determinant from the bosonic fields of 
the system $({H},{H}^\dag,\tchi,\txi,\eta,\kappa,F^2{}_2)$, 
$$
\tilde{Z}^{1-{\rm loop}}_{{\rm H,B}}
={1\over\Det_{(0,1)}\l[{\Delta}^{\N=2}_{{\rm H,B}}\r]}.
$$

Combining the one-loop determinants from both the systems, 
we obtain 
\begin{eqnarray*}
{Z}^{1-{\rm loop}}_{{\rm H}}
&=&{Z}^{1-{\rm loop}}_{{\rm H,F}}\tilde{Z}^{1-{\rm loop}}_{{\rm H,F}}
{Z}^{1-{\rm loop}}_{{\rm H,B}}\tilde{Z}^{1-{\rm loop}}_{{\rm H,B}}
\nn\\
&=&
{\Det_{(0,0)}\l[\Delta^{\N=2}_{{\rm H,B}}\r]
\over\Det_{(0,1)}\l[{\Delta}^{\N=2}_{{\rm H,B}}\r]}
{\Det_{(1,0)}\l[\tilde{\Delta}^{\N=2}_{{\rm H,B}}\r]
\over\Det_{(0,0)}\l[\tilde{\Delta}^{\N=2}_{{\rm H,B}}\r]}
{\Det_{(0,0)}\Big[
\l({\e}^\dag\t^m\e\r)\cD^{(0)}{}_m-g\l[\s,\c\r]-2i{\tr\over{r}^2}\Big]
\over
\Det_{(1,0)}\Big[
\l({\e}^\dag\t^m\e\r)\cD^{(0)}{}_m-g\l[\s,\c\r]+2i{\tr\over{r}^2}\Big]
}
\nn\\
&&\quad
\times
{\Det_{(1,0)}\Big[
\l({\e}^\dag\t^m\e\r)\cD^{(2)}{}_m-g\l[\s,\c\r]
\Big]
\over
\Det_{(0,0)}\Big[
\l({\e}^\dag\t^m\e\r)\cD^{(-2)}{}_m-g\l[\s,\c\r]\Big]}.
\end{eqnarray*}

We may regard the differential operators 
$$
\l({\e}^\dag\t^m\e\r)\cD^{(0)}_m={2i\over\tr}L_3,
\quad
\l({\e^c}^\dag\t^m\e\r)\cD^{(0)}_m={2i\over{r}}L_+,
\quad
\l({\e}^\dag\t^m\e^c\r)\cD^{(0)}_m={2i\over{r}}L_-,
$$
as the generators of the Lie algebra of $SU(2)$ satisfying that 
$$
\l[L_3,\,L_{\pm}\r]={\pm}L_{\pm}, 
\qquad
\l[L_+,\,L_{-}\r]=2L_{3}. 
$$ 
Then, the scalar spherical harmonics $\vp_{l,m,\tm}$ 
$(l=0,1/2,1,3/2,\cdots; -l\le{m},\tm\le{l})$ on the $S^3$ obey 
$$
L_3\vp_{l,m,\tm}=m\vp_{l,m,\tm}, 
\qquad
L_{\pm}\vp_{l,m,\tm}=\sqrt{(l\mp{m})(l\pm{m}+1)}\vp_{l,m\pm{1},\tm}. 
$$

Each of the fluctuations is in the adjoint representation of the gauge 
group $G$, whose Cartan generators we denote as $H_i$ ($i=1,\cdots,r$) 
with r the rank of $G$, and the remaining generators as $E_\a$ with 
$\a$ a root of $G$. We assume that they obey 
$$
\l[H_i,\,E_\a\r]=\a_i E_\a,
\qquad
\l[E_\a,\,E_{-\a}\r]=\sum_{i=1}^{r}\a_iH_i\equiv\a\c{H},
$$
and are normalized as 
$$
\Tr\l[H_iH_j\r]=\dl_{i,j}, 
\qquad
\Tr\l[E_{-\a}E_\a\r]=1.
$$

As explained in \cite{BlauThompson}, 
the Hodge decomposition implys that for the space $\Om^{k,l}(\RS)$ 
of all the ($k,l$)-forms on the Riemann surface $\RS$, 
$$
\Om^{1,0}(\RS)\oplus\Om^{0,1}(\RS)
=\l(\Om^{0,0}(\RS)\ominus{H}^{0}(\RS)\r)\oplus
\l(\Om^{0,0}(\RS)\ominus{H}^{0}(\RS)\r)\oplus{H}^1(\RS),
$$
where $H^p(\RS)$ is the space of all the harmonic $p$-forms on $\RS$. 
It follows from this that for a constant $D$, 
\begin{eqnarray*}
{\Det_{(0,0)}\l[D\r]\over\Det_{(1,0)}\l[D\r]}=D^{b_0(\RS)-\hf{b}_1(\RS)}
=D^{\hf\chi(\RS)}
\end{eqnarray*}
with $b_i(\RS)=\dim{H}_i(\RS)$ the $i$-th Betti number, and 
with the Euler number $\chi(\RS)$ of the surface $\RS$; 
$$
\chi(\RS)=b_0(\RS)-b_1(\RS)+b_2(\RS)=2b_0(\RS)-b_1(\RS),
$$
where we have used the Hodge duality; $b_0(\RS)=b_2(\RS)$. 

The one-loop determinant $\Det_{(k,l)}$ in $Z^{1-{\rm loop}}_{{\rm H}}$ is 
defined over the space with the basis\footnote{
More precisely, the basis consists of 
$\{\vp_{l,m,\tm}\otimes{E}_{\a}\otimes{v}\}$ and 
$\{\vp_{l,m,\tm}\otimes{H}_{i}\otimes{v}\}$, 
with $v\in\Om^{k,l}(\RS)$. However, 
the Cartan part of the Lie algebra of $G$ contributes a constant 
to the determinant, and we will omit them in computing 
the partition function. 
} 
$\{\vp_{l,m,\tm}\otimes{E}_{\a}\otimes{v}\}$, 
where $v\in\Om^{k,l}(\RS)$.

In terms of the basis $\{\vp_{l,m,\tm}\otimes{E}_{\a}\otimes{v}\}$, we 
obtain 
\begin{eqnarray*}
{\Det_{(0,0)}\Big[
\l({\e}^\dag\t^m\e\r)\cD^{(0)}{}_m-g\l[\s,\c\r]-2i{\tr\over{r}^2}\Big]
\over
\Det_{(1,0)}\Big[
\l({\e}^\dag\t^m\e\r)\cD^{(0)}{}_m-g\l[\s,\c\r]+2i{\tr\over{r}^2}\Big]
}
=\l[\prod_{\a\in\La}
\prod_{l\in\hf{\bf Z}_{\ge0}}\prod_{m=-l}^{l}\prod_{\tm=-l}^{l}
\l({2i\over\tr}m-g\sa+2i{\tr\over{r}^2}\r)
\r]^{\hf\chi(\RS)},
\end{eqnarray*}
up to an overall constant, 
where $\a$ is a root of the Lie algebra of the gauge group, and 
$\La$ is the set of all the roots of it.

By the hermitian conjugation, we can see that
$$
\Det_{(0,0)}\Big[
\l({\e}^\dag\t^m\e\r)\cD^{(-2)}{}_m-g\l[\s,\c\r]\Big]
=
\Det_{(0,0)}\Big[
\l({\e}^\dag\t^m\e\r)\cD^{(2)}{}_m-g\l[\s,\c\r]\Big],
$$
and in a simialr way to above, we can compute 
\begin{eqnarray*}
{\Det_{(1,0)}\Big[
\l({\e}^\dag\t^m\e\r)\cD^{(2)}{}_m-g\l[\s,\c\r]
\Big]
\over
\Det_{(0,0)}\Big[
\l({\e}^\dag\t^m\e\r)\cD^{(-2)}{}_m-g\l[\s,\c\r]\Big]}
={1\over\l[\prod_{\a\in\La}
\prod_{l\in\hf{\bf Z}_{\ge0}}\prod_{m=-l}^{l}\prod_{\tm=-l}^{l}
\l({2i\over\tr}m+{2i\over\tr}\l(1-{\tr^2\over{r}^2}\r)-g\sa\r)
\r]^{\hf\chi(\RS)}},
\end{eqnarray*}
upto a constant factor.

After replacing spin $l$ by $n=2l=0,1,2,\cdots$, and shifting $n\to{n-1}$, 
we find that 
\begin{eqnarray*}
&&{\Det_{(0,0)}\Big[
\l({\e}^\dag\t^m\e\r)\cD^{(0)}{}_m-g\l[\s,\c\r]-2i{\tr\over{r}^2}\Big]
\over
\Det_{(1,0)}\Big[
\l({\e}^\dag\t^m\e\r)\cD^{(0)}{}_m-g\l[\s,\c\r]+2i{\tr\over{r}^2}\Big]
}
{\Det_{(1,0)}\Big[
\l({\e}^\dag\t^m\e\r)\cD^{(2)}{}_m-g\l[\s,\c\r]
\Big]
\over
\Det_{(0,0)}\Big[
\l({\e}^\dag\t^m\e\r)\cD^{(-2)}{}_m-g\l[\s,\c\r]\Big]}
\nn\\
&&=\prod_{\a\in\La}\l[\prod_{n=1}^{\infty}
\l({n-1+2{\tr^2\over{r}^2}-i\tr{g}\sa
\over{n+1-2{\tr^2\over{r}^2}+i\tr{g}\sa}}\r)^n\r]^{\hf\chi(\RS)}
=\prod_{\a\in\La}\l[
{1\over{s}_{b=1}\l(i-2i{\tr^2\over{r}^2}-\tr{g}\sa\r)}\r]^{\hf\chi(\RS)},
\end{eqnarray*}
where $s_b(x)$ is a double sine function;
$$
s_b(x)=\prod_{m,n=0}^{\infty}{mb+nb^{-1}+\hf{Q}-ix\over{mb+nb^{-1}+\hf{Q}+ix}},
$$
with $Q=b+b^{-1}$. In particular, when $b=1$, it is reduced to
\begin{equation}
s_{b=1}(x)=\prod_{n=1}^{\infty}\l({n-ix\over{n}+ix}\r)^n. 
\label{doublesineb=1}
\end{equation}
For more details on double sine functions, 
see \cite{HamaHosomichi,Hosomichi,doublesinefunction,Teschner}. 

The remaining factor in the one-loop contribution $Z^{1-{\rm loop}}_{{\rm H}}$ 
is computed in a similar way to yield 
\begin{eqnarray*}
&&{\Det_{(0,0)}\l[\Delta^{\N=2}_{{\rm H,B}}\r]
\over\Det_{(0,1)}\l[{\Delta}^{\N=2}_{{\rm H,B}}\r]}
{\Det_{(1,0)}\l[\tilde{\Delta}^{\N=2}_{{\rm H,B}}\r]
\over\Det_{(0,0)}\l[\tilde{\Delta}^{\N=2}_{{\rm H,B}}\r]}
\nn\\
&&=
\prod_{\a\in\La}\prod_{l\in\hf{\bf Z}_{\ge0}}\prod_{m,\tm=-l}^{l}
\l({{{4\over\tr^2}(m+1-{\tr^2\over{r}^2})^2+{4\over{r}^2}(l-m)(l+m+1)+g^2\sa^2}
\over{{4\over\tr^2}(m+{\tr^2\over{r}^2})^2+{4\over{r}^2}(l-m)(l+m+1)+g^2\sa^2}
}\r)^{\hf\chi(\RS)}
\nn\\
&&=\prod_{\a\in\La}\prod_{l\in\hf{\bf Z}_{\ge0}}\prod_{\tm=-l}^{l}
\l({{{4\over\tr^2}(l+1-{\tr^2\over{r}^2})^2+g^2\sa^2}
\over{{4\over\tr^2}(-l-{\tr^2\over{r}^2})^2+g^2\sa^2}
}\r)^{\hf\chi(\RS)}
=\prod_{\a\in\La}\prod_{n=1}^{\infty}
\l({{(n+1-{2\tr^2\over{r}^2})^2+\tr^2g^2\sa^2}
\over{(n-1+{2\tr^2\over{r}^2})^2+\tr^2g^2\sa^2}
}\r)^{{n\over2}\chi(\RS)}
\nn\\
&&=
\prod_{\a\in\La}
\l[{{s}_{b=1}\l(i-2i{\tr^2\over{r}^2}-\tr{g}\sa\r)
{s}_{b=1}\l(i-2i{\tr^2\over{r}^2}+\tr{g}\sa\r)}\r]^{\hf\chi(\RS)}. 
\end{eqnarray*}

Therefore, wrapping up all the factors, we obtain 
\BBox{500}{60}{
\begin{eqnarray*}
Z^{1-{\rm loop}}_{{\rm H}}
=
\prod_{\a\in\La_+}
\l[{{s}_{b=1}\l(i-2i{\tr^2\over{r}^2}-\tr{g}\sa\r)
{s}_{b=1}\l(i-2i{\tr^2\over{r}^2}+\tr{g}\sa\r)}\r]^{\hf\chi(\RS)}, 
\end{eqnarray*}}
with $\La_+$ the set of all the positive roots in $\La$. 

Let us consider the round limit $\tr\to{r}$ of $Z^{1-{\rm loop}}_{{\rm H}}$. 
To this end, we will derive the formula
\begin{equation}
s_{b=1}(-i+x)s_{b=1}(-i-x)=\Big(2\sinh(\pi{x})\Big)^2,
\label{doublesin^2sinh}
\end{equation}
by using the zeta regularization, 
$$
\prod_{m=1}^{\infty}m=e^{\sum_{m=1}^{\infty}\log{m}}
\quad\to\quad e^{-\zeta'(0)}=\sqrt{2\pi}.
$$
In this regularization, we can prove the above formula as follows:
\begin{eqnarray*}
s_{b=1}(-i+x)s_{b=1}(-i-x)&=&
\prod_{n=1}^{\infty}\l({n-1-ix\over{n+1}+ix}\r)^n\l({n-1+ix\over{n+1}-ix}\r)^n
=x^2\prod_{m=1}^{\infty}\l(m^2+x^2\r)^2
\nn\\
&=&\l(\prod_{m=1}^{\infty}m^4\r)
\l[x\prod_{m=1}^{\infty}\l(1+{x^2\over{m}^2}\r)\r]^2
=\Big(2\sinh(\pi{x})\Big)^2,
\end{eqnarray*}
where we have used the formula 
$$
{1\over\pi}\sinh{\pi{x}}=x\prod_{m=1}^{\infty}\l(1+{x^2\over{m^2}}\r).
$$

We can make use of (\ref{doublesin^2sinh}) to see the round limit 
$\tr\to{r}$ of $Z^{1-{\rm loop}}_{{\rm H}}$, 
\begin{eqnarray*}
Z^{1-{\rm loop}}_{{\rm H}}
\quad\to\quad
\prod_{\a\in\La_+}
\Big[{{s}_{b=1}\l(-i-\tr{g}\sa\r)
{s}_{b=1}\l(-i+\tr{g}\sa\r)}\Big]^{\hf\chi(\RS)} 
=
\prod_{\a\in\La_+}
\Big[2\sinh\l(\pi{rg}\sa\r)\Big]^{\chi(\RS)}.
\end{eqnarray*}

In summary, we have seen that the one-loop constribution 
in the $\N=2$ twisting on the squashed $S^3$ is given by
\BBox{530}{70}{\begin{eqnarray*}
&&Z^{{\rm 1-loop}}=Z^{{\rm 1-loop}}_{{\rm V}}Z^{{\rm 1-loop}}_{{\rm H}}
\nn\\
&&=
\prod_{\a\in\La_+}\Big[2\sinh\l(\pi{rg}\sa\r)\Big]^{\chi(\RS)}
\l[{{s}_{b=1}\l(i-2i{\tr^2\over{r}^2}-\tr{g}\sa\r)
{s}_{b=1}\l(i-2i{\tr^2\over{r}^2}+\tr{g}\sa\r)}\r]^{\hf\chi(\RS)}.
\end{eqnarray*}
}

\section{Localization on the ellipsoid $S^3$} 
\label{EllipsoidLocalization}

We will calculate the partition function by localization 
on the ellipsoid $S^3$ in the background discussed 
in subsection \ref{HosomichiEllipsoid}. 
The calculations we will carry out are quite parallel to 
what we have done for the round and squashed $S^3$'s in the previous section. 
All we have to do is to replace the background gauge field $V$ by the one in 
(\ref{bgV}), and ${\tr/{r}^2}$ by $1/f$. The fixed points discussed 
in the begining of subsection \ref{SquashOne-LoopGaugeMultiplet} 
are the same as for the background on the ellipsoid $S^3$. 

Therefore, we will briefly explain the calculations of 
the one-loop contributions from the $\N=1$ gauge multiplet and 
the $\N=1$ hypermultiplet, separately in the next two subsections.

\subsection{One-Loop Contributions from the $\N=1$ Gauge Multiplet}
\label{SquashOne-LoopGaugeMultiplet}

For the BRST transformations of the $\N=1$ gauge multiplet, 
as discussed in section \ref{localization} 
and done in previous section \ref{SquashLocalization}, 
we will reduce all the component fields in the gauge multiplet 
into scalar fields on the $S^3$.  

As seen in section \ref{SquashLocalization}, upon converting 
the gauge field $A_m$ to $V_0$ and $V_\pm$, the field strength 
$F_{mn}$ and $F_{mz}$ are given, up to the gauge interactions, by
\begin{eqnarray*}
&&\hf\e_{mkl}\l({\e}^\dag\t_m\e\r)F_{kl}
={2\tr\over{r}^2}V_0+i\l({\e^c}^\dag\t_m\e\r)\cD^{(-2)}_mV_{-}
-i\l({\e}^\dag\t_m\e^c\r)\cD^{(2)}_mV_+,
\nn\\
&&\hf\e_{mkl}\l({\e^c}^\dag\t_m\e\r)F_{kl}
={4\tr\over{r}^2}V_++2i\l({\e}^\dag\t_m\e\r)\cD^{(2)}_mV_{+}
-i\l({\e^c}^\dag\t_m\e\r)\cD^{(0)}_mV_0,
\nn\\
&&F_{mz}
=
\l({\e^c}^\dag\t_m\e\r)\l[\hf\l({\e}^\dag\t^n\e^c\r)\cD^{(0)}_nA_z-\cD_zV_-\r]
+\l({\e}^\dag\t_m\e^c\r)\l[\hf\l({\e^c}^\dag\t^n\e\r)\cD^{(0)}_nA_z-\cD_zV_+\r]
\nn\\
&&\hskip 1.5cm
+\l({\e}^\dag\t_m\e\r)\Big[\l({\e}^\dag\t^n\e\r)\cD^{(0)}_nA_z-\cD_zV_0\Big],
\end{eqnarray*}
where we have used (\ref{ellisoidcross}), 
and we will omit the gauge interactions, as before, 
since they have no effects on the partition function in the large $t$ limit.

The BRST transformation of the bosonic fields is given by 
\begin{eqnarray*}
&&\dl_Q\ts=-\qt\txi,
\qquad
\dl_QV_0=-{i\over4}\txi,
\quad
\dl_QV_-=-{i\over4}\teta,
\quad
\dl_QV_+=0,
\qquad
\dl_QA_{\bar{z}}=\qt\tvp, 
\quad
\dl_QA_{z}=0,
\nn\\
&&\dl_Q{D}^1{}_1={i\over4}\l[
\l({\e}^\dag\t^m\e\r)\cD^{(0)}_m\txi-{2i\over{f}}\txi+g\l[\s,\,\txi\r]
+\l({\e^c}^\dag\t^m\e\r)\cD^{(-2)}_m\teta\r],
\nn\\
&&\dl_Q{D}^1{}_2={i\over4}\l[
-\l({\e}^\dag\t^m\e\r)\cD^{(2)}_m\chi+{2i\over{f}}\chi-g\l[\s,\,\chi\r]
+\l({\e^c}^\dag\t^m\e\r)\cD^{(0)}_m\vp-2i\cD_z\eta
\r],
\end{eqnarray*}
where we denote a fixed point of the scalar field $\s$ as the same letter $\s$, 
and the fluctuation about this fixed point $\s$ as $\ts$, as we have done in 
the previous section. 
Henceforth, we will keep this notation until the end of this section. 

The BRST tranformation of the fermionic fields is given by
\begin{eqnarray*}
&&\dl_Q\txi=0, \qquad \dl_Q\teta=0, 
\qquad 
\dl_Q\tvp=0,\qquad \dl_Q\tchi=-D^2{}_1,
\nn\\
&&\dl_Q\xi
=-{2i\over{f}}V_0+g\l[\s,\,V_0\r]+i\l({\e}^\dag\t^m\e\r)\d_m\ts
+\l({\e^c}^\dag\t^m\e\r)\cD^{(-2)}_mV_-
-\l({\e}^\dag\t^m\e^c\r)\cD^{(2)}_mV_++D^1{}_1,
\nn\\
&&\dl_Q\eta
=-{4i\over{f}}V_++2g\l[\s,\,V_+\r]
+2\l({\e}^\dag\t^m\e\r)\cD^{(2)}_mV_+
+i\l({\e^c}^\dag\t^m\e\r)\d_m\ts
-\l({\e^c}^\dag\t^m\e\r)\d_mV_0,
\nn\\
&&\dl_Q\vp=2i\l[\l({\e^\dag}\t^m\e\r)\d_mA_z+g\l[\s,\,A_z\r]
-\cD_zV_0+i\cD_z\ts\r],
\qquad
\dl_Q\chi=2i\l[\l({\e^c}^\dag\t^m\e\r)\d_mA_z-2\cD_zV_+\r],
\end{eqnarray*}
and furthermore, we find that 
\begin{eqnarray}
&&
\dl_Q\l(\dl_Q\xi\r)^\dag
=\hf\l[
{2\over{f}}\txi+ig\l[\s,\,\txi\r]+i\l({\e}^\dag\t^m\e\r)\d_m\txi
+i\l({\e^c}^\dag\t^m\e\r)\cD^{(-2)}_m\teta\r],
\nn\\
&&\dl_Q\l(\dl_Q\eta\r)^\dag
=-{i\over2}\l[
{2i\over{f}}\teta-g\l[\s,\,\teta\r]
+\l({\e}^\dag\t^m\e\r)\cD^{(-2)}_m\teta
-\l({\e}^\dag\t^m\e^c\r)\cD^{}_m\txi
\r],
\nn\\
&&\dl_Q\l(\dl_Q\vp\r)^\dag
=-{i\over2}\l[\l({\e^\dag}\t^m\e\r)\d_m\tvp-g\l[\s,\,\tvp\r]
+2i\cD_{\bar{z}}\txi\r],
\quad
\dl_Q\l(\dl_Q\chi\r)^\dag=-{i\over2}
\l[\l({\e}^\dag\t^m\e^c\r)\d_m\tvp+2i\cD_{\bar{z}}\teta\r], 
\nn\\
&&\dl_Q\l(\dl_Q\tchi\r)^\dag=
-{i\over2}\l[
-\l({\e}^\dag\t^m\e\r)\cD_m\chi+{2i\over{f}}\chi-g\l[\s,\,\chi\r]
+\l({\e^c}^\dag\t^m\e\r)\cD_m\vp-2i\cD_z\eta
\r].
\nn\end{eqnarray}

As we have discussed in the previous section \ref{SquashLocalization}, 
assuming that $\sa=\sum_{i=1}^{r}\s_i\a^i$ is non-zero for a generic 
$(\s^1,\cdots,\s^r)$, we can see that the operator $\l[\s,\,\c\r]$ acting 
on the sector with the basis $\{E_\a\}$ we are interested in is invertible, 
and we will `gauge away' the fluctuation $\ts$ by the shifts 
\begin{eqnarray*}
V_0 \quad\to\quad V_0-i{1\over{g}\l[\s,\,\c\r]}\l({\e}^\dag\t^m\e\r)\d_m\ts,
\qquad
V_+ \quad\to\quad 
V_+-{i\over2}{1\over{g}\l[\s,\,\c\r]}\l({\e^c}^\dag\t^m\e\r)\d_m\ts,
\nn\\
V_- \quad\to\quad 
V_--{i\over2}{1\over{g}\l[\s,\,\c\r]}\l({\e}^\dag\t^m\e^c\r)\d_m\ts,
\qquad
A_z \quad\to\quad 
A_z-{i}{1\over{g}\l[\s,\,\c\r]}\d_z\ts,
\end{eqnarray*}
in the BRST transformation in the large $t$ limit, where we used 
(\ref{CommRelDifOpEllipsoidS}) in Appendix \ref{EllipsoidS^3}. 

Using the remaining gauge transformations, we will `diagonalize' 
the value of the scalar $\s$ at one of the fixed points.  
The latter results in the Fadeev-Popov determinant 
\begin{eqnarray}
Z_{{\rm FP}}=\prod_{\a\in\La}\det_{(0,0)}\l[ig\sa\r]
=\int \l[d\bar{c}(z,\bar{z})dc(z,\bar{z})\r] 
\exp\l[-ig\sum_{\a\in\La}\int_{\RS} d^2z \sqrt{g_{\RS}}
\l(\s\c\a\r)\bar{c}_{-\a}c_{\a}\r],
\label{FPdetEllipsoid}
\end{eqnarray}
with the Fadeev-Popov ghost $c_{\a}(z,\bar{z})$, $\bar{c}_{\a}(z,\bar{z})$ 
$(\a\in\La)$, which are scalar fields on $\RS$. 

This gauge-fixing procedure is quite the same as for the squash $S^3$ 
in section \ref{SquashLocalization}, and  we will set $\ts$ to zero 
in the BRST transformations. 

The bosonic part (\ref{bosongaugeSQ}) of the gauge multiplet 
in the regulator action ${\cal S}_Q$, 
after integrating out the auxiliuary fields $D^{\da}{}_{\db}$ and 
integrating by parts, 
is reduced to the sum of 
\begin{eqnarray}
&&-\int d^5x\sqrt{g}\,\Tr\bigg[
\Big|
-{2i\over{f}}V_0+g\l[\s,\,V_0\r]
+\l({\e^c}^\dag\t^m\e\r)\cD^{(-2)}_mV_-
-\l({\e}^\dag\t^m\e^c\r)\cD^{(2)}_mV_+\Big|^2
\nn\\
&&\hskip3cm
+\Big|
-{4i\over{f}}V_++2g\l[\s,\,V_+\r]
+2\l({\e}^\dag\t^m\e\r)\cD^{(2)}_mV_+
-\l({\e^c}^\dag\t^m\e\r)\d_mV_0\Big|^2
\bigg],  
\label{V0pmSQEllipsoid}
\end{eqnarray}
and 
\begin{eqnarray*}
&&-4\int d^5x\sqrt{g}\,\Tr\Big[
A_z\Delta_0A_{\bar{z}}
+A_z\cD_{\bar{z}}\l(\l({\e}^\dag\t^m\e\r)\d_mV_0+g\l[\s,\,V_0\r]
+2\l({\e^c}^\dag\t^m\e\r)\cD^{(-2)}_mV_-\r)
\nn\\
&&\hskip 1cm
+\cD_{{z}}\l(\l({\e}^\dag\t^m\e\r)\d_mV_0-g\l[\s,\,V_0\r]
+2\l({\e}^\dag\t^m\e^c\r)\cD^{(2)}_mV_+\r){\c}A_{\bar{z}}
+\cD_zV_0\cD_{\bar{z}}V_0+4\cD_zV_+\cD_{\bar{z}}V_-
\Big],
\end{eqnarray*}
where $\Delta_0$ denotes the differential operator
$$
-\l[\l({\e}^\dag\t^m\e\r)\d_m+g\l[\s,\,\c\r]\r]
\l[\l({\e}^\dag\t^n\e\r)\d_n-g\l[\s,\,\c\r]\r]
-\l({\e^c}^\dag\t^m\e\r)\cD^{(-2)}_m\l({\e}^\dag\t^n\e^c\r)\d_n, 
$$
which is potitive and so invertible in the root sector expanded 
in the basis $\{E_{\a}\}$. 

As we have done for the squashed $S^3$ in the previous section 
\ref{SquashOne-LoopGaugeMultiplet}, using 
\begin{eqnarray}
&&D^{(q+2)}_0D^{(q)}_+=D^{(q)}_+\l(D^{(q)}_0+{2i\over{f}}\r)
-i\l({q+2\over2}\r)\l({\e^c}^\dag\t^{mn}\e\r)V_{mn},
\nn\\
&&D^{(q)}_-D^{(q)}_0=\l(D^{(q-2)}_0+{2i\over{f}}\r)D^{(q)}_-
+i\l({q\over2}\r)\l({\e}^\dag\t^{mn}\e^c\r)V_{mn},
\label{D+D0=D0D+}
\end{eqnarray}
with the abbreviations,
$$
D^{(q)}_0=\l({\e}^\dag\t^m\e\r)\cD^{(q)}_m,
\qquad
D^{(q)}_+=\l({\e^c}^\dag\t^m\e\r)\cD^{(q)}_m, 
\qquad
D^{(q)}_-=\l({\e}^\dag\t^m\e^c\r)\cD^{(q)}_m,
$$
derived from (\ref{CommRelDifOpEllipsoidS}) in Appendix \ref{EllipsoidS^3}, 
we will shift $A_z$ and $A_{\bar{z}}$ in the latter integrand to give
\begin{eqnarray}
&&-4\int d^5x\sqrt{g}\,\Tr\Big[
A_z\Delta_0A_{\bar{z}}
+\cD_{{z}}J_+\c{1\over\Delta_{-2}}\l(\cD_{{z}}J_+\r)^\dag
\Big],
\label{AzSQEllipsoid}
\end{eqnarray}
after integrations by parts, 
with 
$$
J_+=2\l(\l({\e}^\dag\t^m\e\r)\d_mV_++g\l[\s,\,V_+\r]-{2i\over{f}}V_+\r)
-\l({\e^c}^\dag\t^m\e\r)\d_mV_0, 
$$
where we have defined the operator $\Delta_{-2}$ by 
\begin{eqnarray*}
-\l[\l({\e}^\dag\t^m\e\r)\cD^{(-2)}_m+{2i\over{f}}+g\l[\s,\,\c\r]\r]
\l[\l({\e}^\dag\t^n\e\r)\cD^{(-2)}_n+{2i\over{f}}-g\l[\s,\,\c\r]\r]
-\l({\e}^\dag\t^n\e^c\r)\cD^{(0)}_n\l({\e^c}^\dag\t^m\e\r)\cD^{(-2)}_m,
\end{eqnarray*}
which is also invertible in the sector we are interested in. 
Here, we have made use of 
\begin{eqnarray*}
{1\over\Delta_0}D^{(0)}_+=D^{(-2)}_+{1\over\Delta_{-2}},
\end{eqnarray*}
which is also deduced from (\ref{CommRelDifOpEllipsoidS}). 

When we shift $V_\pm$ as 
$$
V_\pm \quad\to\quad V_\pm+\hf{1\over{D}^{(\pm2)}_0\mp{2i\over{f}}
\pm{g}\l[\s,\,\c\r]}D^{(0)}_{\pm}V_0,
$$
for a generic $\sa$, the term in $\dl_Q\xi$, 
$$
{D}^{(2)}_-V_+-{D}^{(-2)}_+V_-+\l({2i\over{f}}-g\l[\s,\,\c\r]\r)V_0
$$
is shifted to become 
${D}^{(2)}_-V_+-{D}^{(-2)}_+V_--{\cal K}_0V_0$, with ${\cal K}_0V_0$ denoting 
\begin{eqnarray*}
{g\l[\s,\,\c\r]\over
{\l(D^{(0)}_0+g\l[\s,\,\c\r]\r)\l(D^{(0)}_0-g\l[\s,\,\c\r]\r)}}
\l[{\l(D^{(0)}_0-{2i\over{f}}+g\l[\s,\,\c\r]\r)
\l(D^{(0)}_0-g\l[\s,\,\c\r]\r)}+D^{(-2)}_+D^{(0)}_-\r]V_0,~~~
\end{eqnarray*}
where we have used the formula 
\begin{eqnarray*}
\cD^{(2)}_\pm{1\over{\cD_0^{(2)}\pm{2i\over{f}}\mp{g}\l[\s,\,\c\r]}}
={1\over{\cD_0^{(0)}\mp{g}\l[\s,\,\c\r]}}\cD^{(2)}_\pm, 
\end{eqnarray*}
which follow from (\ref{D+D0=D0D+}) and  
\begin{eqnarray*}
&&\cD^{(q-2)}_0\cD^{(q)}_-
=\cD^{(q)}_-\l(\cD^{(q)}_0-{2i\over{f}}\r)+i\l({q-2\over2}\r)
\l({\e}^\dag\t^{mn}\e^c\r)V_{mn},
\end{eqnarray*}
and the formula 
\begin{eqnarray*}
\cD^{(-2)}_+\cD^{(0)}_--\cD^{(2)}_-\cD^{(0)}_+={4i\over{f}}\cD^{(0)}_0,
\end{eqnarray*}
of (\ref{CommRelDifOpEllipsoidS}), together with (\ref{D0f=0}) 
in Apeendix \ref{EllipsoidS^3}.

Therefore, the integrand of the sum of (\ref{V0pmSQEllipsoid}) and 
(\ref{AzSQEllipsoid}), after integrations by parts, becomes 
\begin{eqnarray*}
&&A_z\Delta_0A_{\bar{z}}
-V_+{1\over\Delta_{-2}}\l[\Delta_{-2}-4\cD_{{z}}\cD_{\bar{z}}\r]
\l(D^{(-2)}_0+{2i\over{f}}+g\l[\s,\,\c\r]\r)
\l(D^{(-2)}_0+{2i\over{f}}-g\l[\s,\,\c\r]\r)V_-
+\qt\l|{\cal K}_0V_0\r|^2,
\end{eqnarray*}
where we have shifted $V_0$ appropriately to eliminate the term 
${D}^{(2)}_-V_+-{D}^{(-2)}_+V_-$, as before. 

Integrating over the remaining fluctuations, 
we obtain the one-loop determinants from 
the bosonic fields of the gauge multiplet,
\begin{eqnarray*}
Z_{{\rm V,B}}^{{\rm 1-loop}}=Z_{{\rm V},0}
{\Det_{(0,0)}\l[\Delta_{-2}\r]\over\Det_{(0,1)}\l[\Delta_0\r]
\Det_{(0,0)}\l[\l[\Delta_{-2}-4\cD_{{z}}\cD_{\bar{z}}\r]
\l(D^{(-2)}_0+{2i\over{f}}+g\l[\s,\,\c\r]\r)
\l(D^{(-2)}_0+{2i\over{f}}-g\l[\s,\,\c\r]\r)
\r]},
\end{eqnarray*}
where $Z_{{\rm V},0}$ denotes the one-loop contribution from $V_0$.  

Taking account of the fact that $V_0$ is a real field; $V_0^\dag=V_0$, 
we find that 
\begin{eqnarray*}
Z_{V,0}&=&\prod_{\a\in\La_+}\det_{(0,0)}\l[
{g\sa\over\l(D^{(0)}_0+{g}\sa\r)
\l(D^{(0)}_0-{g}\sa\r)}\r]^{-2}
\nn\\
&&\qquad\times
\det_{(0,0)}\l[\l(D^{(0)}_0-{2i\over{f}}+g\sa\r)
\Big(D^{(0)}_0-g\sa\Big)+D^{(-2)}_+D^{(0)}_-\r]^{-2}.
\end{eqnarray*}

Therefore, up to an overall constant including the Cartan part, 
$$
Z_{{\rm V},0}={1\over{Z}_{{\rm FP}}}\cdot
{\l(\Det_{(0,0)}\l[D_0^{(0)}-g\l[\s,\,\c\r]\r]\r)^2
\over
\Det_{(0,0)}\l[\l(D^{(0)}_0-{2i\over{f}}+g\l[\s,\,\c\r]\r)
\Big(D^{(0)}_0-g\l[\s,\,\c\r]\Big)+D^{(-2)}_+D^{(0)}_-\r]}.
$$

The computation of the one-loop contributions from the fermionic part 
of the gauge multiplet in the regulator action ${\cal S}_Q$ 
is also parallel to that for the squashed $S^3$. 

Integrating by parts, the fermionic part is reduced to the sum of 
\begin{eqnarray*}
-\l({i\over2}\r)\l[
\raise1.5ex\hbox{$\begin{pmatrix}\txi, \teta\end{pmatrix}$}
\begin{pmatrix}
D^{(0)}_0+{2i\over{f}}+g\l[\s,\c\r]
&
D^{(2)}_-
\cr
D^{(0)}_+
&
-D^{(2)}_0+{2i\over{f}}+g\l[\s,\c\r]
\end{pmatrix}
\begin{pmatrix}\xi\cr\eta\end{pmatrix}\r],
\end{eqnarray*}
and 
\begin{eqnarray*}
\l({i\over2}\r)\bigg[
\raise1.5ex\hbox{$\begin{pmatrix}\tvp, \tchi\end{pmatrix}$}
\begin{pmatrix}
D^{(0)}_0-g\l[\s,\c\r]
&
D^{(2)}_-
\cr
D^{(0)}_+
&
-D^{(2)}_0+{2i\over{f}}-g\l[\s,\c\r]
\end{pmatrix}
\begin{pmatrix}\vp\cr\chi\end{pmatrix}
\bigg]
+
\raise1.5ex\hbox{$\begin{pmatrix}\cD_{\bar{z}}\txi, 
\cD_{\bar{z}}\teta\end{pmatrix}$}
\begin{pmatrix}\vp\cr\chi\end{pmatrix}
+
\raise1.5ex\hbox{$\begin{pmatrix}\tvp,&\tchi\end{pmatrix}$}
\begin{pmatrix}0\cr\cD_z\chi\end{pmatrix}.
\end{eqnarray*}

Integrating over $\vp,\chi,\tvp$, and $\tchi$ 
gives the one-loop determinant 
\begin{eqnarray}
\Det_{(1,0)}\l[
\begin{pmatrix}
D^{(0)}_0-g\l[\s,\c\r]
&
D^{(2)}_-
\cr
D^{(0)}_+
&
-D^{(2)}_0+{2i\over{f}}-g\l[\s,\c\r]
\end{pmatrix}
\r],
\label{vpchideterminant}
\end{eqnarray}
and leaves the integrand 
\begin{eqnarray*}
-{i\over2}\bigg[
\raise1.5ex\hbox{$\begin{pmatrix}\txi, \teta\end{pmatrix}$}
\begin{pmatrix}
D^{(0)}_0-g\l[\s,\c\r]
&
D^{(2)}_-
\cr
D^{(0)}_+
&
-D^{(2)}_0+{2i\over{f}}-g\l[\s,\c\r]
\end{pmatrix}^{-1}
\begin{pmatrix}
D_1
&
0
\cr
D_3
&
D_4
\end{pmatrix}
\begin{pmatrix}\xi\cr\eta\end{pmatrix}
\bigg],
\end{eqnarray*}
after intergation by parts, where the operators $D_1$, $D_3$, and $D_4$ denote 
\begin{eqnarray*}
D_1&=&
\Big(D^{(0)}_0-g\l[\s,\,\c\r]\Big)\l(D^{(0)}_0+{2i\over{f}}
+g\l[\s,\,\c\r]\r)+D^{(2)}_-D^{(0)}_+, 
\nn\\
D_3&=&D^{(0)}_+\l(D^{(0)}_0+{2i\over{f}}+g\l[\s,\,\c\r]\r)
-\l(D^{(2)}_0-{2i\over{f}}+g\l[\s,\c\r]\r)D^{(0)}_+,
\nn\\
D_4&=&
4\cD_{\bar{z}}\cD_z
+\l(D^{(2)}_0-{2i\over{f}}+g\l[\s,\,\c\r]\r)
\l(D^{(2)}_0-{2i\over{f}}-g\l[\s,\,\c\r]\r)
+D^{(0)}_+D^{(2)}_-.
\end{eqnarray*}

Integrating over the remaining $\xi$, $\eta$, $\txi$, and $\teta$, and 
combining the resulting determinant with (\ref{vpchideterminant}),
we obtain the one-loop contributions from the fermionic fields of 
the gauge multiplet, 
\begin{eqnarray*}
Z_{V,F}^{{\rm 1-loop}}=
{\Det_{(1,0)}\l[
\begin{pmatrix}
D^{(0)}_0-g\l[\s,\c\r]
&
D^{(2)}_-
\cr
D^{(0)}_+
&
-D^{(2)}_0+{2i\over{f}}-g\l[\s,\c\r]
\end{pmatrix}
\r]
\over
\Det_{(0,0)}\l[
\begin{pmatrix}
D^{(0)}_0-g\l[\s,\c\r]
&
D^{(2)}_-
\cr
D^{(0)}_+
&
-D^{(2)}_0+{2i\over{f}}-g\l[\s,\c\r]
\end{pmatrix}
\r]}
\Det_{(0,0)}\l[
\begin{pmatrix}
D_1
&
0
\cr
D_2
&
D_4
\end{pmatrix}
\r],
\end{eqnarray*}
with $\Det_{(0,0)}\l[\begin{pmatrix}D_1&0\cr{D_2}&D_4\end{pmatrix}\r]$
evaluated to give 
\begin{eqnarray*}
&&
\Det_{(0,0)}\l[\Big(D^{(0)}_0-g\l[\s,\,\c\r]\Big)\l(D^{(0)}_0
+{2i\over{f}}+g\l[\s,\,\c\r]\r)+D^{(2)}_-D^{(0)}_+\r]
\nn\\
&&\hskip2cm\times
\Det_{(0,0)}\l[
4\cD_{\bar{z}}\cD_z
+\l(D^{(2)}_0-{2i\over{f}}+g\l[\s,\,\c\r]\r)
\l(D^{(2)}_0-{2i\over{f}}-g\l[\s,\,\c\r]\r)
+D^{(0)}_+D^{(2)}_-\r].
\end{eqnarray*}

The determinant 
$$
\Det_{(1,0)}\l[
\begin{pmatrix}
D^{(0)}_0-g\l[\s,\c\r]
&
D^{(2)}_-
\cr
D^{(0)}_+
&
-D^{(2)}_0+{2i\over{f}}-g\l[\s,\c\r]
\end{pmatrix}
\r],
$$
noting the relation
$$
D^{(2)}_-{1\over-D^{(2)}_0+{2i\over{f}}-g\l[\s,\c\r]}
={1\over-D^{(0)}_0-g\l[\s,\c\r]}D^{(2)}_-,
$$
is evaluated to yield 
\begin{eqnarray}
{\Det_{(1,0)}\l[D^{(2)}_0-{2i\over{f}}+g\l[\s,\c\r]\r]
\over
\Det_{(1,0)}\l[D^{(0)}_0+g\l[\s,\c\r]\r]}
\Det_{(1,0)}\l[-\l(D^{(0)}_0+g\l[\s,\c\r]\r)\l(D^{(0)}_0-g\l[\s,\c\r]\r)
-D^{(2)}_-D^{(0)}_+\r].
\label{fermiondeterminantellipsoid}
\end{eqnarray}
Since the determinant of an operator is the same as  the one of its ajoint 
operator, it follows that 
\begin{eqnarray*}
&&\Det_{(1,0)}\l[-\l(D^{(0)}_0+g\l[\s,\c\r]\r)\l(D^{(0)}_0-g\l[\s,\c\r]\r)
-D^{(2)}_-D^{(0)}_+\r]
=\Det_{(0,1)}\l[\Delta_{0}\r],
\nn\\
&&\Det_{(1,0)}\l[D^{(2)}_0-{2i\over{f}}\pm{g}\l[\s,\c\r]\r]
=
\Det_{(0,1)}\l[D^{(-2)}_0+{2i\over{f}}\pm{g}\l[\s,\c\r]\r],
\nn\\
&&\Det_{(1,0)}\l[D^{(0)}_0\pm{g}\l[\s,\c\r]\r]
=
\Det_{(0,1)}\l[D^{(0)}_0\pm{g}\l[\s,\c\r]\r].
\end{eqnarray*}
Using them, (\ref{fermiondeterminantellipsoid}) may be rewritten as
\begin{eqnarray}
{\Det_{(0,1)}\l[D^{(2)}_0+{2i\over{f}}+g\l[\s,\c\r]\r]
\over
\Det_{(0,1)}\l[D^{(0)}_0+g\l[\s,\c\r]\r]}
\Det_{(0,1)}\l[\Delta_0\r].
\label{Det(1,0)fermiellipsoid}
\end{eqnarray}

For the determinant 
$$
\Det_{(0,0)}\l[
\begin{pmatrix}
D^{(0)}_0-g\l[\s,\c\r]
&
D^{(2)}_-
\cr
D^{(0)}_+
&
-D^{(2)}_0+{2i\over{f}}-g\l[\s,\c\r]
\end{pmatrix}
\r],
$$
using the formula
$$
D^{(0)}_+{1\over{D^{(0)}_0-g\l[\s,\c\r]}}
={1\over{D^{(2)}_0-{2i\over{f}}-g\l[\s,\c\r]}}D^{(0)}_+,
$$
and the relation of the determinant of an operator with that of 
the adjoint operator,
\begin{eqnarray*}
&&\Det_{(0,0)}\l[
-\l(D^{(2)}_0-2i{\tr\over{r}^2}-g\l[\s,\c\r]\r)
\l(D^{(2)}_0-2i{\tr\over{r}^2}+g\l[\s,\c\r]\r)-D^{(0)}_+D^{(2)}_-\r]
=\Det_{(0,0)}\l[\Delta_{-2}\r].
\nn\\
&&\Det_{(0,0)}\l[D^{(2)}_0-{2i\over{f}}\pm{g}\l[\s,\c\r]\r]
=
\Det_{(0,0)}\l[D^{(-2)}_0+{2i\over{f}}\pm{g}\l[\s,\c\r]\r],
\nn\\
&&\Det_{(0,0)}\l[D^{(0)}_0\pm{g}\l[\s,\c\r]\r]
=
\Det_{(0,0)}\l[D^{(0)}_0\pm{g}\l[\s,\c\r]\r],
\end{eqnarray*}
we can see that it is reduced to
\begin{eqnarray}
{\Det_{(0,0)}\l[D^{(0)}_0-g\l[\s,\c\r]\r]
\over
\Det_{(0,0)}\l[D^{(-2)}_0+{2i\over{f}}-g\l[\s,\c\r]\r]}
\Det_{(0,0)}\l[\Delta_{-2}\r].
\label{Det(0,0)fermiellipsoid}
\end{eqnarray}

Substituting (\ref{Det(1,0)fermiellipsoid}) and (\ref{Det(0,0)fermiellipsoid}) 
into $Z_{{\rm V, F}}^{{\rm 1-loop}}$, we obtain  
\begin{eqnarray*}
{\Det_{(0,1)}\l[D^{(-2)}_0+{2i\over{f}}+g\l[\s,\c\r]\r]
\Det_{(0,0)}\l[D^{(-2)}_0+{2i\over{f}}-g\l[\s,\c\r]\r]
\over
\Det_{(0,1)}\l[D^{(0)}_0+g\l[\s,\c\r]\r]
\Det_{(0,0)}\l[D^{(0)}_0-g\l[\s,\c\r]\r]}
{\Det_{(0,1)}\l[\Delta_{0}\r]
\over
\Det_{(0,0)}\l[\Delta_{-2}\r]}
\Det_{(0,0)}\l[D_1\r]\Det_{(0,0)}\l[D_4\r].
\end{eqnarray*}

With the same argument about the adjoint operators, we can show that 
\begin{eqnarray*}
&&\Det_{(0,0)}\l[D_1\r]
=\Det_{(0,0)}\l[\l(D^{(0)}_0-{2i\over{f}}+g\l[\s,\,\c\r]\r)
\Big(D^{(0)}_0-g\l[\s,\,\c\r]\Big)+D^{(-2)}_+D^{(0)}_-\r],
\nn\\
&&\Det_{(0,0)}\l[D_4\r]
=\Det_{(0,0)}\l[4\cD_z\cD_{\bar{z}}-\Delta_{-2}\r],
\end{eqnarray*}
and using this, the one-loop contributions 
$Z^{{\rm 1-loop}}_{{\rm V,B}}$, $Z^{{\rm 1-loop}}_{{\rm V,F}}$, 
and $Z_{{\rm FP}}$ from the gauge multiplet are summaried to give 
\BBox{525}{60}{
\begin{eqnarray}
Z^{{\rm 1-loop}}_{{\rm V}}
=Z_{{\rm FP}}Z^{{\rm 1-loop}}_{{\rm V,B}}Z^{{\rm 1-loop}}_{{\rm V,F}}
={\Det_{(0,1)}\l[D^{(-2)}_0+{2i\over{f}}+g\l[\s,\c\r]\r]
\over
\Det_{(0,0)}\l[D^{(-2)}_0+{2i\over{f}}+g\l[\s,\c\r]\r]}
{\Det_{(0,0)}\l[D^{(0)}_0-g\l[\s,\c\r]\r]
\over
\Det_{(0,1)}\l[D^{(0)}_0-g\l[\s,\c\r]\r]}.
\label{Z_Vellipsoid}
\end{eqnarray}
}

The determinant $\Det_{(k,l)}$ can be evaluated by using the basis 
$\{h_{n,m,k}\otimes{v}\otimes{E}_{\a}, 
h_{n,m,k}\otimes{v}\otimes{H}_{i}\}$, for $v$ running over all the 
basis vectors of $\Om^{(k,l)}(\RS)$, the set of all $(k,l)$-forms on $\RS$, 
upon regarding $\Om^{(k,l)}(\RS)$ as a linear space. 
Here, $h_{n,m,k}$ ($n,m=0,1,2,\cdots; k=0,1,\cdots,n+m$) denote 
scalar spherical harmonics on the $S^3$ (See Appedix \ref{EllipsoidS^3} 
for more details) and obey  
\begin{eqnarray*}
&&\cD^{(0)}_0h_{n,m,k}
=\l[{i\over{r}}\l(i\pd{\vp}\r)+{i\over{\tr}}\l(-i\pd{\chi}\r)\r]h_{n,m,k}
=\l[{i(n-k)\over{r}}+{i(m-k)\over{\tr}}\r]h_{n,m,k},
\nn\\
&&\l(\cD^{(-2)}_0+{2i\over{f}}\r)h_{n,m,k}
=\l[{i\over{r}}\l(i\pd{\vp}+1\r)+{i\over{\tr}}\l(-i\pd{\chi}+1\r)\r]h_{n,m,k}
\nn\\
&&\hskip 3.7cm
=\l[{i(n-k+1)\over{r}}+{i(m-k+1)\over{\tr}}\r]h_{n,m,k}.
\end{eqnarray*}

On the basis\footnote{We will again omit the contributions 
from the basis vectors $\vp_{l,m,\tm}\otimes{v}\otimes{H}_{i}$ 
to the determinants, as done for the squashed $S^3$.} 
$\{\vp_{l,m,\tm}\otimes{v}\otimes{E}_{\a}\}$, 
the determinants in $Z^{{\rm 1-loop}}_{{\rm V}}$ are computed to give  
\begin{eqnarray*}
&&\Det_{(k,l)}\l[D^{(0)}_0-g\l[\s,\c\r]\r]
=\prod_{\a\in\La}\prod_{n,m=0}^{\infty}\prod_{k=0}^{m+n}
\dets_{(k,l)}\l[{i(n-k)\over{r}}+{i(m-k)\over\tr}-g\sa\r],
\nn\\
&&\Det_{(k,l)}\l[D^{(-2)}_0+2i{\tr\over{r}^2}+g\l[\s,\c\r]\r]
=\prod_{\a\in\La}\prod_{n,m=0}^{\infty}\prod_{k=0}^{m+n}
\dets_{(k,l)}\l[{i(n-k+1)\over{r}}+{i(m-k+1)\over\tr}
+g\sa\r],
\end{eqnarray*}
where the determinant $\dets_{(k,l)}$ is defined over the space 
$\Om^{(k,l)}(\RS)$.

Taking account of (\ref{Hodge}), we can simplify $Z_{{\rm V}}$,
\begin{eqnarray*}
Z_{{\rm V}}
&=&\prod_{\a\in\La}\prod_{n,m=0}^{\infty}\prod_{k=0}^{m+n}
\l({{i(n-k)\over{r}}+{i(m-k)\over\tr}+g\sa
\over
{i(n-k+1)\over{r}}+{i(m-k+1)\over\tr}+g\sa
}\r)^{\hf\chi(\RS)}
=\prod_{\a\in\La}\prod_{n,m=0}^{\infty}
\l({-{im\over{r}}-{in\over\tr}+g\sa
\over
{i(n+1)\over{r}}+{i(m+1)\over\tr}+g\sa
}\r)^{\hf\chi(\RS)}
\nn\\
&=&\prod_{\a\in\La_+}
\l[\l(ig\sa\r)^2\prod_{n=1}^{\infty}
\l({n^2\over{r}^2}+g^2\sa^2\r)
\l({n^2\over\tr^2}+g^2\sa^2\r)\r]^{\hf\chi(\RS)}.
\end{eqnarray*}

Furthermore, from the formula 
$$
{1\over\pi}\sinh{\pi{x}}=x\prod_{m=1}^{\infty}\l(1+{x^2\over{m^2}}\r),
$$
it follows that 
\BBox{400}{45}{
\begin{eqnarray}
Z^{{\rm 1-loop}}_{{\rm V}}=\prod_{\a\in\La_+}
\l[2\sinh\big(\pi{r}{g}\sa\big)\c2\sinh\big(\pi\tr{g}\sa\big)
\r]^{\hf\chi(\RS)}. 
\label{squashZ_V}
\end{eqnarray}
}

In the round limit $\tr\to{r}$, $Z^{{\rm 1-loop}}_{{\rm V}}$ recovers the 
result for the round $S^3$ in section \ref{SquashLocalization}.

\subsection{One-Loop Contributions from the $\N=1$ Hypermultiplet}

Let us turn to compute the one-loop contributions from the hypermultiplet 
by localization. 

Since the BRST transformations of the scalar fields $H$, $H^\dag$ 
of the hypermultiplet are independent of the background fields, 
they are the same as for the round and squashed $S^3$'s;
\begin{eqnarray*}
\dl_Q\tilde{H}=0, 
\quad
\dl_Q{H}=0, 
\qquad
\dl_Q\tilde{H}^\dag=-{1\over2\sqrt{2}}\tkappa, 
\quad
\dl_Q{H}^\dag=-{1\over2\sqrt{2}}\kappa. 
\end{eqnarray*}

The BRST transformation of the fermions in the hypermultiplet is 
given by
\begin{eqnarray*}
&&\dl_Q\chi=\sqrt{2}i\l[
\l({\e}^\dag\t^m\e\r)\cD^{(-1)}{}_m\tilde{H}+g\l[\s,\,\tilde{H}\r]
+{i\over{f}}\tilde{H}\r],
\quad
\dl_Q\xi=\sqrt{2}i\l({\e^c}^\dag\t^m\e\r)\cD^{(-1)}{}_m\tilde{H},
\nn\\
&&\dl_Q\eta=F^1{}_{1}-2\sqrt{2}\cD_z\tilde{H},
\qquad
\dl_Q\kappa=0, 
\nn\\
&&\dl_Q\tchi=-\sqrt{2}i\l[
\l({\e}^\dag\t^m\e\r)\cD^{(-1)}{}_m{H}+g\l[\s,\,{H}\r]
+{i\over{f}}{H}\r],
\quad
\dl_Q\txi=-\sqrt{2}i\l({\e^c}^\dag\t^m\e\r)\cD^{(-1)}{}_m{H},
\nn\\
&&\dl_Q\teta=F^2{}_{1}+2\sqrt{2}\cD_z{H},
\qquad
\dl_Q\tkappa=0, 
\end{eqnarray*}
and its hermitian conjugate by 
\begin{eqnarray*}
&&\l(\dl_Q\chi\r)^\dag=-\sqrt{2}i\l[
\l({\e}^\dag\t^m\e\r)\cD^{(1)}{}_m\tilde{H}^\dag-g\l[\s,\,\tilde{H}^\dag\r]
-{i\over{f}}\tilde{H}^\dag\r],
\quad
\l(\dl_Q\xi\r)^\dag
=-\sqrt{2}i\l({\e}^\dag\t^m\e^c\r)\cD^{(1)}{}_m\tilde{H}^\dag,
\nn\\
&&\l(\dl_Q\eta\r)^\dag=-F^2{}_{2}-2\sqrt{2}\cD_{\bar{z}}\tilde{H}^\dag,
\qquad
\l(\dl_Q\kappa\r)^\dag=0, 
\nn\\
&&\l(\dl_Q\tchi\r)^\dag=\sqrt{2}i\l[
\l({\e}^\dag\t^m\e\r)\cD^{(1)}{}_m{H}^\dag-g\l[\s,\,{H}^\dag\r]
-{i\over{f}}{H}^\dag\r],
\quad
\l(\dl_Q\txi\r)^\dag=\sqrt{2}i\l({\e}^\dag\t^m\e^c\r)\cD^{(1)}{}_m{H}^\dag,
\nn\\
&&\l(\dl_Q\teta\r)^\dag=F^1{}_{2}+2\sqrt{2}\cD_{\bar{z}}{H}^\dag,
\qquad
\l(\dl_Q\tkappa\r)^\dag=0.
\end{eqnarray*}

Using the BRST transformation of the auxiliuary fields $F^1{}_2$, $F^2{}_2$,
\begin{eqnarray*}
\dl_QF^1{}_2={i\over2}\l[
\l({\e^c}^\dag\t^m\e\r)\cD^{(-1)}{}_m{\chi}
-\l({\e}^\dag\t^m\e\r)\cD^{(1)}{}_m\eta-g\l[\s,\,\eta\r]
+{i\over{f}}\eta-2i\cD_{\bar{z}}\kappa
\r],
\nn\\
\dl_QF^2{}_2={i\over2}\l[
\l({\e^c}^\dag\t^m\e\r)\cD^{(-1)}{}_m{\tchi}
-\l({\e}^\dag\t^m\e\r)\cD^{(1)}{}_m\teta-g\l[\s,\,\teta\r]
+{i\over{f}}\teta-2i\cD_{\bar{z}}\tkappa
\r],
\end{eqnarray*}
where we have omitted the terms 
$g\l(\s^i\r)^{\ta}{}_{\dg}\l[\phi^i,\,\la^{\dg}\r]$ on the right hand sides 
of both the equations, because their contributions vanish 
in the large $t$ limit, 
we find that 
\begin{eqnarray*}
&&\dl_Q\l(\dl_Q\chi\r)^\dag={i\over2}\l[
\l({\e}^\dag\t^m\e\r)\cD^{(1)}{}_m\tkappa-g\l[\s,\,\tkappa\r]
-{i\over{f}}\tkappa\r],
\qquad
\dl_Q\l(\dl_Q\xi\r)^\dag={i\over2}\l({\e}^\dag\t^m\e^c\r)\cD^{(1)}{}_m\tkappa,
\nn\\
&&\dl_Q\l(\dl_Q\eta\r)^\dag
=-{i\over2}\l\{\l({\e^c}^\dag\t^m\e\r)\cD^{(-1)}{}_m\tchi
-\l[\l({\e}^\dag\t^m\e\r)\cD^{(1)}{}_m\txi+g\l[\s,\,\txi\r]
-{i\over{f}}\txi\r]\r\},
\qquad
\dl_Q\l(\dl_Q\kappa\r)^\dag=0, 
\nn\\
&&\dl_Q\l(\dl_Q\tchi\r)^\dag=-{i\over2}\l[
\l({\e}^\dag\t^m\e\r)\cD^{(1)}{}_m\kappa-g\l[\s,\,\kappa\r]
-{i\over{f}}\kappa\r],
\qquad
\dl_Q\l(\dl_Q\txi\r)^\dag=-{i\over2}\l({\e}^\dag\t^m\e^c\r)\cD^{(1)}{}_m\kappa,
\nn\\
&&\dl_Q\l(\dl_Q\teta\r)^\dag
={i\over2}\l\{\l({\e^c}^\dag\t^m\e\r)\cD^{(-1)}{}_m\chi
-\l[\l({\e}^\dag\t^m\e\r)\cD^{(1)}{}_m\xi+g\l[\s,\,\xi\r]
-{i\over{f}}\xi\r]\r\},
\qquad
\dl_Q\l(\dl_Q\tkappa\r)^\dag=0.
\end{eqnarray*}

The system of $(\tilde{H},\tilde{H}^\dag,\chi,\xi,\teta,\tkappa,F^1{}_2)$ 
is identical to the one of $(H,H^\dag,\tchi,\txi,\eta,\kappa,F^2{}_2)$, 
as we have seen for the squashed $S^3$ case 
in subsection \ref{LocalizationN=1Squash}. 
If the former 
contributes the one-loop determinant $Z^{1-{\rm loop}}_{H}$ to the partition 
function, both of the systems contribute $(Z^{1-{\rm loop}}_{H})^2$. 
Therefore, we will focus on the former system only. 

From the fermionic part of the systerm 
$(\tilde{H},\tilde{H}^\dag,\chi,\xi,\teta,\tkappa,F^1{}_2)$ 
of the regulator action ${\cal S}_Q$,
\begin{eqnarray*}
&&-\int \sqrt{g} d^5x \l[
\dl_Q\l(\dl_Q\chi\r)^\dag\c\chi+\dl_Q\l(\dl_Q\xi\r)^\dag\c\xi
+\dl_Q\l(\dl_Q\teta\r)^\dag\c\teta+\dl_Q\l(\dl_Q\tkappa\r)^\dag\c\tkappa \r]
\nn\\
&&={i\over2}\int \sqrt{g} d^5x \l[
\raise1.5ex\hbox{$\begin{pmatrix}\chi,&\xi\end{pmatrix}$}
\begin{pmatrix}
\l({\e}^\dag\t^m\e\r)\cD^{(1)}{}_m-g\l[\s,\c\r]-{i\over{f}}
&
\l({\e^c}^\dag\t^m\e\r)\cD^{(-1)}{}_m
\cr
\l({\e}^\dag\t^m\e^c\r)\cD^{(1)}{}_m
&
-\l({\e}^\dag\t^m\e\r)\cD^{(-1)}{}_m-g\l[\s,\c\r]-{i\over{f}}
\end{pmatrix}
\begin{pmatrix}\tkappa\cr\teta\end{pmatrix}
\r].
\end{eqnarray*}

As we have done in subsection \ref{LocalizationN=1Squash}, 
for four differential operators $D_1,\cdots,D_4$, we have the formula
\begin{eqnarray*}
{\Det}_{(k,l)}\l[\begin{pmatrix}
D_1&D_2\cr{D}_3&D_4
\end{pmatrix}\r]={\Det}_{(k,l)}\l[D_1\r]
{\Det}_{(k,l)}\l[D_4-D_3{1\over{D}_1}D_2\r],
\end{eqnarray*}
for an invertible $D_1$.
In the above case, we have
$$
D_1=\l({\e}^\dag\t^m\e\r)\cD^{(1)}{}_m-g\l[\s,\c\r]-{i\over{f}},
\qquad
D_3=\l({\e}^\dag\t^m\e^c\r)\cD^{(1)}{}_m,
$$
which both act on the spinor $\tkappa$ of negative chirality on $\RS$ and 
of charge $q=1$. Using (\ref{CommRelDifOpEllipsoidS}) in Appendix 
\ref{EllipsoidS^3}, we can identify the operator ${D'}_1$,
\begin{eqnarray*}
D_3D_1
&=&\l({\e}^\dag\t^n\e^c\r)\cD^{(1)}{}_n\l[
\l({\e}^\dag\t^m\e\r)\cD^{(1)}{}_m-g\l[\s,\c\r]-{i\over{f}}
\r]
\nn\\
&=&
\l[\l({\e}^\dag\t^m\e\r)\cD^{(-1)}{}_m
-g\l[\s,\c\r]+{i\over{f}}\r]
\l({\e}^\dag\t^n\e^c\r)\cD^{(1)}{}_n
={D'}_1D_3,
\end{eqnarray*}
and obtain 
\begin{eqnarray*}
{\Det}_{(k,l)}\l[\begin{pmatrix}
D_1&D_2\cr{D}_3&D_4
\end{pmatrix}\r]={{\Det}_{(k,l)}\l[D_1\r]\over{\Det}_{(k,l)}\l[D'_1\r]}
{\Det}_{(k,l)}\l[{D'}_1D_4-D_3D_2\r].
\end{eqnarray*}
We thus find that 
\begin{eqnarray*}
Z^{{\rm 1-loop}}_{{\rm H,F}}=
\Det_{(0,\hf)}\l[\Delta^{\N=1}_{{\rm H,B}}\r]
{\Det_{(0,\hf)}\l[
\l({\e}^\dag\t^m\e\r)\cD^{(1)}{}_m-g\l[\s,\c\r]-{i\over{f}}\r]
\over
\Det_{(0,\hf)}\l[
\l({\e}^\dag\t^m\e\r)\cD^{(-1)}{}_m-g\l[\s,\c\r]+{i\over{f}}\r]},
\end{eqnarray*}
where the differential operator $\Delta^{\N=1}_{{\rm H,B}}$ denotes 
\begin{eqnarray*}
\Delta^{\N=1}_{{\rm H,B}}
&=&
-\l(\l({\e}^\dag\t^m\e\r)\cD^{(-1)}{}_m
-g\l[\s,\c\r]+{i\over{f}}\r)
\l(\l({\e}^\dag\t^m\e\r)\cD^{(-1)}{}_m
+g\l[\s,\c\r]+{i\over{f}}\r)
\nn\\
&&
-\l({\e}^\dag\t^m\e^c\r)\cD^{(1)}{}_m\l({\e^c}^\dag\t^m\e\r)\cD^{(-1)}{}_m.
\end{eqnarray*}

With the same reason as in subsection \ref{LocalizationN=1Squash}, 
we can show that 
$$
{\Det_{(0,\hf)}\l[
\l({\e}^\dag\t^m\e\r)\cD^{(1)}{}_m-g\l[\s,\c\r]-{i\over{f}}\r]
\over
\Det_{(0,\hf)}\l[
\l({\e}^\dag\t^m\e\r)\cD^{(-1)}{}_m-g\l[\s,\c\r]+{i\over{f}}\r]}
=1,
$$
and it follows from this that 
$$
Z^{{\rm 1-loop}}_{{\rm H,F}}=
\Det_{(0,\hf)}\l[\Delta^{\N=1}_{{\rm H,B}}\r].
$$

In the bosonic part of the system 
$(\tilde{H},\tilde{H}^\dag,\chi,\xi,\teta,\tkappa,F^1{}_2)$ 
of the regulator action ${\cal S}_Q$,
\begin{eqnarray*}
-\int d^5x \sqrt{g} \l[
\l(\dl_Q\chi\r)^\dag\c\dl_Q\chi+\l(\dl_Q\xi\r)^\dag\c\dl_Q\xi
+\l(\dl_Q\teta\r)^\dag\c\dl_Q\teta+\l(\dl_Q\tkappa\r)^\dag\c\dl_Q\tkappa 
\r], 
\end{eqnarray*}
we will immediately integrate the auxiliuary fields $F^{\ta}{}_{\cb}$ out,  
and integrate the remaining part of the action by parts to obtain 
\begin{eqnarray*}
-2\int d^5x \sqrt{g}\tilde{H}^\dag\Delta^{\N=1}_{{\rm H,B}}\tilde{H},
\end{eqnarray*}
and see that the one-loop determinant from the bosonic fields of 
the system $(\tilde{H},\tilde{H}^\dag,\chi,\xi,\teta,\tkappa)$ is given by
$$
Z^{1-{\rm loop}}_{{\rm H,B}}={1\over\Det_{(0,\hf)}
\l[\Delta^{\N=1}_{{\rm H,B}}\r]}.
$$

Therefore, the contributions from the hypermultiplet to the partition 
function are trivial; 
$$
\l(Z^{1-{\rm loop}}_{{\rm H}}\r)^2
=\l(Z^{1-{\rm loop}}_{{\rm H,F}}Z^{1-{\rm loop}}_{{\rm H,B}}\r)^2=1.
$$
In the round limit $\tr\to{r}$, the contributions from the hypermultiplet 
reproduce the previous results on the round $S^3$ 
about the hypermultiplet in section \ref{SquashLocalization}.

\section{Summary and Discussions}
\label{SummaryDiscussion}

In this paper, we have seen the effects 
caused by changing the twisting and by deforming a round 3-sphere to 
a squashed and an ellipsoid 3-spheres 
on the partition function on the round $S^3$, which was computed 
in the previous paper \cite{FKM,KM}.

We have discussed the two kinds of twistings - 
the $\N=1$ twisting and the $\N=2$ twising, the former of which 
breaks the $Spin(5)_R$ symmetry group to $U(1)_r\times{SU(2)}_l$, 
which is the subgroup of $SU(2)_{r}\times{SU(2)}_l\simeq{Spin(4)}
\subset{Spin(5)_R}$, while 
the latter breaks the $Spin(5)_R$ to $U(1)_R\times{SU(2)}_R$, 
which is the subgroup ${Spin(2)}_R\times{Spin(3)}_R$ of the $Spin(5)_R$.
In the $\N=1$ twisiting, the only supersymmetry transformation with 
the parameter $\ve^{\da}$ is preserved, and in the $\N=2$ twisting, 
the ones with both $\ve^{\da}$ and $\e^{\ta}$ are available.

The change of the twisting affects on the spin content of the $\N=1$ 
hypermultiplet, when the $\N=2$ gauge multiplet is viewed as the sum of 
the $\N=1$ gauge multiplet and the $\N=1$ hypermultiplet. 

In all the cases we discussed in this paper, the classical action, 
${\cal S}_{\rm cl}$ which is the value of the off-shell action 
(\ref{OffShellAction}) at one $\s^i$ of the fixed-points, can be 
compactly written down\footnote{Recall that upon the localization, 
we rotated $\s^i\to{i}\s^i$ and shifted $D^{\done}{}_{\done}-iF_{45}
-{i\over\a}G_{45}\s$, as discussed at the begining of section 
\ref{localization}.} as 
\begin{eqnarray*}
{\cal S}_{\rm cl}
=\l(\int_{S^3}{2\over{f}}{e}^1\wedge{e}^2\wedge{e}^3\r)
i\sum_{i=1}^{r}\int_{\RS} \,F^i_{45}\,\s^i\,dx^4\wedge{dx}^5,
\end{eqnarray*}
in the zero-area limit of the Riemann surface $\RS$, 
where $1/f$ is replaced by $\tr/^2r$ for the squashed $S^3$ and 
by $1/r$ for the round $S^3$. 
Recall that $\s^i$ is a constant at the fixed point, and 
notice that the scalar curvature $R(\RS)$ of $\RS$ disappears in 
the mass parameter ${\cal M}_{\s}$. 
The integration in the prefactor 
can be easily done to give 
$$
\int_{S^3}{2\over{f}}{e}^1\wedge{e}^2\wedge{e}^3
=\begin{cases}
(2\pi{r})^2 & {\rm for~the~round~}S^3,
\\
{(2\pi{\tr})^2} & {\rm for~the~squashed~}S^3,
\\
(2\pi{r})(2\pi{\tr}) & {\rm for~the~ellipsoid~}S^3.
\end{cases}
$$
Therefore, combining the classical action ${\cal S}_{\rm cl}$ with 
the one-loop contributions 
$Z^{{\rm 1-loop}}=Z^{{\rm 1-loop}}_{{\rm V}}Z^{{\rm 1-loop}}_{{\rm H}}$, 
we obtain the partition function 
$$
Z^{{\cal N}}_{~S^3}=\sum_{m}\prod_{i=1}^{r}
\int\,d\s^i\exp\l[{\cal S}_{\rm cl}\r]\,Z^{{\rm 1-loop}}, 
$$
where the intergers $m_i$  are the `monopole' charges, 
which will be explained below for, just for brevity, the gauge group $G=SU(2)$. 

In the $\N=1$ twisting, we have seen that 
the one-loop contributions from the hypermultiplet are trivial to 
the partition functions. 
\begin{eqnarray*}
Z^{{\rm 1-loop}}_{{\rm H}}=1.
\end{eqnarray*}

Furthermore, on the squashed $S^3$, 
the one-loop determinants from the $\N=1$ gauge multiplet remains the same 
as on the round $S^3$, when we replace $r$ by $\tr$. 
In fact, the partition function $Z^{{\cal N}=1}_{\rm squashed}$ 
for the squashed $S^3$, 
\begin{eqnarray*}
\sum_{m}\prod_{i=1}^{r}\int\,d\s^i\exp\l[{\cal S}_{\rm cl}\r]
Z^{{\rm 1-loop}}_{{\rm V}}Z^{{\rm 1-loop}}_{{\rm H}}
=\sum_{m}\prod_{i=1}^{r}\int\,d\s^i\exp\l[{\cal S}_{\rm cl}\r]
\prod_{\a\in\La_+}\l[2\sinh\big(\pi\tr{g}\sa\big)\r]^{\chi(\RS)}
\end{eqnarray*}
is reduced to the one for the round $S^3$, in the round limit $\tr\to{r}$.

More specifically, let us take the gauge group $G$ to be $SU(2)$, and then 
the generators $\{H,\,E_{\pm}\}$ obey
$$
\l[H,\,E_{\pm}\r]=\pm\sqrt{2}E_{\pm}, 
\qquad
\l[E_{+},\,E_{-}\r]=\sqrt{2}H, 
$$
with our normalization, implying that the positive root $\a=\sqrt{2}$. 
In our convention, we have 
$$
\int_{\RS} \,F_{45}\,dx^4\wedge{dx}^5={2\pi\over{g}}\sqrt{2}m, 
$$
where $m$, which was refered to above as the `monopole charge', 
runs over all the integers. Substituting it to the classical action 
${\cal S}_{\rm cl}$, we see that the path integral gains the contributions 
only from the configurations 
$$
\s={1\over\sqrt{2}}{g\over(2\pi\tr)^2}n,
\qquad {\rm for}~n\in{\bf Z}. 
$$
Therefore, the partition function on the squashed $S^3$ is turned to
$$
Z^{{\cal N}=1}_{\rm squashed}
=2\sum_{n=1}^{\infty}
\l(e^{-{g^2\over4\pi\tr}n}-e^{{g^2\over4\pi\tr}n}\r)^{\chi(\RS)}
=2\l(q^{\hf}-q^{-\hf}\r)^{\chi(\RS)}\sum_{n=1}^{\infty}
\l[\chi_n(q)\r]^{\chi(\RS)},
$$
with $q=e^{-g^2/(2\pi\tr)}$, 
where the character $\chi_n(q)$ is defined by 
$$
\chi_n(q)={q^{{n\over2}}-q^{-{n\over2}}\over{q^{\hf}-q^{-\hf}}}. 
$$

Especially when we consider the round $S^3$ and replace $\tr$ by $r$ 
in the above $Z^{{\cal N}=1}_{\rm squashed}$, the result in the previous 
paper \cite{FKM} is recovered;
\begin{eqnarray}
Z^{{\cal N}=1}_{\rm round}
=2\l(q^{\hf}-q^{-\hf}\r)^{\chi(\RS)}\sum_{n=1}^{\infty}
\l[\chi_n(q)\r]^{\chi(\RS)},
\label{N=1roundIndex}
\end{eqnarray}
with $q=e^{-g^2/(2\pi{r})}$, and we can see that 
it is consistent with the superconformal index computed in \cite{Schur}. 
Using the $2(g-1)$ structure constants and $3(g-1)$ propagators 
in \cite{Schur} to compute the index for the surface $\RS$ of 
genus\footnote{This $g$ is {\it not} the gauge coupling constant $g$.} $g$ 
(therefore, $\chi(\RS)=2-2g$) with no punctures, one obtains the above 
$Z^{{\cal N}=1}_{\rm round}$ up to an factor\footnote{
In the terminology of \cite{Schur}, the factor is given by 
$\l({\cal N}_{222}\r)^{-\chi(\RS)}$. }. 
In the review article \cite{Tachikawa}, it has been elucidated\footnote{
We thank Yuji Tachikawa for elucidating this point.} that 
the discrepancy is attributed to the difference of the renormalization 
prescriptions used here and there, 
and that it can be improved by the requirement of the S-duality 
(a.k.a. the bootstrap).

From the point of view of the number of supersymmetries, 
this result seems puzzling. The partition function 
$Z^{{\cal N}=1}_{\rm round}$ computed under the $\N=1$ twisting 
is supposed to be the index of a four-dimensional 
$\N=1$ supersymmetric theory, while the index in \cite{Schur} 
was computed for a four-dimensional $\N=2$ superconformal theory. 
In \cite{GaddeBeem}, the superconformal index of of $\N=1$ class {\cal S} 
fixed points has been calculated in four dimensions. Among their results, 
the mixed Schur index carrys two fugacities $p$ and $q$ in their notations. 
When we take $p=q$, the index takes the same form as the Schur index of 
the $\N=2$ fixed points given in \cite{Schur}.

The partition function $Z^{{\cal N}=1}_{\rm squashed}$ on the squashed $S^3$ 
is essentially the same as $Z^{{\cal N}=1}_{\rm round}$ on the round $S^3$. 
However, the partition function $Z^{{\cal N}=1}_{\rm ellipsoid}$
on the ellipsoid $S^3$ is deformed from the one on the round $S^3$.  
\begin{eqnarray*}
Z^{{\cal N}=1}_{\rm ellipsoid}
&=&\sum_{m}\prod_{i=1}^{r}\int\,d\s^i
\exp\l[{\cal S}_{\rm cl}\r]Z^{{\rm 1-loop}}_{{\rm V}}Z^{{\rm 1-loop}}_{{\rm H}}
\nn\\
&=&\sum_{m}\prod_{i=1}^{r}\int\,d\s^i
\exp\l[{\cal S}_{\rm cl}\r]
\prod_{\a\in\La_+}
\l[2\sinh\big(\pi{r}{g}\sa\big)\c2\sinh\big(\pi\tr{g}\sa\big)
\r]^{\hf\chi(\RS)}. 
\end{eqnarray*}
This is a similar situation to the three-dimensional case in \cite{Hosomichi}. 
As we have done just before, taking the gauge group $G=SU(2)$ and 
summing over the monopole charge $m$, we can see that the only configurations 
$$
\s={1\over\sqrt{2}}{g\over(2\pi\tr)(2\pi{r})}n,
\qquad {\rm for}~n\in{\bf Z}
$$
contribute to the partition function, and therefore the summation 
of $n$ over intergers yields 
$$
Z^{{\cal N}=1}_{\rm ellipsoid}
=2\l[\l(q^{\hf}-q^{-\hf}\r)\l(p^{\hf}-p^{-\hf}\r)\r]^{\hf\chi(\RS)}
\sum_{n=1}^{\infty}\l[\chi_n(q)\chi_n(p)\r]^{\hf\chi(\RS)},
$$
where $q=e^{-g^2/(2\pi{r})}$ and $p=e^{-g^2/(2\pi{\tr})}$.
It is also consistent with the mixed Schur index\footnote{
More specifically, our result corresponds to the case with $l_1=l_2$
in their notations of \cite{GaddeBeem}, and to the $\N=1$ twist 
in \cite{Sicilian}, as may be seen from the background $R$-symmetry gauge 
field.} 
of $\N=1$ rank one class ${\cal S}$ fixed points in \cite{GaddeBeem}, 
up to an factor from the renoramlization mentioned above.

Let us turn to the $\N=2$ twisting. 
On the round $S^3$, we have seen that 
the hypermultiplet contributes the same one-loop determinants to 
the partition fucntion as the $\N=1$ gauge multiplet does. 
Deforming the round $S^3$ to the squashed $S^3$, we have observed that 
the one-loop contributions from the hypermultiplet are deformed by the 
deformation paramter of the $S^3$. 

In fact, in the partition function $Z^{{\cal N}=2}_{\rm squashed}$ on the 
squashed $S^3$
\begin{eqnarray*}
Z^{{\cal N}=2}_{\rm squashed}
&=&
\sum_{m}\prod_{i=1}^{r}\int\,d\s^i\exp\l[{\cal S}_{\rm cl}\r]
Z^{{\rm 1-loop}}_{{\rm V}}Z^{{\rm 1-loop}}_{{\rm H}},
\end{eqnarray*}
we have seen that the one-loop contributions 
$Z^{{\rm 1-loop}}_{{\rm V}}Z^{{\rm 1-loop}}_{{\rm H}}$ are given by
\begin{eqnarray*}
\prod_{\a\in\La_+}
\l[
\Big(2\sinh\big(\pi\tr{g}\sa\big)\Big)^{2}
{{s}_{b=1}\l(i-2i{\tr^2\over{r}^2}-\tr{g}\sa\r)
{s}_{b=1}\l(i-2i{\tr^2\over{r}^2}+\tr{g}\sa\r)}\r]^{\hf\chi(\RS)}. 
\end{eqnarray*}
Therefore, for the gauge group $G=SU(2)$, upon summing the magnetic charge 
$m$ over intergers, we obtain 
\begin{eqnarray*}
Z^{{\cal N}=2}_{\rm squashed}
=2\l(q^{\hf}-q^{-\hf}\r)^{\chi(\RS)}\sum_{n=1}^{\infty}
\l[\big(\chi_n(q)\big)^2
{{s}_{b=1}\l(i-2i{\tr^2\over{r}^2}-{g^2\over4\pi^2\tr}n\r)
{s}_{b=1}\l(i-2i{\tr^2\over{r}^2}+{g^2\over4\pi^2\tr}n\r)}
\r]^{\hf\chi(\RS)},
\end{eqnarray*}
with $q=e^{-g^2/(2\pi\tr)}$, 
where the double sine functions may be rewritten as
\begin{eqnarray*}
{{s}_{b=1}\l(i-2i{\tr^2\over{r}^2}-{g^2\over4\pi^2\tr}n\r)
{s}_{b=1}\l(i-2i{\tr^2\over{r}^2}+{g^2\over4\pi^2\tr}n\r)}
=\prod_{m=1}^{\infty}\l[{
\l(m+1-2{\tr^2\over{r}^2}\r)^2+\l({g^2\over4\pi^2\tr}n\r)^2
\over
\l(m-1+2{\tr^2\over{r}^2}\r)^2+\l({g^2\over4\pi^2\tr}n\r)^2
}\r]^m,
\end{eqnarray*}
and in the round limit $\tr\to{r}$, they reduce to
$$
\l[\l(q^{\hf}-q^{-\hf}\r)\c\chi_n(q)\r]^2, 
$$
with $q=e^{-g^2/(2\pi{r})}$. When recognizing $\chi_n(q)$ as the $q$-deformed 
number\footnote{This definition of the $q$-deformed number $[n]_q$ slightly 
differs from the one in \cite{FKM} by the factor $1/(q^\hf-q^{-\hf})$.} 
$[n]_q$, we may regard the square root of 
the double sine functions as a deformation of $[n]_q$. 

We thus find that the partition function 
$Z^{{\cal N}=2}_{\rm round}$ on the round $S^3$ is given for $G=SU(2)$ by 
\begin{eqnarray*}
Z^{{\cal N}=2}_{\rm round}
&=&
\sum_{m}\int\,d\s\exp\l[{\cal S}_{\rm cl}\r]
\l[2\sinh\big(\pi{r}{g}\sa\big)\r]^{2\chi(\RS)}
\nn\\
&=&2\l(q^{\hf}-q^{-\hf}\r)^{2\chi(\RS)}\sum_{n=1}^{\infty}
\l[\chi_n(q)\r]^{2\chi(\RS)}.
\end{eqnarray*}
This result suggests that the partition function $Z^{{\cal N}=2}_{\rm round}$ 
does not corresponds to the Schur limit of the superconformal index 
discussed in \cite{Schur}. We expect that it gives another simple limit 
of the superconformal index of $\N=2$ rank one class ${\cal S}$ fixed points, 
where the index can be calculated by the two-dimensional $q$-deformed 
Yang-Mills theory but with the measure 
$\l[2\sinh\big(\pi{r}{g}\sa\big)\r]^{\chi(\RS)}$ squared.

\vskip 3cm

\centerline{\bf Acknowledgement}

\medskip
We would like to thank Fumitaka Fukui for collaborations 
at the early stage of this work. 
We are grateful to Kazuo Hosomichi, Yosuke Imamura, and  Hiroki Matsuno 
for heplful discussions. 
We would like to thank Yuji Tachikawa for helpful discussions on many crucial 
points. 
The work of T.~K. was supported in part 
by a Grant-in-Aid \#23540286 from the MEXT of Japan.

\clearpage
\appendix
\noindent
{\Large \bf Appendix}

\section{Our Conventions of (Anti-)Symmetrization of Indices and 
Differential Forms}
\label{conventions}

The convention of the antisymmetrization and symmetrization\footnote{
The reference \cite{BSvP2} gives weights to each of the terms 
on the right hand side; for example, 
$A_{[\m}B_{\n]}=(1/2)(A_{\m}B_{\n}-A_{\n}B_{\m})$, 
$X_{(\m}Y_{\n)}=(1/2)(X_{\m}Y_{\n}+X_{\n}Y_{\m})$.
}
may be seen from 
$$
A_{[\m}B_{\n]}=A_{\m}B_{\n}-A_{\n}B_{\m}, 
\qquad
X_{(\m}Y_{\n)}=X_{\m}Y_{\n}+X_{\n}Y_{\m}.
$$
However, for the six-dimensional gamma matrice 
(for the definition of them, see the next appendix), we define 
$$
\u{\G}^{\u{ab}}=\hf\u{\G}^{[\u{a}}\u{\G}^{\u{b}]}, 
\qquad
\u{\G}^{\u{a_1\cdots{a}_n}}
={1\over{n}}\u{\G}^{[\u{a_1}}\u{\G}^{\u{a_2\cdots{a}_n}]}
={1\over{n!}}\u{\G}^{[\u{a_1}}\u{\G}^{\u{a_2}}\cdots\u{\G}^{\u{a}_n]}.
$$
For the five-dimensional gamma matrices, 
$\g^{a_1\cdots{a}_n}$ is defined in the same way.

For an $n$-form $\u{A}_n$ in six dimensions, we define 
$$
\u{A}_n
={1\over{n}!}\,\u{A}{}_{\u{\m_1\cdots\m_n}}\,
d\u{X}^{\u{\m}_1}\wedge\cdots\wedge{}d\u{X}^{\u{\m}_n}
={1\over{n}!}\,\u{A}{}_{\u{{a}_1\cdots{a}_n}}\,
\u{E}^{\u{a}_1}\wedge\cdots\wedge{}\u{E}^{\u{a}_n},
$$
where $X^\m$ ($\m=0,1,\cdots,5$) denote the local coordinates and 
$\u{E}^{\u{a}}$ ($a=0,1,\cdots,5$) are the sechsbein one-form. 
We also define the Hodge dual of the form $\u{A}_n$ by
$$
\u{*}\,\u{A}_n
={1\over(6-n)!}\l[{1\over{n}!}
\ve_{\u{{a}_1\cdots{a}_{6-n}{b}_1\cdots{b}_n}}
\u{A}{}^{\u{{b}_1\cdots{b}_n}}\,
\r]\u{E}^{\u{a}_1}\wedge\cdots\wedge{}\u{E}^{\u{a}_{6-n}},
$$
with $\ve_{01\cdots5}=1$. Because of the property 
$\u{*}\u{*}\,\u{A}_n=(-)^{n+1}\u{A}_n$, one may define 
the self-dual and the anti-self dual 
parts, respectively, of a three-form $\u{A}_3$ by
$$
\u{A}_{3}^{\pm}=\hf\l(\u{A}_3\pm\u{*}\,\u{A}_3\r)=\pm\u{*}\,\u{A}_{3}^{\pm}.
$$
We define the external derivative $d\u{A}_n$ of the $n$-form 
$$
d\u{A}_n
={1\over{n}!}\,\d{}_{\u{\m}{}_1}\u{A}{}_{\u{\m_2\cdots\m_{n+1}}}\,
d\u{X}^{\u{\m}{}_1}\wedge\cdots\wedge{}d\u{X}^{\u{\m}{}_{n+1}}
={1\over(n+1)!}\,\d{}_{[\u{\m}{}_1}\u{A}{}_{\u{\m_2\cdots\m_{n+1}}]}\,
d\u{X}^{\u{\m}{}_1}\wedge\cdots\wedge{}d\u{X}^{\u{\m}{}_{n+1}}.
$$

For an $n$-form $a$ in five dimensions, 
\begin{eqnarray*}
a={1\over{n}!}\,a{}_{{\m_1\cdots\m_n}}\,
d{x}^{{\m}_1}\wedge\cdots\wedge{}d{x}^{{\m}_n} 
={1\over{n}!}\,a{}_{{a_1\cdots{}a_n}}\,
{e}^{{a}_1}\wedge\cdots\wedge{}{e}^{{a}_n}, 
\end{eqnarray*}

similarly, the external derivative $da$ and the Hodge dual $*a$ are
\begin{eqnarray*}
da&=&{1\over(n+1)!}\,\d_{[\m_1}a{}_{{\m_2\cdots\m_{n+1}}]}\,
d{x}^{{\m}_1}\wedge\cdots\wedge{}d{x}^{{\m}_{n+1}}, 
\nn\\
*a&=&{1\over(5-n)!}
\l[{1\over{n}!}\ve_{a_1\cdots{a}_{5-n}b_1\cdots{}b_n}a{}^{{b_1\cdots{}b_n}}\r]
d{e}^{{a}_1}\wedge\cdots\wedge{}d{e}^{{a}_{5-n}}. 
\end{eqnarray*}

\section{Gamma Matices of the 6-Dimensional Lorentz Group}
\label{LorentzGamma}

We define the six-dimensional gamma matrices $\u{\G}^{\u{a}}$ 
($\u{a}=0,1,\cdots,5$) such that they satisfy
$$
\l\{\u{\G}^{\u{a}},\,\u{\G}^{\u{b}}\r\}=2\u{\eta}^{\u{ab}}\,{\bf 1}_{8},
$$
with ${\bf 1}_{8}$ the $8\times8$ unit matrix and the 
Lorentz metric $({\eta}^{\u{ab}})={\rm diag.}(-1,+1,\cdots,+1)$.
While only $\u{\G}^{0}$ is anti-hermition, the others are hermition. 

The chirality is defined by the matrix 
$$
\u{\G}^{7}=\u{\G}^{0}\u{\G}^{1}\cdots\u{\G}^{5},
\qquad
\l(\u{\G}^{7}\r)^2={\bf 1}_{8}, 
$$
and it enjoys the properties
\begin{eqnarray}
\u{\G}^{\u{a_1\cdots{a}_6}}=-\ve^{\u{a_1\cdots{a}_6}}\,\u{\G}^7,
\qquad
\u{\G}^{\u{abc}}=-{1\over3!}\ve^{\u{abcdef}}\,\u{\G}{}_{\u{def}}\,\u{\G}^7,
\label{topgamma}
\end{eqnarray}
with the convetions $\ve^{01\cdots5}=-1$ ($\ve_{01\cdots5}=+1$).

The charge conjugation matrix $\u{C}$ is a unitary matrix satisfying 
\begin{eqnarray}
\u{C}^T=\u{C}, 
\qquad
\l(\u{\G}_{\u{a}}\r)^T=-\u{C}\u{\G}_{\u{a}}\u{C}^{-1},
\label{C6}
\end{eqnarray}
with $T$ denoting the transpose of the matrices,  
and thus $\l(\u{\G}^{7}\r)^T=-\u{C}\u{\G}^{7}\u{C}^{-1}$.

On the reduction along the time direction from the six-dimensional Minkowski 
space to the five-dimensional Euclidean space, where the gamma matrices are 
five $4\times4$ hermitian matrices $\g^a$ ($a=1,\cdots,5$) satisfying 
$$
\big\{\g^a,\,\g^b\big\}=2{\dl}^{{ab}}\,{\bf 1}_{4}, 
\qquad
\g^1\cdots\g^5={\bf 1}_{4},
$$
with ${\bf 1}_{4}$ the $4\times4$ unit matrix,
we define 
\begin{eqnarray*}
\u{\G}^{0}={\bf 1}_{4}\otimes{i\t_2}
=\begin{pmatrix}
0& {\bf 1}_4\\ 
-{\bf 1}_4&0
\end{pmatrix}, 
\quad
\u{\G}^{a}=\g^a\otimes{\t_1} 
=\begin{pmatrix}
0 & \g^a \\ 
\g^a & 0
\end{pmatrix}, 
\quad
\u{\G}^{7}={\bf 1}_{4}\otimes{\t_3}
=\begin{pmatrix}
{\bf 1}_4 & 0\\ 
0&-{\bf 1}_4
\end{pmatrix}, 
\end{eqnarray*}
for $a=1,\cdots,5$, with the Pauli matrices $\t_1$, $\t_2$, $\t_3$.

The property (\ref{topgamma}) is reduced to 
$$
\g^{abcde}\equiv{1\over5!}\g^{[a}\g^{b}\g^{c}\g^{d}\g^{e]}=\ve^{abcde},
$$
with $\ve^{12345}=\ve_{12345}=1$.

The six-dimensional charge conjugate matrix $\u{C}$ is related to 
the five-dimensional charge conjugation matrix $C$ by
$$
\u{C}=C\otimes{i\t_2}
=\begin{pmatrix}
0&C\\-C&0
\end{pmatrix},
$$
and one can see that the charge conjugation matrix $C$ enjoys
the properties
$$
C^T=-C, 
\qquad
\l(\g_a\r)^T=C\g_a{C}^{-1}.
$$
It follows from them that 
$$
\l(C\g^{a_1\cdots{a}_n}\r)^T=-(-)^{{n(n-1)\over2}}\l(C\g^{a_1\cdots{a}_n}\r),
\qquad
\l(\g^{a_1\cdots{a}_n}C^{-1}\r)^T
=-(-)^{{n(n-1)\over2}}\l(\g^{a_1\cdots{a}_n}C^{-1}\r). 
$$

A more explicit form of the five-dimensional gamma matrices $\g^\m$ takes 
\begin{eqnarray*}
\g^a=\t_a\otimes\t_2, 
\quad (a=1,2,3)
\qquad
\g^4={\bf 1}_2\otimes\t_1,
\qquad
\g^5={\bf 1}_2\otimes\t_3,
\end{eqnarray*}
with the charge conjugation matrix $C=C_3\otimes{\bf 1}_2$, 
where $C_3=i\t_2$.

\section{Gamma Matices of the $R$-Symmetry Group Spin(5)$_R$}
\label{RGamma}

We give the explicit form of 
the gamma matrices of the $R$-symmetry group Spin(5)$_R$
\be
\p^1=\t^1\otimes\t^2,
\quad
\p^2=\t^2\otimes\t^2,
\quad
\p^3=\t^3\otimes\t^2,
\quad
\p^4={\bf 1}\otimes\t^1,
\quad
\p^5={\bf 1}\otimes\t^3=\p^1\cdots\p^4,
\ne
with the Pauli matrices $\t_1$, $\t_2$, $\t_3$,
satisfying that
$$
\l\{\p^I,\,\p^J\r\}=2\,\dl^{IJ},
$$
where $I,\,J$ run from 1 to 5. 
We use them to define
$$
\p^{I_1\cdots{I}_n}={1\over{n}!}\,\p^{[I_1}\cdots\p^{{I}_n]}, 
\qquad
\p^{I_1\cdots{I}_5}=\ve^{I_1\cdots{I}_5},
$$
with $\ve^{12345}=1$.

We also explicitly give 
the charge conjugation matrix $\Om$ of the Spin(5)$_R$
$$
\Om=i\t_2\otimes{\bf 1}=-\Omega^\dag=-\Omega^T,
$$
where $\Omega^T$ is the transpose of the matrix $\Omega$,
which satisfies 
$$
\Omega\l(\p^I\r)^T\Omega^{-1}=\p^I,
\qquad
(I=1,2,\cdots,5).
$$
It follows from these properties that
$$
\l(\Om\p^{I_1\cdots{I}_n}\r)^T=-(-){}^{{n(n-1)\over2}}
\l(\Om\p^{I_1\cdots{I}_n}\r),
\qquad
\l(\p^{I_1\cdots{I}_n}\Om^{-1}\r)^T=-(-){}^{{n(n-1)\over2}}
\l(\p^{I_1\cdots{I}_n}\Om^{-1}\r).
$$

Given the components
$$
\Omega=\Big(\Omega_{\a\b}\Big), 
\quad
\Omega^{-1}=-\Big(\Omega^{\a\b}\Big),
\qquad
(\a,\b=1,\cdots,4)
$$
one has 
$$
\Om^{\a\g}\Om_{\b\g}=\dl^{\a}{}_{\b}.
$$

The index $\a$ of a spinor $\e^\a$ of the Spin(5)$_R$ is lowered by $\Omega$ 
as 
$$
\e_\a=\e^\b\Omega_{\b\a}, 
\qquad
\e^\a=\Omega^{\a\b}\e_\b,
$$
and this convention is consistent with 
$$
\Om^{\a\g}\,\Om^{\b\dl}\,\Om_{\g\dl}=\Om^{\a\b}.
$$

Since the components $\Omega_{\a\b}$ are real, 
$$
\Big(\Omega_{\a\b}\Big)^*=\Omega^{\a\b},
$$
where $\Big(\Omega_{\a\b}\Big)^*$ denotes the complex conjugate of
$\Omega_{\a\b}$.

The Fierz tranformation of two matrices ($M_{\a\b}$), ($N_{\a\b}$) 
$$
M_{\a\b}\,N_{\g\dl}
={1\over4}\l[
\l(M\Om^{-1}N\r){}_{\a\dl}\Om_{\g\b}
+\l(M\p^I\Om^{-1}N\r){}_{\a\dl}\l(\Om\p_I\r){}_{\g\b}
-\hf\l(M\p^{IJ}\Om^{-1}N\r){}_{\a\dl}\l(\Om\p_{IJ}\r){}_{\g\b}
\r]
$$
may be useful to verify some of the calculations in the text.

\subsection{Spin(5)$_R$ $\longrightarrow
$ Spin(4)$_R~\simeq~SU(2)_l\times{SU(2)_r}$}

A vector $v^I$ ($I=1,\cdots,4,5$) of the Spin(5)$_R$ group is decomposed 
into irreducible representations of the subgroup Spin(4)$_R$ as 
one vector $v^i$ ($i=1,\cdots,4$) and one singlet $v^5$. 
A spinor $\psi^\a$ ($\a=1,\cdots,4$) of the Spin(5)$_R$ group is decomposed into 
$$
\psi^\a=
\begin{pmatrix}
\psi^{\ta} \\ \psi^{\da}
\end{pmatrix},
\qquad
(\ta=1,2;~\da=1,2),
$$
with $\psi^{\ta}$ in $({\bf 2},{\bf 1})$ 
and 
$\psi^{\da}$ in $({\bf 1},{\bf 2})$ of the 
$SU(2)_l\times{}SU(2)_r\simeq$ Spin(4)$_R$ group. 

The gamma matrices of the Spin(5)$_R$ group are reduced into
\begin{eqnarray*}
\p^i=
\begin{pmatrix}
& \s^i \\ \bar{\s}^i &
\end{pmatrix},
\quad (i=1,\cdots,4),
\qquad
\p^5=
\begin{pmatrix}
{\bf 1}_{2} &  \\  & -{\bf 1}_{2}
\end{pmatrix}=\p^1\cdots\p^4,
\end{eqnarray*}
where
$$
(\s^i)=(-i\overrightarrow{\t}, 1)=(\s^i{}^{\ta}{}_{\db}), 
\qquad
(\bar{\s}^i)=(i\overrightarrow{\t}, 1)=(\bar{\s}^i{}^{\da}{}_{\tb}),
$$
with $\t^a$ ($a=1,2,3$) the Pauli matrices. 
The matrices $\s^i$, $\bar{\s}^i$ ($i=1,\cdots,4$) obey the relations
\begin{eqnarray*}
&&\hskip2.5cm
\bar{\s}^i{}^{\db}{}_{\ta}=\ve_{\ta\tg}\ve^{\db\ddl}\s^i{}^{\tg}{}_{\ddl},
\qquad
{\s}^i{}^{\tg}{}_{\ddl}=\ve^{\tg\ta}\ve_{\ddl\db}\bar{\s}^i{}^{\db}{}_{\ta},
\nn\\
&&\s^i{}^{\ta}{}_{\db}\,\bar{\s}^i{}^{\dg}{}_{\tdl}
=2
\dl{}^{\ta}{}_{\tdl}\,\dl{}^{\dg}{}_{\db},
\qquad
\s^i{}^{\ta}{}_{\db}\,{\s}^i{}^{\tg}{}_{\ddl}
=2
\ve{}^{\ta\tg}\,\ve{}_{\db\ddl},
\qquad
\bar{\s}^i{}^{\da}{}_{\tb}\,\bar{\s}^i{}^{\dg}{}_{\tdl}
=2
\ve{}^{\da\dg}\,\ve{}_{\tb\tdl},
\nn\\
&&\hskip4cm
\Tr\l[\s^i\bar{\s}^j\r]=\s^i{}^{\ta}{}_{\db}\bar{\s}^j{}^{\db}{}_{\ta}=2\dl^{ij}.
\end{eqnarray*}

The generators of the Spin(4)$_R$ group in the spinor representation 
become the direct sum 
$$
\p^{ij}=\hf
\begin{pmatrix}
\s^i\bar{\s}^j-\s^j\bar{\s}^i & \\  & \bar{\s}^i\s^j-\bar{\s}^j\s^i
\end{pmatrix}
\equiv
\begin{pmatrix}
\s^{ij} & \\ & \bar{\s}^{ij} 
\end{pmatrix}, 
\quad
(i,j=1,\cdots,4)
$$
obeying 
$$
\s^{ij}=-\hf\ve^{ijkl}\s_{kl},
\qquad
\bar{\s}^{ij}=\hf\ve^{ijkl}\bar{\s}_{kl}.
$$
For example, one has 
$$
\bar{\s}^{12}=\bar{\s}^{34}=i\t^3, 
\qquad
{\s}^{12}=-{\s}^{34}=i\t^3. 
$$

The charge conjugation matrix gives 
$$
\Om=
\begin{pmatrix}
\ve_{\ta\tb} & \\ & \ve_{\da\db}
\end{pmatrix}
=
\begin{pmatrix}
i\t_2 & \\ & i\t_2
\end{pmatrix},
\qquad
\Om^{-1}=
\begin{pmatrix}
-\ve^{\ta\tb} & \\ & -\ve^{\da\db}
\end{pmatrix}
=
\begin{pmatrix}
-i\t_2 & \\ & -i\t_2
\end{pmatrix},
$$
with $\ve_{12}=\ve^{12}=1$.

The spinor indices $\ta$, $\da$ are raised or lowered as 
\begin{eqnarray*}
\psi_{\ta}=\psi^{\tb}\ve_{\tb\ta}, 
\quad
\psi^{\ta}=\ve^{\ta\tb}\psi_{\tb};
\qquad
\psi_{\da}=\psi^{\db}\ve_{\db\da}, 
\quad
\psi^{\da}=\ve^{\da\db}\psi_{\db}.
\end{eqnarray*}

\section{Symplectic Majorana-Weyl Spinors}
\label{SMW}

For a six-dimensional Dirac spinor $\u{\psi}$, 
we define $\overline{\u{\psi}}=\l(\u{\psi}\r)^\dag\u{\G}^0$. 
In six-dimensional Minkowski space, the symplectic Majorana condition 
on an even number of spinors can be imposed. In our case, all the spinors 
in the Weyl multiplet and the tensor multiplet of the supergravity 
carry the spinor indices of the Spin(5)$_R$ symmetry group, 
and the dimension of the spinor representation is four - an even number. 
Let us take one of such spinors, say $\u{\psi}^\a$, and it obeys 
the symplectic Majorana condition 
$$
\l(\u{\psi}^\a\r)^\dag\u{\G}^0=\l(\u{\psi}^\b\r)^T\u{C}\Om_{\b\a},
$$ 
and the other spinors in the multiplets obey the same condition. 

In the Minkowski space, the Weyl condition 
$\u{\G}^7\u{\psi}=\pm\u{\psi}$
and the symplectic Majorana condition can be imposed on spinors 
at the same time. 
In fact, all the spinors of the multiplets are symplectic Majorana-Weyl 
spinors, and also so are the parameters of the supersymmetry and the conformal 
supersymmetry transformations, as explained in the text. 

After the dimensional reduction, the spinors in 
the supergravity multiplets give rise to symplectic Majorana spinors 
in the five-dimensional Euclidean space. If $\u{\psi}^\a$ 
is a symplectic Majorana-Weyl spinor of positive chirality, 
it takes the form 
$$
\u{\psi}^\a=\begin{pmatrix}\psi^\a\\0\end{pmatrix},
$$
and is reduced to the symplectic Majorana spinor $\psi^\a$ obeying 
\begin{eqnarray}
\l(\psi^\a\r)^\dag=\l({\psi}^\b\r)^T{C}\Om_{\b\a},
\label{fiveMajorana}
\end{eqnarray}
in the five-dimensional Euclidean space. 
If it is of negative chirality, one can see from 
$$
\u{\psi}^\a=\begin{pmatrix}0\\\psi^\a\end{pmatrix},
$$
that $\psi^\a$ also obeys the same condition (\ref{fiveMajorana}).

It is convenient to introduce the notations for the conjugate of 
a five-dimensional spinor $\e^\a$
$$
\bar\e^\a\equiv\l(\e^\a\r)^TC,
$$
and the abbreviation of the spinor bilinear 
$$
\l(\bar\e\c\p_{I_1\cdots{I}_n}\g^{a_1\cdots{a}_m}\eta\r)
\equiv
\l(\Om\p_{I_1\cdots{I}_n}\r)_{\a\b}{\c}\l(\e^\a\r)^TC\g^{a_1\cdots{a}_m}\eta^\b,
$$
of two five-dimensional spinors $\e^\a$, $\eta^\a$.

The Fierz transformation of five-dimensional spinors $\e^\a$, $\eta^\a$ gives 
$$
\eta^\a\bar\e^\b
=\eta^\a\l(\e^\b\r)^TC
=-{1\over4}\bigg[
\l(\e^\b{C}\eta^\a\r){\bf 1}_4+\l(\e^\b{C}\g^a\eta^\a\r)\,\g_a
-\hf\l(\e^\b{C}\g^{ab}\eta^\a\r)\,\g_{ab}
\bigg],
$$
and the following formula is repeatedly used in the calculations in the text: 
\begin{eqnarray*}
\eta^\a\bar\e^\b-\e^\a\bar\eta^\b
&=&-{1\over8}\bigg[
\l(\bar\e\c\eta\r)\l(\Om^{-1}\r)^{\a\b}
+\l(\bar\e\c\p_I\eta\r)\l(\p^I\Om^{-1}\r)^{\a\b}
+\l(\bar\e\c\g^a\eta\r)\l(\Om^{-1}\r)^{\a\b}\g_a
\nn\\
&&\qquad\qquad\qquad
+\l(\bar\e\c\p_I\g^a\eta\r)\l(\p^I\Om^{-1}\r)^{\a\b}\g_a
+{1\over4}\l(\bar\e\c\p_{IJ}\g^{ab}\eta\r)\l(\p^{IJ}\Om^{-1}\r)^{\a\b}\g_{ab}
\bigg]. 
\end{eqnarray*}

The following abbreviation for the bilinears of spinors 
$\psi^{\ta}$, $\chi^{\ta}$ in ({\bf 2}, {\bf 1}) 
and 
$\la^{\da}$, $\e^{\da}$ in ({\bf 1}, {\bf 2}) 
of the $SU(2)_l\times{SU(2)_r}$ $R$-symmetry group is used:
\begin{eqnarray*}
&&\bar\psi\c\s_{i_1\cdots{i}_{2n}}\g^{a_1\cdots{a}_m}\chi
\equiv\ve_{\ta\tg}(\s_{i_1\cdots{i}_{2n}})^{\tg}{}_{\tb}\c
\l(\psi^{\ta}\r)^T{C}\g^{a_1\cdots{a}_m}\chi^{\tb}, 
\nn\\
&&\bar\psi\c{\s}_{i_1\cdots{i}_{2n+1}}\g^{a_1\cdots{a}_m}\la
\equiv\ve_{\ta\tg}({\s}_{i_1\cdots{i}_{2n+1}})^{\tg}{}_{\db}\c
\l(\psi^{\ta}\r)^T{C}\g^{a_1\cdots{a}_m}\la^{\db}, 
\nn\\
&&\bar\e\c\bar{\s}_{i_1\cdots{i}_{2n+1}}\g^{a_1\cdots{a}_m}\chi
\equiv\ve_{\da\dg}(\bar{\s}_{i_1\cdots{i}_{2n+1}})^{\dg}{}_{\tb}\c
\l(\e^{\da}\r)^T{C}\g^{a_1\cdots{a}_m}\chi^{\tb}, 
\nn\\
&&\bar\e\c\bar{\s}_{i_1\cdots{i}_{2n}}\g^{a_1\cdots{a}_m}\la
\equiv\ve_{\da\dg}(\bar{\s}_{i_1\cdots{i}_{2n}})^{\dg}{}_{\db}\c
\l(\e^{\da}\r)^T{C}\g^{a_1\cdots{a}_m}\la^{\db}. 
\end{eqnarray*}

For the spinors $\ve^{\da}$, $\eta^{\da}$, the Fierz transformation 
\begin{eqnarray*}
&&\ve^{\da}\bar{\eta}_{\db}=\ve^{\da} {\eta}^{\dg}C\ve_{\dg\db}
=-\qt\l[
\l(\bar{\eta}_{\db}\,\ve^{\da}\r)\,{\bf 1}_4
+\l(\bar{\eta}_{\db}\g^a\ve^{\da}\r)\,\g_a
-\hf\l(\bar{\eta}_{\db}\g^{ab}\ve^{\da}\r)\,\g_{ab}\r],
\end{eqnarray*}
is also useful to verify computations in the text such as 
the algebras of the supersymmetry transformations. 
In particular, we have often made use of the formula
\begin{eqnarray*}
\eta^{\da}\bar{\ve}_{\db}-\ve^{\da}\bar{\eta}_{\db}
=-\qt\l[
\l(\bar{\ve}\c\eta\r)\dl^{\da}{}_{\db}
+\l(\bar{\ve}\c\g^a\eta\r)\dl^{\da}{}_{\db}\,\g_a
+{1\over8}\l(\bar{\ve}\c\bar{\s}_{ij}\g^{ab}\eta\r)
\l(\bar{\s}^{ij}\r)^{\da}{}_{\db}\,\g_{ab}\r].
\end{eqnarray*}

\section{The Supersymmetry Condition from the Spinor $\chi^{\a\b}{}_\g$}
\label{SUSYchi}

One of the supersymmetry conditions which the supersymmetric backgrounds 
in the five-dimensional supergravity should obey is the requirement that 
$\dl\chi^{\a\b}{}_{\g}\big(\Om\p^I\big)_{\b\a}$ should vanish. 
It yields the condition
\begin{eqnarray}
&&
{4\over5\a}\l[
G^{ab}t^I{}_{ab}\Om_{\g\dl}
-{1\over4}G^{ab}t^J{}_{ab}\l(\Om\p^{I}{}_{J}\r){}_{\g\dl}
\r]\e^{\dl}
-2\l[
t^I{}_{ab}t^J{}^{ab}\l(\Om\p_J\r){}_{\g\dl}
-{1\over5}t^J{}_{ab}t_J{}^{ab}\l(\Om\p^I\r){}_{\g\dl}
\r]\e^{\dl}
\nn\\
&&+{4\over15}M^I{}_{J}\l(\Om\p^J\r){}_{\g\dl}\e^{\dl}
+\ve^{abcde}\bigg[t^I{}_{ab}t^J{}_{cd}\l(\Om\p_J\r)_{\g\dl}
-{1\over5}\,t_J{}_{ab}t^J{}_{cd}\l(\Om\p^I\r)_{\g\dl}
\bigg]\g_{e}\e^\dl
\nn\\
&&+{8\over5}\,\l[
\cD^a{t}^I{}_{ab}\Om_{\g\dl}-\qt\cD^a{t}^K{}_{ab}\l(\Om\p^I{}_K\r){}_{\g\dl}
\r]\g^{b}\e^\dl
-{3\over5}\,\l[
\cD_a{S}^I{}_{J}\l(\Om\p^J\r){}_{\g\dl}
-{1\over3}\,\cD_a{S}_{KL}\l(\Om\p^{IKL}\r){}_{\g\dl}
\r]\g^{a}\e^\dl
\nn\\
&&
+{2\over5}\ve^{abcde}\l[
\cD_at^I{}_{bc}\Om_{\g\dl}-{1\over4}\cD_at^J{}_{bc}\l(\Om\p^I{}_{J}\r){}_{\g\dl}
\r]\g_{de}\e^{\dl}
+{3\over10}\l[
F_{ab}{}^{IJ}\l(\Om\p_J\r){}_{\g\dl}
-{1\over3}\,F_{ab}{}^{KL}\l(\Om\p^I{}_{KL}\r){}_{\g\dl}
\r]\g^{ab}\e^\dl
\nn\\
&&
-{8\over5\a}\l[
G_a{}^ct^I{}_{bc}\Om_{\g\dl}-\qt\,G_a{}^ct_J{}_{bc}\l(\Om\p^{IJ}\r){}_{\g\dl}
\r]\g^{ab}\e^\dl
-{12\over5}\l[
t^I{}_{ac}t^J{}_{b}{}^{c}\l(\Om\p_J\r)_{\g\dl}
-{1\over3}t^K{}_{ac}t^L{}_{b}{}^{c}\l(\Om\p^I{}_{KL}\r)_{\g\dl}
\r]\g^{ab}\e^\dl
\nn\\
&&
+{3\over10\a}\l[G_{ab}S^I{}_J\l(\Om\p^J\r){}_{\g\dl}
-{1\over3}\,G_{ab}S_{KL}\l(\Om\p^{IKL}\r){}_{\g\dl}
\r]\g^{ab}\e^\dl
\nn\\
&&
+{4\over5}\,\bigg[
t^I{}_{ab}S_{KL}\l(\Om\p^{KL}\r){}_{\g\dl}
-\hf{t}^{K}{}_{ab}S_{IJ}\l(\Om\p^{J}{}_{K}\r){}_{\g\dl}
\nn\\
&&\qquad\qquad
-\qt{t}_{J}{}_{ab}S_{KL}\l(\Om\p^{IJKL}\r){}_{\g\dl}
-\hf{t}^{J}{}_{ab}S^{I}{}_{J}\Om{}_{\g\dl}
+{3\over4}{t}^{K}{}_{ab}S_{JK}\l(\Om\p^{IJ}\r){}_{\g\dl}
\bigg]\g^{ab}\e^\dl=0.
\label{SUSYchieq}
\end{eqnarray}

Here, for convenience, we will write once again the covariant derivatives 
and the field strength
\begin{eqnarray*}
\cD_\m{t}^I{}_{ab}
&=&\d_\m{t}^I{}_{ab}-b_\m{t}^I{}_{ab}
+\l(\Om_\m\r){}_a{}^{c}\,{t}^I{}_{cb}
+\l(\Om_\m\r){}_b{}^{c}\,{t}^I{}_{ac}
-A_\m{}^I{}_J\,{t}^J{}_{ab},
\nn\\
\cD_\m{S}_{IJ}
&=&\d_\m{S}_{IJ}-b_\m{S}_{IJ}-A_\m{}_I{}^{K}\,{S}_{KJ}
-A_\m{}_J{}^{K}\,{S}_{IK},
\nn\\
F_{\m\n}{}^I{}_J
&=&\d_\m{A}_\n{}^I{}_J-\d_\n{A}_\m{}^I{}_J
-A_\m{}^{I}{}_{K}\,A_\n{}^{K}{}_{J}
+A_\n{}^{I}{}_{K}\,A_\m{}^{K}{}_{J}. 
\end{eqnarray*}

\section{The SUSY Transform of the Mass Term of the Scalars}
\label{SUSYMassPhi}

When the interested readers attempt to ensure the supersymmetry invariance of 
the actions $L$ and $S$ in sections \ref{TensorSUGRA} and 
\ref{NonAbelianTensor}, 
respectively, it may be convenient to show how the mass term 
${\cal M}_B{}_{IJ}\phi^I\phi^J$ in the actions transforms under a supersymmetry 
transformation\footnote{
For the abelian case, we need to regard 
the matrices $\phi^I$ and $\chi^\a$ as $1\times1$ matrices
and to forget the trace $\Tr$ in the formulas here. 
}. 
\begin{eqnarray}
&&\dl\l(\Tr\l[\hf\,{\cal M}_{B}{}_{IJ}\phi^I\phi^J\r]\r)
=-{i\over4}\Tr\bigg[
\Big(S^I{}_KS^{K}{}_{J}\,\phi^J+{1\over20\a^2}G^{ab}G_{ab}\phi^I
-4t^I{}_{ab}t_J{}^{ab}\phi^J
\nn\\
&&\hskip8cm
+{4\over15}M^I{}_J\phi^J
+{1\over5}R(\Om)\phi^I\Big)(\bar\chi\c\p_I\e)\bigg].
\label{SUSYtrfmasstermPhi}
\end{eqnarray}

The two last terms on the right hand side of (\ref{SUSYtrfmasstermPhi}) 
depend on $M^I{}_J$ and $R(\Om)$. If they are given 
in terms of the backgrounds $S_{IJ}$, $G_{ab}$, $t^I{}_{ab}$, 
they may cancel the supersymmetry variation of the other terms in the actions. 
In fact, this is the case, if one uses the supersymmetry condition 
(\ref{SUSYchieq}) and the Killing spinor equation (\ref{KSE}), 
as will seen below. 

Using the supersymmetry condition (\ref{SUSYchieq}), the term 
$$
-{i\over4}\Tr\l[\phi_I\,\bar\chi^\a\c{4\over15}M^I{}_J(\Om\p^J)_{\a\b}\e^\b\r]
$$ 
on the right hand side of (\ref{SUSYtrfmasstermPhi}) can be straightforwardly 
replaced by terms depending on the backgrounds $S_{IJ}$, $G_{ab}$, $t^I{}_{ab}$.

The commutation relation of the covariant derivatives gives
$$
\hf\g^{ab}\l[\cD_a,\,\cD_b\r]\e^\a
=
-{1\over4}R(\Om)\e^\a
-{1\over8}F_{ab}{}^{IJ}\l(\p_{IJ}\r){}^\a{}_\b\g^{ab}\e^\b,
$$
and on the other hand, using the Killing spinor equation (\ref{KSE}), 
one obtains 
\begin{eqnarray*}
&&\hf\g^{ab}\l[\cD_a,\,\cD_b\r]\e^\a
=\g^{ab}\cD_a\cD_b\e^\a
\nn\\
&&=\cD_aS^{IJ}\l(\p_{IJ}\r){}^\a{}_\b\g^a\e^\b
-{1\over2\a}\cD_aG_{bc}\g^{ab}\g^{c}\e^\a
-{1\over8\a}\cD_aG_{bc}\g^{ad}\g_{d}{}^{bc}\e^\a
+{1\over2}\cD_at^I{}_{bc}\l(\p_I\r){}^\a{}_\b\g^{ad}\g_{d}{}^{bc}\e^\b
\nn\\
&&\quad
+S^{IJ}\l(\p_{IJ}\r){}^\a{}_\b\g^a\cD_a\e^\b
-{1\over2\a}G_{bc}\g^{ab}\g^{c}\cD_a\e^\a
-{1\over8\a}G_{bc}\g^{ad}\g_{d}{}^{bc}\cD_a\e^\a
+{1\over2}t^I{}_{bc}\l(\p_I\r){}^\a{}_\b\g^{ad}\g_{d}{}^{bc}\cD_a\e^\b.
\end{eqnarray*}
Comparing them, one obtains the formula 
\begin{eqnarray*}
&&-{2\over5}i\c\l(\Om\p_I\r){}_{\a\b}
\Tr\bigg[\phi^I\c\bar\chi^\a\l(-{1\over4}R(\Om)\e^\b\r)\bigg]
\nn\\
&&=-{i\over20}
\Tr\bigg[\Big(
F_{ab}{}^{KL}(\bar\chi\c\p_I\p_{KL}\g^{ab}\e)
+8\,\cD_aS_{KL}(\bar\chi\c\p_I\p^{KL}\g^{a}\e)
-4\,\cD_aG_{bc}(\bar\chi\c\p_I\g^{ab}\g^{c}\e)
\nn\\
&&\quad
-\,\cD_aG_{bc}(\bar\chi\c\p_I\g^{ad}\g_{d}{}^{bc}\e)
+4\,\cD_at^J{}_{bc}(\bar\chi\c\p_I\p_J\g^{ad}\g_{d}{}^{bc}\e)
+8\,S_{KL}(\bar\chi\c\p_I\p^{KL}\g^{a}\cD_a\e)
\nn\\
&&\quad
-4\,G_{bc}(\bar\chi\c\p_I\g^{ab}\g^{c}\cD_a\e)
-G_{bc}(\bar\chi\c\p_I\g^{ad}\g_{d}{}^{bc}\cD_a\e)
+4\,t^J{}_{bc}(\bar\chi\c\p_I\p_J\g^{ad}\g_{d}{}^{bc}\cD_a\e)
\Big)\phi^I\bigg].
\end{eqnarray*}
With the help of (\ref{KSE}) once more, the last four terms on the right hand 
of the above equation yield 
\begin{eqnarray*}
&&-{2\over5}iS_{KL}(\bar\chi\c\p_I\p^{KL}\g^{a}\cD_a\e)\phi^I
+{i\over5\a}G_{bc}(\bar\chi\c\p_I\g^{ab}\g^{c}\cD_a\e)\phi^I
+{i\over20\a}G_{bc}(\bar\chi\c\p_I\g^{ad}\g_{d}{}^{bc}\cD_a\e)\phi^I
\nn\\
&&
-{i\over5}t^I{}_{bc}\phi^J(\bar\chi\c\p_J\p_I\g^{ad}\g_{d}{}^{bc}\cD_a\e)
\nn\\
&&=-{i\over2}S_{IJ}S_{MN}\phi^K(\bar\chi\c\p_K\p^{IJ}\p^{MN}\e)
+{9\over20\a}iS_{IJ}G_{ab}\phi^K(\bar\chi\c\p_K\p^{IJ}\g^{ab}\e)
\nn\\
&&\quad
-{3\over5}iS_{IJ}t^{L}{}_{ab}\phi^K(\bar\chi\c\p_K\p^{IJ}\p_L\g^{ab}\e)
-{i\over40\a^2}G_{ab}G_{cd}\phi^I
\big(\bar\chi\c\p_I(5\g^{abcd}+\dl^{ab}\dl^{cd})\e\big)
\nn\\
&&\quad
-{i\over10\a}G_{ab}t^{I}{}_{cd}\phi^J\big(\bar\chi\c\p_J\p_{I}
(8\dl^{ac}\g^{bd}+6\dl^{ab}\dl^{bd})\e\big)
\nn\\
&&\quad
+{2\over5}it^{I}{}_{ab}t^{J}{}_{cd}\phi^K\big(\bar\chi\c\p_K\p_I\p_J
(\g^{abcd}+\dl^{ac}\g^{bd}+3\dl^{ac}\dl^{bd})\e\big).
\end{eqnarray*}

Using these formulas, it may be more accessible to verify 
the supersymmetry invariance of both the actions $L$ and $S$.

\section{Round, Squashed, and Ellipsoid 3-Spheres}
\label{3-sphere}

A 3-sphere $S^3$ is given by the set of 
solutions of $(x_1,x_2,x_3,x_4)\in{\bf R}^4$ to 
\begin{eqnarray}
x_1^2+x_2^2+x_3^2+x_4^2=1.
\label{algebraicS3}
\end{eqnarray}
If we describe it in terms of complex variables $(z,w)\in{\bf C}^2$
as $z=x_4+ix_3$, $w=x_2+ix_1$, since the defining equaiton becomes 
$|z|^2+|w|^2=1$, the two by two matrix 
$$
\begin{pmatrix}z&w\\-w^*&z^*\end{pmatrix}
$$
yields an element of a Lie group of $SU(2)$. 
Conversely, any element of the $SU(2)$ group may take the form of 
the two by two matrix in the fundamental representation. 
More explicitly, if we introduce polar coordinates ($\psi,\th,\phi$) 
and identify 
\begin{eqnarray*}
z=e^{{i\over2}(\psi+\phi)}\cos\l({\th\over2}\r),
\qquad
w=e^{{i\over2}(\phi-\psi)}\sin\l({\th\over2}\r),
\end{eqnarray*}
where $0\le\psi\le4\pi$, $0\le\th\le\pi$, $0\le\phi\le2\pi$, 
the equivalence of the 3-sphere $S^3$ to the Lie group  $SU(2)$ is 
understood by the mapping 
\begin{eqnarray}
U(\psi,\th,\phi)
=e^{{i\over2}\phi\t_3}e^{{i\over2}\th\t_2}e^{{i\over2}\psi\t_3}
=\begin{pmatrix}z&w\\-w^*&z^*\end{pmatrix}, 
\label{mappingU}
\end{eqnarray}
with the Pauli matrices $\t_a$ ($a=1,2,3$) 
\begin{eqnarray*}
\t_1=\begin{pmatrix}0&1\\1&0\end{pmatrix}, 
\quad
\t_2=\begin{pmatrix}0&-i\\i&0\end{pmatrix}, 
\quad
\t_3=\begin{pmatrix}1&0\\0&-1\end{pmatrix}. 
\end{eqnarray*}
This is a convenient parametrization for a round and a squashed $S^3$, 
as will be seen soon. On the other hand, for an ellipsoid $S^3$, 
we use another set of polar coordinates ($\phi$, $\chi$, $\th$) as
\begin{eqnarray*}
z=e^{{i}\vp}\cos{\th},
\qquad
w=e^{{i}\chi}\sin{\th},
\end{eqnarray*}
where $0\le\vp\le2\pi$, $0\le\chi\le2\pi$, $0\le\th\le\pi/2$, and therefore, 
the former coordinates are related to the latter as
$$
\th\l|_{\rm former}\r.= 2\th\l|_{\rm latter}\r.,
\quad
\phi\l|_{\rm former}\r.= (\vp+\chi)\l|_{\rm latter}\r.,
\quad
\psi\l|_{\rm former}\r.= (\vp-\chi)\l|_{\rm latter}\r..
$$

The mapping $U$ gives the vielbeins $\m^{(0)}_m$ ($m=1,2,3$) 
on a unit round sphere, 
\begin{eqnarray*}
\m^{(0)}&=&\sum_{a=1}^{3}\m^{(0)}_a\t^a
=\l({1\over{i}}\r)U^{-1}dU
\nn\\
&=&\t_1\l({\sin\th\cos\psi{d\phi}-\sin\psi{d\th}\over2}\r)
+\t_2\l({\sin\th\sin\psi{d\phi}+\cos\psi{d\th}\over2}\r)
+\t_3\l({\cos\th{d\phi}+{d\psi}\over2}\r).
\end{eqnarray*}

In terms of the vielbeins, the metric of a round sphere of radius $r$ is given by 
$$
ds^2=-r^2\Tr\l[\l(\m^{(0)}\r)^2\r]
=r^2\l[\l(\m^{(0)}_1\r)^2+\l(\m^{(0)}_2\r)^2+\l(\m^{(0)}_3\r)^2\r]
=\l(\m_1\r)^2+\l(\m_2\r)^2+\l(\m_3\r)^2
$$
with $\m_m=r\m^{(0)}_m$, ($m=1,2,3$) and the spin connection, 
$$
\om^{(0)}_{mn}=\ve_{mnk}\mu^{(0)}_k={1\over{r}}\ve_{mnk}\mu_k,
$$
for $m,n,k=1,2,3$. The coframe $\nu^m$ ($m=1,2,3$) is the inverse of the 
vielbein $\mu_m$. 

The isometry group of the round sphere is 
$SO(4)\simeq\l[{SU(2)}_L\times{SU(2)}_R\r]/{\bf Z}_2$, and it acts on 
the matrix $U(\psi,\th,\phi)$ as
$$
U(\psi,\th,\phi) \quad\to\quad g_L\c{U(\psi,\th,\phi)}\c{g}_R^{-1}
$$
for $g_L\in{SU(2)}_L$ and $g_R\in{SU(2)}_R$, and one can see that 
the vielbeins are transformed as 
$$
\m^{(0)} \quad\to\quad {g}_R\c\m^{(0)}\c{g}_R^{-1}.
$$

\subsection{Killing Spinors on a Round 3-Sphere}
\label{KSroundS}

The Killing spinor equation on a round 3-sphere is given by 
\begin{equation}
\l(d+\qt\om^{(0)}_{mn}\t^{mn}\r)\e={i\over2r}\m_m\t^m\e. 
\label{roundKSE}
\end{equation}
When $\e$ satisfies the Killing equation, 
the spinor $C_3^{-1}\e^*$ gives another solution to the equation. 
One solution to the equation is a constant spinor $\e_0$; $d\e_0=0$.

Another Killing equation 
$$
\l(d+\qt\om^{(0)}_{mn}\t^{mn}\r)\e=-{i\over2r}\m_m\t^m\e,
$$
is rewritten into 
$$
d\e=-{i}\m^{(0)}\e=-U^{-1}dU\c\e, 
$$
which is solved by $\e=U^{-1}\e_0$, since one has 
$$
d(U^{-1}\e_0)=-U^{-1}dU\c\l(U^{-1}\e_0\r).
$$

In the text, we make frequent use of the constant Killing spinor $\e=\e_0$ 
and its charge conjugate $\e^c=C_3^{-1}{\e_0}^*$. 
We normalize them such that $\e^\dag\e={\e^c}^\dag\e^c=1$, and then 
the Fierz transformation gives 
\begin{equation}
\e\,\e^\dag+\e^c\,{\e^c}^\dag={\bf 1}_2.
\label{Fierz3DKillingSpinor}
\end{equation}

We can make three Killing vectors ${\e}^\dag\t^m\e$, ${\e^c}^\dag\t^m\e$, 
and ${\e}^\dag\t^m\e^c$ out of $\e$ and $\e^c$; 
$\l({\e^c}^\dag\t_m\e\r)^*=\l({\e}^\dag\t_m\e^c\r)$, 
and they obeys 
$$
\cD_m\l({\e}^\dag\t_n\e\r)={1\over{r}}\e_{mnk}\l({\e}^\dag\t_k\e\r),
\qquad
\cD_m\l({\e^c}^\dag\t_n\e\r)={1\over{r}}\e_{mnk}\l({\e^c}^\dag\t_k\e\r),
\qquad
\cD_m\l({\e}^\dag\t_n\e^c\r)={1\over{r}}\e_{mnk}\l({\e}^\dag\t_k\e^c\r),
$$
where $\e_{mnk}$ is the constant antisymmetric tensor with $\e_{123}=1$. 
From the norm of $\e$ and its Fierz transformation, we can deduce that 
\begin{eqnarray}
\l({\e}^\dag\t^m\e\r)\l({\e}^\dag\t^m\e\r)=1,
\qquad
\l({\e}^\dag\t^m\e\r)\l({\e^c}^\dag\t^m\e\r)=0,
\qquad
\l({\e}^\dag\t^m\e\r)\l({\e}^\dag\t^m\e^c\r)=0,
\nn\\
\l({\e^c}^\dag\t^m\e\r)\l({\e}^\dag\t^m\e^c\r)=2,
\qquad
\l({\e^c}^\dag\t^m\e\r)\l({\e^c}^\dag\t^m\e\r)=0, 
\qquad
\l({\e}^\dag\t^m\e^c\r)\l({\e}^\dag\t^m\e^c\r)=0,
\label{KillingVectorOrthonormality}
\end{eqnarray}
and thus they span the three-dimensional space;
$$
\l({\e}^\dag\t^m\e\r)\l({\e}^\dag\t^n\e\r)
+\hf\l({\e^c}^\dag\t^m\e\r)\l({\e}^\dag\t^n\e^c\r)
+\hf\l({\e}^\dag\t^m\e^c\r)\l({\e^c}^\dag\t^n\e\r)
=\dl^{mn},
$$
so that we can expand a vector $A_m$ in terms of the Killing vectors,
\begin{eqnarray*}
A_m
&=&\l({\e}^\dag\t_m\e\r)\l({\e}^\dag\t^n\e\r)A_n
+\hf\l({\e^c}^\dag\t_m\e\r)\l({\e}^\dag\t^n\e^c\r)A_n
+\hf\l({\e}^\dag\t_m\e^c\r)\l({\e^c}^\dag\t^n\e\r)A_n
\nn\\
&=&\l({\e}^\dag\t_m\e\r)V_0+\l({\e^c}^\dag\t_m\e\r)V_-
+\l({\e}^\dag\t_m\e^c\r)V_+.
\end{eqnarray*}
Similarly, the differential operator $\d_m=\sum_{\m=1}^{3}\nu_m{}^\m\d_\m$ 
is expanded as
$$
\d_m
=\l({\e}^\dag\t_m\e\r)\l({\e}^\dag\t^n\e\r)\d_n
+\hf\l({\e^c}^\dag\t_m\e\r)\l({\e}^\dag\t^n\e^c\r)\d_n
+\hf\l({\e}^\dag\t_m\e^c\r)\l({\e^c}^\dag\t^n\e\r)\d_n.
$$

Since it satisfies the commutation relation
$$
\l[\d_m,\,\d_n\r]=-{2\over{r}}\e_{mnk}\d_k,
$$
the covariant derivatives on a scalar field $\Phi$ commute with each other,
\begin{eqnarray*}
\l[\cD_m,\,\cD_n\r]\Phi=\l[\d_m,\,\d_n\r]\Phi
+\l[\l(\om_m\r)_n{}^k-\l(\om_n\r)_m{}^k\r]\d_k\Phi=0. 
\end{eqnarray*}
Using the properties
\begin{eqnarray*}
&&\e^{mnk}\l({\e}^\dag\t_n\e\r)\l({\e^c}^\dag\t_k\e\r)
=-i\l({\e^c}^\dag\t^m\e\r), 
\qquad
\e^{mnk}\l({\e}^\dag\t_n\e\r)\l({\e}^\dag\t_k\e^c\r)=i\l({\e}^\dag\t^m\e^c\r), 
\nn\\
&&\hskip2cm
\e^{mnk}\l({\e^c}^\dag\t_n\e\r)\l({\e}^\dag\t_k\e^c\r)
=-2i\l({\e}^\dag\t^m\e\r), 
\end{eqnarray*}
we can deduce the commutation relations among the differential operators
$\l({\e}^\dag\t^n\e\r)\cD_n$, $\l({\e}^\dag\t^n\e^c\r)\cD_n$, and 
$\l({\e^c}^\dag\t^n\e\r)\cD_n$ on a scalar field,
\begin{eqnarray}
&&\l[\l({\e}^\dag\t^m\e\r)\cD_m,\,\l({\e^c}^\dag\t^n\e\r)\cD_n\r]
\nn\\
&&=\l[\l({\e}^\dag\t^n\e\r)\cD_n\l({\e^c}^\dag\t^m\e\r)
-\l({\e^c}^\dag\t^n\e\r)\cD_n\l({\e}^\dag\t^m\e\r)\r]\d_m
+\l({\e}^\dag\t^m\e\r)\l({\e^c}^\dag\t^n\e\r)\l[\cD_m,\,\cD_n\r]
\nn\\
&&=-{2\over{r}}\e_{mnk}\l({\e}^\dag\t^n\e\r)\l({\e^c}^\dag\t^k\e\r)\cD_m
={2i\over{r}}\l({\e^c}^\dag\t^m\e\r)\cD_m,
\nn\\
&&\l[\l({\e}^\dag\t^m\e\r)\cD_m,\,\l({\e}^\dag\t^n\e^c\r)\cD_n\r]
=-{2\over{r}}\e_{mnk}\l({\e}^\dag\t^n\e\r)\l({\e}^\dag\t^k\e^c\r)\cD_m
=-{2i\over{r}}\l({\e}^\dag\t^m\e^c\r)\cD_m,
\label{CommRelDifOpRoundS}\\
&&\l[\l({\e^c}^\dag\t^m\e\r)\cD_m,\,\l({\e}^\dag\t^n\e^c\r)\cD_n\r]
=-{2\over{r}}\e_{mnk}\l({\e^c}^\dag\t^n\e\r)\l({\e}^\dag\t^k\e^c\r)\cD_m
={4i\over{r}}\l({\e}^\dag\t^m\e\r)\cD_m,
\nn
\end{eqnarray}

Therefore, when we regard them as
$$
L_3=-i{r\over2}\l({\e}^\dag\t^m\e\r)\cD_m, 
\qquad
L_+=-i{r\over2}\l({\e^c}^\dag\t^m\e\r)\cD_m, 
\qquad
L_-=-i{r\over2}\l({\e}^\dag\t^m\e^c\r)\cD_m, 
$$
they form the $SU(2)$ algebra,
$$
\l[L_3,\,L_{\pm}\r]=\pm{L_{\pm}},
\qquad
\l[L_+,\,L_-\r]=2\,L_3.
$$

\subsection{Killing Spinors on a Squashed 3-Sphere}

Let us turn to a squashed 3-sphere. In terms of the Hopf fibration of 
a 3-sphere, the circle fiber in a squashed $S^3$ has 
the different radius $\tr$ from the radius $r$ of the 2-sphere base, 
while $\tr=r$ for a round 3-sphere. 
The metric of the squashed $S^3$ is thus given by
$$
ds^2=r^2\l[\l(\m^{(0)}_1\r)^2+\l(\m^{(0)}_2\r)^2\r]+\tr^2\l(\m^{(0)}_3\r)^2
=\l(e_1\r)^2+\l(e_2\r)^2+\l(e_3\r)^2,
$$
where $e_1=r\m^{(0)}_1$, $e_2=r\m^{(0)}_2$, and $e_3=\tr\m^{(0)}_3$. 

Since the vielbeins $\m^{(0)}$ are still invariant under the $SU(2)_L$ 
transformations, it shows that the isometry group $SO(4)$ is broken to 
$\l[SU(2)_L\times{U(1)_R}\r]/{\bf Z}_2$.

Here, we assume that $\tr\ge{r}$.
Solving the equation $de^m+\om^m{}_n\wedge{e}^n=0$, one obtains 
the Levi-Civita spin connection $\om^{ab}$, 
$$
\l(\om_1\r)_{23}=\l(\om_2\r)_{31}={\tr\over{r}^2},
\qquad
\l(\om_3\r)_{12}={\tr\over{r}^2}+{2\over\tr}\l(1-{\tr^2\over{r}^2}\r). 
$$

The curvature tensor of the spin connection $\om^{mn}$ on the squashed 
3-sphere $S^3$
$$
R^m{}_{n}=d\om^m{}_n+\om^m{}_{k}\wedge\om^k{}_n,
$$
is computed to give 
$$
R^1{}_2={\tr^2\over{r}^4}\,e^1\wedge{e}^2
+{4\over{r}^2}\l(1-{\tr^2\over{r}^2}\r)\,e^1\wedge{e}^2,
\quad
R^2{}_3={\tr^2\over{r}^4}\,e^2\wedge{e}^3,
\quad
R^3{}_1={\tr^2\over{r}^4}\,e^3\wedge{e}^1,
$$
and the scalar curvature 
$$
R=R_{mn}{}^{mn}={6\tr^2\over{r}^4}
+{8\over{r}^2}\l(1-{\tr^2\over{r}^2}\r)
$$

\subsection{Constant Killing Spinors on a Squashed 3-Sphere}
\label{constKSSquashS}

The constant spinors $\e=\e_0$ and $\e^c=C_3^{-1}\e^*$ obey
\begin{eqnarray*}
&&\l(d+\qt\om_{mn}\t^{mn}\r)\e
=+{i\over2}{\tr\over{r}^2}e_m\t^m\e
+{i\over{\tr}}\l(1-{\tr^2\over{r}^2}\r)e_3\t_3\,\e,
\nn\\
&&
\l(d+\qt\om_{mn}\t^{mn}\r)\e^c
=+{i\over2}{\tr\over{r}^2}e_m\t^m\e^c
+{i\over{\tr}}\l(1-{\tr^2\over{r}^2}\r)e_3\t_3\e^c,
\end{eqnarray*}
respectively, with the spin connection $\om^{mn}$ on the squashed 3-sphere. 
Therefore, if we impose the condition
\begin{equation}
\t_3\e=\e, \qquad \t_3\e^c=-\e^c,
\label{HosomichiKillingSpinor}
\end{equation}
we may regard the second terms on the right hand sides of the above two 
Killing spinor equations as the background gauge field;
\begin{eqnarray*}
&&\l[d+\qt\om_{mn}\t^{mn}-{i\over{\tr}}\l(1-{\tr^2\over{r}^2}\r)e_3\r]\e
=+{i\over2}{\tr\over{r}^2}e_m\t^m\e,
\nn\\
&&
\l[d+\qt\om_{mn}\t^{mn}+{i\over{\tr}}\l(1-{\tr^2\over{r}^2}\r)e_3\r]\e^c
=+{i\over2}{\tr\over{r}^2}e_m\t^m\e^c,
\end{eqnarray*}
and we may further regard the gauge field as the $R$-symmetry gauge field, 
when the spinors are embedded into a single five-dimensional spinor. 

On the other hand, regarding the gauge field as a $U(1)$ gauge field,
$$
V=-{1\over{\tr}}\l(1-{\tr^2\over{r}^2}\r)e_3
$$
we can see that the Killing spinors $\e$ and $\e^c$ carry charge 1 and 
-1, respectively, and so obey
\begin{eqnarray*}
&&\cD\e=\l[d+\qt\om_{mn}\t^{mn}+iV\r]\e
=+{i\over2}{\tr\over{r}^2}e_m\t^m\e,
\qquad
\cD\e^c=\l[d+\qt\om_{mn}\t^{mn}-iV\r]\e^c
=+{i\over2}{\tr\over{r}^2}e_m\t^m\e^c.
\end{eqnarray*}

It follows from the Killing spinor equations that 
\begin{eqnarray*}
\l[\cD_m,\,\cD_n\r]\e&=&\l[\qt{R}_{mn}{}^{kl}\t_{kl}+iV_{mn}\r]\e
={i\over2}\l({\tr\over{r}^2}\r)^2\e_{mnk}\t_k\e
={\tr\over{r}^2}\e_{mnk}\cD_k\e,
\nn\\
\l[\cD_m,\,\cD_n\r]\e^c&=&\l[\qt{R}_{mn}{}^{kl}\t_{kl}-iV_{mn}\r]\e^c
={i\over2}\l({\tr\over{r}^2}\r)^2\e_{mnk}\t_k\e^c
={\tr\over{r}^2}\e_{mnk}\cD_k\e^c,
\end{eqnarray*}
where $V_{mn}$ are the components of the field strength of the gauge field $V$, 
$$
dV=\hf{V}_{mn}e^m\wedge{e}^n
=-{2\over{r}^2}\l(1-{\tr^2\over{r}^2}\r)e^1\wedge{e}^2.
$$
Multiplying $\t^{mn}$ from the left on both the left and right sides, 
we obtain 
\begin{eqnarray*}
\l[\hf{R}(S^3)-iV_{mn}\t^{mn}\r]\e=3\l({\tr\over{r}^2}\r)^2\e,
\qquad
\l[\hf{R}(S^3)+iV_{mn}\t^{mn}\r]\e^c=3\l({\tr\over{r}^2}\r)^2\e^c.
\end{eqnarray*}

From the conditions (\ref{HosomichiKillingSpinor}), we take 
$$
\e=\begin{pmatrix}1\cr0\end{pmatrix},
\qquad
\e^c=\begin{pmatrix}0\cr1\end{pmatrix},
$$
and the Fierz transformation of $\e$ and $\e^c$ remains the same as in 
(\ref{Fierz3DKillingSpinor}).

Three vectors ${\e}^\dag\t^m\e$, ${\e^c}^\dag\t^m\e$, and ${\e}^\dag\t^m\e^c$ 
are of charge 0, 2, and -2, respectively under the gauge field $V$, and obey 
$$
\cD_m\l({\e}^\dag\t_n\e\r)={\tr\over{r}^2}\e_{mnk}\l({\e}^\dag\t_k\e\r),
\qquad
\cD_m\l({\e^c}^\dag\t_n\e\r)={\tr\over{r}^2}\e_{mnk}\l({\e^c}^\dag\t_k\e\r),
\qquad
\cD_m\l({\e}^\dag\t_n\e^c\r)={\tr\over{r}^2}\e_{mnk}\l({\e}^\dag\t_k\e^c\r),
$$
where the covariant derivatives contain the gauge field $V$ as a connection 
according to their charges.
They satisfy (\ref{KillingVectorOrthonormality}) and form 
an orthonormal basis 
$$
\l({\e}^\dag\t^m\e\r)\l({\e}^\dag\t^n\e\r)
+\hf\l({\e^c}^\dag\t^m\e\r)\l({\e}^\dag\t^n\e^c\r)
+\hf\l({\e}^\dag\t^m\e^c\r)\l({\e^c}^\dag\t^n\e\r)
=\dl^{mn},
$$
and obey the relations of the cross product,
\begin{eqnarray}
&&\e^{mnk}\l({\e}^\dag\t_n\e\r)\l({\e^c}^\dag\t_k\e\r)
=-i\l({\e^c}^\dag\t^m\e\r), 
\qquad
\e^{mnk}\l({\e}^\dag\t_n\e\r)\l({\e}^\dag\t_k\e^c\r)=i\l({\e}^\dag\t^m\e^c\r), 
\nn\\
&&\hskip2cm
\e^{mnk}\l({\e^c}^\dag\t_n\e\r)\l({\e}^\dag\t_k\e^c\r)
=-2i\l({\e}^\dag\t^m\e\r). 
\label{squashcross}
\end{eqnarray}

We will denote the covariant derivative $\cD$ on a scalar field $\Phi$ of 
charge $q$ under the gauge field $V$ as 
$$
\cD^{(q)}{}_m\Phi=\d_m\Phi+iqV_m\Phi.
$$
The commutation relations of the differential operators 
$\l({\e}^\dag\t^m\e\r)\cD^{(q)}{}_m$, 
$\l({\e^c}^\dag\t^m\e\r)\cD^{(q)}{}_m$, and 
$\l({\e}^\dag\t^m\e^c\r)\cD^{(q)}{}_m$ essentially remains the same as 
in (\ref{CommRelDifOpRoundS}), if $1/r$ is replaced by $\tr/r^2$;
\begin{eqnarray}
&&\l({\e}^\dag\t^n\e^c\r)\cD^{(q)}{}_n\,\l({\e}^\dag\t^m\e\r)\cD^{(q)}{}_m
=\l({\e}^\dag\t^m\e\r)\cD^{(q-2)}{}_m\,\l({\e}^\dag\t^n\e^c\r)\cD^{(q)}{}_n
+2i{\tr\over{r}^2}\l({\e}^\dag\t^n\e^c\r)\cD^{(q)}{}_n,
\label{CommRelDifOpSquashS}\\
&&\l({\e^c}^\dag\t^n\e\r)\cD^{(q)}{}_n\,\l({\e}^\dag\t^m\e\r)\cD^{(q)}{}_m
=\l({\e}^\dag\t^m\e\r)\cD^{(q+2)}{}_m\,\l({\e^c}^\dag\t^n\e\r)\cD^{(q)}{}_n
-2i{\tr\over{r}^2}\l({\e^c}^\dag\t^n\e\r)\cD^{(q)}{}_n,
\nn\\
&&\l({\e^c}^\dag\t^n\e\r)\cD^{(q-2)}{}_n\,\l({\e}^\dag\t^m\e^c\r)\cD^{(q)}{}_m
=\l({\e}^\dag\t^m\e^c\r)\cD^{(q+2)}{}_m\,\l({\e^c}^\dag\t^n\e\r)\cD^{(q)}{}_n
\nn\\
&&\hskip7cm
+4i{\tr\over{r}^2}\l({\e}^\dag\t^n\e\r)\cD^{(q)}{}_n
+q\e_{mnk}\l({\e}^\dag\t^m\e\r)V_{nk},
\nn\end{eqnarray}
except for the last term on the right hand side in the last equation.

\subsection{Non-Constant Killing Spinors on a Squashed 3-Sphere}

Similarly to the constant spinors $\e=\e_0$ and $C_3^{-1}\e_0^*$, 
the spinor $\e=U^{-1}\e_0$ on the squashed 3-sphere gives the solution 
to the differential equation
$$
\l(d+\qt\om_{mn}\t^{mn}\r)\e
=-{i\over2r}\l(2-{\tr\over{r}}\r)e_m\t^m\e
+{i\over{r}}\l(1-{\tr\over{r}}\r)e_3\t_3\,\e.
$$
However, the computation of the partion function with this Killing spinor 
would give the same result as with the constant spinor $\e_0$, and  
this Killing spinor isn't used in this paper.

As discussed in \cite{Imamura}, there is another Killing spinor which 
is given by a Lorentz tramsform of $U^{-1}\e_0$. 
As explained in subsection \ref{ImamuraSquash}, the slant periodic boundary 
condition is rotated to the time direction upon the dimensional reduction 
from the six-dimensional theory. The rotation induces the Lorentz 
transformation on the coframe $e^m$. 
This is quite parallel to the three-dimensional stroy 
in \cite{Imamura}. Therefore, one obtains the Lorentz trasformed 
Killing spinors on the squashed 3-sphere with the supersymmetry 
background in subsection \ref{ImamuraSquash}, 
which is five-dimensional analogues of a three-dimensional Killing spinor 
given in the paper \cite{Imamura}. 

It has been discussed in \cite{Imamura} that 
the spinor $\e=e^{(1/2)\xi\t_3}U^{-1}\e_0$ 
with $\cosh\xi=\tr/{r}$ and $\sinh\xi=\sqrt{(\tr/{r})^2-1}$ 
is the solution to 
$$
\l(d+\qt\om_{mn}\t^{mn}\r)\e
=-{i\over2}{\tr\over{r}^2}e_m\t^m\e
+{1\over{r}}\sqrt{{\tr^2\over{r}^2}-1}\,\ve_{3mn}e^m\t^n\e,
$$
and its charge conjugate $C^{-1}_3\e^*$ satisfies
$$
\l(d+\qt\om_{mn}\t^{mn}\r)(C_3^{-1}\e^*)
=-{i\over2}{\tr\over{r}^2}e_m\t^m(C_3^{-1}\e^*)
-{1\over{r}}\sqrt{{\tr^2\over{r}^2}-1}\,\ve_{3mn}e^m\t^n(C_3^{-1}\e^*).
$$



We have explained in subsection \ref{ImamuraSquash} that 
the slant boundary condition with the parameter $u$ on the round 3-sphere 
is transformed via the change of coordinates to 
the periodic boundary condition on the squashed 3-sphere 
with the fiber radius $\tr$ and the base radius $r$, 
as have been discussed in \cite{Imamura}.  
We have introduced the intermediate parameter $\xi$ 
in subsection \ref{ImamuraSquash}, 
which explicitly appears in the Killing spinors. 
It could be convenient to summarize the relations of the parameters 
$u$, $\xi$ and the ratio $\tr/r$ as 
\begin{eqnarray*}
&&\cosh\xi={\tr\over{r}}={1\over\sqrt{1-u^2}},
\quad
\sinh\xi
={u\over\sqrt{1-u^2}},
\quad
e^{\pm\xi/2}=\sqrt{{\tr+r\over2r}}\pm\sqrt{{\tr-r\over2r}}
=\l({{1\pm{u}\over1\mp{u}}}\r)^{{1\over4}}.
\end{eqnarray*}

\subsection{Killing Spinors on an Ellipsoid 3-Sphere}
\label{EllipsoidS^3}

In order to obtain an ellipsoid $S^3$, following \cite{Hosomichi}, 
we will deform the defining equation (\ref{algebraicS3}) to give 
$$
{x_1^2+x_2^2\over{\tr}^2}+{x_3^2+x_4^2\over{r}^2}=1, 
$$
with the flat metric 
$$
ds^2=dx_1^2+dx_2^2+dx_3^2+dx_4^2.
$$
Substituting the polar coordinates 
\begin{eqnarray*}
z=x_4+ix_3={r}e^{{i}\vp}\cos{\th},
\qquad
w=x_2+ix_1={\tr}e^{{i}\chi}\sin{\th},
\end{eqnarray*}
where $0\le\vp\le2\pi$, $0\le\chi\le2\pi$, $0\le\th\le\pi/2$,
into the metric, one obtains
$$
ds^2=\l(e^1\r)^2+\l(e^2\r)^2+\l(e^3\r)^2,
$$
where the dribeins $e^{1,2,3}$ are 
$$
e^1=r\cos\th\,d\vp,
\quad
e^2=\tr\sin\th\,d\chi,
\quad
e^3=f(\th)d\th, 
$$
with $f(\th)=\sqrt{{\tr}^2\cos^2\th+r^2\sin^2\th}$. 
The coframes, which are the dual basis to the dribeins, are given by
$$
\th_1={1\over{r}\cos\th}\pd{\vp}, 
\qquad
\th_2={1\over{\tr}\sin\th}\pd{\chi}, 
\qquad
\th_3={1\over{f(\th)}}\pd{\th}. 
$$

The compatible spin connection is given by
$$
\om^{12}=0, 
\quad
\om^{23}={1\over{f(\th)}\tan\th}e^2,
\quad
\om^{31}={\tan\th\over{f}(\th)}e^1,
$$
and the curvature tensor 
$R^m{}_n=d\om^{m}{}_{n}+\om^{m}{}_{k}\wedge\om^{k}{}_{n}$ 
by
$$
R^{12}={1\over{f(\th)}^2}e^1\wedge{e}^2,
\quad
R^{23}={r^2\over{f(\th)}^4}e^2\wedge{e}^3,
\quad
R^{31}={\tr^2\over{f(\th)}^4}e^3\wedge{e}^1, 
$$
and therefore one obtains the scalar curvature 
$$
R(S_{ell}^3)={2\over{f(\th)}^4}\l({f(\th)}^2+r^2+\tr^2\r). 
$$

When $r=\tr=f$, the ellipsoid $S^3$ becomes a round one, and then 
spinors 
$$
\e=
{1\over\sqrt{2}}
\begin{pmatrix}
e^{{i\over2}\l(\chi-\vp+\th\r)}
\\
-e^{{i\over2}\l(\chi-\vp-\th\r)}
\end{pmatrix},
\qquad
\e^c=C^{-1}_3\e^*=
{1\over\sqrt{2}}
\begin{pmatrix}
e^{-{i\over2}\l(\chi-\vp-\th\r)}
\\
e^{-{i\over2}\l(\chi-\vp+\th\r)}
\end{pmatrix},
$$
obey the Killing spinor equations on the round $S^3$
$$
\l(d+{1\over4}\om^{mn}\t_{mn}\r)\e={i\over2{r}}\m^m\t_m\e,
\qquad
\l(d+{1\over4}\om^{mn}\t_{mn}\r)\e^c={i\over2{r}}\m^m\t_m\e^c,
$$
In order to keep these Killing spinors on the ellipsoid $S^3$ with $\tr\not=r$, 
turing on a background $U(1)$ gauge field 
\begin{equation}
V=\hf\l[-{1\over{r}\cos\th}\l({r\over{f}}-1\r)e^1
+{1\over\tr\sin\th}\l({\tr\over{f}}-1\r)e^2\r],
\label{bgV}
\end{equation}
as in \cite{Hosomichi}, one can see that they satisfy 
$$
D\e\equiv\l(d+{1\over4}\om^{mn}\t_{mn}+iV\r)\e={i\over2{f}}e^m\t_m\e,
\qquad
D\e^c\equiv\l(d+{1\over4}\om^{mn}\t_{mn}-iV\r)\e^c={i\over2{f}}e^m\t_m\e^c.
$$
The use of the Killing spinor equations twice for $DD\e$ and $DD\e^c$ leads to 
\begin{eqnarray}
\l(\hf{R}-i\t^{mn}V_{mn}\r)\e
=\l({3\over{f}^2}+{2i\over{f}^2}\t^m\d_m{f}\r)\e,
\quad
\l(\hf{R}+i\t^{mn}V_{mn}\r)\e^c
=\l({3\over{f}^2}+{2i\over{f}^2}\t^m\d_m{f}\r)\e^c,
\label{DDeDDec}
\end{eqnarray}
where $V_{mn}$ is the field strength of $V$,
$$
dV=\hf{V}_{mn}\,e^m\wedge{e}^n,
$$
and the partial differential $\d_m{f}$ is defined by
$df=e^m\d_m{f}$.

As on the squashed $S^3$ in subsection \ref{constKSSquashS},
three vectors ${\e}^\dag\t^m\e$, ${\e^c}^\dag\t^m\e$, and ${\e}^\dag\t^m\e^c$ 
are of charge 0, 2, and -2, respectively under the gauge field $V$. 
They obey 
$$
\cD_m\l({\e}^\dag\t_n\e\r)={1\over{f}}\e_{mnk}\l({\e}^\dag\t_k\e\r),
\qquad
\cD_m\l({\e^c}^\dag\t_n\e\r)={1\over{f}}\e_{mnk}\l({\e^c}^\dag\t_k\e\r),
\qquad
\cD_m\l({\e}^\dag\t_n\e^c\r)={1\over{f}}\e_{mnk}\l({\e}^\dag\t_k\e^c\r),
$$
where the covariant derivatives contain the gauge field $V$ as a connection 
according to their charges. 
They satisfy (\ref{KillingVectorOrthonormality}) and give 
an orthonormal basis 
$$
\l({\e}^\dag\t^m\e\r)\l({\e}^\dag\t^n\e\r)
+\hf\l({\e^c}^\dag\t^m\e\r)\l({\e}^\dag\t^n\e^c\r)
+\hf\l({\e}^\dag\t^m\e^c\r)\l({\e^c}^\dag\t^n\e\r)
=\dl^{mn},
$$
and obey the same relations of the cross product,
\begin{eqnarray}
&&\e^{mnk}\l({\e}^\dag\t_n\e\r)\l({\e^c}^\dag\t_k\e\r)
=-i\l({\e^c}^\dag\t^m\e\r), 
\qquad
\e^{mnk}\l({\e}^\dag\t_n\e\r)\l({\e}^\dag\t_k\e^c\r)=i\l({\e}^\dag\t^m\e^c\r), 
\nn\\
&&\hskip2cm
\e^{mnk}\l({\e^c}^\dag\t_n\e\r)\l({\e}^\dag\t_k\e^c\r)
=-2i\l({\e}^\dag\t^m\e\r), 
\label{ellisoidcross}
\end{eqnarray}
as in (\ref{squashcross}) for the squashed $S^3$.  

More concretely, we have
\begin{eqnarray*}
\Big(\l({\e}^\dag\t^1\e\r),\,\l({\e}^\dag\t^2\e\r),\,\l({\e}^\dag\t^3\e\r)\Big)
&=&\l(-\cos\th,\,\sin\th,\,0\r),
\nn\\
\Big(\l({\e^c}^\dag\t^1\e\r),\,\l({\e^c}^\dag\t^2\e\r),\,
\l({\e^c}^\dag\t^3\e\r)\Big)
&=&ie^{i\l(\chi-\vp\r)}\l(\sin\th,\,\cos\th,\,-i\r),
\nn\\
\Big(\l({\e}^\dag\t^1\e^c\r),\,\l({\e}^\dag\t^2\e^c\r),\,
\l({\e}^\dag\t^3\e^c\r)\Big)
&=&-ie^{-i\l(\chi-\vp\r)}\l(\sin\th,\,\cos\th,\,i\r),
\end{eqnarray*}
and it implys that 
\begin{eqnarray}
\l({\e}^\dag\t^m\e\r)\d_m{f}=0,
\label{D0f=0}
\end{eqnarray}
where $\d_m$ is defined such that $d=e^m\d_m$.

We will denote the covariant derivative $\cD$ on a scalar field $\Phi$ of 
charge $q$ under the gauge field $V$ as 
$$
\cD^{(q)}{}_m\Phi=\d_m\Phi+iqV_m\Phi.
$$
The commutation relations of the differential operators 
$\l({\e}^\dag\t^m\e\r)\cD^{(q)}{}_m$, 
$\l({\e^c}^\dag\t^m\e\r)\cD^{(q)}{}_m$, and 
$\l({\e}^\dag\t^m\e^c\r)\cD^{(q)}{}_m$ 
slightly differ from (\ref{CommRelDifOpSquashS}) 
by the terms with the field strength $V_{mn}$, 
even if $\tr/r^2$ is replaced by $1/f$, and 
using (\ref{DDeDDec}), we will replace them by the terms including $f$ 
to give 
\begin{eqnarray}
&&\l({\e}^\dag\t^n\e^c\r)\cD^{(q)}{}_n\,\l({\e}^\dag\t^m\e\r)\cD^{(q)}{}_m
=\l({\e}^\dag\t^m\e\r)\cD^{(q-2)}{}_m\,\l({\e}^\dag\t^n\e^c\r)\cD^{(q)}{}_n
+{2i\over{f}}\l({\e}^\dag\t^n\e^c\r)\cD^{(q)}{}_n
+iq\l({\e}^\dag\t^n\e^c\r)\d_n{1\over{f}},
\nn\\
&&\l({\e^c}^\dag\t^n\e\r)\cD^{(q)}{}_n\,\l({\e}^\dag\t^m\e\r)\cD^{(q)}{}_m
=\l({\e}^\dag\t^m\e\r)\cD^{(q+2)}{}_m\,\l({\e^c}^\dag\t^n\e\r)\cD^{(q)}{}_n
-{2i\over{f}}\l({\e^c}^\dag\t^n\e\r)\cD^{(q)}{}_n,
+iq\l({\e^c}^\dag\t^n\e\r)\d_n{1\over{f}},
\nn\\
&&\l({\e^c}^\dag\t^n\e\r)\cD^{(q-2)}{}_n\,\l({\e}^\dag\t^m\e^c\r)\cD^{(q)}{}_m
=\l({\e}^\dag\t^m\e^c\r)\cD^{(q+2)}{}_m\,\l({\e^c}^\dag\t^n\e\r)\cD^{(q)}{}_n
\label{CommRelDifOpEllipsoidS}\\
&&\hskip8cm
+{4i\over{f}}\l({\e}^\dag\t^n\e\r)\cD^{(q)}{}_n
+q\e_{mnk}\l({\e}^\dag\t^m\e\r)V_{nk},
\nn\end{eqnarray}
except for the last term on the right hand side in the last equation.

Let us consider the scalar spherical harmonics in the coordinates 
($\chi,\th,\vp$) in the round limit $\tr\to{r}$ and set the radius 
to be unity; $r=1$. The mapping from ($\chi,\th,\vp$) to the 3-sphere is 
$$
U=e^{{i\over2}(\chi+\vp)\t_3}\,e^{{i}\th\t_2}\,e^{{i\over2}(\vp-\chi)\t_3},
$$
and the $SU(2)_{{\rm L}}\times{SU(2)}_{{\rm R}}$ isometry of 
the round 3-sphere is given by $g_{{\rm L}}\in{SU(2)}_{{\rm L}}$ and 
$g_{{\rm R}}\in{SU(2)}_{{\rm R}}$ as
$$
U \quad\to\quad g_{{\rm L}}{\c}U{\c}g^{-1}_{{\rm R}}.
$$
The left- and the right-invariant one-forms are given by
\begin{eqnarray*}
{1\over{i}}U^{-1}dU
=\l(\t_1+i\t_2\r)e^-{}_{({\rm L})}+\l(\t_1-i\t_2\r)e^+{}_{({\rm L})}
+\t_3e^3{}_{({\rm L})},
\nn\\
{1\over{i}}dU{\c}U^{-1}
=\l(\t_1+i\t_2\r)e^-{}_{({\rm R})}+\l(\t_1-i\t_2\r)e^+{}_{({\rm R})}
+\t_3e^3{}_{({\rm R})},
\end{eqnarray*}
where $e^a_{({\rm L})}$, $e^a_{({\rm R})}$ ($a=+,-,3$) play the role of 
the dreibeins, respectively. Their coframes which are dual to 
$e^a_{({\rm L,R})}$, respectively, are 
\begin{eqnarray*}
&&\th^{({\rm L})}_{\pm}
=e^{\mp{i}(\vp-\chi)}\l(\th_1\sin\th+\th_2\cos\th\mp{i}\th_3\r),
\qquad
\th^{({\rm L})}_{3}=\l(\th_1\cos\th-\th_2\sin\th\r),
\nn\\
&&\th^{({\rm R})}_{\pm}
=e^{\pm{i}(\vp+\chi)}\l(-\th_1\sin\th+\th_2\cos\th\mp{i}\th_3\r),
\qquad
\th^{({\rm R})}_{3}=\l(\th_1\cos\th+\th_2\sin\th\r).
\end{eqnarray*}
Then, identifying these coframes as
\begin{eqnarray*}
L_{\pm}=-{1\over2i}\th^{{\rm L}}_{\pm},
\quad
L_{3}=-{1\over2i}\th^{{\rm L}}_{3}={1\over2i}\l(\pd{\chi}-\pd{\vp}\r);
\qquad
\tilde{L}_{\pm}={1\over2i}\th^{{\rm R}}_{\pm},
\quad
\tilde{L}_{3}={1\over2i}\th^{{\rm R}}_{3}={1\over2i}\l(\pd{\chi}+\pd{\vp}\r),
\end{eqnarray*}
we can see that they form the Lie algebras of $SU(2)$;
\begin{eqnarray*}
\l[L_3,\,L_{\pm}\r]=\pm{L}_{\pm},
\quad
\l[L_+,\,L_{-}\r]=2{L}_{3};
\qquad
\l[\tilde{L}_3,\,\tilde{L}_{\pm}\r]=\pm\tilde{L}_{\pm},
\quad
\l[\tilde{L}_+,\,\tilde{L}_{-}\r]=2\tilde{L}_{3}.
\end{eqnarray*}
Furthermore, a simple calculation shows that 
$$
L^2=\hf\l(L_+L_-+L_-L_+\r)+L_3L_3=\hf\l(\tilde{L}_+\tilde{L}_-
+\tilde{L}_-\tilde{L}_+\r)+\tilde{L}_3\tilde{L}_3=\tilde{L}^2.
$$

A spherical harmonics $f_n$ corresponding to one of the highest weight states 
defined by 
$$
L_+{f}_n=\tilde{L}_+{f}_n=0,
$$
is obtained by 
$$
f_n=c_ne^{in\chi}\sin^n\th, 
\quad {\rm with~~} c_n={n+1\over2\pi^2},
$$
where $n$ is an integer required by the periodicity under $\chi\to\chi+2\pi$
and is non-negative required by the normalizability of $f_n$; 
$n\in{\bf Z}_{\ge0}$. It has the eigenvalues ($l=n/2,m=n/2,\tm=n/2$)
for $L^2=\tilde{L}^2$, $L_3$ and $\tilde{L}_3$,
$$
L^2f_n=\tilde{L}^2f_n=\l({n\over2}\r)\l({n\over2}+1\r)f_n, 
\qquad 
L_3f_n={n\over2}f_n, 
\qquad
\tilde{L}_3f_n={n\over2}f_n,
$$
which corresponds to the spherical harmonics $\vp_{l,m,\tm}=\vp_{n/2,n/2,n/2}$.

Therefore, a state corresponding to $\vp_{l,l,\tm}$ is given by 
$$
\tilde{L}_-^nf_{n+m}={(m+n)!\over{m}!}c_{m+n}e^{im\chi-in\vp}
\cos^n\th{\c}\sin^m\th,
$$
with\footnote{Note that the integer $m$ here is not the eigenvalue of $L_3$.} 
$m=l+\tm$ and $n=l-\tm$. The integers $m$ and $n$ run from 0 to $\infty$.
A further calculation yields 
$$
L_-^k\l(\tilde{L}_-^nf_{n+m}\r)=
(-)^k{(m+n)!\over{m}!}c_{m+n}e^{i(m-k)\chi-i(n-k)\vp}
{1\over\cos^{n-k}\th{\c}\sin^{m-k}\th}
\l({1\over\sin2\th}{d~\over{d\th}}\r)^k
\cos^{2n}\th{\c}\sin^{2m}\th,
$$
where the integer $k$ runs from 0 to $n+m$. 
Multiplying it by an appropriate normalization constant, we refer to it 
as $h_{n,m,k}$ ($n,m=0,1,2,3,\cdots;k=0,1,2,\cdots,n+m$).

\subsection{From Killing Spinors on 3-Spheres to 5-Dimensional Killing Spinors}

Introducing the basis of two-dimensional spinors 
$$
\zeta_{\pm}={1\over\sqrt{2}}\begin{pmatrix}1\cr\pm{i}\end{pmatrix}
=\l(\zeta_{\mp}\r)^*,
\qquad
\t_2\,\zeta_{\pm}=\pm\zeta_{\pm},
$$
and combining them with the Killing spinors $\e$, $C^{-1}\e^*$ on the 3-sphere 
to give the five-dimensional spinors 
$$
\e\otimes\zeta_{\pm}, 
\qquad 
C_3^{-1}\e^*\otimes\zeta_{\pm}.
$$

The action of $i\g^{45}={\bf 1}_2\otimes\t_2$ on these spinors can be seen as 
\begin{eqnarray*}
&&\g^{45}\l(\e\otimes\zeta_{\pm}\r)
=\mp{i}\l(\e\otimes\zeta_{\pm}\r),
\qquad
\g^{45}(C_3^{-1}\e^*\otimes\zeta_{\pm})
=\mp{i}(C_3^{-1}\e^*\otimes\zeta_{\pm}),
\end{eqnarray*}
and for the gamma matrices $\g_a$ ($a=1,2,3$), 
$$
\g_a\l(\e\otimes\zeta_{\pm}\r)=\pm\t_a\e\otimes\zeta_{\pm},
\quad
\g_a\l(C_3^{-1}\e^*\otimes\zeta_{\pm}\r)
=\pm\t_aC_3^{-1}\e^*\otimes\zeta_{\pm}
=\mp{C}_3^{-1}\l(\t_a\r)^*\e^*\otimes\zeta_{\pm}, 
\quad
(a=1,2,3)
$$


In the text, we assume that the Killing spinors $\ve^{\da}$ obey 
$$
i\g_{45}\,\ve^{\da}=\l(\t_3\r)^{\da}{}_{\db}\ve^{\db},
$$
or more explicitly, we make the ansatz 
$$
\ve^{\dot{1}}=\e\otimes\zeta_{+},
\qquad
\ve^{\dot{2}}=C_{3}^{-1}\e^*\otimes\zeta_{-}. 
$$



\subsection{Auxiliuary $SU(2)$ Flavor Spinors}
\label{AuxiliuarySpinor}

In order to construct the off-shell supersymmetry of the five-dimensional theory, 
besides the supersymmetry parameter $\ve^{\da}$, 
we need to introduce an additional supersymmetry parameter $\ve^{\ca}$ 
($\ca=1,2$), with the index $\ca$ of an additional $SU(2)$ flavor 
group which are not an subgroup of the $Spin(5)_R$ $R$-symmetry group. 

In this paper, we take the supersymmetry parameter $\ve^{\da}$ in the form
$$
\ve^{\dot{1}}=\e\otimes\zeta_{+},
\qquad
\ve^{\dot{2}}=C_{3}^{-1}\e^*\otimes\zeta_{-}, 
$$
where $\e$ is an two-dimensional spinor. By a simple examination, one can see 
that $\e$ and $C^{-1}_3\e^*$ are linearly independent. 
For this parameter $\ve^{\da}$, we take $\ve^{\ca}$ to be 
$$
\ve^{\check{1}}=\e\otimes\zeta_{-},
\qquad
\ve^{\check{2}}=C_{3}^{-1}\e^*\otimes\zeta_{+};
\qquad
\ve^{\ca}=\g^{5}\ve^{\da}
$$
so that they obey
$$
i\g^{45}\ve^{\ca}=-\l(\t_3\r)^{\ca}{}_{\cb}\ve^{\cb}.
$$

Since the two-dimensional spinors $\zeta_{\pm}$ are linearly independent, 
these supersymmetry parameters $\ve^{\da}$, $\ve^{\ca}$ form the basis of 
the four-dimensional linear space of five-dimensional spinors. 

The covariant derivative $\cD_\m\ve^{\ca}$ is defined by 
$$
\cD_\m\ve^{\ca}=\d_\m\ve^{\ca}+\qt\om_\m{}^{bc}\g_{bc}\,\ve^{\ca}
-\qt\check{A}_\m{}^{ij}\l(\bar{\s}_{ij}\r){}^{\ca}{}_{\cb}\ve^{\cb},
$$
where we assume that 
the gauge field $\check{A}_\m{}^{ij}$ takes the same as ${A}_\m{}^{ij}$; 
$\check{A}_\m{}^{ij}={A}_\m{}^{ij}$.

\section{Dictionary among Notations}
\label{Dictionary}

\begin{table}
\begin{center}
\begin{tabular}{|c|c|c|}\hline
\makebox[40mm]{Bergshoeff {\it et\,al.} \cite{BSvP2}}
&\makebox[40mm]{Cordova-Jafferis \cite{CJ}}
&\makebox[40mm]{ours}
\\ \hline\hline\hline
\multicolumn{3}{|c|}{the Lorentz vector indices} 
\\ \hline
{$a,b$}&{${\u{a}},\u{b}$}&{${\u{a}},\u{b}$}
\\ \hline
\multicolumn{3}{|c|}{the Spin(5)$_R\simeq$USp(4)$\simeq$sp(4)$_R$ 
spinor indices} 
\\ \hline
{$i,j$}&{$m,n$}&{${{\a}},{\b}$}
\\ \hline
\multicolumn{3}{|c|}{the Spin(5)$_R\simeq$USp(4)$\simeq$sp(4)$_R$ 
vector indices} 
\\ \hline
&&{$I,J$}
\\ \hline\hline
\multicolumn{3}{|c|}{the Spin(5)$_R$ charge conjugation matrix $\Leftrightarrow$
the USp(4)$\simeq$sp(4)$_R$ invariant metric} 
\\ \hline
{$\Om_{ij}$}&{$\Om_{mn}$}&{$\Om_{\a\b}$}
\\ \hline
\multicolumn{3}{|c|}{the gamma matrices of the Lorentz group} 
\\ \hline
{$\g^a$}&{$\G^{\u{a}}$}&{$\u{\G}^{\u{a}}$}
\\ \hline
\multicolumn{3}{|c|}{the charge conjugation matrix of the Lorentz group} 
\\ \hline
{$C$}&{$\u{C}$}&$i\u{C}$
\\ \hline\hline
\multicolumn{3}{|c|}{the spinor indices of Spin(5)$_R$} 
\\ \hline
{$\psi{}^{i}=\Om^{ij}\psi{}_{j}$,}
&{$\psi{}^{m}=\Om^{mn}\psi{}_{n}$,}
&{$\psi{}^{\a}=\Om^{\a\b}\psi{}_{\b}$,}
\\ 
{$\psi{}_{i}=\psi{}^{j}\Om_{ji}$}
&{$\psi{}_{m}=\psi{}^{n}\Om_{nm}$}
&{$\psi{}_{\a}=\psi{}^{\b}\Om_{\b\a}$}
\\ \hline
\multicolumn{3}{|c|}{the conjugate of a spinor $\u{\psi}^\a$} 
\\ \hline
\multirow{2}{*}{$\bar\psi{}^{i}=i\l(\psi_i\r)^\dag\g^0=\l(\psi^i\r)^TC$}
&\multirow{2}{*}{$\bar\psi{}^{m}=i\l(\psi_m\r)^\dag\G^0=\l(\psi^m\r)^T\u{C}$}
&\multirow{2}{*}{$\bar{\u{\psi}}{}_{\a}
=\l(\u{\psi}^\a\r)^\dag\u{\G}{}^0
=\l(\u{\psi}^\b\r)^T\u{C}\Om_{\b\a}$}
\\ 
& &  \\ \hline
\multicolumn{3}{|c|}{the local Lorentz (dependent) gauge field 
$\u{\Om}{}_{\u{\m}}{}^{\u{ab}}$} 
\\ \hline
\multirow{2}{*}{${\om}{}_{{\m}}{}^{{ab}}$}
&\multirow{2}{*}{${\om}{}_{\u{\m}}{}^{\u{ab}}$}
&\multirow{2}{*}{$\u{\Om}{}_{\u{\m}}{}^{\u{ab}}$}
\\ 
& &  \\ \hline
\multicolumn{3}{|c|}{the Spin(5)$_R$ gauge field $\u{V}{}_{\u{\m}}{}^{IJ}$} 
\\ \hline
\multirow{3}{*}{$V_\m{}^{ij}=V_\m{}^{ji}$}
&\multirow{3}{*}{$V_{\u{\m}}{}^{mn}=V_{\u{\m}}{}^{nm}$}
&\multirow{2}{*}{$\u{V}_{\u{\m}}{}^{\a}{}_{\b}
=(1/2){\u{V}}_{\u{\m}}{}^{IJ}\big(\p_{IJ}\big){}^{\a}{}_{\b}$,}
\\ 
& &  \\ 
& &$ {\u{V}}_{\u{\m}}{}^{IJ}=-{\u{V}}_{\u{\m}}{}^{JI}$ 
\\ \hline
\multicolumn{3}{|c|}{the auxiliuary field $\u{M}{}^{\a\b}{}_{\g\dl}$} 
\\ \hline
{$D^{ij}{}_{kl}$}
&{$D{}^{mn}{}_{rs}$}
&{$\u{M}{}^{\a\b}{}_{\g\dl}$}
\\ \hline\hline
\multicolumn{3}{|c|}{the scalar field $\u{\phi}{}^{\a\b}$ of the tensor 
multiplet} 
\\ \hline
{$\phi^{ij}$}
&{$\Phi^{mn}$}
&{$\u{\phi}{}^{\a\b}$}
\\ \hline
\multicolumn{3}{|c|}{the scalar spinor $\u{\chi}{}^{\a}$ of 
the tensor multiplet} 
\\ \hline
{$\psi^{i}$}
&{${\varrho}^m$}
&{$\u{\chi}{}^{\a}$}
\\ \hline
\end{tabular}
\label{Dictionarynotation}
\caption{Dictionary of notations in \cite{BSvP2,CJ} and ours.}
\end{center}
\end{table}

In this paper, we follow the same procedure as in \cite{CJ} 
to obtain five-dimensional $\N=2$ supersymmetric Yang-Mills theory 
in the supergravity background, but our notations are different from 
the ones in \cite{CJ}. The procedure is based on the dimensional reduction 
of the off-shell supergravity in \cite{BSvP2}, where different notations 
from \cite{CJ} and ours however, are used. Therefore, the list of the 
diffrent notations among \cite{BSvP2,CJ} and ours may be convenient for 
the readers.  

We use the common Lorentz metric with the signature 
$(-,+,\cdots,+)$. However, the indices of Lorentz vectors are different; 
for a Lorentz vector $\u{V}^{\u{a}}$ in our notations, it is 
$V^a$ in \cite{BSvP2} and ${V}^{\u{a}}$ in \cite{CJ}. 
Note here that the underline $\underline{\quad}$ indicates that 
it is a six-dimensional one in our paper. That's why 
the above $V$ carries the underline $\u{V}$ in ours, 
but not in \cite{BSvP2,CJ}. 
However, the underline $\underline{\quad}$ carried by the Lorentz 
indices or the coordinate frame indices means that they are 
six-dimensional in \cite{CJ} and ours, but not in \cite{BSvP2}. 

We use the common algebra of the gamma matrices 
of the Lorentz group, but the different notations of the gamma matrices 
are used, as listed in Table \ref{Dictionarynotation}. 
The properties (\ref{C6}) of our charge conjugation matrix $\u{C}$ 
are also enjoyed by $C$ in \cite{BSvP2} and  $\u{C}$ in \cite{CJ}.
However, since our definition of the conjugate $\u{\bar{\psi}}_{\a}$ 
of spinors is different from the ones in \cite{BSvP2,CJ} by $i=\sqrt{-1}$;
$\bar{\psi}^i\to{i}\bar{\u{\psi}}^{\a}$, 
as is seen in Table \ref{Dictionarynotation}, our charge conjugation matrix 
$\u{C}$ multiplied by $i$ is equal to the ones in \cite{BSvP2,CJ}.

The spinor indices $\a,\b$ of the Spin(5)$_R$ symmetry in our paper are 
$i,j$ in \cite{BSvP2} and $m,n$ in \cite{CJ}, and they run from 1 to 4. 
Only in our paper, the vector indices $I,J$ of the Spin(5)$_R$ are introduced. 
Since there is the equivalence Spin(5) $\simeq$ USp(4), the Spin(5)$_R$ 
group is referred to as USp(4) in \cite{BSvP2}, and its Lie algebra 
as sp(4)$_R$ in \cite{CJ}. The invariant metric $\Om_{ij}$ of USp(4) 
in \cite{BSvP2} is $\Om_{mn}$ in \cite{CJ} and is 
the charge conjugation matrix $\Om_{\a\b}$ of Spin(5)$_R$ in ours.

As shown in Table \ref{Dictionarynotation}, the gauge field 
$\u{\Om}{}_{\u{\m}}{}^{\u{ab}}$ of the local Lorentz symmetry 
is referred to as $\om{}_{\m}{}^{ab}$ in \cite{BSvP2} and 
$\om{}_{\u{\m}}{}^{\u{ab}}$ in \cite{CJ}. 
The curvature tensor of the gauge field $\om{}_{\m}{}^{ab}$ is 
defined in \cite{BSvP2} by 
$$
R{}_{\m\n}{}^{ab}(M)=\d_\m\om_\n{}^{ab}-\d_\n\om_\m{}^{ab}
+\om_\m{}^{ac}\om_\n{}_{c}{}^{b}
-\om_\n{}^{ac}\om_\m{}_{c}{}^{b}
-\underline{2\l(f_\m{}^ae_\n{}^b-f_\m{}^be_\n{}^a
-f_\n{}^ae_\m{}^b+f_\n{}^be_\m{}^a\r)}+\cdots,
$$
with the ellipses $\cdots$ denoting the contribution from the fermionic fields, 
following the usual construction of the curvature tensors in the conformal 
tensor calculus \cite{Supergravity}. Upon the deformation of the superconformal 
algebra, the gauge field $f_\m{}^a$ becomes dependent \cite{BSvP2};
$$
f_\m{}^a=-{1\over8}{R'}_\m{}^a(M)+{1\over80}e_\m{}^a{R'}(M)
+{1\over32}T^{ij}{}_{{\m}cd}T_{ij}{}^{acd},
$$
in their notations, 
where the curvature ${R'}{}_{\m\n}{}^{ab}(M)$ is $R{}_{\m\n}{}^{ab}(M)$ 
with the underlined terms omitted in the above. 
The Ricci curvature ${R'}_\m{}^a(M)=e^\n{}_b{R'}_{\m\n}{}^{ba}(M)$ and 
the scalar curvature ${R'}(M)=e^\m{}_a{R'}_\m{}^a(M)$ are also defined 
in \cite{BSvP2}. 

In the notation of \cite{CJ}, $\u{R}$ is the scalar curvature of 
${\om}{}_{\u{\m}}{}^{\u{ab}}$ and is equal to ${R'}(M)$ with 
the fermionic contributions omitted.

The curvature tensor $\u{R}{}_{\u{\m\n}}{}^{\u{ab}}(\u{\Om})$ 
of $\u{\Om}{}_{\u{\m}}{}^{\u{ab}}$ in our conventions is the same as 
the bosonic contribution of ${R'}{}_{\m\n}{}^{ab}(M)$ in \cite{BSvP2}; 
$\u{R}{}_{\u{\m\n}}{}^{\u{ab}}(\u{\Om})={R'}{}_{\m\n}{}^{ab}(M)\big|$, 
with $\big|$ denoting the omission of the terms including ferminic fields. 
Our convention of the Ricci curvature is $\u{R}{}_{\u{\m}}{}^{\u{a}}(\u{\Om})=
\u{\Th}^{\u{\n}}{}_{\u{b}}\u{R}{}_{\u{\n\m}}{}^{\u{ba}}(\u{\Om})
=-{R'}_\m{}^a(M)\big|$, and $\u{R}(\u{\Om})=-{R'}(M)\big|$ in \cite{BSvP2} 
and $\u{R}(\u{\Om})=-\u{R}$ in \cite{CJ}.

\subsection{Difference in Notations between the Previous Paper and this paper}
\label{oldbackground}

\begin{table}
\begin{center}
\begin{tabular}{|c|c|}\hline
\makebox[40mm]{the previous paper \cite{FKM}}
&\makebox[40mm]{this paper}
\\ \hline\hline
\multicolumn{2}{|c|}{the gamma matrices of the Lorentz group} 
\\ \hline
\multirow{2}{*}{$\G^M$, $C_5$,\quad ($M=1,\cdots,5$)}
&\multirow{2}{*}{$\g^\m$, $C$,\quad ($\m=1,\cdots,5$)}
\\ 
& \\ \hline
\multicolumn{2}{|c|}{the supersymmetry parameter $\ve^{\da}$} 
\\ \hline
\multirow{2}{*}{$\Sigma^{\da}$}&\multirow{2}{*}{
$\ve^{\da}/\sqrt{2}$}
\\ 
& \\ \hline
\multicolumn{2}{|c|}{the non-abelian gauge field $A_\m$} 
\\ \hline
{$v_M$}&{$A_\m$}
\\ \hline
\multicolumn{2}{|c|}{the gaugino field $\la^{\da}$} 
\\ \hline
\multirow{2}{*}{$\Psi^{\da}$}&\multirow{2}{*}{
$\la^{\da}/(2\sqrt{2})$}
\\ 
 & \\ \hline
\multicolumn{2}{|c|}{the scalar field $\s$} 
\\ \hline
{$\s$}&{$-\s$}
\\ \hline
\multicolumn{2}{|c|}{the scalar fields $\phi^i$~~ ($i=1,\cdots,4$)} 
\\ \hline
{$H_{\da}$} & 
{$\pm{i}(\s_i)^1{}_{\da}\,\phi^i/\sqrt{2}$}
\\ 
\multirow{2}{*}{$\bar{H}^{\da}=\l({H}_{\da}\r)^*$} & 
\multirow{2}{*}{$\pm{i}(\bar{\s}_i)^{\da}{}_{1}\,\phi^i/\sqrt{2}$}
\\ 
&
\\ \hline
\multicolumn{2}{|c|}{the spinor field $\psi^{\ta}$~~ ($i=1,\cdots,4$)} 
\\ \hline
{$\Xi$} & 
{$\mp({i}/2)\,\psi^{\ta=1}$}
\\ 
{$\Xi^\dag$} & 
{$\pm({i}/2)\l(\psi^{\ta=2}\r)^TC$}
\\ \hline\hline
\multicolumn{2}{|c|}{the auxiliuary $SU(2)$ indices ($\ca=1,2$)} 
\\ \hline
$\a$,$\b$ & $\ca$, $\cb$
\\ \hline 
$F_{H\a}$ & $F_{\ca}$ 
\\ \hline 
\end{tabular}
\label{Diffrencenotation}
\caption{Difference of notations between \cite{FKM} and this paper.}
\end{center}
\end{table}

Before estimating the partition function, 
we don't need to use the complex conjugate of the scalar fields $\phi^I$ 
and the hermition conjugate of the spinor $\chi^\a$ to 
obtain the supersymmetry transformations, the supersymmetry algebra, 
and the supersymmetric action. 
However, the scalar fields $\phi^I$ have the negative signature 
in the kinetic terms, and therefore, it is necessary to perform 
the ``Wick's rotation'' of the scalar fields $\phi^I$ to define 
the partition function. We will regard them as pure imaginary. 

However, the four scalars $\phi^i$ ($i=1,\cdots,4$) in the previous paper 
have the positive signature, and we can regard them as real scalars. 
Thus, they form the complex scalars $H^{\da}$, $\bar{H}_{\da}$, and 
the scalars $H^{\da}$, $\bar{H}_{\da}$ in the previous paper can be 
identified as 
$$
H_{\da}=\pm{i\over\sqrt{2}}\l(\s_i\r){}^1{}_{\da}\phi^i,
\qquad
\bar{H}^{\da}=\l(H_{\da}\r)^*
=\pm{i\over\sqrt{2}}\l(\bar{\s}_i\r){}^{\da}{}_{1}\phi^i, 
$$
in terms of $\phi^i$ ($i=1,\cdots,4$) in this paper. 
\begin{eqnarray*}
\mp{i\over\sqrt{2}}\l(\s_i\r){}^2{}_{\da}\phi^i=\ve_{\da\db}\bar{H}^{\db},
\qquad
\mp{i\over\sqrt{2}}\l(\bar{\s}_i\r){}^{\da}{}_{2}\phi^i
=\ve^{\da\db}{H}_{\db}, 
\qquad
\phi^i=\mp{i\over\sqrt{2}}\l[\l(\bar{\s}^i\r){}^{\da}{}_{1}H_{\da}
+\l({\s}^i\r){}^{1}{}_{\da}\bar{H}^{\da}\r],
\nn\\
\phi^1=\pm{1\over\sqrt{2}}\l(H_2-\bar{H}^2\r),~
\phi^2=\pm{i\over\sqrt{2}}\l(H_2+\bar{H}^2\r),~
\phi^3=\pm{1\over\sqrt{2}}\l(H_1-\bar{H}^1\r),~
\phi^4=\mp{i\over\sqrt{2}}\l(H_1+\bar{H}^1\r).
\end{eqnarray*}

In order for the supersymmetry transformation and the action 
in the previous paper to be consistent with the ones in this paper, 
we need to identify 
$$
\Xi=\mp{i\over2}\psi^{\ta=1}, 
\qquad
\Xi^\dag=\pm{i\over2}\l(\psi^{\ta=2}\r)^TC.
$$
But, the symplectic Majorana condition of $\psi^{\ta}$ 
is inconsistent with the hermitian conjugation of $\Xi$. 
Since we don't use the symplectic Majorana condition of $\psi^{\ta}$ 
to ensure the supersymmetry transformation, the supersymmetry algebra, 
and the supersymmetric action, this identification never causes any troubles. 

Through the dictionary, the on-shell supersymmetry transformation (\ref{SUSY}) 
may be rewritten in terms of the notations in the previous paper as 
\begin{eqnarray*}
\dl{v}_M&=&-i\bar{\Sigma}_{\da}\G_M\Psi^{\da},
\qquad
\dl{\s}=i\bar{\Sigma}_{\da}\Psi^{\da},
\qquad
\dl{H}_{\da}=-i\bar{\Sigma}_{\da}\Xi,
\qquad
(\dl\bar{H}^{\da}=i\bar\Xi\Sigma^{\da}),
\nn\\
\dl\Psi^{\da}&=&-\hf\l[\hf{F}_{MN}\G^{MN}+\G^M\cD_M\s
-{1\over2\a}G_{MN}\G^{MN}\s\r]\Sigma^{\da}
\nn\\
&&-\hf{S}_{ij}\s\l(\bar{\s}^{ij}\r){}^{\da}{}_{\db}\Sigma^{\db}
-ig\bigg(\l[\bar{H}^{\da},\,H_{\db}\r]
-\hf\dl^{\da}{}_{\db}\l[\bar{H}^{\dg},\,H_{\dg}\r]\bigg)\Sigma^{\db},
\nn\\
\dl{\Xi}&=&\bigg[
\G^M\cD_M{H}_{\da}+ig\l[\s,\,H_{\da}\r]-{1\over2\a}G_{MN}\G^{MN}H_{\da}
-t_{MN}\G^{MN}H_{\da}
\nn\\
&&\quad
-\hf\l(S^{ij}+\ve^{ijkl}S_{kl}\r)\l[\l(\bar{\s}_j\r)^{\db}{}_{1}H_{\db}
+\l(\s_j\r)^1{}_{\db}\bar{H}^{\db}\r]\l(\s_i\r)^1{}_{\da}
\bigg]\Sigma^{\da},
\end{eqnarray*}
and the equations of motion (\ref{chieom}) of $\la^{\da}$, $\psi^{\ta}$ as
\begin{eqnarray*}
&&\G^M\cD_M\Psi^{\da}+ig\l[\s,\,\Psi^{\da}\r]
+ig\l[\bar{H}^{\da},\,\Xi\r]
-ig\ve^{\da\db}\l[H_{\db},\,C^{-1}\Xi^*\r]
\nn\\
&&\qquad
={1\over8\a}G_{MN}\G^{MN}\Psi^{\da}
+{1\over4}S_{ij}\l(\bar{\s}^{ij}\r)^{\da}{}_{\db}\Psi^{\db}
+{1\over2}t_{MN}\G^{MN}\Psi^{\da},
\nn\\
&&\G^M\cD_M\Xi-ig\l[\s,\,\Xi\r]-2ig\l[H_{\da},\,\Psi^{\da}\r]
\nn\\
&&\qquad
={1\over8\a}G_{MN}\G^{MN}\Xi
+{1\over4}S_{ij}\l(\l(\s^{ij}\r)^1{}_1\Xi+\l(\s^{ij}\r)^2{}_2C^{-1}\Xi^{*}\r)
-\hf{t}_{MN}\G^{MN}\Xi.
\end{eqnarray*}
The Killing spinor equation (\ref{KSE}) in the old notations becomes 
$$
\cD_M\Sigma^{\da}
={1\over4}S_{ij}\l(\bar{\s}^{ij}\r){}^{\da}{}_{\db}\G_M\Sigma^{\db}
-{1\over2\a}G_{MN}\G^N\Sigma^{\da}-{1\over8\a}G_{KL}\G_M{}^{KL}\Sigma^{\da}
-{1\over2}t_{KL}\G_M{}^{KL}\Sigma^{\da},
$$
with the covariant derivative
$$
\cD_M\Sigma^{\da}
\equiv
\d_M\Sigma^{\da}+\hf\,b_M\Sigma^{\da}+\qt\Om_M{}^{KL}\G_{KL}\Sigma^{\da}
-\qt\,A_M{}^{ij}\l(\bar{\s}_{ij}\r)^{\da}{}_{\db}\G_M\Sigma^{\db}.
$$

Under the background
$$
\a=1
\quad
(b_M=0), 
\qquad
{1\over\a}G_{45}=-1,
\qquad 
S_{12}=S_{34}=\hf, 
\qquad
t_{45}={1\over4},
\qquad
A^{12}=A^{34}=-\hf\om^{45},
$$
the Killing spinor equation is reduced to the one in the previous paper 
\begin{eqnarray*}
\cd_M\Sigma^{\da}=-\hf\G_M{}^{45}\Sigma^{\da}.
\end{eqnarray*}
The Killing spinor in the previous paper has the additional property
\begin{eqnarray*}
&&\hskip2cm
\G^{45}\Sigma^{\da}=-i\l(\t_3\r)^{\da}{}_{\db}\Sigma^{\db}.
\nn\\
&&(\G^4\Sigma^{\da}=i\l(\t_3\r)^{\da}{}_{\db}\G^5\Sigma^{\db},
\qquad
\G^5\Sigma^{\da}=-i\l(\t_3\r)^{\da}{}_{\db}\G^4\Sigma^{\db}). 
\end{eqnarray*}
The equations of motion are also reduced to the ones in the previous paper.

In order to lift the on-shell supersymmetry transformation to the off-shell one, 
introducing the auxiliuary field $D^{\da}{}_{\db}$ to replace 
$$
D{}^{\da}{}_{\db}=i\s\l(\t_3\r){}^{\da}{}_{\db}
+2ig\l(
\l[\bar{H}^{\da},\,H_{\db}\r]-\hf\dl^{\da}{}_{\db}\l[\bar{H}^{\dg},\,H_{\dg}\r]
\r), 
$$
and using the equations of motion of the spinors with the background, 
the supersymmetry transformation of $D^{\da}{}_{\db}$ reproduces
the previous one 
$$
\dl{D}^{\da\db}
=i\bar{\Sigma}^{(\da}\bigg(\G^M\cD_M\Psi^{\db)}+ig\l[\s,\,\Psi^{\db)}\r]\bigg).
$$
It yields the off-shell supersymmetry of 
the vector multiplet ($v_M,\la^{\da},\s,D^{\da}{}_{\db}$) 
in the previous paper. 

For the self-dual $S_{ij}$; $\ve_{ij}{}^{kl}S_{kl}=2S_{ij}$, 
the term in $\dl\Xi$,
$$
-\hf\l(S^{ij}+\ve^{ijkl}S_{kl}\r)\l[\l(\bar{\s}_j\r)^{\db}{}_{1}H_{\db}
+\l(\s_j\r)^1{}_{\db}\bar{H}^{\db}\r]\l(\s_i\r)^1{}_{\da}\Sigma^{\da}
=\mp{i\over\sqrt{2}}\l(S^{ij}+\ve^{ijkl}S_{kl}\r)\phi_j\l(\s_i\r)^1{}_{\db}
\Sigma^{\db},
$$
can be rewritten by using the formulas
$$
\s^k\bar{\s}_{ij}=\dl^k{}_{[i}\s_{j]}+\ve_{ij}{}^{kl}\s_l,
\qquad
{\s}_{ij}\s^k=\dl^k{}_{[i}\s_{j]}-\ve_{ij}{}^{kl}\s_l,
$$
as
$$
\pm{i\over\sqrt{2}}\c{3\over4}S_{ij}\l(\s^k\r)^1{}_{\da}\phi_k
\l(\bar{\s}^{ij}\r)^{\da}{}_{\db}\Sigma^{\db}
={3\over4}S_{ij}\l(\bar{\s}^{ij}\r)^{\da}{}_{\db}H_{\da}\Sigma^{\db}.
$$
It facilitates the computation to verify that the off-shell supersymmetry 
transformation of the hypermultiplet ($H_{\da},\Xi,F_{H\a}$) 
with the auxiliuary field $F_{H\a}$ gives rise to the one in the previous paper. 
In fact, by modifying the on-shell supersymmetry transformation as
$$
\dl\Xi \quad\to\quad \dl\Xi+F_{H\a}\check{\Sigma}^{\a},
\qquad
\dl{F}_{H\a}=i\bar{\check{\Sigma}}_{\a}({\rm the~e.o.m.~of~}\Xi),
$$
one can lift it to the off-shell supersymmetry transformation.

In the previous paper, we have given the action\footnote{
The mass term of $H_{\da}$ in the action ${\cal L}$ didn't include 
the curvature term $R(\RS)\bar{H}^{\da}H_{\da}$ in the previous paper \cite{FKM}. 
}
\begin{eqnarray*}
&&{\cal L}=\int \sqrt{g}\,d^5x\,\Tr\bigg[
-\qt{v}_{MN}v^{MN}+\hf\,D_M\s{D}^M\s
-\qt\,D^{\da}{}_{\db}D^{\db}{}_{\da}
-D^M\bar{H}^{\da}D_MH_{\da}
+\bar{F}_H{}^{\a}F_{H\a}
\nn\\
&&\hskip3.3cm
+\hf\om_{{\rm c.s.}}
+\s^2
-\l(1+\qt{R}(\RS)\r)\bar{H}^{\da}H_{\da}
+\l({i\over2}(\t_3)^{\da}{}_{\db}D^{\db}{}_{\da}-v_{45}\r)\s
\nn\\
&&\hskip3.3cm
+i\bar{\Psi}_{\da}\G^MD_M\Psi^{\da}
-i\bar\Xi\G^MD_M\Xi
+\hf\l(\t_3\r)^{\da}{}_{\db}\bar\Psi_{\da}\Psi^{\db}
-{i\over2}\bar\Xi\G_{45}\Xi
\nn\\
&&\hskip3.3cm
+g^2\l[\s,\,\bar{H}^{\da}\r]\l[\s,\,{H}_{\da}\r]
+igD^{\da}{}_{\db}\l[\bar{H}^{\db},\,H_{\da}\r]
\nn\\
&&\hskip3.3cm
-g\bar{\Psi}_{\da}\l[\s,\,\Psi^{\da}\r]
-g\bar\Xi\l[\s,\,\Xi\r]
-2g\bar\Xi\l[H_{\da},\,\Psi^{\da}\r]
+2g\l[\bar{H}^{\da},\,\bar{\Psi}_{\da}\r]\Xi
\bigg], 
\end{eqnarray*}
where the term $\om_{{\rm c.s.}}$ denotes 
$$
\int\,\sqrt{g}\,d^5x~\Tr\l[\ve^{mnk}v_m\l(\d_nv_k+{i\over3}g\l[v_n,\,v_k\r]\r)\r]
=\int v\wedge\l(dv+{2\over3}(ig)v\wedge{v}\r)\wedge\vol({\RS})
=\int\,\sqrt{g}\,d^5x~\Tr[\om_{{\rm c.s.}}],
$$
with $\vol(\RS)$ the volume form of the Riemann surface $\RS$.

Integrating out the auxiliuary fields $D^{\da}{}_{\db}$, $F_{H\a}$ 
in the action, one obtains
\begin{eqnarray*}
&&{\cal L}\big|=\int \sqrt{g}\,d^5x\,\Tr\bigg[
-\qt{v}_{MN}v^{MN}
-v_{45}\s
+\hf\,D_M\s{D}^M\s
-D^M\bar{H}^{\da}D_MH_{\da}
\nn\\
&&\hskip3.5cm
+\hf\om_{{\rm c.s.}}
+\hf\s^2
-\l(1+\qt{R}(\RS)\r)\bar{H}^{\da}H_{\da}
-g(\t_3)^{\da}{}_{\db}\s\l[\bar{H}^{\db},\,H_{\da}\r]
\nn\\
&&\hskip3.5cm
+g^2\l[\s,\,\bar{H}^{\da}\r]\l[\s,\,{H}_{\da}\r]
-g^2\l[\bar{H}^{\da},\,H_{\db}\r]\l[\bar{H}^{\db},\,H_{\da}\r]
+\hf{g}^2\l[\bar{H}^{\da},\,H_{\da}\r]\big[\bar{H}^{\db},\,H_{\db}\big]
\nn\\
&&\hskip3.5cm
+i\bar{\Psi}_{\da}\G^MD_M\Psi^{\da}
-i\bar\Xi\G^MD_M\Xi
+\hf\l(\t_3\r)^{\da}{}_{\db}\bar\Psi_{\da}\Psi^{\db}
-{i\over2}\bar\Xi\G_{45}\Xi
\nn\\
&&\hskip3.5cm
-g\bar{\Psi}_{\da}\l[\s,\,\Psi^{\da}\r]
-g\bar\Xi\l[\s,\,\Xi\r]
-2g\bar\Xi\l[H_{\da},\,\Psi^{\da}\r]
+2g\l[\bar{H}^{\da},\,\bar{\Psi}_{\da}\r]\Xi
\bigg].  
\end{eqnarray*}
It enables us to compare it easily with the action $S$ in 
(\ref{5DLF}-\ref{5DLB}), and one can see that the substitution of 
the background on the round 3-sphere 
in subsection \ref{oldR3Sbg} into the action $S$ exactly 
reproduces the above action.

\clearpage

\end{document}